\documentstyle[aps,prl,floats,graphicx]{revtex}
\begin{document}

\def\topfraction{1} \def\bottomfraction{1} \def\textfraction{0}

\draft
\title{Superfluid analogies of cosmological phenomena}
\author{G.E. Volovik\\
Low Temperature Laboratory, Helsinki
  University of Technology\\
 Box 2200, FIN-02015 HUT, Finland\\
and\\  
L.D. Landau Institute for Theoretical Physics, 117334 Moscow,
  Russia}
\maketitle
\begin{abstract}
In a modern viewpoint the relativistic quantum field theory is the emergent phenomenon
arising in the low energy corner of the physical fermionic vacuum -- the medium, whose nature
remains unknown. The same phenomenon occurs in the condensed matter systems: In the extreme limit
of low energy the condensed matter system of special universality class acquires all the
symmetries, which we know today in high energy physics: Lorentz invariance, gauge invariance,
general covariance, etc.  The chiral fermions as well as gauge bosons and gravity field arise
as fermionic and bosonic collective modes of the system. The inhomogeneous states of the condensed matter ground state -- vacuum --
induce nontrivial effective metrics of the space, where the free quasiparticles move along
geodesics.  This conceptual similarity between condensed matter and quantum vacuum allows us to
simulate many phenomena in high energy physics and cosmology, including axial anomaly,
baryoproduction and magnetogenesis, event horizon and Hawking radiation, cosmological
constant and rotating vacuum, etc., probing these phenomena in
ultra-low-temperature superfluid helium, atomic Bose condensates and superconductors. Some of
the experiments have been already conducted.
\end{abstract}

\tableofcontents

\section{Introduction. Physical vacuum as condensed matter.} 

The traditional Grand Unification view is that the low-energy symmetry of our
world is the remnant of a larger symmetry, which exists at high energy and
is broken, when the energy is reduced. According to this philosophy the higher
the energy the higher is the symmetry: $U(1)\times SU(3) \rightarrow
U(1)\times SU(2)\times SU(3) \rightarrow  SO(10) \rightarrow$supersymmetry,
etc. The less traditional view is quite opposite: it is argued that starting
from some energy scale one probably finds that the higher the energy the
poorer are the symmetries of the physical laws, and finally even the Lorentz
invariance and gauge invariance will be smoothly violated
\cite{FrogNielBook,Chadha}.  From this point of view the relativistic quantum
field theory is an  effective theory \cite{Weinberg,Jegerlehner}. It is an
emergent phenomenon arising as a fixed point in the low energy corner of the
physical vacuum. In the vicinity of the fixed point the system acquires new
symmetries which it did not have at higher energy. And it is quite possible
that even such symmetries as Lorentz symmetry and gauge invariance are not
fundamental, but gradually appear when the fixed point is approached. 

Both scenaria are supported by the condensed matter systems. In
particular, superfluid $^3$He-A provides an instructive example of both ways
of the behavior of the symmetry.  At high temperature the $^3$He gas
and at lower temperature the $^3$He liquid have all the symmetries,
which the ordinary condensed matter can have: translational invariance, global
$U(1)$ group and global $SO(3)$ symmetries of spin and orbital rotations.
When the temperature decreases further the liquid
$^3$He reaches the superfluid transition temperature
$T_c$, below which it spontaneously looses  all its
symmetries except for the translational one -- it is still liquid. This
breaking of symmetry at low temperature and thus at low energy reproduces the
Grand Unification scheme, where the symmetry breaking is the most important
component. 

However, this is not the whole story. When the temperature is reduced further
the opposite ``anti-grand-unification'' scheme starts to work: in the limit
$T\rightarrow 0$  the superfluid $^3$He-A gradually acquires from
nothing almost all the symmetries, which we know today in high energy physics:
(analog of) Lorentz invariance, local gauge invariance, elements of general
covariance, etc.  It appears that such an enhancement of symmetry in the limit
of low energy  happens because $^3$He-A belongs to a special universality
class of Fermi systems \cite{parallel}. For the condensed matter of such class
the chiral fermions as well as gauge bosons and gravity field arise as
fermionic and bosonic collective modes together with the chirality itself and
with corresponding symmetries. The inhomogeneous deformations of the condensed
matter ground state -- quantum vacuum -- induce nontrivial effective metrics of
the space, where the free quasiparticles move along geodesics.  This conceptual
similarity between condensed matter and quantum vacuum gives some hint on the
origin of symmetries and also allows us to simulate many phenomena in high
energy physics and cosmology.

The quantum field theory, which we have now, is incomplete due to ultraviolet
diveregences at small scales. The crucial example is provided by the quantum
theory of gravity, which after 70 years of research is still far from
realization in spite of numerous beautiful achievements \cite{Rovelli}. This is
a strong indication that the gravity, both classical and quantum, is not
fundamental: It is effective field theory  which is not applicable at small
scales, where the ``microscopic'' physics of vacuum becomes important and
according to the ``anti-grandunification'' scenario some or all of the known
symmetries in Nature are violated. The analogy between quantum vacuum and
condensed matter could  give an insight into this transPlanckian physics since
it provides examples of the physically imposed deviations from Lorentz and other
invariances at higher energy. This is important in  many different areas of
high energy physics and cosmology, including possible CPT violation and black
holes, where the infinite red shift at the horizon opens the route to the
transPlanckian physics. 

The condensed matter teaches us that the low-energy properties of different
condensed matter vacua (magnets, superfluids, crystals, superconductors,
etc.)  are robust, i.e. they do not depend much on the details of microscopic
(atomic) structure of these substances. The main role is played by symmetry and
topology of condensed matter: they determine the soft (low-energy)
hydrodynamic variables, the effective Lagrangian describing the low-energy
dynamics, topological defects and quantization of physical parameters. The
microscopic details provide us only with the ``fundamental constants'', which
enter the effective phenomenological Lagrangian, such  as speed of ``light''
(say, the speed of sound), superfluid density, modulus of elasticity, magnetic
susceptibility, etc.  Apart from these ``fundamental constants'', which can be
rescaled, the systems behave similarly in the infrared limit if they belong to
the same universality and symmetry classes, irrespective of their microscopic
origin. 

The detailed information on the system is lost in such acoustic or
hydrodynamic limit \cite{LaughlinPines}. From the properties of the low energy
collective modes of the system -- acoustic waves in case of crystals -- one
cannot reconstruct the atomic structure of the crystal since all the crystals
have similar acoustic waves described by the same equations of the same
effective theory, in crystals it is the classical theory of elasticity. The
classical fields of collective modes can be quantized to obtain quanta of
acoustic waves -- the phonons. This quantum field remains the effective
field which is applicable only in the long-wave-length limit, and does not
give a detailed information on the real quantum structure of the underlying
crystal (exept for its symmetry class). In other words one cannot construct the
full quantum theory of real crystal using the quantum theory of elasticity.
Such theory would always contain divergencies on atomic scale, which cannot be
regularized. 

The same occurs in other effective theories of condensed matter.
In particular the naive approach to calculate the ground state (vacuum) energy
of superfluid liquid $^4$ using the zero point energy of phonons gives even
the wrong sign of the vacuum energy, as we shall see in Sec.
\ref{VacuumEnergyAnd}. 

It is quite probable that in the same way the quantization of
classical gravity, which is one of the infrared collective modes of quantum
vacuum, will not add more to our understanding of the ``microscopic''
structure of the vacuum \cite{Hu96,Padmanabhan,LaughlinPines}. Indeed,
according to this ``anti-grandunification'' analogy,  such properties of our
world, as gravitation, gauge fields, elementary chiral fermions, etc., all
arise in the low energy corner  as a low-energy soft modes of the underlying
``condensed matter''. At high energy (of the Planck scale) these modes merge
with the continuum of the all high-energy degrees of freedom of the ``Planck
condensed matter'' and thus cannot be separated anymore from each other. Since
the gravity is not fundamental, but appears as an effective field in the
infrared limit, the only output of its quantization would be the quanta of the
low-energy gravitational waves -- gravitons. The more deep quantization of
gravity makes no sense in this phylosophy. In particular, the effective theory
cannot give any prediction for the vacuum energy and thus for the cosmological
constant.

The main advantage of the condensed matter analogy is that in principle we
know the condensed matter structure at any relevant scale, including the
interatomic distance,  which plays the part of one of the Planck length
scales in the hierarchy of scales. Thus the condensed matter can suggest
possible routes from our present low-energy corner of ``phenomenology''  to the
``microscopic'' physics  at Planckian and trans-Planckian energies.  It can
also show the limitation of the effective theories: what quantities can be
calculated within the effective field theory using, say, renormalization group
approach, and what qantities depend essentially on the details of the
transPlanckian physics. 
 
In the main part of the review we consider superfluid $^3$He in its A-phase,
which belongs the special class of Fermi liquids, where the effective gravity,
gauge fields and chiral fermions appear in the low-energy corner together with
Lorentz and gauge invariance \cite{parallel,LammiTalk}, and discuss the
correspondence between the phenomena in superfluid $^3$He-A and that in
relativistic particle physics. However, some useful analogies can be provided
even by Bose liquid -- superfluid $^4$He, where a sort of the effective
gravitational field appears in the low energy corner. That is why it is
instructive to start with the simplest effective field theory of Bose
superfluid which has a very restricted number of effective fields. 

\section{Landau-Khalatnikov two-fluid hydrodynamics as effective theory of
gravity.}
\label{LandauKhalatnikovSection}

\subsection{Superfluid vacuum and quasiparticles.} 
 
According to Landau and Khalatnikov \cite{Khalatnikov} a weakly excited state
of the collection of interacting $^4$He atoms  can be considered as a small
number of elementary excitations -- quasiparticles (phonons and rotons). In
addition, the state without excitation -- the ground state or vacuum -- can
have collective degrees of freedom. The superfluid vacuum can move without
friction, and  inhomogeneity of the flow serves as the gravitational and/or 
other effective fields. The matter propagating  in the presence of this
background is represented by fermionic (in Fermi superfluids) or bosonic (in
Bose superfluids) quasiparticles, which form the so called normal component of
the liquid. Such two-fluid hydrodynamics introduced by Landau and Khalatnikov
\cite{Khalatnikov} is the example of the effective field theory which
incorporates the motion of both the superfluid background (gravitational field)
and its excitations (matter). This is the counterpart of the Einstein equations,
which incorporate both gravity and matter.

One must distinguish between the bare particles and quasiparticles in
superfluids. The particles are the elementary objects of the system on a
microscopic ``transPlanckian'' level, these are the atoms of the underlying
liquid ($^3$He or $^4$He atoms). The many-body system of the interacting atoms
form the quantum vacuum -- the ground state. The nondissipative collective
motion of the superfluid vacuum with zero entropy is determined by the
conservation laws experienced by the atoms and by their quantum coherence in
the superfluid state. The quasiparticles are the particle-like excitations
above this vacuum state. The bosonic excitations in superfluid
$^4$He and fermionic and bosonic excitations in superfluid $^3$He form the
viscous normal component of these liquids, which correspond to matter in our
analogy.  The normal component is responsible for the thermal and kinetic
low-energy properties of superfluids. 

\subsection{Dynamics of superfluid vacuum.} \label{DynamicsSuperfluidVacuum}

In the simplest superfluid the coherent motion of the superfluid vacuum
is characterized by two collective (hydrodynamic) variables: the particle
number density $n({\bf r},t)$  of  atoms comprising the liquid and superfluid
velocity ${\bf v}_{\rm s}({\bf r},t)$ of their coherent motion. In superfluid
$^4$He the superfluid velocity is the gradient of the phase of the order
parameter (${\bf v}_{\rm s}=(\hbar/m)\nabla \Phi$, where $m$ is the bare mass
of particle -- the mass of
$^4$He atom) and thus the flow of vacuum is curl-free: $\nabla\times {\bf
v}_{\rm s}=0$. This is not however a rule: as we shall see in Sec.
\ref{RelativisticLevel} the superfluid vacuum flow of
$^3$He-A can have a continuous vorticity,  $\nabla\times {\bf
v}_{\rm s}\neq 0$. 

The particle number conservation provides one of the equations of the effective
theory of superfluids -- the continuity equation:
\begin{equation}
{\partial n\over \partial t}+ \nabla\cdot{\bf J}=0~.
\label{ContinuityEquation}
\end{equation}
In a strict microscopic theory of monoatomic lquid, $n$ and the particle
current ${\bf J}$ are given by the particle distribution function 
$n({\bf p})$: 
\begin{equation}
n=\sum_{\bf p}  n({\bf
p})~~,~~{\bf J}={1\over m} \sum_{\bf p} {\bf p} n({\bf p})~.
\label{TotalCurrent1}
\end{equation}
The liquids considered here are nonrelativistic and obeying the
Galilean transformation law. In the Galilean system the momentum of particles
and the particle current are related by the second Eq.(\ref{TotalCurrent1}). 

The particle distribution function  $n({\bf p})$ is typically rather
complicated function of momentum even at $T=0$ because of the strong
interaction between the bare atoms in a real liquid.  $n({\bf p})$ can be
determined only in a fully microscopic theory and thus never enters the
effective theory of superfluidity. The latter instead is determined by
quasiparticle distribution function  $f({\bf p})$, which is simple because at
low $T$ the number of quasiparticles is small and their interaction can be
neglected. That is why in equilibrium $f({\bf p})$ given by the thermal  Bose
distribution (or by the Fermi distribution for fermionic quasiparticles) and in
nonequilibrium it can be found from the conventional kinetic equation for
quasiparticles.  

In the effective theory the  particle current has two contributions
\begin{equation}
{\bf J}= n {\bf v}_{\rm s}+
{\bf J}_{\rm q}~,~{\bf J}_{\rm q}={1\over m}{\bf P}~,~{\bf P} \sum_{\bf p} {\bf
p} f({\bf p})~.
\label{TotalCurrent2}
\end{equation}
 The  first term $n {\bf v}_{\rm s}$ is the current
transferred coherently by the collective motion of superfluid vacuum with the superfluid
velocity ${\bf v}_{\rm s}$. In equilibrium at $T=0$ this is the only current,
but if quasiparticles are excited above the ground state, their momentum ${\bf
P}$ gives an additional contribution to the particle current providing the
second term in Eq.(\ref{TotalCurrent2}).    Note that
under the Galilean transformation to the coordinate system moving with the
velocity
${\bf u}$, at which the superfluid velocity transforms as ${\bf v}_{\rm
s}\rightarrow {\bf v}_{\rm s} + {\bf u}$, the momenta of particle and
quasiparticle transform differently:    
${\bf p} \rightarrow {\bf p}  + m {\bf u}$ for microscopic particles (atoms) and ${\bf p}
\rightarrow {\bf p}$ for quasiparticles. The latter occurs because the
quasiparticle in effective low-energy theory has no information on such
characteristic of the transPlanckian world as the mass of the bare atoms
comprising the vacuum state.

The second equation for the collective variables is the London
equation for the superfluid velocity, which is curl-free in superfluid $^4$He
($\nabla\times{\bf v}_{\rm s}=0$):
\begin{equation}
m{\partial  {\bf v}_{(s)}\over \partial t}   +   \nabla {\delta {\cal E}\over \delta
n} =0~.
\label{LondonEquation}
\end{equation} 
Together with the kinetic equation for the quasiparticle   distribution
function $f({\bf p})$, the Eqs.(\ref{LondonEquation}) and (\ref{ContinuityEquation}) for
collective fields ${\bf v}_{\rm s}$ and $n$ give the complete effective theory for the
kinetics  of quasiparticles (matter) and coherent motion of vacuum (gravitational field) if
the energy functional
${\cal E}$ is known. In the limit of low temperature, where the density
if thermal quasiparticles are small, the interaction
between quasiparticles can be neglected. Then the simplest Ansatz satisfying the
Galilean invariance is
\begin{equation}
{\cal E}=\int d^3r \left( {m\over 2}n{\bf v}_{\rm s}^2 + \epsilon(n) -\mu n+ 
\sum_{\bf p} \tilde E({\bf p},{\bf r}) f({\bf p},{\bf r})\right)~.
\label{Energy}
\end{equation} 
Here $\epsilon(n)$ (or $\tilde\epsilon(n)=\epsilon(n) -\mu n$) is the 
vacuum energy density as a function of the particle density; $\mu$ is the overall constant
chemical potential, which is the Lagrange multiplier responsible for the conservation of the
total number $N=\int d^3x~n$ of the $^4$He atoms;
$\tilde E({\bf p},{\bf r})=E({\bf p},n({\bf r}))+ {\bf
p}\cdot{\bf v}_{\rm s}({\bf r})$ is the  Doppler shifted quasiparticle energy in the
laboratory frame with  $E({\bf p},n({\bf r}))$ being the quasiparticle energy
measured in the frame comoving with the superfluid vacuum. 

\subsubsection{Absence of canonical Lagrangian formalism in effective
theories.}
\label{AbsenceLagrangian}

The Eqs. (\ref{ContinuityEquation}) and
(\ref{LondonEquation}) can be  obtained from the Hamiltonian formalism using
the energy in Eq.(\ref{Energy}) as Hamiltonian and the following Poisson
brackets 
\begin{equation}
 \left\{{\bf v}_{\rm s}({\bf r}_1),n({\bf r}_2)\right\}={1\over m}\nabla \delta({\bf
r}_1-{\bf r}_2)~,~ \left\{n({\bf r}_1),n({\bf r}_2)\right\}= \left\{{\bf v}_{\rm s}({\bf
r}_1),{\bf v}_{\rm s}({\bf r}_2)\right\}=0~.
\label{PoissonBrackets}
\end{equation} 
The Poisson brackets between components of superfluid velocity are
zero only for curl-free superfluidity. In a general case it is
\begin{equation}
 \left\{ v_{{\rm s}i}({\bf r}_1),v_{{\rm s}j}({\bf r}_2)\right\}=-{1\over mn}e_{ijk}
 (\nabla\times {\bf
v}_{\rm s})_k \delta( {\bf r}_1-{\bf r}_2)~.
\label{PoissonBracketsVelocity}
\end{equation}
In this case even at $T=0$, when the quasiparticles are absent, the
Hamiltonian description of the hydrodynamics is only possible: There is no
Lagrangian, which can be expressed in terms of the hydrodynamic variables
${\bf v}_{\rm s}$ and $n$. The absence of the Lagrangian in many condensed
matter systems is one of the consequences of the reduction of the degrees of
freedom in effective field theory, as compared with the fully microscopic
description \cite{WessZumFerro}. In ferromagnets, for example, the number of
the hydrodynamic variables is odd: 3 components of the magnetization vector
${\bf M}$. They thus cannot form the canonical pairs of conjugated variables.
As a result one can use either the Hamiltonian description or introduce the
effective action with the Wess-Zumino term, which contains an extra coordinate
$\tau$:
\begin{equation}
S_{\rm WZ}\propto \int d^3x~dt~d\tau ~ {\bf M}\cdot(\partial_t{\bf
M}\times\partial_\tau{\bf M})~.
\label{WessZumino}
\end{equation}
According to the analogy the presence of the Wess-Zumino term in the
relativistic quantum field theory would indicate that such theory is effective.

\subsection{Normal component -- ``matter''.}\label{NormalComponentSection}

 In a local thermal equilibrium the distribution of quasiparticles is characterized by
local temperature $T$ and by local velocity of the quasiparticle gas ${\bf
v}_{\rm n}$, which is called the normal component velocity:
\begin{equation}
f_{\cal T}({\bf p})=\left(\exp { \tilde E({\bf p})- {\bf p}  {\bf v}_n\over
T} \pm 1\right)^{-1}~,
\label{Equilibrium}
\end{equation}
where the sign + is for the fermionic quasiparticles in Fermi superfluids and the   sign -
is for the bosonic quasiparticles in Bose superfluids. Since $\tilde E({\bf p})=E({\bf
p}) +{\bf p}\cdot {\bf v}_{\rm s}$, the equilibrium distribution is determined by the
Galilean invariant quantity ${\bf v}_{\rm n} - {\bf v}_{\rm s}\equiv {\bf w}$, which is the 
normal component velocity measured in the frame comoving with superfluid vacuum. It is called
the counterflow velocity. In the limit when the conterflow velocity ${\bf v}_{\rm n}-{\bf
v}_{\rm s}$ is small, the quasiparticle
(``matter'') contribution to the liquid momentum and thus to the particle
current is proportional to the counterflow velocity:
\begin{equation}
 J_{{\rm q} i} =n_{{\rm n}ik}(v_{{\rm n}k}-v_{{\rm s}k})~,~n_{{\rm
n}ik}=-\sum_{\bf p}{p_ip_k\over m}~{\partial f_{\cal T}\over \partial E}~~,
\label{EquilibriumCurrent}
\end{equation}
where the tensor $n_{{\rm n}ik}$ is the so called density of the normal component. In this
linear regime the total current in Eq.(\ref{TotalCurrent2}) can be represented
as the sum of the currents carried by the normal and superfluid components
\begin{equation}
J_{i} =n_{{\rm s}ik}v_{{\rm s}k}+n_{{\rm n}ik}v_{{\rm n}k} ~,
\label{TotalCurrent3}
\end{equation}
where tensor $n_{{\rm s}ik}=n\delta_{ik}-n_{{\rm n}ik}$ is the so called  
density of superfluid component.
In the isotropic
superfluids,
$^4$He and $^3$He-B, the normal component density is an isotropic tensor,
$n_{{\rm n}ik}=n_{{\rm n}}\delta_{ik}$, while in anisotropic superfluid 
$^3$He-A the normal component density is a
uniaxial tensor \cite{VollhardtWolfle}.  At
$T=0$ the quasiparticles are frozen out and one has
$n_{{\rm n}ik}=0$ and $n_{{\rm s}ik}=n\delta_{ik}$ in all monoatomic
superfluids.

\subsection{Quasiparticle spectrum and effective metric}

The structure of the quasiparticle spectrum in superfluid $^4$He becomes more and more
universal the lower the energy. In the low energy corner the spectrum of these
quasiparticles, phonons, can be obtained in the framework of the effective theory. Note that
the effective theory is unable to describe the high-energy part of the spectrum -- rotons,
which can be determined in a fully microscopic theory only.  On the contrary, the spectrum of
phonons is linear, $E({\bf p},n)\rightarrow c(n)|{\bf p}|$, and only the ``fundamental
constant'' -- the speed of ``light''  $c(n)$ -- depends on the physics of the higher energy
hierarchy rank. Phonons represent the quanta of the collective modes of the
superfluid vacuum, sound waves, with the speed of sound obeying 
$c^2(n)=(n/m)(d^2\epsilon/dn^2)$. All other information on the microscopic atomic nature of
the liquid is lost. Note that for the curl-free superfluids the sound waves represent the only
``gravitational'' degree of freedom. The Lagrangian for these ``gravitational
waves'' propagating above the smoothly varying background is obtained from
equations  (\ref{ContinuityEquation}) and  (\ref{LondonEquation}) at $T=0$ by
decomposition of the superfluid velocity and density into the smooth and
fluctuating parts:   
 ${\bf v}_{\rm s}=
{\bf v}_{\rm s~~smooth} +\nabla \alpha$ \cite{unruh,vissersonic}.  The
quadratic part of the Lagrangian for the scalar field $\alpha$ is
\cite{StoneIordanskii}:
\begin{equation}
{\cal L}=  {m\over 2}n \left( (\nabla \alpha)^2- {1\over c^2}\left(\dot\alpha    + 
({\bf v}_{\rm s}\cdot\nabla)\alpha\right)^2\right)\equiv {1\over
2}\sqrt{-g}g^{\mu\nu}\partial_\mu\alpha \partial_\nu\alpha~.
\label{LagrangianSoundWaves}
\end{equation} 
The quadratic Lagrangian for sound waves has necessarily the Lorentzian form,
where the effective Riemann metric experienced by the sound wave, the so
called acoustic metric, is simulated by the smooth parts of the hydrodynamic
fields:
\begin{equation}
 g^{00}=-{1\over mnc} ~,~ g^{0i}=-{v_{\rm s}^i\over mnc} ~,~ g^{ij}=
{c^2\delta^{ij} -v_{\rm s}^i v_{\rm s}^j\over mnc} ~,
\label{ContravarianAcousticMetric}
\end{equation}  
\begin{equation}
 g_{00}=-{mn\over  c}(c^2-{\bf v}_{\rm s}^2) ~,~
g_{0i}=-{mnv_{{\rm s} i}\over c} ~,~ g_{ij}= {mn\over  c}\delta_{ij} ~,~
\sqrt{-g}={m^2n^2\over c}~.
\label{CovarianAcousticMetric}
\end{equation}  
Here and further ${\bf v}_{\rm s}$ and $n$ mean the smooth parts of the velocity and density
fields.  Phonons in superfluids and crystals provide a typical example of how
an enhanced symmetry and effective Lorentzian metric appear in condensed matter
in the low energy corner.

The energy spectrum of sound wave quanta, phonons, which represent the ``gravitons''
in this effective gravity, is determined by 
\begin{equation}
 g^{\mu\nu}p_\mu p_\nu=0~,~~{\rm or} ~~ (\tilde E-{\bf p}\cdot {\bf v}_{\rm
s})^2=c^2p^2  ~.
\label{PhononEnergySpectrum}
\end{equation}  

\subsection{Effective metric for bosonic collective modes in other systems.} 

The effective action in Eq.(\ref{LagrangianSoundWaves}) is typical for the low
energy collective modes in ordered systems. The more general case is provided
by the Lagrangian for the Goldstone bosons in antiferromegnets -- the spin waves.
The spin wave dynamics in $x-y$ antiferromagnets and in $^3$He-A is governed by
the Lagrangian for the Goldstone variable $\alpha$, which is the angle of the
antiferromagnetic vector: 
\begin{equation}
{\cal L}=  {1\over 2}\eta^{ij}  \nabla_i \alpha\nabla_j \alpha- {1\over
2}\chi\left(\dot\alpha    +  ({\bf v}\cdot\nabla)\alpha\right)^2\equiv
{1\over 2}\sqrt{-g}g^{\mu\nu}\partial_\mu\alpha \partial_\nu\alpha~.
\label{LagrangianSpinWaves}
\end{equation}
Here the matrix $\eta^{ij}$ is the spin rigidity; $\chi$ is the spin
susceptibility; and
${\bf v}$ is the local velocity of crystall in antiferromagnets and
superfluid velocity, ${\bf v}={\bf v}_{\rm s}$, in
$^3$He-A. In antiferromagnets these 10 coefficients give rise to all ten
components of the effective  Riemann metric:
\begin{equation}
 g^{00}=-(\eta\chi)^{1/2} ~,~ g^{0i}=-(\eta\chi)^{1/2}v^i ~,~
g^{ij}=\left({\eta\over \chi}\right)^{1/2} (\eta^{ij} -\chi v^i v^j)
~,~\eta^{-1} ={\rm det}(\eta^{ij})~,
\label{ContravarianSpinWaveMetric}
\end{equation}  
\begin{equation}
 g_{00}=-(\eta\chi)^{-1/2} (1 -\chi\eta_{ij} v^i v^j) ~,~
g_{0i}=-\left({\chi\over \eta}\right) ^{1/2}\eta_{ij} v^j ~,~ g_{ij}=
\left({\chi\over \eta}\right) ^{1/2}\eta_{ij} ~,~
\sqrt{-g}=\left({\chi\over \eta}\right) ^{1/2}~.
\label{CovarianSpinWaveMetric}
\end{equation}  
The effective interval is 
\begin{equation}
ds^2=-  {1\over (\eta\chi)^{1/2} }dt^2+\left({\chi\over \eta}\right) ^{1/2}
\eta_{ij}(dx^i-v^idt)(dx^j-v^jdt)~.
\label{SpinWaveInterval}
\end{equation}  
This form of the interval corresponds to the Arnowitt-Deser-Misner
decomposition of the space-time metric, where the function
\begin{equation}
N= {1\over (\eta\chi)^{1/4} }~,
\label{LapseFunction}
\end{equation} 
is known as lapse function;   $g_{ij}=\left( \chi/ \eta\right) ^{1/2}\eta_{ij}$
gives the three-metric describing the geometry of space; and the velocity
vector ${\bf v}$ plays the part of the so-called shift function  (see e.g. the
book
\cite{VisserBook}). 

\subsection{Effective quantum field and effective action} 
 
The effective action in  Eq.(\ref{LagrangianSoundWaves}) for phonons and
in Eq.(\ref{LagrangianSpinWaves}) for spin waves (magnons) formally obeys the
general covariance. In addition,  in the classical limit of
Eq.(\ref{PhononEnergySpectrum}) corresponding to geometrical optics (in our
case this is geometrical acoustics) the propagation of phonons is invariant
under the conformal transformation of metric,
$g^{\mu\nu}\rightarrow \Omega^2 g^{\mu\nu}$. This symmetry is lost at the quantum level: the
Eq.(\ref{LagrangianSoundWaves}) is not invariant under general conformal
transformations, however the reduced symmetry is still there:
Eq.(\ref{LagrangianSoundWaves}) is invariant under scale transformations with
$\Omega={\rm Const}$. 

As we shall see further, in the superfluid $^3$He-A the  other effective
fields and new symmetries appear in the low energy corner, including also the
effective $SU(2)$ gauge fields and gauge  invariance. The symmetry of
fermionic Lagrangian induces, after integration over the quasiparticles
degrees of freedom, the corresponding symmetry of the effective action for the
gauge fields. Moreover, in addition to superfluid velocity field  there are
appear the other gravitational degrees of freedom with the spin-2 gravitons.
However, as distinct from the effective gauge fields, whose effective action
is very similar to that in particle physics, the effective gravity cannot
reproduce in a full scale the Einstein theory: the effective action for the
metric is contaminated by the noncovariant terms, which come from the
``transPlanckian'' physics
\cite{parallel}.  The origin of difficulties with effective gravity in condensed matter is
probably the same as the source of the problems related to quantum gravity and cosmological
constant. 

The quantum quasiparticles interact with the classical collective fields ${\bf v}_{\rm s}$
and $n$, and with each other. In Fermi superfluid $^3$He the fermionic quasiparticles
interact with many collective fields describing the multicomponent order parameter and with
their quanta. That is why one obtains the interacting Fermi and Bose quantum fields,
which are in many respect similar to that in particle physics. However, this
field theory can be applied to a lowest orders of the perturbation theory
only.  The higher order diagrams are divergent and nonrenormalizable, which
simply means that the effective theory is valid when only the low
energy/momentum quasiparticles are involved even in their virtual states. This
means that only those terms in the effective action can be derived by
integration over the quasiparticle degrees of freedom, whose integral are
concentrated solely in the low-energy region. For the other processes one must
go beyond the effective field theory and consider the higher levels of
description, such as Fermi liquid theory, or further the microscopic level of
the underlying liquid with atoms and their interactions. In short, all the
terms in effective action come from the microscopic ``Planck'' physics, but only
some fraction of them can be derived in a self-consistent way within the
effective field theory itself. 

In Bose supefluids the fermionic degrees of freedom are absent, that is why the quantum field
theory there is too restrictive, but nevertheless it is useful to consider it since it
provides the simplest example of the effective theory. On the other hand the
Landau-Khalatnikov scheme is rather universal and is easily extended to superfluids with
more complicated order parameter and with fermionic degrees of freedom (see
the book \cite{VollhardtWolfle}).

\subsection{Vacuum energy and cosmological constant. Nullification of
vacuum energy.}\label{VacuumEnergyAnd}

The vacuum energy densities $\epsilon(n)$ and $\tilde\epsilon(n)=\epsilon(n)
-\mu n$, and also the parameters which characterize the quasparticle energy
spectrum cannot be determined by the effective theory: they are provided solely
by the higher (microscopic) level of description. The vacuum is characterized
by the equilibrium value of the particle number density
$n_0(\mu)$ at given chemical potential $\mu$, which is determined by the minimization of 
the energy $\tilde\epsilon(n)$ which enters the
functional in Eq.(\ref{Energy}): $d\tilde\epsilon/dn = 0$. This energy is related to
the pressure in the liquid created by external sources provided by the environment. From the
definition of the pressure, $P=-d(V\epsilon(N/V))/dV$  where $V$ is the volume of the system
and $N$ is the total number of the $^4$He atoms, one obtains that the  energy density
$\tilde\epsilon(n)=\epsilon(n) -\mu n$ of the vacuum in equilibrium and the vacuum pressure
are related in the same way as in the Einstein cosmological term:
\begin{equation}
 \tilde \epsilon_{\rm vac~equilibrium} = -P_{\rm vac}~.
\label{VacuumEnergyPressure}
\end{equation}  
Close to the equilibrium state one can expand the vacuum energy in terms of
deviations of particle density from its equilibrium value. Since the linear
term disappears due to the stability of the superfluid vacuum, one has
\begin{equation}
\tilde\epsilon(n)\equiv \epsilon(n)-\mu n=-P_{\rm vac} +{1\over 2} {m c^2\over n_0(\mu)}
(n-n_0(\mu))^2~.
\label{VacuumCloseToEquil} 
\end{equation} 

It is important that our vacuum is liquid, i.e. it can be in
equilibrium without interaction with the environment. In this equilibrium
state the pressure in the liquid is absent, $P_{\rm vac}=0$, and thus the
vacuum energy density $\tilde\epsilon$ is zero: 
\begin{equation}
\tilde\epsilon_{\rm vacuum~of~self-sustaining~system} \equiv 0~.
\label{Zer0VacuumEnergy} 
\end{equation} 
This can be the possible route to the solution of the problem of the vacuum
energy in quantum field theory. From the only assumption that the underlying
physical vacuum is liquid, i.e. the self-sustaining system, it follows that
the energy of the vacuum in its equilibrium state at $T=0$ is identically
zero and thus does not depend on the microscopic details.  The nullification of
the relevant vacuum energy $\tilde\epsilon$ in Eq.(\ref{Zer0VacuumEnergy})
remains even after the phase transition to the broken symmetry state occurs. At
first glance, the vacuum energy must decrease in a phase transition, as is
usually follows from the Ginzburg-Landau description of the phase transition.
But in the isolated system the chemical potential $\mu$ will be automatically
ajusted to preserve the zero external pressure and thus the zero energy of the
vacuum.

As distinct from the energy density $\tilde\epsilon$ which enters the
action and thus corresponds to the energy density of the quantum vacuum, the
value of the energy $\epsilon$, which is the proper energy of the liquid,
is not zero in equilibrium. It does depend on the microscopic (transPlanckian)
details and can be found in microscopic calculations only.  At zero external
pressure the vacuum energy per one atom of the liquid
$^4$He  coincides with the chemical potential $\mu$. From numerical
simulations of the many-body problem it was obtained that  
$\mu=\epsilon(n_0(\mu))/n_0(\mu) \sim -7$K \cite{Woo}.  The negative value of
the chemical potential is the property of the liquid.

Let us now compare these two vacuum energies $\tilde\epsilon=0$ and
$\epsilon <0$ with what the effective theory can tell us on the vacuum
energy. In the effective theory the vacuum energy is given by the zero point
energy of (in our case) the phonon modes
\begin{equation}
\epsilon_{\rm eff} = (1/2)\sum_{E({\bf p})<\Theta} cp ={1\over 16\pi^2}
{\Theta^4\over \hbar ^3 c^3} ={1\over 16\pi^2}\sqrt{-g} \left(g^{\mu\nu}\Theta_{\mu}
\Theta_{\nu}\right)^2  ~.
\label{VacuumEnergyEffective}
\end{equation}  
Here $c$ is the speed of sound; $\Theta \sim \hbar c/a $ is the
Debye characteristic temperature with $a$ being an interatomic space;
 $\Theta_{\mu}=(-\Theta,0,0,0)$.  $\Theta$ plays the part of the ``Planck''
cut-off energy scale $E_{\rm P}$. In superfluid $^4$He this cut-off is of the
same order of magnitude as
$mc^2$, i.e. the ``Planck mass'' $E_{\rm P}/c^2$ appears to be of order of the
mass of $^4$He atom $m$.  Thus the effective theory gives for the vacuum energy
density the value of order $E_{\rm P}^4/c^3$, while the stability condition
which comes from the microscopic ``transPlanckian'' physics gives an exact
nullification of the vacuum energy at
$T=0$. Being mapped to the cosmological constant problem, the estimation in
Eq.(\ref{VacuumEnergyEffective}), with $c$ being the speed of light and
$\Theta$ being the real Planck energy $E_{\rm P}$, gives the cosmological term
by 120 orders of magnitude higher than its upper experimental limit
\cite{Weinberg2}. This certainly confirms that the effective field theory is
unable to predict the relevant energy of the vacuum.

We wrote the Eq.(\ref{Zer0VacuumEnergy}) in the form which is different from
the conventional cosmological term $\Lambda \sqrt{-g}$. This is to show that 
both forms (and the other possible forms too) have the similar drawbacks. The
Eq.(\ref{VacuumEnergyEffective}) is conformal invariant due to conformal
invariance experienced by the quasiparticle energy spectrum in
Eq.(\ref{PhononEnergySpectrum}) (actually, since this term does not depend on
derivatives, the conformal invariance is equivalent to invariance under
multiplication of $g_{\mu\nu}$ by constant factor). However, in
Eq.(\ref{VacuumEnergyEffective}) the general covariance is violated by the
cut-off. On the contrary, the conventional cosmological term  
$\Lambda \sqrt{-g}$ obeys the general covariance, but it is not invariant under
 transformation $g_{\mu\nu}\rightarrow \Omega^2g_{\mu\nu}$ with constant
$\Omega$.  Thus both forms of the vacuum energy violate one or the other
symmetry of the low-energy effective Lagrangian Eq.(\ref{LagrangianSoundWaves})
for phonons, which means that the vacuum energy cannot be determined
exclusively within the low-energy domain. 

The estimation within the effective theory cannot resolve between
vacuum energies $\epsilon$ and $\tilde\epsilon=\epsilon-\mu n$, since in the
effective theory there is no notion of the conserved number of the $^4$He
atoms of the underlying liquid. And in both cases the effective theory gives a
wrong answer. It certainly violates the zero condition
(\ref{Zer0VacuumEnergy}) for
$\tilde\epsilon$. Comparing it with the liquid energy $\epsilon$, one finds that
the magnitude of $\epsilon_{\rm eff}(n_0)/n_0\sim 10^{-2}\Theta\sim 10^{-1}$K
(as follows from Eq.(\ref{VacuumEnergyEffective})) is  smaller than the result
obtained for $\epsilon$ in the microscopic theory. Moreover it has an opposite
sign. This means again that the effective theory must be used with great
caution, when one calculates those quantities, which crucially
(non-logarithmically) depend on the ``Planck'' energy scale. For them the
higher level ``transPlanckian'' physics must be used only. In a given case the
many-body wave function of atoms of the underlying quantum liquid has been
calculated to obtain the vacuum energy \cite{Woo}. The quantum fluctuations of
the phonon degrees of freedom in Eq.(\ref{Zer0VacuumEnergy}) are already
contained in this microscopic wave function. To add the energy of this zero
point motion of the effective field to the microscopically calculated energy
$\epsilon$ would be the double counting. 

Consideration of the equilibrium condition Eq.~(\ref{VacuumCloseToEquil}) shows that the
proper regularization of the equilibrium vacuum energy in the effective action must by
equating it to exact zero. In addition, from the Eq.~(\ref{VacuumCloseToEquil}) it
follows that the variation of the vacuum energy over the metric determinant must be also zero
in equilibrium: $d\tilde\epsilon/dg|_{n=n_0(\mu)}=(d\tilde\epsilon/d n)_{n=n_0(\mu)}/(d g/d
n)_{n=n_0(\mu)}= 0$. This apparently shows that the vacuum energy in $^3$He-A can be neither of
the form of Eq.(\ref{VacuumEnergyEffective}) nor in the form
$\Lambda\sqrt{-g}$. The metric dependence of the vacuum energy  consistent with the
Eq.(\ref{VacuumCloseToEquil}) could be only of the type
$\Lambda(g-g_0)^2$, so that the cosmological term in Einstein equation would be  $\propto
\Lambda(g-g_0)g_{\mu\nu}$. This means that in equilibrium, i.e. at $g=g_0$, the cosmological
term is zero and thus only the nonequilibrium vacuum is ``gravitating''. 

Thus the condensed matter analogy suggests two ways how to resolve the
cosmological constant puzzle. Both
are based on the notion of the stable equilibrium state of the quantum vacuum,
which is determined by the ``microscopic'' transPlanckian physics.
 
(1) If one insists that the cosmological term must be
$\Lambda\sqrt{-g}$, then for the self-sustaining vacuum the absence of the
external pressure requires that $\Lambda=0$ in equilibrium at $T=0$. At
nonzero $T$ the vacuum energy (and thus the vacuum gravitating mass) must be of
order of the energy density of matter (see
Sec.\ref{GlobalThermodynamicEquilibrium} and Sec.\ref{VacuumPressureSection}),
which agrees with the modern experimental estimation of the cosmological
constant \cite{Riess}.
 
This however does not exclude the Casimir effect, which appears if the vacuum is not
homogeneous and describes the change in the zero-point oscillations due to, say, boundary
conditions. The smooth deviations from the homogeneous equilibrium vacuum are within the
responsibility of the low-energy domain, that is why these deviations can be successfully
described by the effective field theory, and their energy can gravitate.  

(2) The cosmological term has a form $\Lambda(g-g_0)^2$ with the preferred background
metric $g_0$. The equilibrium vacuum with this background metric is not gravitating, while in
nonequilibrium, when $g\neq g_0$, the perturbations of the vacuum are grivitating. In
relativistic theories  such dependence of the Lagrangian on
$g$ can occur in the models where the determinant of the metric is the
dynamical variable which is not transformed under coordinate transformations,
i.e. the ``fundamental'' symmetry in the low-energy corner is not the general
covariance, but the the invariance under coordinate transformations with unit
determinant.

In conclusion of this Section, the gravity is the low-frequency,
and actually the classical output of all the quantum degrees of freedom of the
``Planck condensed matter''. So one should not quantize the gravity again,
i.e. one should not use the low energy quantization for construction of the
Feynman diagrams technique with diagrams containing the integration over high
momenta. In particular, the effective field theory is not appropriate for the
calculation of the vacuum energy and thus of the cosmological constant.
Moreover, one can argue that, whatever the real ``microscopic'' structure of
the vacuum is, the energy of the equilibrium vacuum is not gravitating: The
diverging energy of quantum fluctuations of the effective fields and thus the
cosmological term must be regularized to zero as we discussed above, since (i)
these fluctuations are already contained in the ``microscopic wave function''
of the vacuum; (ii) the stability of this ``microscopic wave function'' of the
vacuum requires the absence of the terms linear in $g_{\mu\nu}-
g_{\mu\nu}^{(0)}$ in the effective action; (iii) the self-sustaining
equilibrium vacuum state requires the nullification of the vacuum energy in
equilibrium at $T=0$.

\subsection{Einstein action and higher derivative terms}
\label{EinsteinActionAnd}

In principle, there are the higher order nonhydrodynamic terms in the effective
action, which are not written in Eq.(\ref{Energy}) since they contain space
and time  derivatives of the hydrodynamics variable, $n$ and  ${\bf v}_{\rm
s}$,  and  thus are relatively small. Though they are determined by the
microscopic ``transPlanckian'' physics, some part of them can be obtained
using the effective theory. The standard procedure, which was first used
by Sakharov to obtain the effective action for gravity
\cite{Sakharov}, is the integration over the fermionic or bosonic fields in
the gravitational background. In our case we must integrate over  the massless
scalar field $\alpha$ propagating in inhomogeneous $n$ and  ${\bf v}_{\rm s}$
fields, which provide the effective metric. The integration gives the
curvature term in Einstein action, which can also be written in two ways.  The
form which respects the general covariance of the Lagrangian for
$\alpha$ field in Eqs. (\ref{LagrangianSoundWaves}) and
(\ref{LagrangianSpinWaves}) is
\begin{equation}
{\cal L}_{\rm  Einstein}=-{1 \over 16\pi G }  
\sqrt{-g} R    ~~,
\label{EinsteinAction} 
\end{equation}
This form does not obey the invariance under multiplication of $g_{\mu\nu}$ by constant
factor, which shows its dependence on the  ``Planck'' physics. The
gravitational Newton constant $G$  is expressed in terms of the ``Planck'' cutoff:
$G^{-1}\sim \Theta^2$. Another form, which explicitly contains the ``Planck'' cutoff,
\begin{equation}
{\cal L}_{\rm  Einstein}=-{1 \over 16\pi }  
\sqrt{-g} R g^{\mu\nu}\Theta_{\mu}\Theta_{\nu}   ~~,
\label{EinsteinActionModified} 
\end{equation}
is equally bad: the action is invariant under the scale transformation of the metric, but the
general covariance is violated since the cut-off four-vector provides the
preferred reference frame. Such incompatibility of different low-energy
symmetries is the hallmark of the effective theories. 

To give an impression on the relative magnitude of the Einstein action
let us express the Ricci scalar in terms of the superfluid velocity field
only, keeping $n$ and $c$ fixed:
\begin{equation}
\sqrt{-g} R  ={mn\over c^2}\left(2\partial_t \nabla\cdot  {\bf v}_{\rm s} + \nabla^2(v_{\rm
s}^2)\right) ~.
\label{RicciScalar} 
\end{equation}
In superfluids the Einstein action is small compared to the dominating kinetic
energy term $mn{\bf v}_{\rm s}^2/2$ in Eq.(\ref{Energy})   by   factor
$a^2/l^2$, where $a$ is again the atomic (``Planck'') length scale and $l$ is the
characteristic macroscopic length at which the velocity field changes. That is why it can be
neglected in the hydrodynamic limit, $a/l \rightarrow 0$. Moreover, there are many  terms of
the same order in effective actions which do not display the general covariance, such as
 $(\nabla\cdot  {\bf v}_{\rm s})^2$. They are provided by microscopic physics, and there is
no rule in superfluids according to which these noncovariant terms  must be smaller
than the Eq.(\ref{EinsteinAction}).  But in principle, if the gravity field as
collective field arises from the other degrees of freedom, different from the
superfluid condensate motion, the Einstein action can be dominating. We shall
discuss this on example of the ``improved'' $^3$He-A in Sec.
\ref{Improved3He}. 
 
The effective action for the gravity field must also contain the higher order
derivative terms, which are quadratic in the Riemann tensor,  
\begin{equation}
  \sqrt{-g}(q_1R_{\mu\nu\alpha\beta}R^{\mu\nu\alpha\beta}+
q_2R_{\mu\nu}R^{\mu\nu}+q_3 R^2 )~\ln \left(
{g^{\mu\nu}\Theta_{\mu}\Theta_{\nu}\over R}\right)~.
\label{SquareRicciScalar} 
\end{equation} 
The parameters $q_i$ depend on the matter content of the
effective field theory. If the ``matter'' consists of scalar fields, phonons or
spin waves,  the integration over these collectives modes gives
$q_1=-q_2= (2/5)q_3= 1/(180\cdot 32\pi^2)$ (see e.g. \cite{FrolovFursaev}).
These terms logarithmically depend on the cut-off and thus their calculation in
the framework of the effective theory is justified.  Because of the logarithmic
divergence (they are of the relative order
$(a/l)^4~\ln (l/a)$) these terms dominate over the noncovariant terms of order
$(a/l)^4$,  which can be obtained only in fully microscopic calculations. Being
determined essentially by the phononic Lagrangian in
Eq.(\ref{LagrangianSoundWaves}), these terms respect (with logarithmic
accuracy) all the symmetries of this Lagrangian including the general
covariance and the invariance under rescaling the metric. That is why they are
the most appropriate terms for the self-consistent effective theory of gravity. 

This is the general rule: the logarithmically divergent terms in
action play a special role, since they always can be obtained within the
effective theory and with the logarithmic accuracy they are dominating over
the nonrenormalizable terms. As we shall see below the logarithmic terms 
arise in the effective action for the effective gauge fields, which appear 
in superfluid $^3$He-A in a low energy
corner (Sec.\ref{RunningCouplingConstant}). These terms in superfluid $^3$He-A
have been obtained first in microscopic calculations, however it appeared that
their physics can be completely determined by the low energy tail and thus
they can be calculated within the effective theory. This is well known in
particle physics as running coupling constants, zero charge effect and
asymptotic freedom.

Unfortunately in effective gravity of superfluids
the logarithmic terms as well as Einstein term are small compared with the main
terms -- the vacuum energy and the kinetic energy of the vacuum flow, which
depend on the 4-th power of cut-off parameter. This means that the superfluid
liquid is not the best condensed matter for simulation of Einstein gravity. In
$^3$He-A there are other components of the order parameter, which also give
rise to the effective gravity, but superfluidity of
$^3$He-A remains to be an obstacle. To fully simulate the Einstein gravity,
one must try to construct the non-superfluid condensed matter system which
belongs to the same universality class as $^3$He-A, and thus contains the 
effective Einstein gravity as emergent phenomenon, which is not contaminated
by the superfluidity. Such a system with suppressed superfluidity is discussed
in Sec. \ref{Improved3He}. 

\section{``Relativistic'' energy-momentum tensor for ``matter'' moving in
``gravitational'' superfluid background in two fluid
hydrodynamics}\label{SectionEnergyMomentum}

\subsection{Kinetic equation for quasiparticles (matter)}

Now let us discuss the dynamics of ``matter'' (normal component) in the
presence of the ``gravity field''  (superfluid motion). It is determined by
the kinetic equation for the distribution function
$f$ of the quasiparticles:
\begin{equation}
\dot f - {\partial \tilde E\over \partial {\bf r}} \cdot {\partial f\over \partial
{\bf p}}+ {\partial \tilde E\over \partial {\bf p}} \cdot {\partial f\over \partial
{\bf r}}={\cal J}_{coll}~.
\label{KineticEq}
\end{equation}
The collision integral conserves the momentum and the energy   of
quasiparticles,  i.e. 
\begin{equation}
\sum_{\bf p}  {\bf p}  {\cal J}_{coll}=\sum_{\bf
p}  \tilde E ({\bf p}) {\cal J}_{coll}=\sum_{\bf
p}  E({\bf p})  {\cal J}_{coll}=0~, 
\label{ConservationCollision}
\end{equation}  
but not necessarily the number of quasiparticle: the quasiparticle
number is not conserved in superfluids, though in the low-energy limit there
can arise an approximate conservation law. 

\subsection{Momentum exchange between superfluid vacuum and quasiparticles}

From the Eq.(\ref{ConservationCollision}) and from the two equations for the superfluid
vacuum, Eqs.(\ref{ContinuityEquation},\ref{LondonEquation}),  one obtains the time evolution
of the momentum density for each of two subsystems: the superfluid background (vacuum) and
quasiparticles (matter). The momentum evolution of the superfluid vacuum is
\begin{equation}
m\partial_t (n {\bf v}_{{\rm s}})  =- m\nabla_i(J_i{\bf v}_{\rm s}) -n
 \nabla\left( {\partial \epsilon\over \partial n}  +\sum_{\bf p} f
{\partial E \over \partial n} \right) + P_i  \nabla  v_{{\rm s}i}~.
\label{SuperfluidMomentumEq}
\end{equation} 
where ${\bf P}=m{\bf J}_{\rm q}$ is the momentum of lquid carried by quasiparticles (see
Eq.(\ref{TotalCurrent2})),
while the evolution
of the  momentum density of quasiparticles:
\begin{equation}
\partial_t  {\bf P}   =\sum_{\bf p} {\bf p}  \partial_t f   =  -
\nabla_i(v_{{\rm s}i}{\bf P})    -\nabla_i \left(  \sum_{\bf p} {\bf p}f
{\partial   E\over
\partial p_i}\right)  -\sum_{\bf p}f 
 \nabla   E  - P_i
 \nabla  v_{{\rm s}i}~.
\label{QuasiparticleMomentumEq}
\end{equation} 

Though the momentum of each subsystem is not conserved because of the
interaction with the other subsystem, the  total momentum density of the
system, superfluid vacuum + quasiparticles, must be conserved because of the
fundamental principles of the underlying microscopic physics. This can be easily
checked by summing two equations, (\ref{SuperfluidMomentumEq}) and 
(\ref{QuasiparticleMomentumEq}),
\begin{equation}
m\partial_t J_i   = \partial_t (mn  v_{{\rm s}i} +P_i)=  
-\nabla_i \Pi_{ik}~,
\label{TotalMomentumEq}
\end{equation} 
where the stress tensor
\begin{equation}
\Pi_{ik} = m J_i v_{{\rm s}k} + v_{{\rm s}i} P_k+  \sum_{\bf p} p_k f {\partial E\over
\partial p_i}  + \delta_{ik} \left(n\left({\partial \epsilon\over \partial
n}+ \sum_{\bf p} f{\partial E\over \partial n}\right) - 
\epsilon\right)  ~.
\label{StressTensor}
\end{equation} 

\subsection{Covariance vs conservation.}

The same happens with the energy. The total energy of the two subsystems is
conserved, while there is an energy exchange between the two subsystems of
quasiparticles and superfluid vacuum.  It appears that in the low
energy limit the momentum and energy exchange between the subsystems occurs in
the same way as the exchange of energy and momentum between matter and the
gravitational field. This is because in the low energy limit the
quasiparticles are ``relativistic'', and thus this exchange must be described
in the general relativistic covariant form. The
Eq.(\ref{QuasiparticleMomentumEq}) for the  momentum density of quasiparticles
as well as the corresponding equation for the quasiparticle energy density can
be represented as
\begin{equation}
T^\mu{}_{\nu;\mu}=0 ~~,~~{\rm or}~~ {1\over
\sqrt{-g}}\partial_\mu \left(T^\mu{}_\nu \sqrt{-g}\right) - {1\over
2}T^{\alpha\beta} \partial_\nu g_{\alpha\beta}= 0
 \, ,
\label{CovariantConservation}
\end{equation}
where $T^\mu{}_\nu$ is the usual ``relativistic'' energy-momentum tensor of
``matter'', which will be discussed in the
next Sec.\ref{EnergyMomentumTensorMatter}. This result does not depend on the
dynamic equations for the superfluid condensate (gravity field); the latter are
even not covariant in our case. The Eq.(\ref{CovariantConservation}) follows
solely from the ``relativistic'' spectrum of quasiparticles. As is known from
the general relativity, the Eq.(\ref{CovariantConservation}) does not represent
any conservation in a strict sense, since the covariant derivative is not a
total derivative
\cite{LandauLifshitz2}. The extra term, the second term in
Eq.(\ref{CovariantConservation}) which is not the total derivative, describes
the force acting on quasiparticles  (matter) from the superfluid condensate (an
effective gravitational field). For $\nu=i$ this extra term represents two last
terms in Eq.(\ref{QuasiparticleMomentumEq}) for quasiparticle momentum (see 
Sec.\ref{EnergyMomentumTensorMatter}).

The covariant form of the energy and momentum ``conservation'' for matter in
Eq.(\ref{CovariantConservation}) cannot be extended to the ``gravity'' field.
In the conservation law  $\partial_\mu T^\mu{}_\nu({\rm total})=0$ the total
energy-momentum tensor of superfluid and quasiparticles is evidently
noncovariant, as is seen from Eq.(\ref{TotalMomentumEq}). This happens partly
because the dynamics of the superfluid background is not covariant. However,
even for the fully covariant dynamics of gravity in Einstein theory the
problem of the energy-momentum tensor remains.  It is impossible to construct
such total energy momentum tensor, $T^\mu{}_\nu({\rm
total}) = T^\mu{}_\nu({\rm matter}) +T^\mu{}_\nu({\rm gravity})$,
which could have a covariant form and simultaneously satisfy the real
conservation law $\partial_\mu T^\mu{}_\nu({\rm total})=0$. Instead one
has the noncovariant energy momentum pseudotensor for the gravitational
background \cite{LandauLifshitz2}. 

From the condensed-matter point of view, this failure to construct the fully
covariant conservation law is a clear indication that the  Einstein gravity is
really an effective theory. As we mentioned above, effective theories in
condensed matter are full of such  contradictions related to incompatible
symmetries. In a given case the general covariance is incompatible with the
conservation law; in  cases of the vacuum energy
term (Sec.\ref{VacuumEnergyAnd}) and the Einstein term 
(Sec.\ref{EinsteinActionAnd}), obtained within the effecive theory the
general covariance is incompatible with the scale invariance; in the case of an
axial anomaly, which is also reproduced in condensed matter
(Sec.\ref{ChiralAnomaly}), the conservation of the baryonic charge is
incompatible with quantum mechanics; the  action of the Wess-Zumino type, which
cannot be written in 3+1 dimension in the covariant form (as we discussed at
the end of Sec.\ref{DynamicsSuperfluidVacuum}, Eq.(\ref{WessZumino})), is
almost typical phenomenon in various condensed matter systems whose low-energy
dynamics cannot be described by theeverywhere deterimined Lagrangian; the
momentum density determined as  variation of the hydrodynamic energy over
${\bf v}_{\rm s}$ does not coincide with the canonical momentum in most of the
condensed matter systems; etc. There are many other examples of apparent
inconsistencies in the effective theories of condensed matter. All such
paradoxes are naturally built in the effective theory; they necessarily arise
when the fully microsopic description is reduced to the effective theory with
restricted number of collective degrees of freedom. 

The paradoxes disappear completely (together with the effective symmetries of
the low-energy physics) on the fundamental level, i.e. in a fully  microscopic
description  where all degrees of freedom are taken into account. In an atomic
level of description the dynamics of $^4$He atoms is fully determined by the
well defined microscopic Lagrangian which respects all the symmetries of atomic
physics, or by canonical Hamiltonian formalism for pairs of canonically
conjugated variables, coordinates and momenta of atoms. Though this ``Theory of
Everything'' does not contain the paradoxes, in most case it fails to describe
the low-energy physics just because of the enormous amount of degrees of
freedom. The effective theory is to be constructed to incorporate the phenomena
of the low-energy physics, which sometimes are too exotic (the Quantum Hall
Effect is an example) to be predicted  by ``The Theory of Everything''
\cite{LaughlinPines}. 

\subsection{Energy-momentum   tensor for ``matter''.}
\label{EnergyMomentumTensorMatter}

Let us specify the tensor $T^\mu{}_\nu$ for quasiparticles, which enters
Eq.(\ref{CovariantConservation}), for  the simplest case, when the gravity is
simulated by the superflow only. If we neglect the space-time dependence of
the density $n$ and of the speed of sound $c$, then the constant factor $mnc$
can be removed from the metric in
Eqs.(\ref{ContravarianAcousticMetric}-\ref{CovarianAcousticMetric}) and the
effective metric is simplified:
\begin{equation}
 g^{00}=-1 ~,~ g^{0i}=-v_{\rm s}^i  ~,~ g^{ij}=
 c^2\delta^{ij} -v_{\rm s}^i v_{\rm s}^j  ~,
\label{ContravarianAcousticMetricReduced}
\end{equation}  
\begin{equation}
 g_{00}=-\left(1-{{\bf v}_{\rm s}^2\over c^2}\right) ~,~
g_{0i}=-{v_{{\rm s} i}\over c^2} ~,~ g_{ij}= {1\over  c^2}\delta_{ij} ~,~
\sqrt{-g}={1\over c^3}~.
\label{CovarianAcousticMetricReduced}
\end{equation}  
In this case the energy-momentum tensor of quasiparticles can be represented as
\cite{FischerVolovik} 
\begin{equation}
\sqrt{-g} T^\mu{}_\nu=\sum_{\bf p} f v_G^\mu p_\nu\,,\qquad
v_G^\mu v_{G\mu} = -1 +\frac1{c^2}
\frac{\partial E }{\partial p_i}
\frac{\partial E }{\partial p_i}\,,
\end{equation}
where $p^0= E$; $p_0= - \tilde E=-E-{\bf p}\cdot{\bf v}_{\rm s}$; $v_G^\mu$
is the group four velocity of quasiparticle defined as
\begin{eqnarray}
v_G^i& =& \frac{\partial \tilde E}{\partial p_i}\,,\quad v_G^0=1\,,\quad
v_{Gi}= \frac1{c^2} \frac{\partial  E }{\partial p_i}\,,
\quad v_{G0}=-\left(
1+\frac{{v_{\rm s}^i }}{c^2}  \frac{\partial E }{\partial
p_i}\right)\,.
\end{eqnarray}
Space-time  indices are throughout
assumed to be raised and lowered by the   metric in
Eqs.(\ref{ContravarianAcousticMetricReduced}-\ref{CovarianAcousticMetricReduced}). The group
four velocity is null in the relativistic domain of the spectrum only: $v_G^\mu v_{G\mu}=0$ if
$E =cp$.
The relevant components of the energy-momentum tensor are:
\begin{eqnarray}
\sqrt{-g} T^0{}_i & = & \sum_{\bf p} f {p}_i
= P_i\qquad \mbox{\sf momentum density in either frame},
 \nonumber\\
\quad -\sqrt{-g} T^0{}_0& = & \sum_{\bf p} f \tilde E\qquad
\mbox{\sf   energy density in laboratory frame},
\nonumber\\
\sqrt{-g} T^k{}_i & = &  \sum_{\bf p} f p_i v_G^k \qquad
\mbox{\sf   momentum flux in laboratory frame}, \nonumber\\
-\sqrt{-g} T^i{}_0 & = & -\sum_{\bf p} f \tilde E\frac{\partial E}{\partial p_i}
= \sum_{\bf p} f \tilde E v_G^i\qquad  \mbox{\sf  
energy flux in laboratory frame}, \nonumber\\
 \sqrt{-g} T^{00} & = & \sum_{\bf p} f  p^0 = \sum_{\bf p} f  E 
 \qquad  \mbox{\sf  energy density in comoving frame}.
\end{eqnarray}
With this definition of the momentum-energy tensor the covariant conservation law in 
Eq.(\ref{CovariantConservation}) acquires the form:
\begin{equation}
(\sqrt{-g}T^\mu{}_\nu)_{,\mu}=\sum_{\bf p} f \partial_\nu \tilde E=  P_i \partial_\nu v_{\rm
s}^i + 
\sum_{\bf p} f |{\bf p}|\partial_\nu c 
 \,.
\label{CovariantConservation2}
\end{equation}
The right-hand side represents ``gravitational'' forces acting on the
``matter'' from the superfluid vacuum (for $\nu=i$ this is just the two last
terms in Eq.(\ref{QuasiparticleMomentumEq})).

\subsection{Local thermodynamic
equilibrium.}

Local thermodynamic equilibrium is characterized by the local
temperature $T$ and local normal component velocity ${\bf v}_{\rm n}$  in
Eq.(\ref{Equilibrium}).  In local thermodynamic equilibrium the components of energy-momentum
for the quasiparticle system (matter) are determined by the generic
thermodynamic potential (the pressure), which has the form
\begin{equation}
\Omega=\mp T\frac1{(2\pi\hbar)^3} \sum_{s}  \int d^3p~{\rm ln}(1\mp f)\,,
\label{Pressure}
\end{equation}
with the upper sign for fermions and lower sign for bosons.
For phonons one has
\begin{equation}
\Omega=  \frac{\pi^2}{30\hbar^3}  T_{\rm eff}^4\sqrt{-g}\,,\qquad
T_{\rm eff} = \frac{T}{\sqrt{1-w^2}}~\,,
\label{EffectiveT}
\end{equation}
where the renormalized effective temperature $T_{\rm eff}$ absorbs all the dependence on
two velocities of liquid. The components of the energy momentum tensor are given as
\begin{equation}
T^{\mu\nu}  =    (\varepsilon  + \Omega )u^\mu
u^\nu+\Omega g^{\mu\nu}\,,\qquad \varepsilon=-\Omega +T{\partial \Omega
\over\partial T}=3\Omega\,,\qquad T^\mu{}_\mu=0~.
\label{QuasipStressTensorRel2}
\end{equation}
The four velocity of the ``matter'', $u^\alpha$ and
$u_\alpha=g_{\alpha\beta}u^{\beta}$, which satisfies the normalization
equation $u_\alpha u^\alpha=-1$, is expressed in terms of superfluid and normal component
velocities as
\begin{equation}
u^0={1\over \sqrt{ 1 - w^2}}\,,\qquad u^i={v_{(n)}^i\over \sqrt{  1 -
w^2}}\,,\qquad u_i= { w_i \over \sqrt{{ 1 - w^2}}}\,,\qquad
u_0=-{1+{{\bf w} \cdot{\bf v}_{{\rm s}}} \over
\sqrt{ { 1 - w^2}}}\,.
\label{4Velocity}
\end{equation}

\subsection{Global thermodynamic
equilibrium.  Tolman temperature. Pressure of ``matter'' and
``vacuum'' pressure.}\label{GlobalThermodynamicEquilibrium}

The distribution of quasiparticles in local equilibrium in
Eq.(\ref{Equilibrium}) can be expressed via the
temperature four-vector $\beta^\mu $ and thus via the effective temperature $T_{\rm eff}$:
\begin{equation}
f_{\cal T} = {1\over
1+\exp[-\beta^\mu p_\mu]}\,,\qquad \beta^\mu ={u^\mu\over  T_{\rm eff}}
=\left({1\over T}, {{\bf v}_n \over  T
}\right)\,,\qquad \beta^\mu\beta_\mu=-T_{\rm eff}^{-2}~.
\label{4Temperature}
\end{equation}

For the relativistic system, the true equilibrium with vanishing entropy
production is established if $\beta^\mu$ is a timelike Killing vector
satisfying
\begin{equation}
\beta_{\mu;\nu}+
\beta_{\nu;\mu}=0~,\;\;{\rm or}\qquad
\beta^\alpha\partial_\alpha g_{\mu\nu}+
(g_{\mu\alpha}\partial_\nu +g_{\nu\alpha}\partial_\mu)\beta^\alpha=0
~.
\label{EquilibriumConditions}
\end{equation}
For a time-independent, space-dependent situation the  condition
$0=\beta_{0;0}=\beta^i\partial_i g_{00}$ gives
$\beta^i=0$, while the other conditions are satisfied when
$\beta^0={\rm constant}$. Hence the true equilibrium requires that
${\bf v}_{\rm n}=0$ in the frame where the superfluid velocity field is time independent
(i.e. in the frame where $\partial_t{\bf v}_{\rm s}=0$), and
$T={\rm constant}$. These are just the global equilibrium conditions in
superfluids, at which no dissipation occurs.  From the   equilibrium conditions $T={\rm
constant}$ and ${\bf v}_{\rm n}=0$ it follows that under the global  equilibrium the effective
temperature in Eqs.(\ref{EffectiveT}) is space dependent according to
\begin{equation}
T_{\rm eff}({\bf r})={T\over \sqrt{ 1 - v_{\rm s}^2({\bf r})}}={T\over
\sqrt{-g_{00}({\bf r})}}\,.
\label{TolmanLaw}
\end{equation}
According to Eq.(\ref{4Temperature})  the
effective temperature $T_{\rm eff}$ corresponds to the
``covariant relativistic'' temperature in general relativity. It is an apparent
temperature as measured by the local observer, who ``lives'' in superfluid
vacuum and uses sound for communication as we use the light signals. The
Eq.(\ref{TolmanLaw}) is exactly the Tolman's law in general relativity
\cite{Tolman}, which shows how the local temperature ($T_{\rm eff}$) changes
in the gravity field in equilibrium. The role of the constant Tolman
temperature is played by the real constant temperature $T$ of the liquid.

Note that $\Omega$ is the pressure created by quasiparticles (``matter''). In superfluids this
pressure is supplemented by the  pressure of the superfluid component -- the vacuum
pressure discussed in Sec.\ref{VacuumEnergyAnd}-- so that the total pressure 
in equilibrium is 
\begin{equation}
P=P_{\rm vac} +P_{\rm matter}= P_{\rm vac}+ \frac{\pi^2}{30\hbar^3}  T_{\rm eff}^4\sqrt{-g}~,
\label{TotalPressure}
\end{equation}
For the liquid in the absence of the interaction with
environment the total pressure of the liquid is zero in equilibrium, which
means that the the vacuum pressure compensates the pressure of matter. In
noneqilibrium situation this compensation is not complete, but the two
pressures are of the same order of magnitude. Maybe this can provide the
natural solution of the cosmological constant problem: $\Lambda$ appears to be
almost zero without fine tuning.

In conclusion of this Section, the normal part of the superfluid $^4$He fully
reproduces the dynamics of the relativistic matter in the presence of the
gravity field. Though the ``gravity'' itself is not determined by Einstein
equations, using the proper superflow fields we can simulate many phenomena
related to the classical and quantum behavior of matter in a curved space-time,
including the black-hole physics.

\section{Universality classes of fermionic vacua.} \label{UniversalityClassesOf}

Now we proceed to the Fermi systems, where the effective theory involves both
bosonic and fermionic fields. What kind of the effective fields
arises depends on the universality class of the Fermi systems, which determines
the behavior of the fermionic quasiparticle spectrum at low energy.

\subsubsection{Classes of fermionic quasiparticle spectrum} 

There are three generic classes of the fermionic spectrum in condensed matter.

\begin{figure}[t]
\centerline{\includegraphics[width=\linewidth]{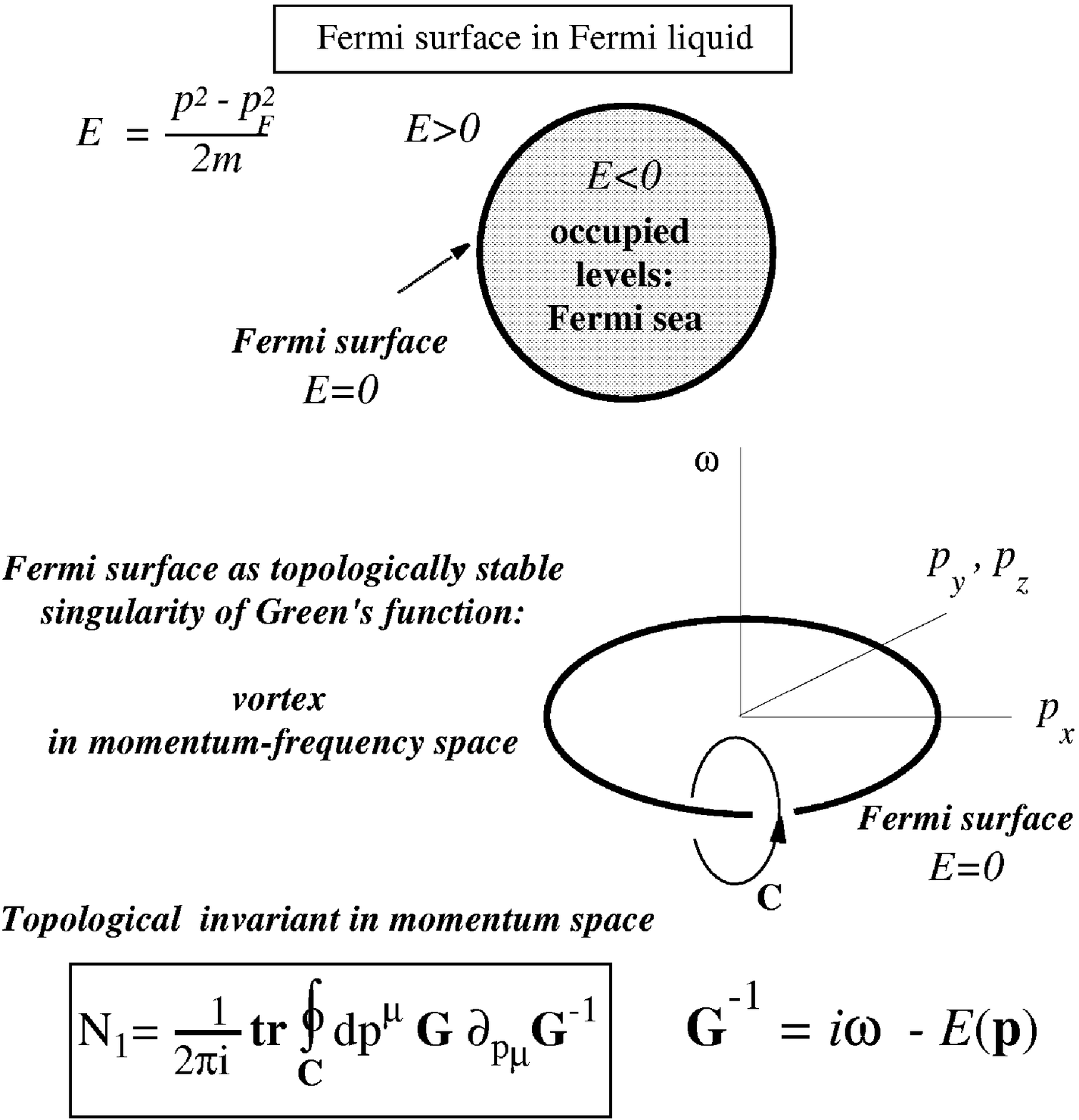}}
\medskip
\caption{Fermi surface as a topological object in momentum space. {\it Top}: In
the Fermi gas the Fermi surface bounds the solid Fermi sphere of the occupied
negative energy states. {\it Bottom}: Fermi surface survives even if the
interaction between the particles is introduced. The reason for that is that
the Fermi surface is the topologically stable object: it is the vortex in the 4D
momentum-frequency space $(\omega,{\bf p}$.}
\label{FermiSurfaceFig}
\end{figure}

In (isotropic) Fermi liquid the spectrum of fermionic quasiparticles approaches at low energy
the universal behavior 
\begin{equation}
E({\bf p},n)\rightarrow v_F(n)\left(|{\bf
p}|-p_F(n)\right)~,~~{\rm class~~~(i)}~~,
\label{Spectrum1}
\end{equation}
 with two ``fundamental constants'', the Fermi velocity
$v_F$ and Fermi momentum $p_F$. The values of these parameters are governed by
the microscopic physics, but in the effective theory of Fermi liquid they are
the fundumental constants. The energy of the fermionic quasiparticle in
Eq.(\ref{Spectrum1}) is zero on a two dimensional manifold
$|{\bf p}|=p_F(n)$ in 3D momentum space, called the Fermi surface (Fig.
\ref{FermiSurfaceFig}). 

\begin{figure}[t]
\centerline{\includegraphics[width=\linewidth]{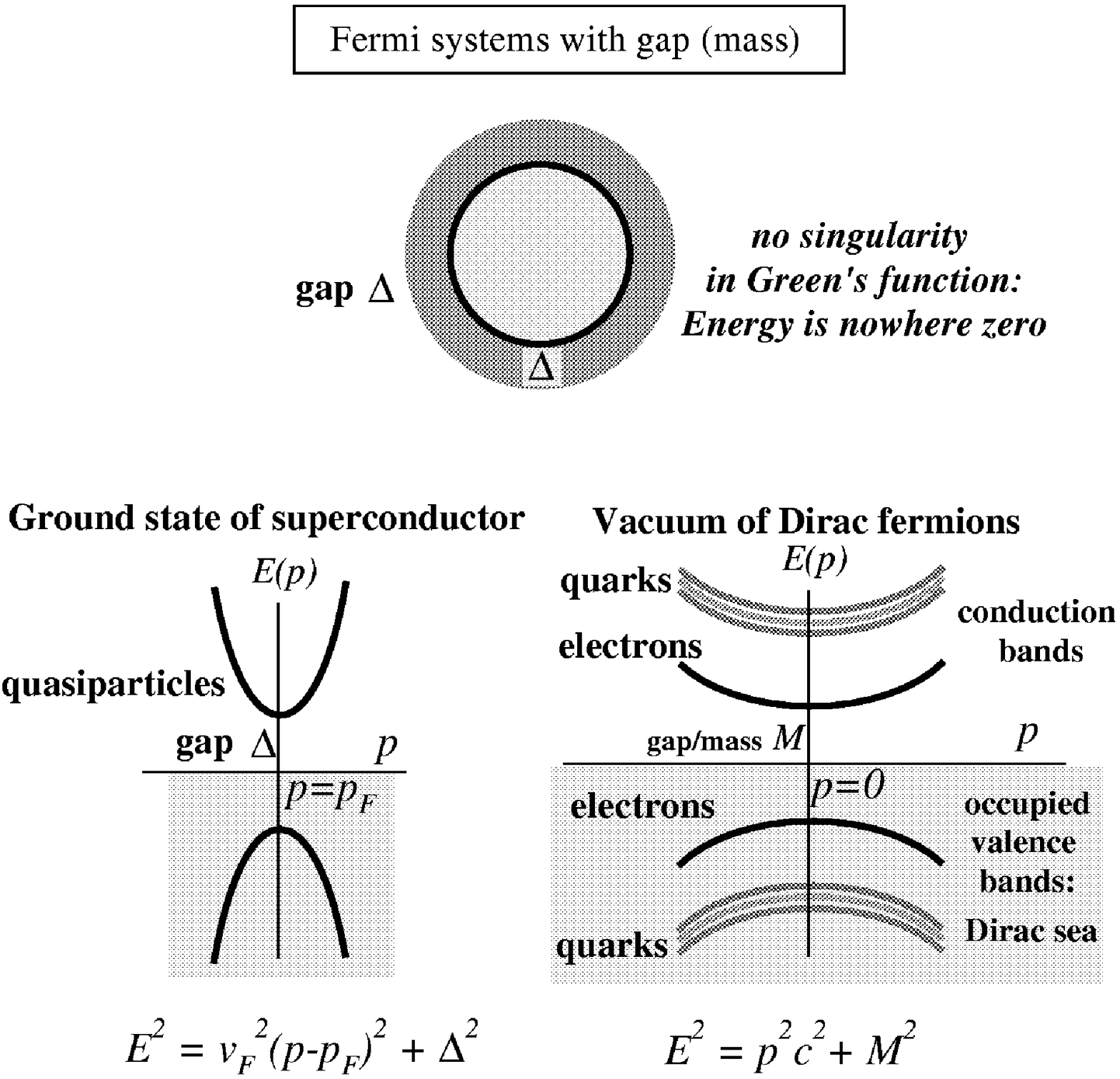}}
\medskip
\caption{Fermi systems with gap or mass. {\it Top}: The gap which appears on
the Fermi surface in conventional superconductors and in $^3$He-B. {\it Bottom
left}: Quasiparticle spectrum in conventional superconductors and $^3$He-B. {\it Bottom
right}: The spectrum of Dirac particles and quasiparticle spectrum in
semiconductors.}
\label{GappedSystemsFig}
\end{figure}

In isotropic
superconductor and in superfluid
$^3$He-B the energy of quasipartilce is nowhere zero
(Fig.\ref{GappedSystemsFig}), the gap $\Delta_0$ in the spectrum appears as an
additional ``fundamental constant''
\begin{equation}
E^2({\bf p})\rightarrow \Delta_0^2 +  v_F^2 \left(|{\bf
p}|- p_F\right)^2~,~~{\rm class~~~(ii)}~~.
\label{Spectrum2}
\end{equation}

\begin{figure}[t]
\centerline{\includegraphics[width=\linewidth]{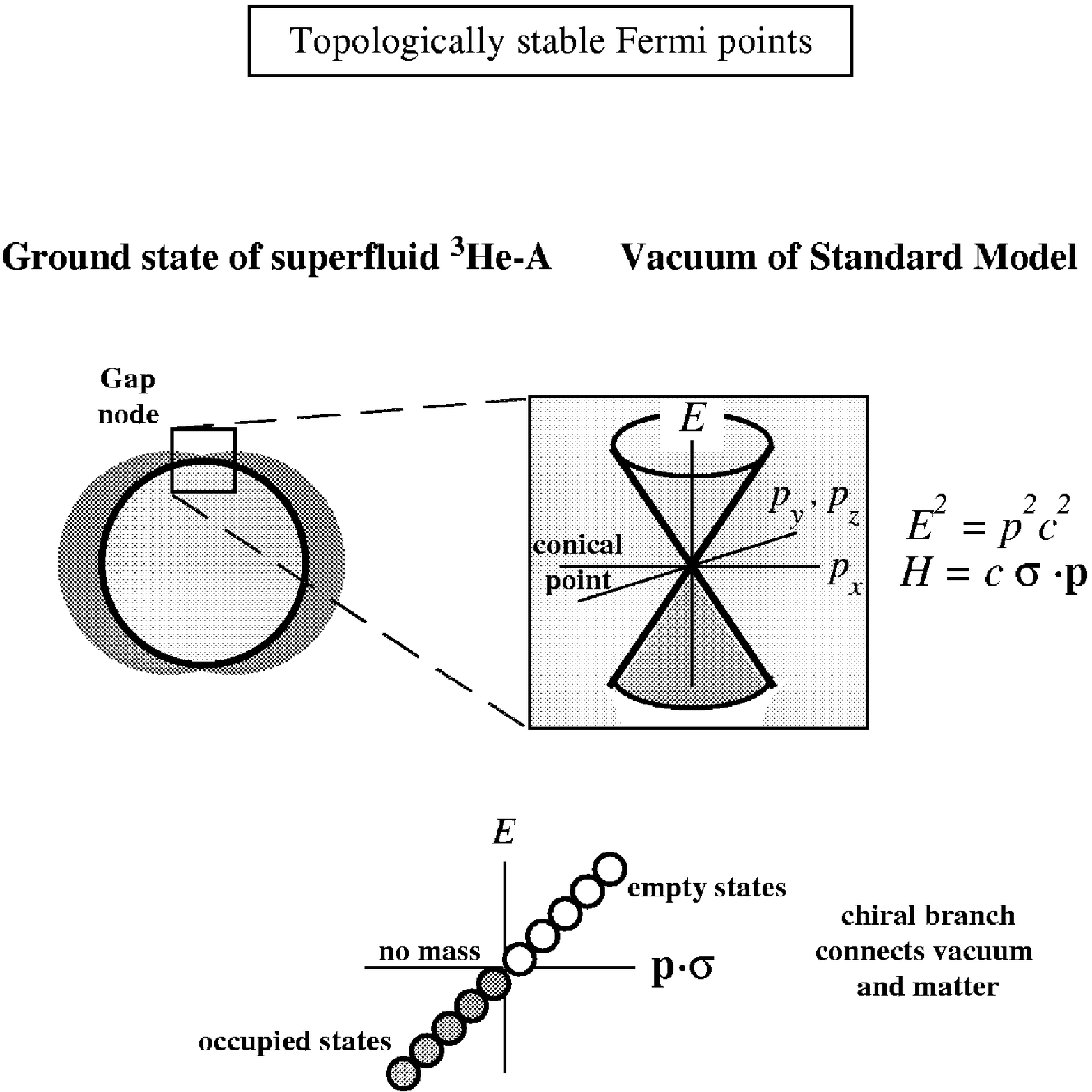}}
\medskip
\caption{{\it Top}: Gap node in superfluid $^3$He-A is the conical point in the
energy-momentum space. {\it Bottom}:  The spectrum of the right-handed
chiral particle particle: its spin ${\bf \sigma}$ is oriented along the
momentum ${\bf p}$. Quasiparticles in the vicinity of the
nodes in $^3$He-A and elementary particles in the Standard Model above the
electroweak transition are chiral fermions.}
\label{FermiPointFig}
\end{figure}

In $^3$He-A the gap $\Delta({\bf p})$ depends on the direction of the
momentum
${\bf p}$: $\Delta^2({\bf p})=\Delta_0^2({\bf p}\times {\hat{\bf l}})^2/p_F^2$.
It becomes zero in two opposite directions called the gap nodes -- along and
opposite to the unit vector ${\hat{\bf l}}$. As a result the
quasiparticle energy is zero at two isolated points ${\bf p}=\pm p_F{\hat{\bf
l}}$ in 3D momentum space (Fig.
\ref{FermiPointFig}). Close to the gap node at
${\bf p}={\bf p}_0$ the spectrum has a form 
\begin{equation}
E^2({\bf p})\rightarrow g^{ik}(p_i -p_{0i})(p_i -p_{0k})~, ~~{\rm
class~~~(iii)}~~.
\label{Spectrum3}
\end{equation}

These three spectra represent three topologically distinct universality
classes of the fermionic vacuum in 3+1 dimension:  (i) Systems with Fermi
surface (Fig. \ref{FermiSurfaceFig}); (ii) Systems with gap or mass (Fig.
\ref{GappedSystemsFig}); and (iii) Systems with Fermi points (Fig.
\ref{FermiPointFig}).
Systems with the Fermi lines in the spectrum are topologically unstable and by
small perturbations can be transformed to one of the three classes. The same
topological classification is applicable to the fermionic vacua in high energy
physics. The vacuum of Dirac fermions, with the excitation spectrum 
$E^2({\bf p})\rightarrow M^2 + c^2  |{\bf p}|^2$, belongs to the class (ii).  The vacuum of
the Weyl fermions in the Standard Model,  with the excitation spectrum $E^2({\bf
p})\rightarrow  c^2  |{\bf p}|^2$, belongs to the class (iii). As we shall see
below, the latter class is very special, since in this class the relativistic
quantum field theory with chiral fermions emerges in the low energy corner,
while the collective fields form the gauge fields and gravity. 

$^3$He liquids present examples of all 3 classes.  The normal $^3$He liquid at
$T>T_c$  and also the ``high energy physics'' of superfluid $^3$He phases
(with energy $E\gg \Delta_0$) are representative of the class (i). Below the
superfluid transition temperature $T_c$ one has either an isotropic superfluid
$^3$He-B of the class (ii) or superfluid $^3$He-A, which belongs to the class 
(iii), where the relativistic quantum field theory with chiral fermions
gradually arises  at low temperature.  The great advantage of superfuid
$^3$He-A is that it can be described by the BCS theory, which incorporates all
the hierarchy of the energy scales:
The
``transPlanckian'' scale of energies $E\gg
\Delta_0$ which in the range $v_F
p_F \gg E \gg \Delta_0$ is described by the effective theory of universality
class (i); the ``Planck'' scale physics at $E\sim \Delta_0$; and the low-energy
physics of energies $ E \ll \Delta_0^2/v_Fp_F$ which is described by the
effective relativistic theory of universality class (iii).   Let us start with
the universality class (i).

\subsection{Fermi surface as topological object} \label{FermiSurface}

The Fermi surface (Fig.\ref{FermiSurfaceFig}) naturally appears in the
noninteracting Fermi gas, where the energy spectrum of fermions is
\begin{equation}
E(p)={p^2\over 2m}-\mu~,
\label{FermiGasEnergySpectrum}
\end{equation}
and  $\mu >0$ is as before the chemical potential. The Fermi surface
bounds the volume in the momentum space where the energy is
negative, $E(p)<0$, and where the particle states are all occupied at
$T=0$. In this isotropic model the Fermi surface
is a sphere of radius $p_F=\sqrt{2m\mu}$. Close to the Fermi surface the
energy spectrum is $E(p)\approx v_F(p-p_F)$, where $v_F=\partial_p E|_{p=p_F}$
is the Fermi velocity.

It is important that the Fermi surface
survives even if interactions between particles are introduced. Such stability
of the Fermi surface comes from the
topological property of the Feynman quantum
mechanical propagator -- the one-particle Green's function
\begin{equation}
{\cal G} =(z-{\cal H})^{-1}~.
\label{Propagator1}
\end{equation}
Let us write the propagator for a given momentum ${\bf p}$ and for the
imaginary frequency, $z=ip_0$. The imaginary frequency is introduced to
avoid the conventional singularity of the propagator ``on the mass shell'',
i.e. at $z=E(p)$. For noninteracting particles the propagator has the form
\begin{equation}
G={1\over ip_0 -v_F(p-p_F)}~.
\label{Propagator2}
\end{equation}
Obviously there is still a singularity:  On the 2D hypersurface $(p_0=0,
p=p_F)$ in the 4-dimensional space $(p_0, {\bf p})$ the propagator is not
well defined. This singularity is stable, i.e. it cannot be eliminated by
small perturbations. The reason is that the phase $\Phi$ of the Green's
function $G=|G|e^{i\Phi}$ changes by $2\pi$ around the path $C$ embracing this
2D hypersurface in the 4D-space (see the bottom of Fig.\ref{FermiSurfaceFig},
where one dimension is skipped, so that the Fermi surface is presented as a
closed line in 3D space). The phase winding number
$N_1=1$ cannot change continuously, that is why it is robust towards any
perturbation. Thus the singularity of the Green's function on the 2D-surface in
the momentum space is preserved, even when interactions between particles are
introduced.

Exactly the same topological conservation of the winding number leads to the
stability of the quantized vortex in superfluids and superconductors,
the only difference being that, in the case of vortices, the phase winding
occurs in the real space, instead of the momentum space. The complex order
parameter $\Psi=|\Psi|e^{i\Phi}$ changes by $2\pi n_1$ around the path embracing
the vortex line in 3D space or vortex sheet in 3+1 space-time.
The connection between the space-time topology and the energy-momentum space
topology  is, in fact, even deeper (see {\it 
e.g.} Ref.\cite{VolovikMineev1982}). If the order parameter depends on
space-time, the propagator in semiclassical
aproximation depends both on 4-momentum and on
space-time
coordinates $G(p_0,{\bf p}, t, {\bf r})$.
The topology
in the 4+4 dimensional space describes:
the momentum space topology
of the homogeneous system; topological defects of the order parameter in
space-time;
topology
of the energy spectrum within the topological 
defects \cite{Grinevich1988}; and quantization of physical parameters (see Section
\ref{Gapped2D}).

In the more complicated cases, when the Green's function is the matrix with
spin and band indices,  the phase of the Green's function becomes
meaningless. In this case one should use  a general analytic expression for
the integer momentum-space topological
invariant which is
responsible for the stability of the Fermi surface:
\begin{equation}
N_1={\bf Tr}~\oint_C {dl\over 2\pi i} {\cal G}(p_0,p)\partial_l {\cal
G}^{-1}(p_0,p)~.
\label{InvariantFor FS}
\end{equation}
Here the integral is taken over an arbitrary contour $C$ in the momentum
space  $({\bf p},p_0)$, which encloses the Fermi  hypersurface (Fig.
\ref{FermiSurfaceFig} {\it bottom}); and
${\bf Tr}$ is the trace over the spin and band indices.

\subsubsection{Landau Fermi liquid
}

\begin{figure}[t]
\centerline{\includegraphics[width=\linewidth]{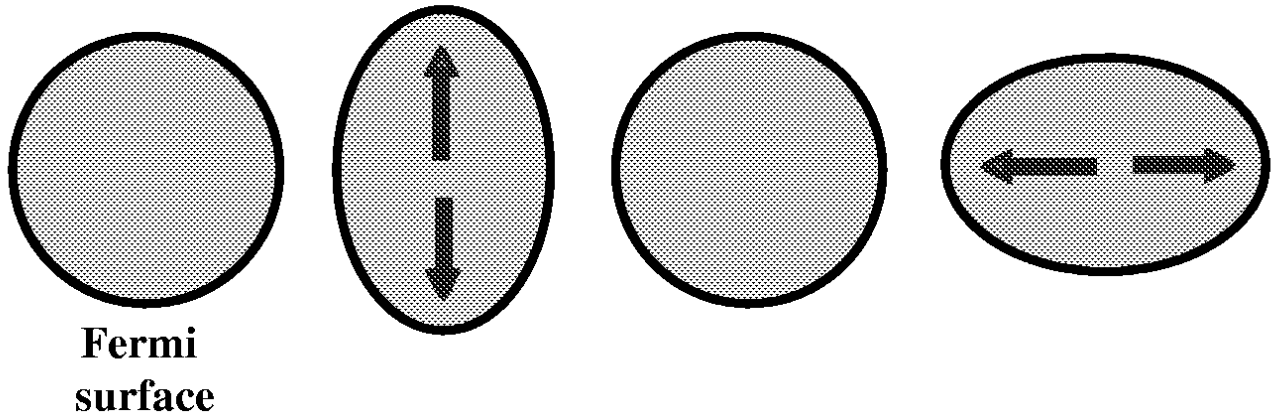}}
\medskip
\caption{Bosonic collective modes in the fermionic vacuum of the
Fermi-surface universality class.  Collective motion of particles
comprising the vacuum is seen by an individual quasiparticle as dynamical modes
of the  Fermi surface. Here the propagating elliptical deformations of the
Fermi surface are drawn.}
\label{CollectiveModesFSFig}
\end{figure}

The topological
class of systems with Fermi
surface
is rather broad. In particular it
contains conventional Landau Fermi-liquids,
in
which the propagator preserves the pole. Close to the pole the propagator is
\begin{equation}
G={Z\over ip_0 -v_F(p-p_F)}~.
\label{PropagatorWithZ}
\end{equation}
Evidently the residue $Z\neq 1$ does not change the
topological
invariant for the propagator,
Eq.(\ref{InvariantFor FS}), which remains $N_1=1$. This is essential for
the Landau theory of an interacting Fermi liquid;
it confirms the assumption that there is one to one correspondence between the low
energy quasiparticles  in Fermi liquids and
particles in a Fermi gas. It is also important for the consideration of the
bosonic collective modes
of the Landau
Fermi-liquid.
The interaction between the fermions
cannot not change the topology
of the fermionic
spectrum,  but it produces the effective field
acting on a given particle by the other moving particles. This effective
field
cannot destroy the  Fermi
surface
owing to its 
topological
stability, but it can locally shift the
position of the  Fermi surface.
Therefore a collective motion of the particles is seen by an individual
quasiparticle as dynamical modes of the  Fermi surface (Fig.
\ref{CollectiveModesFSFig}. These bosonic modes are known as the
different harmonics of the zero sound \cite{Khalatnikov}.

Note that the Fermi hypersurface exists for any spatial dimension. In the
2+1 dimension the Fermi hypersurface is a line in 2D momentum space, which
cooresponds to the vortex loop in the 3D frequency-momentum space in Fig.
\ref{FermiSurfaceFig}.

Topological stability also means that any adiabatic change of the system
will leave the system within the same class. Such adiabatic perturbation can
include the change of the interaction strength between the particles,
deformation of the Fermi surface, etc. Under adiabatic perturbation no
spectral flow across the Fermi surface occurs (of course, if the deformation is
slow enough), so the state without excitations transforms to the other state, in
which  excitations are also absent, i.e. the vacuum transforms to the vacuum.
The absence of the spectral flow leads in particular to the  Luttinger's theorem
which states that the volume of the Fermi surface is invariant under adiabatic
deformations, if the number of particles is kept constant
\cite{LuttingerTheorem}. Since the isotropic Fermi liquid can be obtained from
the Fermi gas by adiabatical switching on the interaction between the
particles, the relation between the particle density and the Fermi momentum
remains the same as in the Fermi gas, 
\begin{equation}
n={p_F^3\over 3\pi^2\hbar^3}~.
\label{LuttingerTheorem}
\end{equation}
 Topological approach
to Luttinger's theorem has been recently discussed in
\cite{LuttingerTheoremTopology}. The processes related to the spectral flow of
quasiparticle energy levels will be considered in Sections \ref{ChiralAnomaly}
and \ref{FermionZeroModesOnVort}  in connection with the phenomenon of axial
anomaly. 

\subsubsection{Non-Landau Fermi liquids}

In the 1+1 dimension, the Green's function looses its pole but nevertheless the
Fermi surface
is still there
\cite{NewClass,Blagoev}. Though the Landau Fermi liquid transforms to another
states, this occurs within the same topological class with given $N_1$. An
example is provided by the Luttinger liquid.
Close to the Fermi surface the Green's function for the Luttinger liquid
can be approximated as (see \cite{Wen,NewClass,LuttingerLiquidReview})
\begin{eqnarray}
\nonumber G(z,p)\sim \\
(ip_0 -v_1\tilde p)^{g-1\over 2}(ip_0+v_1\tilde p)^{g\over
2}(ip_0 -v_2\tilde p)^{g-1\over 2}(ip_0+v_2\tilde p)^{g\over 2}
\label{Propagator3}
\end{eqnarray}
where $v_1$ and $v_2$ correspond to Fermi velocities of spinons and holons
and $\tilde p=p-p_F$. The above equation is not exact but reproduces the
momentum space topology
of the Green's
function in Luttinger Fermi liquid.
If $g\neq
0$ and
$v_1\neq v_2$, the singularity in the $(\tilde p, z=ip_0)$ momentum space
occurs on the Fermi surface,
 i.e. at  $(p_0=0,
~\tilde p=0)$.  The momentum space topological
invariant in Eq.(\ref{InvariantFor FS}) remains the same
$N_1=1$, as for the conventional Landau Fermi-liquid.
The
difference from Landau Fermi liquid
occurs only at
real frequency $z$: The quasiparticle pole is absent and one has the branch
cut singularities instead of the mass shell, so that the quasiparticles are
not well defined. The population of the particles has no jump on the Fermi
surface,
but has a power-law singularity in the
derivative \cite{Blagoev}.

Another example of the non-Landau Fermi liquid
is
the  Fermi liquid
with exponential behavior of the
residue \cite{Yakovenko}. It also has the Fermi surface
with the same topological
invariant, but the
singularity at the Fermi surface
is exponentially weak.

\subsection{Fully gapped systems: ``Dirac particles'' in superconductors   
and in superfluid $^3$He-B} \label{FullyGapped}

Although the systems we have discussed in Sec.\ref{FermiSurface} contain fermionic and bosonic
quantum fields, this is not the relativistic quantum field theory which we need for the
simulation of quantum vacuum: There is no Lorentz invariance and the oscillations of the
Fermi surface do not resemble the gauge field even remotely. The situation is somewhat better
for superfluids and superconductors  with fully gapped spectra. For example, the
Nambu-Jona-Lasinio model in particle physics provides a parallel with conventional
superconductors \cite{NambuJona-Lasinio}; the symmetry breaking scheme in superfluid $^3$He-B
was useful for analysis of the color superconductivity in quark matter
\cite{ColorSuperfluidity}. 

In $^3$He-B the Hamiltonian of free Bogoliubov quasiparticles is the $4\times 4$ matrix (see
Eq.(\ref{BogoliubovNambuHamBPhase}) below):
\begin{eqnarray}
{\cal H}=\left( \matrix {M({\bf p}) & c{\bf\sigma}\cdot{\bf p}  \cr c{\bf  \sigma}\cdot{\bf
p}  & -M({\bf p})\cr }
\right) =
{\check \tau}_3 M({\bf p})+c{\check \tau}_1 {\bf  \sigma}\cdot{\bf p}~,\\
M({\bf p})={p^2\over
2m}-\mu\approx v_F(p-p_F)~,~ c={\Delta_0 \over p_F}~~~,~~~{\cal
H}^2=E^2\approx M^2({\bf p}) +
\Delta_0^2~,~ 
\label{PropagatorFullyGapped}
\end{eqnarray}
where the Pauli $2\times 2$   matrices ${\bf\sigma}$ describe  the conventional spin of
fermions and $2\times 2$   matrices 
$\check {\bf\tau}$ describe the Bogoliubov-Nambu isospin in the
particle-hole space (see Sec.\ref{BCS}).  The  Bogoliubov-Nambu  Hamiltonian
becomes ``relativistic'' in the limit 
$mc^2 \gg \mu$, where it asymptotically approaches 
the Dirac Hamiltonian for relativistic particles of mass $\mu$. However in a real   $^3$He-B
one has an opposite limit
$mc^2 \ll \mu$ and the energy spectrum is far from being ``relativistic''. Nevertheless 
$^3$He-B also serves as a model system for simulations of phenomena in particle physics  and
cosmology. In particular,  
vortex nucleation in nonequilibrium phase transition has been observed   
\cite{MiniBigBang} as experimental verification of the Kibble mechanism describing formation
of cosmic strings in expanding Universe
\cite{Kibble}; the global vortices in  $^3$He-B were also used for 
experimental simulation of the production of baryons by cosmic strings mediated
by spectral flow \cite{BevanNature} (see
Sec.\ref{FermionZeroModesOnVort}).

\subsection{Systems with Fermi points}\label{SystemsWithFermiPoints}  

\subsubsection{Chiral particles and Fermi point} 

In particle physics the energy spectrum $E({\bf p})=cp$ is characteristic of the
massless chiral fermion, lepton or quark, in the Standard Model  with $c$ being
the speed of light. As distinct from the case of Fermi surface, where the energy of
quasiparticle is zero at the surface in 3D momentum space, the energy of a chiral particle is
zero at the point ${\bf p}=0$. We call such point the Fermi point.  The Hamiltonian for the
massless spin-1/2 particle is a
$2\times 2$ matrix
\begin{equation}
 {\cal H}=\pm 
c{\bf \sigma}\cdot {\bf p}
\label{Neutrino}
\end{equation}  
 which is expressed in terms of the Pauli spin matrices ${\bf \sigma}$. The sign $+$ is for a
right-handed particle  and $-$  for a left-handed one: the spin of the particle is oriented
along or opposite to its momentum, respectively. 

\subsubsection{Topological invariant for Fermi point} 

\begin{figure}[t]
\centerline{\includegraphics[width=\linewidth]{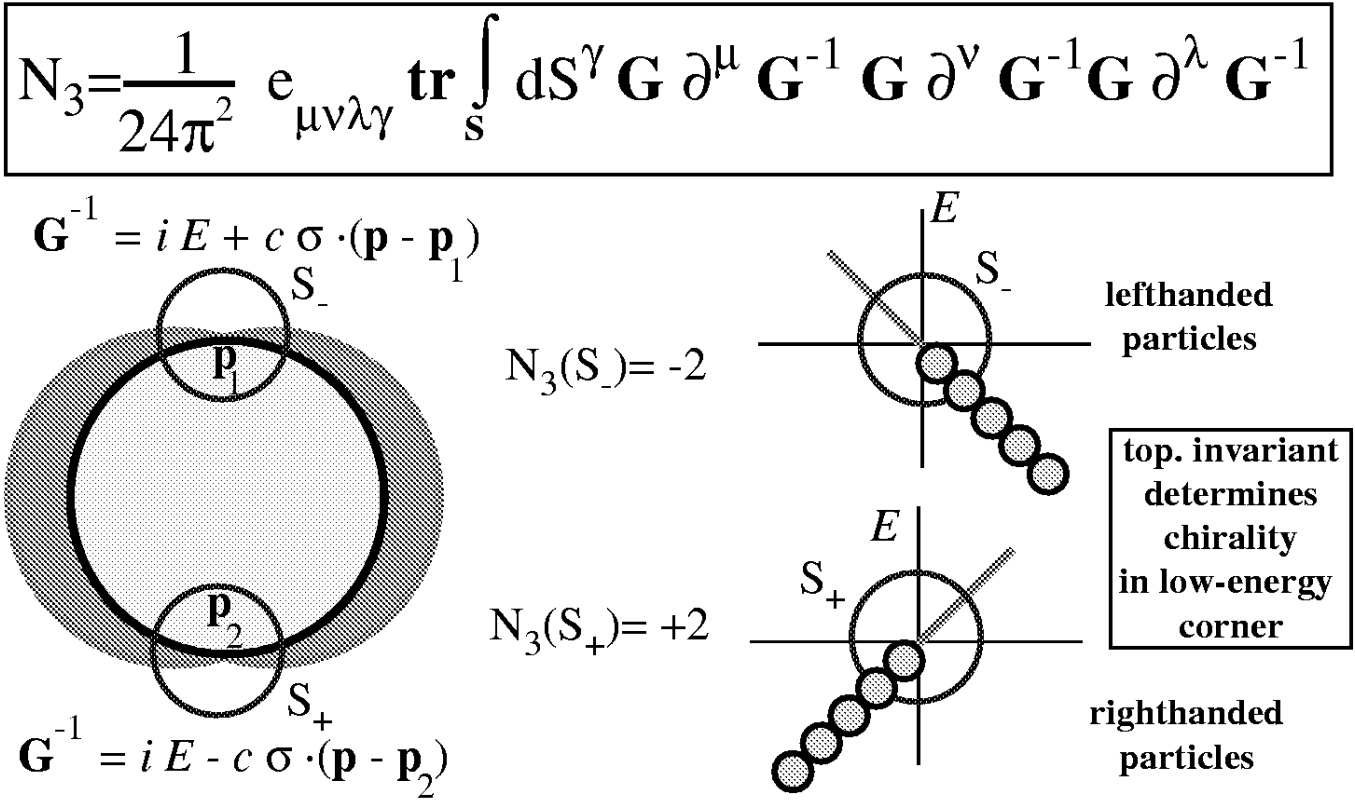}}
\medskip
\caption{The Green's function for fermions in $^3$He-A and in the
Standard Model have point singularity in the 4D momentum-frequency space,
which is described by the integer-valued topological invariant $N_3$. The
Fermi points in $^3$He-A at ${\bf p}=\pm p_F{\hat{\bf l}}$ have $N_3=\mp 2$.
The Fermi point at ${\bf p}$ for the chiral relativistic particle in Eq.
(\protect\ref{Neutrino}) has $N_3=C_a$, where $C_a=\pm 1$ is the chirality.
The chirality, however, appears only in the low energy corner together with
the Lorentz invariance. Thus the topological index $N_3$ is the generalization
of the chirality to the Lorentz noninvariant case.}
\label{FermiPointTopologyFig}
\end{figure}

Even if the Lorentz symmetry is violated far from the Fermi point, the Fermi
point will survive. The stability of the Fermi point is prescribed by the
mapping of the surface $S_2$ surrounding the degeneracy point in 3D-momentum
space into the complex projective space $CP^{N-1}$ of the eigenfunction
$\omega({\bf p})$ of $N\times N$ Hamiltonian describing the fermion with $N$
components \cite{NielsenNinomiya} ($N=2$ for one Weyl spinor). The topological
invariant can be written analytically in terms of the Green's function ${\cal
G} =(ip_0-{\cal H})^{-1}$ determined on the imaginary  frequency axis, $z=ip_0$
\cite{Exotic} (Fig.\ref{FermiPointTopologyFig}). One can see that this
propagator has a singularity at the point in the 4D momentum-frequency space:
$(p_0=0,{\bf p}=0)$.  The invariant is represented as the integral
around  the 3-dimensional surface $\sigma$ embracing such singular point
\begin{equation} 
N_3 = {1\over{24\pi^2}}e_{\mu\nu\lambda\gamma}~
{\bf tr}\int_{\sigma}~  dS^{\gamma}
~ {\cal G}\partial_{p_\mu} {\cal G}^{-1}
{\cal G}\partial_{p_\nu} {\cal G}^{-1} {\cal G}\partial_{p_\lambda}  {\cal
G}^{-1}~.
\label{TopInvariant}
\end{equation}
Here ${\cal G}$ is an arbitrary matrix function of ${\bf p}$ and $p_0$, which is
continuous and differentiable outside the singular point. One can check that
under continuous variation of the matrix function the integrand changes by the
full derivative. That is why the integral over the closed 3-surface does not
change, i.e. $N_3$ is invariant under continuous deformations of the Green's
function and also of the closed 3-surface. The possible values of the invariant
can be easily found: if one chooses the matrix  function which changes in
$U(2)$ space one obtains the integer values of $N_3$. They describe the
mapping  of the $S^3$ sphere surrounding the degeneracy point in 4-space of the
energy-momentum
$(p_0,{\bf p})$ into the $SU(2)=S^3$ space.  The same integer values $N_3$ are
preserved for any Green's function matrix, if it is well determined, i.e. if
${\rm det} ~G^{-1}\neq 0$. The index $N_3$ thus represent topologically
different Fermi points -- the singular points in 4D momentum-frequency space.

\begin{figure}[t]
\centerline{\includegraphics[width=\linewidth]{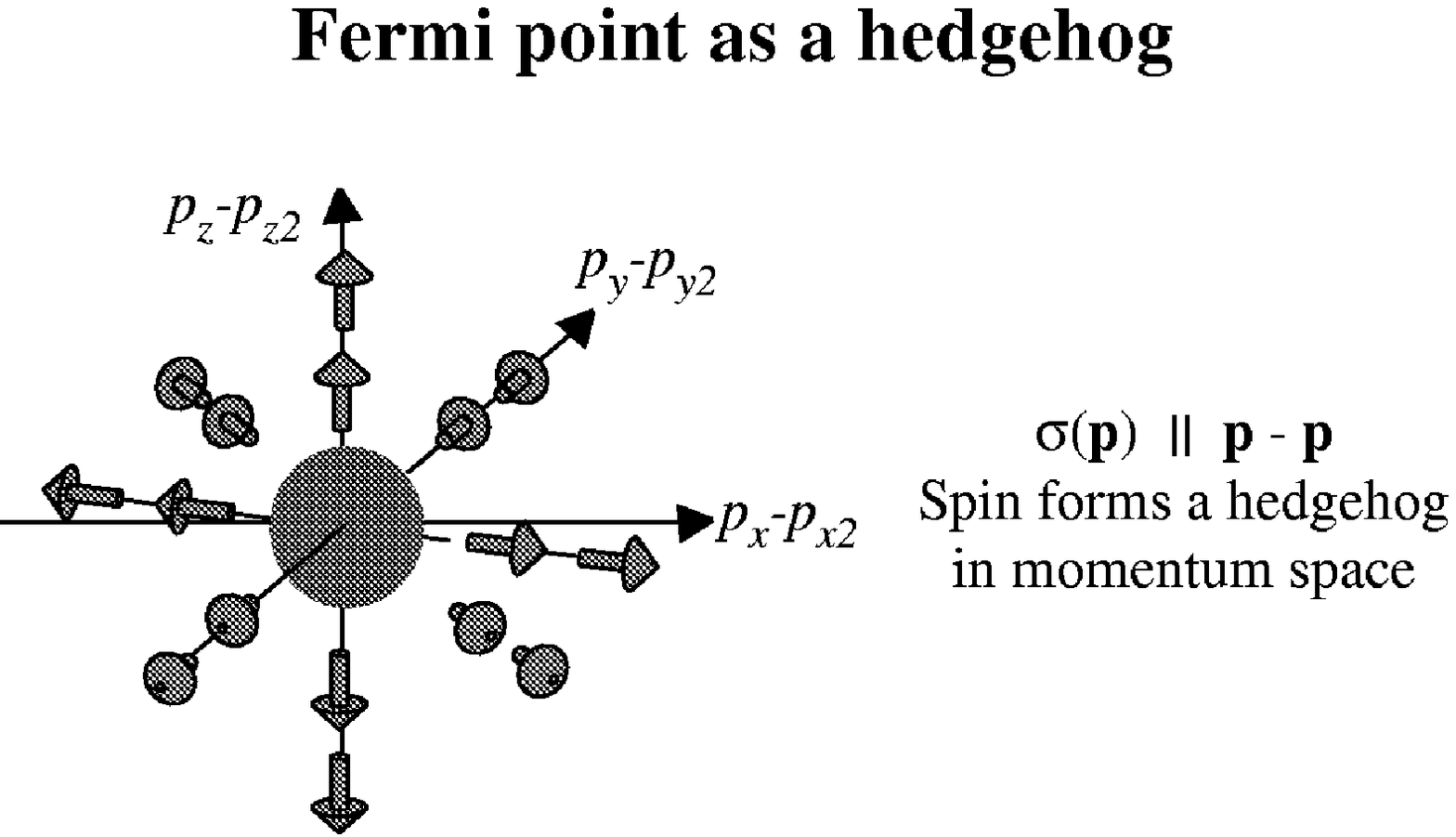}}
\medskip
\caption{Illustration of the
meaning of the topological invariant for the simplest case: Fermi point as a
hedgehog in 3D momentum space. For each momentum ${\bf p}$ we draw the
direction of the quasiparticle spin, or its equivalent in
$^3$He-A -- the Bogoliubov spin. Topological invariant for the hedgehog is
the mapping $S^\rightarrow S^2$ with integer index $N_3$ which is $N_3=+1$
for the drawn case of right-handed particle. The topological invariant $N_3$ is
robust to any deformation of the spin field ${\bf\sigma}({\bf p})$: one cannot
comb the hedgehog smooth.}
\label{HedgehogFig}
\end{figure}

\subsubsection{Topological invariant as the generalization of chirality.}
\label{GeneralizationChirality}

For the chiral fermions in Eq.(\ref{Neutrino}) this invariant has values 
$N_3=C_a$, where $C_a=\pm 1$ is the chirality of the fermion. For this case the
meaning of this topological invariant can be easily visualized (Fig.
\ref{HedgehogFig}). Let us consider the behavior of the
particle spin
${\bf s}({\bf p})$ as a function the particle momentum ${\bf p}$ in the
3D-space  
${\bf p}=(p_x,p_y.p_z)$. For the right-handed particle, whose spin is parallel
to the momentum, one has
${\bf s}({\bf p})={\bf p}/2p$, while for left-handed ones ${\bf s}({\bf p})=-  {\bf
p}/2p$. In both cases the spin distribution in the momentum space looks like a
hedgehog, whose spines are represented by spins. Spines point outward for the
right-handed particle producing the mapping of the sphere $S^2$  in 3D
momentum space onto the sphere
$S^2$ of the spins with
the index $N_3=+1$. For the left-handed particle the spines of the hadgehog
look  inward and one has the mapping with $N_3=-1$.  In the 3D-space the
hedgehog is topologically stable.  

What is important here that the Eq.(\ref{TopInvariant}), being the
topological invariant, does not change under any (but not very large)
perturbations. This means that even if the interaction between the particles
is introduced and the Green's functions changes drastically, the result remains
the same: $N_3=1$ for the righthanded particle and $N_3=-1$ for the lefthanded one.
The singularity of the Green's function remains, which means that the
quasiparticle spectrum remain gapless: fermions are massless even in the
presence of interaction.  

Above we considered the relativistic fermions. However, the topological invariant
is robust to any deformation, including those which violate the Lorentz
invariance. This means that the topological description is far
more general than the description in terms of chirality, which is valid only when the Lorentz
symmetry is obeyed. In paticular, the notion of the Fermi point can be extended to the
nonrelativistic condensed matter, such as  superfluid $^3$He-A (Fig.
\ref{FermiPointTopologyFig}), while the chirality of quasiparticles is not
determined in the nonrelativistic system. This means that the charge $N_3$ is
the topological generalization of chirality. 

From the topological point of view the Standard Model and the Lorentz noninvariant ground
state of $^3$He-A belong to the same universality class of systems with Fermi points, though
the underlying ``microscopic'' physics can be essentially different.

\subsubsection{Relativistic massless chiral fermions emerging near Fermi point.}
\label{RelativisticMassless}

\begin{figure}[t]
\centerline{\includegraphics[width=\linewidth]{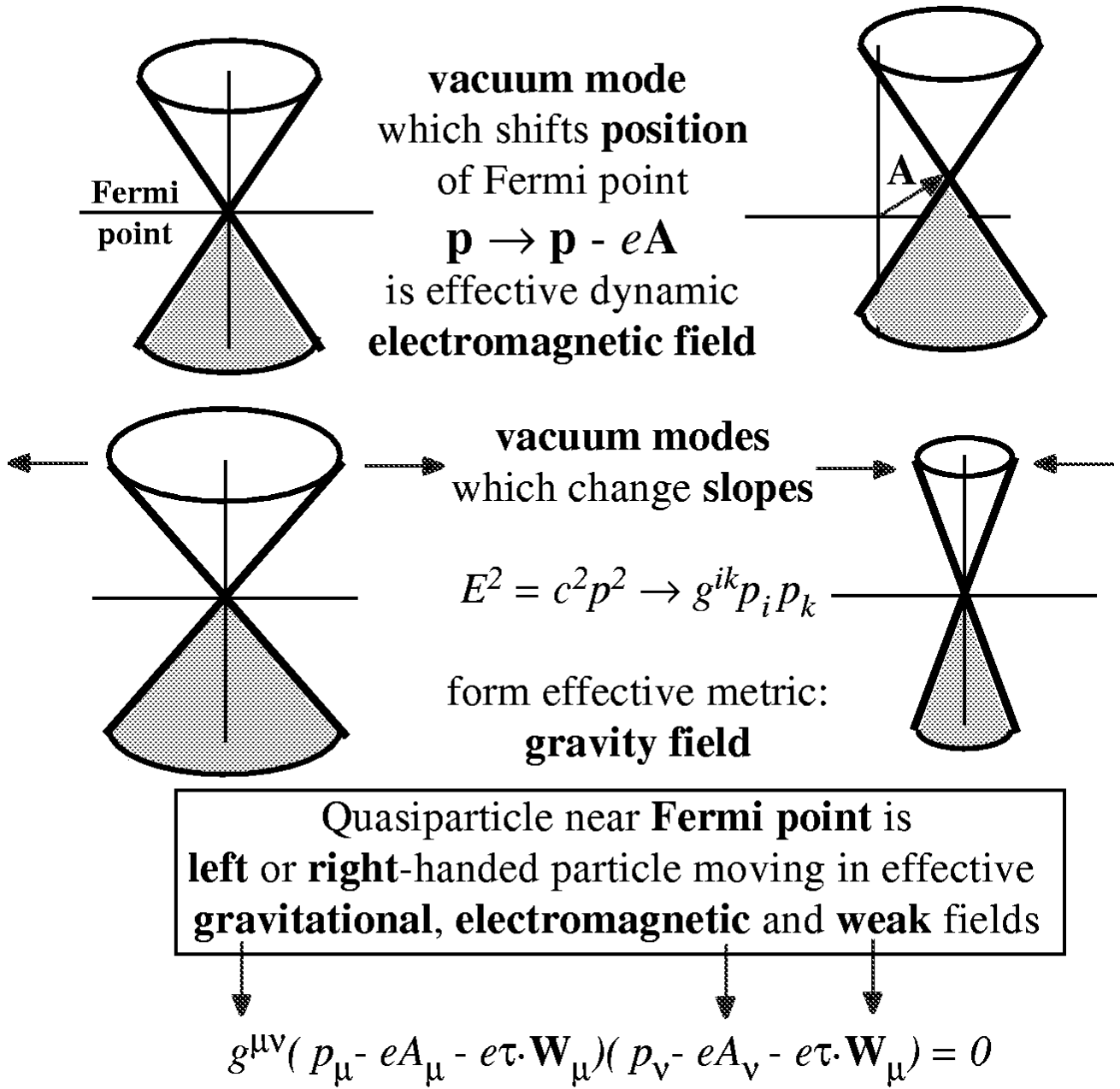}}
\medskip
\caption{Bosonic collective modes of the fermionic vacuum which belongs to the
Fermi-point universality class. The slow (low-energy) vacuum motion cannot
destroy the topologically stable Fermi point, it can only shift the point and/or
change its slopes. The shift corresponds to the change of the gauge field ${\bf
A}$, while the slopes (``speeds of light'') form  the metric tensor  $g^{ik}$. 
are oscillating Collective motion of particles comprising the vacuum is thus
seen by an individual quasiparticle as gauge and gravity fields. Thus the chiral
fermions, gauge fields and gravity  appear in low-energy corner together with 
physical laws: Lorentz, gauge, conformal invariance and general covariance.}
\label{CollectiveModesFPFig}
\end{figure}

The most remarkable property of systems with the Fermi point is that the
relativistic invariance always emerges at low energy. Close to the Fermi point 
$p_{\mu}^{(0)}$ in the 3+1 momentum-energy space one can expand the propagator in terms of
the deviations from this Fermi point, $p_\mu -p_{\mu}^{(0)}$.  If the Fermi
point is not degenerate, i.e. $N_3=\pm 1$, then close to the Fermi
point only the linear deviations survive. As a result the general form of the
inverse propagator is 
\begin{equation}
{\cal G}^{-1}=\sigma^a e^\mu_a (p_{\mu} - p_{\mu}^{(0)})
~.
\label{GeneralPropagator}
\end{equation}
Here we returned back from the imaginary frequency axis to the real energy, so that $z=E=-p_0$
instead of  $z=ip_0$; and $\sigma^a=(1,{\bf \sigma})$. The quasiparticle spectrum $E({\bf p})$
is given by the poles of the propagator, and thus by equation
\begin{equation}
g^{\mu\nu}(p_{\mu} - p_{\mu}^{(0)})(p_{\nu} - p_{\nu}^{(0)})=0~,~g^{\mu\nu}=\eta^{ab}e^\mu_a
e^\nu_b ~,
\label{GeneralEnergy}
\end{equation}
where $\eta^{ab}=diag(-1,1,1,1)$.  Thus in the vicinity of the Fermi point the
massless quasiparticles are always described by the Lorentzian metric
$g^{\mu\nu}$, even if the underlying Fermi system is not Lorentz invariant;
superfluid $^3$He-A is an example (see next Section).  On this example we shall
also see that the quantities $g^{\mu\nu}$ and  $p_{\mu}^{(0)}$ are dynamical
variables, related to the collective modes of $^3$He-A, and they play the
part of the effective gravity and gauge fields correspondingly (Fig.
\ref{CollectiveModesFPFig}).

In conclusion, from the condensed matter point of view, the classical (and
quantum) gravity is not a fundamental interaction. Matter (chiral particles and
gauge fields) and gravity (vierbein or metric field) inevitably appear together
in the low energy corner as collective fermionic and  bosonic  zero modes of the
underlying system, if the system belongs to the universality class with Fermi
points.

\subsection{Gapped systems with nontrivial topology in 2+1 dimensions}\label{Gapped2D} 

Even for the fully gapped systems there can exist the momentum-space
topological invariants, which characterize the vacuum states. Typically this
occurs in 2+1 systems,  e.g. in the 2D electron system exhibiting quantum Hall
effect \cite{Ishikawa}; in thin film of $^3$He-A (see \cite{VolovikYakovenko}
and Sec.9 of Ref.\cite{Exotic}); in the 2D superconductors with broken time
reversal symmetry
\cite{VolovikEdgeStates}; in the 2+1 world of fermions living within the
domain walls. The quantum (Lifshitz) transition between the states with
different topological invariants occurs through the intermediate gapless regime
\cite{Exotic}.

The ground states, vacua, in 2D systems or in quasi-2D thin films are characterized by 
invariants of the type 
\begin{equation} 
N_3 = {1\over{24\pi^2}}e_{\mu\nu\lambda}~
{\bf tr}\int   d^2pdp_0
~ {\cal G}\partial_{p_\mu} {\cal G}^{-1}
{\cal G}\partial_{p_\nu} {\cal G}^{-1} {\cal G}\partial_{p_\lambda}  {\cal
G}^{-1}~,
\label{3DTopInvariant}
\end{equation} 
The integral now is over the whole 3D momentum-energy space
$p_\mu=(p_0,p_x,p_y)$, or if the crystalline system is considered the
integration over $p_x,p_y$ is bounded by Brillouin zone. The integrand  is
determined everywhere in this space since the system is fully gapped and thus
the Green's function is nowhere singular. In thin films, in addition to spin
indices, the Green's function matrix ${\cal G}$ contains the indices of the
transverse levels, which come from quantization of motion along the normal to
the film \cite{Exotic}. Fermions on different transverse levels represent
different families of fermions with the same properties. This would correspond
to generations of fermions in the Standard Model, if our 3D world
is situated within the soliton wall in 4D space. 

The topological invariants of the type in Eq.(\ref{3DTopInvariant}) determine
the anomalous properties of the film. In particular, they are responsible for
quantization of physical parameters, such as Hall conductivity \cite{Ishikawa}
and spin Hall conductivity
\cite{VolovikYakovenko,Senthill,ReadGreen}. The Eq.(\ref{3DTopInvariant}) leads to
quantization of the $\theta$-factor  \cite{VolovikYakovenko}
\begin{equation} 
  \theta ={\pi\over 2}~ N_3 
 ~ ,
\label{Theta}
\end{equation}
in front of the Chern-Simons term
\begin{equation} 
 S_\theta =  { \hbar \theta\over{32\pi^2}}\int 
d^2x\hskip1mm dt\hskip1mm e^{\mu\nu\lambda}
  A_\mu   F_{\nu\lambda} ~,~F_{\nu\lambda}=\partial _\nu A_\lambda - 
\partial_\lambda A_\nu=
\hat {\bf d}\cdot\left(\partial_\nu\hat {\bf d}\times\partial_\lambda\hat {\bf d}\right) ,
\label{ChernSimonsFilm}
\end{equation}
where $\hat {\bf d}$ is unit vector characterizing the fermionic spectrum.
The $\theta$-factor determines the quantum statistics of the skyrmions. They
are fermions or bosons depending on the thickness of the film and their
statistics abruptly changes when the film grows \cite{Exotic}. For more general
2+1 condensed matter systems with different types of momentum-space invariants,
see \cite{Yakovenko2}.
   
The action in Eq.(\ref{ChernSimonsFilm}) represents the product of two
topological invariants: $N_3$ in momentum space and Hopf invariant in 2+1
coordinate space-time. This is an example of topological term
in action characterized by combined momentum-space/real-space topology. 

\section{Fermi points: $^3$He-A vs Standard Model}\label{FermiPoints}

The reason why all the attributes of the relativistic quantum field theory
arise from nothing in $^3$He-A is that both systems, the Standard Model
and $^3$He-A, have the same topology in momentum space.  The energy spectrum of
fermions in $^3$He-A also contains point zeroes, the gap nodes, which are
described by the same topological invariant in the momentum space in
Eq.(\ref{TopInvariant}) (see Fig. \ref{FermiPointTopologyFig}). For one
isolated Fermi point the nonzero topological invariant  
gives singularity in the Green's function and thus the gapless
spectrum, which for the relativistic system means the absence of the fermion
mass. 

It appears, however, that in both systems the total
topological charge of all the Fermi points in the momentum space is zero. In
$^3$He-A one has $\sum_a N_{3a}=2-2=0$. Nevertheless, the separation of zeroes
in momentum space prohibits masses for fermions.  The mass can appear when the
Fermi points merge. But even in this case the absence of the fermionic mass can
be provided by the symmetry of the system. This happens in the so called
planar phase of $^3$He \cite{VollhardtWolfle} and also in the Standard Model,
where the for each Fermi point one has $\sum_a
N_{3a}=\sum_aC_a=0$, but the discrete symmetry in the planar phase
of $^3$He-A and the electroweak symmetry in the Standard Model prohibit masses
for fermions. Fermions become massive when this symmetry is broken. This will
be discussed in more detail in Sec.\ref{StandardModelAnd} (see also ref.
\cite{MomentumSpaceTopology}).

\subsection{Superfluid $^3$He-A} 

\subsubsection{Fermi liquid level.} 

In $^3$He-A, the number of Fermi points and thus the number of Fermionic
species is essentially smaller than in the
Standard Model of strong ane electroweak interactions. In place of the various
quarks and leptons, there are only four species occurring as left and right
``weak'' doublets. One way to write these might be
\cite{VolovikVachaspati}
\begin{equation}
\chi_L = \pmatrix{\nu\cr e\cr}_L \ ,
\chi_R = \pmatrix{\nu\cr e\cr}_R \ .
\label{LeftRightDoublet}
\end{equation}
In this section we discuss how this is obtained.

The pair-correlated systems (superconductors and $ ^3$He superfluids) in
their unbroken symmetry state above $T_c$ belong to the class of Fermi systems
with Fermi surface. In terms of the field operator $\psi_{\alpha}$ for $^3$He
atoms the action is
\begin{equation}
S=\int dt~d^3x~ \left[ \psi^\dagger_{\alpha}
  \left ( i\partial_t - {{{\bf 
p}^{2}}
\over {2m}}+\mu\right)
                       \psi_{\alpha} \right ]  ~+~S_{int}
               \ ,
\label{He3Action}
\end{equation}
where $S_{int}$ includes the time-independent interaction of two
atoms (the quartic term), $m$ is the mass of $^3$He atom,
$ {\bf p}=-i \nabla$ is  the momentum operator and $\mu$ is the chemical
potential.  In general this system is described by the large number of
fermionic degrees of freedom. However, in the low-temperature limit, the number
of degrees of freedom is effectively reduced and the system is well described as
a system of noninteracting quasiparticles (dressed particles). Since the Fermi
liquid belongs to the same universality class as the weakly interacting Fermi
gas, at low energy one can map it to the Fermi gas degrees of freedom. This is
the essence of the Landau theory of Fermi liquids.   The
particle-particle interaction renormalizes the effective mass of 
quasiparticle: $m\rightarrow m^*$. The residual interaction is reduced at low
$T$ because of the small number of thermal quasiparticles above the
Fermi-surface and can be neglected. Thus the effective action for
quasiparticles becomes
\begin{equation}
S=\int dt~d^3x~\psi^\dagger_{\alpha}[i\partial_t -
M ( {\bf  p} ) ]\psi_{\alpha}~~,
\label{He3Action2}
\end{equation}
where $M ( {\bf  p} )$ is the quasiparticle energy spectrum. In a
Fermi-liquid this description is valid in the so called degeneracy limit,
when the temperature $T$ is much smaller than the effective Fermi
temperature,     $\Theta\sim \hbar^2 /(m a^2)$, which plays the part of the Planck energy in
the Fermi liquid. Here
$a$ is again the interparticle distance in the liquid.  Further for simplicity
we use the following Ansatz for the quasiparticle energy in Fermi liquid
\begin{equation}
M ( {\bf  p} )={p^2-p_F^2\over 2m^*}\approx v_F(p-p_F)
\label{FermiGasEnergy}
\end{equation}
where $m^*$ is the renormalized mass of the quasiparticle, and the
Fermi velocity is now; the last expression
is the most general form of the low-energy spectrum of excitations in the
isotropic Landau Fermi liquids in the vicinity of the Fermi surface.

\subsubsection{BCS level.} \label{BCS}

Below the superfluid/superconducting transition temperature $T_c$, new
collective degrees of freedom appear, which are the order parameter fields,
corresponding to the Higgs field in particle physics.  In
superconductors the order parameter is the vacuum expectation value of the
product of two annihilation operators (Cooper pair wave function)
\begin{equation}
\Delta \propto <{\rm vac}\vert  \psi \psi
\vert {\rm vac}>~~.
\label{GapFunction}
\end{equation}
The order parameter $\Delta$ is a $2\times 2$ matrix in conventional spin
space. It breaks the global $U(1)_N$ symmetry $\psi_{\mu}\rightarrow
e^{i\alpha}{\bf
\psi}_{\mu}$, which is responsible for the conservation of the particle number
$N$ of
$^3$He atoms, since under
$U(1)_N$ the order parameter transforms as
$\Delta(x,y)
\rightarrow e^{2i\alpha}\Delta(x,y)$. If $\Delta$ has nontrivial spin and
orbital structure, it also breaks the $SO(3)_L$ and $SO(3)_S$
symmetries under rotations in orbital and spin spaces correspondingly.

The interaction of the fermionic degrees of freedom with the order parameter 
can be obtained in BCS model using the Hubbard-Stratonovich procedure.  The
essence of this procedure, which can be easily visualized if one omits all the
coordinate dependence  and spin indices, is
the decomposition of pair interaction $S_{int}=\int  g
\psi^\dagger\psi^\dagger \psi \psi$, where $g$ is the generalized interaction
potential.  The formal way is to introduce the constant Gaussian term in the
path integral $\int d\Delta ~d\Delta^* \exp (i\Delta ~ \Delta^*/g)$ and
shift the argument $\Delta \rightarrow \Delta - \psi \psi$. In this way the
quartic term in action is cancelled  and one has the BCS action with only
quadratic forms:
\begin{eqnarray}
S=\int dt~d^3x~\psi^\dagger_{\alpha}\left[i\partial_t -M({\bf p})
\right]\psi_{\alpha}
\label{BCSAction1}
\\
+\int dt~d^3x~d^3y~[\psi^\dagger_{\alpha}(x){\bf
\psi}^\dagger_{\beta}(y) \Delta_{\alpha\beta}(x,y) +
\Delta_{\alpha\beta}^\star (x,y)\psi_{\beta}(y)
\psi_{\alpha}(x)]
\label{BCSAction2}\\
-\int {|\Delta|^2 \over g}~.
\label{BCSAction3}
\end{eqnarray}
The last term is the symbolic form of the quadratic form of the order
parameter. 

The interaction of fermions with the bosonic  order
parameter field in Eq.(\ref{BCSAction2}) allows transitions between states
differing by two atoms,  $N$ and $N\pm 2$. The order parameter 
$\Delta$ serves as the matrix element of such transition. This means that the
particle number $N$ is not conserved in the broken symmetry state and
the single-fermion elementary excitation of this broken symmetry vacuum
represents the mixture of the $N=1$ (particle) and $N=-1$ (hole) states. 

In electroweak theory the interactions corresponding to those in 
Eq.(\ref{BCSAction2}) are the Yukawa coupling which appear in the
broken symmetry state between the left-handed
$SU(2)_L$ doublets and the right-handed fermion singlets. An example of such
an interaction is the term:
\begin{equation}
G_d ({\bar u}, {\bar d})_L \Phi d_R
\label{YukawaInteraction}
\end{equation}
in the electroweak Lagrangian.
When $\Phi$ acquires a vacuum expectation value during the electroweak
phase transition, this gives rise to the nonconservation of the isospin
and hypercharge in the same manner as the charge $N$ is not conserved in
the broken symmetry action Eq.(\ref{BCSAction2}).  Such hybridization of left
and right particles leads to the lepton and quark masses. Similarly, in
superfluids and superconductors these terms give rise to the gap on the
Fermi surface.

However, the more close link to the BCS has the color superconductivity in
quark matter \cite{ColorSuperfluidity}, where the order parameter is the matrix
element between the states  differing by two quarks. Among the different phases
of the color superconductivity there is a representative of the Fermi point
universality class, the  phase where the gap hass point node like in $^3$He-A
\cite{ColorSuperconductivity1}.

For the  $s$-wave spin-singlet pairing in superconductors and
the  $p$-wave spin-triplet pairing in superfluid $^3$He the matrix element
has the following general form:
\begin{equation}
\Delta_{ s-{\rm wave}} ({\bf  r},{\bf  p})=  i \sigma ^{(2)} \Psi({\bf  r})
~~,
\label{SWave}
\end{equation}
\begin{equation}
\Delta_{ p-{\rm wave}} ({\bf  r},{\bf  p})=  i \sigma ^{(\mu )}\sigma ^{(2)}
A_{\mu \,i}({\bf  r})   p_i~~.
\label{PWave}
\end{equation}
Here ${\bf \sigma}$ are the Pauli matrices in spin space,
${\bf  r}=({\bf  x}+{\bf  y})/2$ is the center of mass coordinate of the
Cooper pair, while the momentum ${\bf  p}$
describes the relative motion of the two fermions
within the Cooper pair: it is the Fourier transform of  the coordinates
$({\bf  x}-{\bf  y})$. The complex scalar function $\Psi({\bf  r})$ and
$3\times 3$ matrix $ A_{\mu \,i}({\bf  r})$ are the order parameters for $s$-
and $p$-wave pairing respectively, which are space independent in
global equilibrium. 

Two superfluid states with $p$-wave pairing are realized in $^3$He. The state
which belongs to the Fermi point universality class is the $^3$He-A phase. In
$^3$He-A the matrix order parameter is factorized into the product of the spin
part described by the real unit vector $\hat{\bf d}$ and the  orbital part
described by the complex vector $\hat{\bf e}_1 + i\hat{\bf e}_2$
\begin{equation}
A_{\mu \,i}=\Delta_0\hat d_\mu \left(\hat e_{1i} + i \hat e_{2i}\right)~~,
~~\hat{\bf e}_1\cdot \hat{\bf e}_1=\hat{\bf e}_2\cdot \hat{\bf e}_2=1~~,~~\hat{\bf e}_1\cdot
\hat{\bf e}_2=0~.
\label{APhase}
\end{equation}
The $^3$He-B belongs to the universality class with gap, its the matrix order
parameter is  
\begin{equation}
A_{\mu \,i}=\Delta_0 R_{\mu \,i}e^{i\theta}~,
\label{BPhase}
\end{equation}
where $R_{\mu \,i}$ is the real orthogonal matrix. These two broken
symmetry vacua correspond to two different routes of the symmetry breaking:
$U(1)_N\times SO(3)_L\times SO(3)_S \rightarrow U(1)_{N-L}\times U(1)_{S}$
for $^3$He-A and 
$U(1)_N\times SO(3)_L\times SO(3)_S \rightarrow SO(3)_{S+L}$ for $^3$He-B.

The easiest way to treat the action in
Eq.(\ref{BCSAction1}-\ref{BCSAction3}), in which the states with 1 particle
and -1 particle are hybridized by the order parameter, is to double the number
of degrees of freedom introducing the antiparticle  (hole) for each particle by
introducing the  Bogoliubov-Nambu field operator
$\chi$. It is the spinor in a new  particle-hole space (Nambu
space):
\begin{equation}
{\bf  \chi}=\pmatrix{{\bf u}  \cr
                {\bf v} \cr}=\pmatrix{\Phi  \cr
                i\sigma ^{(2)}\Phi^\dagger\cr}~~;~~
{\bf  \chi}^\dagger =({\bf u}^\dagger~, ~ {\bf v}^\dagger )~~.
\label{BogoliubovNambuField}
\end{equation}
Under $U(1)_N$ symmetry  operation $\Phi_{\alpha}\rightarrow
e^{i\phi}\Phi_{\alpha}$ this spinor transforms as
\begin{equation}
{\bf  \chi}\rightarrow
e^{i\check \tau _3 \phi} {\bf  \chi}~~.
\label{GaugeTransfBogoliubovNambu}
\end{equation}
Here $\check \tau _i$ are Pauli matrices in the Nambu space,
such that $\check \tau_3$ is the particle number operator ${\bf N}$ with 
eigenvalues $+1$ for the particle component of the quasiparticle and $-1$ for
its hole component.

The eigenvalue equation for the quasiparticle spectrum is
\begin{equation}
{\cal H}  \chi =E  \chi~~ ,
\label{BogoliubovNambuEquation}
\end{equation}
where the Hamiltonian for quasiparticles in $s$-wave superconductors:
\begin{equation}
{\cal H}_{ s-{\rm wave}}= M({\bf p}) \check \tau
_3 + \pmatrix{0 &\Psi({\bf  r}) \cr
              \Psi^*( {\bf  r}) & 0 \cr }~~;
\label{BogoliubovNambuHamSWave}
\end{equation}
in the B-phase of $^3$He:
\begin{equation}
{\cal H}_{\rm B-phase}= M({\bf p})\check \tau
_3 + {{\Delta_0}\over {p_F}}({\bf \sigma}\cdot {\bf p} )\check \tau
_1~ ;
\label{BogoliubovNambuHamBPhase}
\end{equation}
and in $^3$He-A: 
\begin{equation}
{\cal H}_{\rm A-phase}= M({\bf p})\check \tau
_3 + {{\Delta _0}\over {p_F}}({\bf \sigma}\cdot\hat{\bf d}({\bf  r}))(\check \tau
_1\hat e_1({\bf  r})\cdot    {\bf p} -\check \tau
_2\hat e_2({\bf  r})\cdot    {\bf p}  ) ~ .
\label{BogoliubovNambuHamAPhase}
\end{equation}
It is also instructive to consider the so called planar phase, where:
\begin{equation}
{\cal H}_{\rm planar}= M({\bf p})\check \tau
_3 + {{\Delta _0}\over {p_F}} \check \tau
_1((\hat e_1({\bf  r})\cdot    {\bf p})\sigma_x  + (\hat e_2({\bf  r})\cdot   
{\bf p}) \sigma_y ) ~ .
\label{BogoliubovNambuHamPlanarPhase}
\end{equation}

The square of the fermionic energy in these pair-correletaed states is
correspondingly
\begin{equation}
E_{ s-{\rm wave}}^2({\bf  p})=
{\cal H}^2=M^2({\bf p})+\vert \Psi\vert^2~~ ,
\label{BogoliubovNambuEnergySWave}
\end{equation}
\begin{equation}
E_{{\rm B-phase}}^2({\bf  p})=
{\cal H}^2=M^2({\bf p})+{\Delta_0^2\over p_F^2}p^2 \approx
\left({p^2-p_F^2\over 2m^*}\right)^2+ \Delta_0^2 ~~ ,
\label{BogoliubovNambuEnergyBPhase}
\end{equation}
\begin{equation}
E_{\rm A-phase}^2({\bf  p})=E_{\rm planar}^2({\bf  p})={\cal H}^2=M^2({\bf p})
+\left({{\Delta _0}\over {p_F}}\right)^2({\bf  p} \times {\bf \hat  l})^2~~
,~~{\bf \hat  l}={\bf \hat  e}_1\times {\bf \hat  e}_2~.
\label{BogoliubovNambuEnergyAPhase}
\end{equation}
Here we took into accont that $\Delta_0\ll v_Fp_F$ in $^3$He. These equations
show that fermionic quasiparticles in the
$s$-wave superfluids/superconductors and in $^3$He-B have a gap in the spectrum
and thus belong to the unviersality class of Dirac vacuum. In the A-phase of
$^3$He the quasiparticle spectrum has two zeros at ${\bf  p}=\pm p_F\hat{\bf
l}$ (Fig. \ref{FermiPointFig}). Each of these nodes is topologically stable and
is described by the topological invariant in Eq.(\ref{TopInvariant}) which
characterizes the Fermi point. One obtains that
$N_3=-2$ for the Fermi point at ${\bf p}=+p_F\hat{\bf l}$, which means that
there are two fermionic species living near this node, and
similarly $N_3=+2$ for the Fermi point at ${\bf p}=-p_F\hat{\bf l}$, where the
other two fermionic species live. The nonzero values of the invariant, show that
the $^3$He-A belongs to the universality class of Fermi points. 

\subsubsection{``Relativistic'' level.}\label{RelativisticLevel}

As we already discussed in
Sec.\ref{RelativisticMassless}, for the system of this universality class, in
the vicinity of each Fermi point
$a$ the energy spectrum becomes ``relativistic''.   In our case of
Bogoliubov-Nambu fermions,  the expansion of Eq.(\ref{BogoliubovNambuHamAPhase})
in terms of the deviations from the Fermi points gives the following
``relativistic'' Hamiltonian
\begin{equation}
{\cal H}_a=  \sum_b(e^j_b)_a {\bf{\check \tau}}^b (p_j-e_a A_j) \ .
\label{RelatBogoliubovNambuEnergy}
\end{equation}
Here ${\bf A}=p_F\hat{\bf l}$ plays the role of a vector potential of
``electromagnetic'' field, $e_a$ is the corresponding ``electric''
charge of the $a$-th fermion. Altogether, there are four chiral fermionic
species $a$ in $^3$He-A. The two left fermions living in
the vicinity of the node
${\bf  p}=p_F\hat{\bf l}$ have ``electric'' charge $e_a=+1$; the quantum
number which distinguishes between these two fermions is  $S_3=\pm 1/2$ 
--  the projection of the conventional spin $(1/2)\sigma$ of the $^3$He atom
on the axis $\hat{\bf d}$. The chirality is determined by the
projection of Bogoliubov-Nambu spin, not by the conventional spin (see
discussion in Sec. \ref{SpinIsospin}).   For the two right fermions living in
the vicinity of the opposite node at
${\bf  p}=-p_F\hat{\bf l}$ the ``electric'' charge is $e_a=-1$. Thus one has
$e_a=-C_a$, where $C_a$ is the chirality of the fermion.  

The coefficients $(e^i_b)_a$ for each fermion $a$
form the so called $dreibein$, or triad, the local coordinate frames for the
fermionic particles are
\begin{equation}
{\bf  e}_1=2S_3c_\perp\hat {\bf  e}_1 \ , \
{\bf  e}_2=-2S_3c_\perp\hat {\bf  e}_2 \ , \
{\bf  e}_3=-C_a c_\parallel\hat{\bf l}  \ ,\  
\label{Dreibein}
\end{equation} 
where 
\begin{equation}
c_\perp={{\Delta_0}\over{p_F}}~~,~~
c_\parallel=v_F={p_F\over m^*}
\label{SpeedsOfLight}
\end{equation}
play the part of the speed of light propagating in directions perpendicular and parallel to
$\hat{\bf l}$ respectively. At zero external pressure one has  $c_\perp \sim
3~$cm/sec and $v_F\sim 55~$m/sec.

Eq.(\ref{Dreibein}) represents the 3-dimensional version of the
$vierbein$ or tetrads, which are used to describe gravity
in the tetrad formalism of general relativity. Eq.(\ref{RelatBogoliubovNambuEnergy}) is the
Weyl Hamiltonian for charged chiral particles. From Eq.(\ref{Dreibein}) it follows that $det[
e^i_b ]$ has the same  sign as the topological invariant $N_3$, which reflects the connection
between the chirality $C_a$ and topological invariant: $N_3=2C_a$. The factor 2 comes from
two spin projections $S_3$ of each fermion to the axis $\hat{\bf d}$ (see below
Sec.\ref{SpinIsospin}).

The superfluid velocity
${\bf v}_{\rm s}$ in superfliud $^3$He-A is determined by the twist of the triad
${\hat{\bf e}}_1,{\hat{\bf e}}_2,{\hat{\bf e}}_3={\hat{\bf l}}$ and corresponds to  
torsion in the tetrad formalism of  gravity  (the space dependent
rotation of vectors
${\hat{\bf e}}_1$ and ${\hat{\bf e}}_2$ about
axis ${\hat{\bf l}}$):
\begin{equation}
{\bf v}_{\rm s} ={\hbar\over 2 m} \hat e_1^i{\bf \nabla} \hat
e_2^i ~.
\label{v_s}
\end{equation}
One can check that the superfluid velocity is properly transformed under the
Galilean transformation: ${\bf v}_{\rm s}\rightarrow {\bf v}_{\rm s}+{\bf
u}$. As distinct from the curl-free superfluid velocity in superfluid
$^4$He the vorticity in $^3$He-A can be nonzero and continuous as
follows from Eq.(\ref{v_s}) \cite{Mermin-Ho}:
\begin{equation}
  \nabla\times{\bf v}_{\rm s} =
      {\hbar\over 4m~}e_{ijk} \hat l_i{\bf \nabla} \hat l_j\times{\bf
\nabla}
\hat l_k ~~.
\label{Nermin-Ho}
\end{equation}

The conventional 3+1 metric tensor expressed in terms of the
triad $e^i_b$ is
\begin{equation}
g^{ij} =\sum_{b=1}^3   e^j_b e^i_b=c_\parallel^2 \hat l^i\hat l^j
+c_\perp^2(\delta^{ij} -\hat l^i\hat l^j)  ~~,~~g^{00}=-1~~,~~g^{0i}=0~,
\label{MetricAPhase}
\end{equation}
so the energy spectrum of fermions in the vicinity of each of the
nodes is
\begin{equation}
E_a^2({\bf  p})=g^{ij}(p_i-e_a A_i)(p_j- e_a A_j)  .
\label{SquareSpectrumAPhase}
\end{equation}
The fully ``relativistic''  equation 
\begin{equation}
g^{\mu\nu}(p_\mu-e_a A_\mu)(p_\nu- e_a A_\nu) =0 ,
\label{SquareSpectrumAPhaseGeneral}
\end{equation}
includes also the nonzero nondiagonal metric $g^{0i}$ and scalar potential $A_0$, both being
induced by superfluid velocity ${\bf v}_{\rm s}$ which produces the Doppler
shift
\begin{equation}
\tilde E_a({\bf  p})=E_a ({\bf  p}) + {\bf  p}\cdot {\bf
v}_{\rm s}=E_a ({\bf  p}) + ({\bf  p}-e_a  {\bf A})\cdot {\bf v}_{\rm s}+ e_a
p_F
\hat{\bf l}\cdot {\bf v}_{\rm s}~.
\label{ScalarPotAndMixedMetricEl}
\end{equation}
This gives finally all the components of the effective metric tensor and
effective electromagnetic field in terms of the obsrevables in $^3$He-A:
\begin{eqnarray}
g^{ij}  =c_\parallel^2 \hat l^i\hat l^j +c_\perp^2(\delta^{ij}
-\hat l^i\hat l^j) - v_{\rm s}^i v_{\rm s}^j,~g^{00}=-1,~g^{0i}=-v_{\rm s}^i,
~\sqrt{-g}={1\over c_\parallel
c_\perp^2} ,\label{MetricAPhaseGeneral}
\\g_{ij}  ={1\over c_\parallel^2} \hat l^i\hat l^j +{1\over c_\perp^2}(\delta^{ij}
-\hat l^i\hat l^j),~g^{00}=-(1-g_{ij}v_{\rm s}^i v_{\rm
s}^j),~g^{0i}=-g_{ij}v_{\rm s}^j,\label{MetricAPhaseGeneralContra}
\\ds^2=-dt^2+g_{ij} (dx^i- v_{\rm s}^i)(dx^j- v_{\rm s}^j)~,
\label{MetricAPhaseGeneralInterval}
\\
A_0=p_F\hat{\bf l}\cdot {\bf v}_{\rm s}~,~{\bf  A}=p_F\hat{\bf l}~.
\label{4Potential}
\end{eqnarray}

Ironically the enhanced symmetry of the ``relativistic''
equation (\ref{SquareSpectrumAPhaseGeneral}) arises due to the spontaneous
symmetry breaking in $^3$He-A, i.e. the symmetry emerges from the symmetry
breaking. In $^3$He-A the anisotropy of space along the $\hat{\bf l}$ direction
appears in previously isotropic liquid giving rise to the Fermi points at ${\bf
p}_a=-C_ap_F\hat{\bf l}$ and to all phenomena following from the existence of
the Fermi points. The propagating oscillations of the anisotropy axis ${\bf 
A}=p_F\hat{\bf l}$ are the Goldstone bosons arisng in $^3$He-A due to
the symmetry breaking. They are viewed by quasiparticles living near the Fermi
points as the propagating electromagnetic waves. However, the spontaneous
symmetry breaking is not the necessary condition for the Fermi points to exist.
For example, the Fermi point can naturally appear in semicondictors without any
symmetry breaking \cite{Abrikosov}, and if it is there it is difficult to
destroy the Fermi point because of its topological stability.  Moreover, it is
the symmetry breaking, which is the main reason why the effective action for the
metric $g^{\mu\nu}$ and gauge field $A_\nu$ is contaminated by the
non-covariant terms.  The latter come from the gradients of Goldstone fields,
and in most cases they dominate over the natural Maxwell and Einstein terms in
the effective action. An example is the $n{\bf v}_{\rm s}^2$ term in Eq.
\ref{Energy}: it dominates over the Einstein action for $g^{\mu\nu}$, which
contains derivatives of ${\bf v}_{\rm s}$ (see Sec. \ref{EinsteinActionAnd}). 
That is why the condensed matter, where the analogy with the
relativistic fiedl theory is realized in full, would be such system where the
Fermi point exists without the symmetry breaking. Unfortunately, at the moment
we have no such condensed matter.

\subsubsection{Hierarchy of energy scales.}\label{HierarchyEnergyScales} 

The expansion in Eq.(\ref{RelatBogoliubovNambuEnergy}) and thus the
relativistic description of quasiparticles dynamics occurs at low enough
energy and is violated at higher (``trans-Planckian'') energies where the
Lorentz and other symmetries disappear. There is a hierarchy of energy scales
of the ``trans-Planckian'' physics in
$^3$He-A:
\begin{equation}
{\Delta_0^2\over v_Fp_F}\ll \Delta_0 \ll v_Fp_F\ll E_{\rm el}~,
\label{EnergyScales} 
\end{equation}
where $E_{\rm el}$ is the energy of excitations of the electronic states in
the $^3$He atom. These energy scales are correspondingly $\sim~10^{-10}$ eV,
$10^{-7}$ eV, $10^{-4}$ eV and 1 eV. 

The first ``Planckian'' energy scale $\Delta_0^2/v_Fp_F=m^*c_\perp^2$ serves as
the Lorentzian cut-off: At energies $E\ll \Delta_0^2/v_Fp_F$  the quasiparticle
spectrum determined by Eq.(\ref{BogoliubovNambuHamAPhase}) can be  expanded in
the vicinity of the Fermi point to give the Lorentz invariant form of the Weyl
Hamiltonian in Eq.(\ref{RelatBogoliubovNambuEnergy}). In this region
$g^{ik}(p_i -p_{0i})(p_k -p_{0k})\ll (m^*c_\perp^2)^2$, and the nonlinear
corrections to the energy spectrum are small and in many cases can be neglected.
These corrections become important when the physics of the black hole
horizon and other exotic space-times are discussed (see Secs.
\ref{VierbeinDomainWall} and \ref{HorizonsErgoregions}). 

The region between two ``Planckian'' energy scales,
$\Delta_0^2/v_Fp_F \ll E\ll \Delta_0$, is the so called quasiclassical region,
in this nonrelativistic region we can use the quasiclassical approximation to
calculate, say, the quasiparticle bound states in the core of the vortex (Sec.
\ref{FermionZeroModesOnVort}). 

In the region between the second and third ``Planck'' scale, $\Delta_0  \ll E\ll
v_Fp_F$, the effect of pairing can be neglected and one effectively has the
degenerate Fermi-system described by the Fermi surface universality class.
The BCS theory in the Bogoliubov-Nambu
representation is appropriate for the whole range up to third ``Planck'' scale,
$E\ll  p_Fv_F$, and thus represents the self-consistent quantum field theory
describing simultaneously  the ``relativistic'' low-energy physics of the
Fermi point scale and two high-energy levels of the  ``transPlanckian''
physics. This is a great advantage of the BCS theory of superfluid $^3$He-A. 

Finally, at $ v_Fp_F \ll E\ll E_{\rm el}$  the system can be described as that
of classical weakly interacting atoms.

\subsubsection{Spin vs isospin.}\label{SpinIsospin} 

The nonunit values of  $N_3=2C_a$ show that each Fermi point
is doubly degenerate due to the spin degree of freedom of the $^3$He atom. For each
projection $S_3$ of spin one has $N_3=+ 1$ at ${\bf p}=-p_F\hat{\bf l}$ and
$N_3=- 1$ at ${\bf p}=+p_F\hat{\bf l}$.  Thus we have two doublets, one left at ${\bf
p}=+p_F\hat{\bf l}$ and one right at ${\bf p}=-p_F\hat{\bf l}$, 
\begin{equation}
\chi_L = \pmatrix{\chi_L^{S_3=+1/2}\cr
\chi_L^{S_3=-1/2}\cr}  \equiv \pmatrix{\nu\cr e\cr}_L  \ ,
~~\chi_R = \pmatrix{\nu\cr e\cr}_R \equiv \pmatrix{\chi_R^{S_3=+1/2}\cr
\chi_R^{S_3=-1/2}\cr}\equiv \pmatrix{\nu\cr e\cr}_R  \ .
\label{LeftRightDoublet2}
\end{equation}
From Eq.(\ref{LeftRightDoublet2}) it follows that spin ${\bf \sigma}$ of quasiparticle  plays
the part of the isospins in the extended Standard Model with $SU(2)_{L}\times
SU(2)_{R}$ symmetry.  The conventional spin of the $^3$He atom is thus responsible for the
$SU(2)$ degeneracy, but not for chirality. On the other hand the Bogoliubov spin
$\check{\bf\tau}$ is responsible for chirality in
$^3$He-A and thus plays the same role as the conventional spin $\bf\sigma$ of chiral fermions
in Standard Model in Eq.(\ref{Neutrino}).

Such interchange of spin and isospin shows that that the only origin of the chirality of the
(quasi)particle is the nonzero value of the topological invariant $N_3$. What kind of spin is
related to the chirality depends on the details of the matrix structure of the Fermi point. In
this sense there is no principle difference between  spin and isospin: changing the matrix
structure one can gradually convert isospin to spin, while the topological
charge $N_3$ of the Fermi point remains invariant under such a rotation.

\subsection{Standard Model and its Momentum Space Topology}
\label{StandardModelAnd}

\subsubsection{Fermions in Standard Model}\label{FermionsStandardModel}

In Standard Model of electroweak and strong interactions (if the righthanded neutrino is
present, as follows from the Kamiokande experiments) each family of quarks and leptons
contains 8 lefthanded and 8 righthanded fermions transforming under the gauge group
$G(213)= SU(2)_{L}\times U(1)_Y\times SU(3)_C$ of weak, hypercharge and  strong
interactions correspondingly. The 16 fermions with their diverse hypercharges and electric
charges can be organized in a  simple way, if one assumes the higher symmetry
group, such as $SO(10)$ at a grand unification scale. We use here a type of
Pati-Salam model \cite{PatiSalam,Foot} with the symmetry group
$G(224)=SU(2)_{L}\times SU(2)_{R} \times SU(4)_C$. This group  $G(224)$ is the
minimal subgroup of $SO(10)$ group which preserves all its important
properties \cite{PatiNew}. The advantage of the group $G(224)$ as compared
with the $SO(10)$ group, which is important for the correct  definition of the
topological charge, is that $G(224)$ organizes 16 fermions not into one
multiplet of 16 left fermions as the $SO(10)$ group does, but into the left
and right baryon-lepton octets with the left-right symmetry on the fundamental
level:
\begin{equation}
\matrix{ ~&SU(2)_L & SU(2)_R \cr 
SU(4)_C&\left(\matrix{
u_L&d_L\cr
u_L&d_L\cr
u_L&d_L\cr
\nu_L&e_L\cr
}\right)
&
\left(\matrix{
u_R&d_R\cr
u_R&d_R\cr
u_R&d_R\cr
\nu_R&e_R\cr
}\right)\cr
}
\label{SU4}
\end{equation}
Here the $SU(3)_C$ colour group is extended to $SU(4)_C$  colour group by introducing as a
charge the  difference between baryonic and leptonic numbers $B-L$; and the $SU(2)_{R}$
group for the right particles is added.  When the energy is reduced the $G(224)$ group
transforms to the intermediate subgroup $G(213)$ of the electroweak and strong
interactions with the hypercharge given by $Y={1\over 2} (B-L) +W_{R3}$. At   energy below
about 200 GeV the electroweak symmetry is violated and we have the group $SU(3)_C
\times U(1)_Q$ of strong and electromagnetic interactions with the electric charge
$Q=Y+W_{L3}={1\over 2} (B-L)+W_{R3}+W_{L3}$. These charges are:
\begin{equation}
\matrix{
Fermion&W_{L3}&W_{R3}&B-L&\rightarrow&Y&\rightarrow&Q\cr
u_L(3)&+{1\over 2}&0 &{1\over 3}&~&{1\over 6}&~&{2\over 3}\cr
u_R(3)&0 &+{1\over 2}  &{1\over 3}&~&{2\over 3}&~&{2\over 3}\cr
d_L(3) & -{1\over 2}&0 &{1\over 3}&~&{1\over 6}&~&-{1\over 3}\cr
d_R(3) &0 &-{1\over 2}  &{1\over 3}&~&-{1\over 3}&~&-{1\over 3}\cr
\nu_L &+{1\over 2} &0  &-1&~&-{1\over 2}&~&0\cr
\nu_R &0 &+{1\over 2}  &-1&~&0&~&0\cr
e_L &-{1\over 2} &0  &-1&~&-{1\over 2}&~&-1\cr
e_R &0 &-{1\over 2}  &-1&~&-1&~&-1\cr
} 
\label{SU42}
\end{equation}

In the above $G(224)$ model, 16 fermions of one generation can be represented as
the product
$Cw$ of 4 bosons and 4 fermions \cite{Terazawa}. This scheme is similar to the
slave-boson approach in condensed matter, where the particle is considered as a
product of the spinon and holon. Spinons are fermions which carry spin, while
holons are ``slave''-bosons which carry electric charge \cite{Marchetti}.  In the
Terazawa scheme \cite{Terazawa} the ``holons''
$C$ form the 
$SU(4)_C$ quartet of spin-0
$SU(2)$-singlet particles which carry baryonic and leptonic charges, their
$B-L$ charges  of the
$SU(4)_C$ group are $( {1\over 3},  {1\over 3}, {1\over 3}, -1)$.  The ``spinons''
are spin-${1\over 2}$ particles
$w$, which are
$SU(4)_C$ singlets and $SU(2)$-isodoublets; they carry spin and isospin.
\begin{equation}
\left(\matrix{
u_L&d_L&u_R&d_R\cr
u_L&d_L&u_R&d_R\cr
u_L&d_L&u_R&d_R\cr
\nu_L&e_L&\nu_R&e_R\cr
}\right)=\left(\matrix{
C_{1/3}\cr
C_{1/3}\cr
C_{1/3}\cr
C_{-1}\cr
}\right)\times
\left(\matrix{w_L^{+1/2}&
w_L^{-1/2}&w_R^{+1/2}&w_R^{-1/2}\cr}\right)
\label{Terazawa}
\end{equation}

Here $\pm 1/2$  means the charge $W_{L3}$ for the left spinons and  $W_{R3}$
for the right spinons, which coincides with the  electric charge of
spinons: $Q={1\over 2}(B-L) +W_{L3}+W_{R3}=W_{L3}+W_{R3}$. In Terazawa notations
$w_1=(w_L^{+1/2},w_R^{+1/2})$ forms the doublet of spinons with $Q=+1/2$ and 
$w_2=(w_L^{-1/2},w_R^{-1/2})$ --  with $Q=-1/2$.
These 4 spinons, 2 left and 2 right, 
transform under
$SU(2)_{L}
\times SU(2)_{R}$ symmetry group.

\subsubsection{Momentum-space topological invariants}\label{Momentum-space}

In the case of one chiral fermion the massless (gapless) character of its
energy spectrum in Eq.(\ref{GeneralEnergy}) is protected by the
momentum-space topological invariant. However, in case of the equal number of left and right
fermions the total topological charge $N_3$ in Eq.(\ref{TopInvariant}) is zero for the Fermi
point at ${\bf p}=0$, if the trace is over all the fermionic species.  Thus the mass
protection mechanism does not work and in principle an arbitrary small interaction between
the fermions can provide the Dirac masses for all 8 pairs of fermions. The mass, however,
does not appear for some or all fermions, if the interaction has some symmetry
elements. This situation occurs in the  planar phase of superfluid
$^3$He  in Eq.(\ref{BogoliubovNambuHamPlanarPhase}) and in
the Standard Model.   In both cases the weighted momentum-space topological
invariants can be determined. They do not change under perturbations, which
conserve given symmetry, and thay are the functions of parameters of this
symmetry group. In the Standard Model the relevant symmetries are the
electroweak symmetries  $U(1)_Y$ and $SU(2)_L$  generated by the hypercharge
and by the weak charge.
 
\subsubsection{Generating function for topological invariants constrained by symmetry}

Let us introduce the matrix ${\cal N }$ whose trace gives the invariant $N_3$ in
Eq.(\ref{TopInvariant}):
\begin{equation} 
{\cal N} = {1\over{24\pi^2}}e_{\mu\nu\lambda\gamma}~
 \int_{\sigma}~  dS^{\gamma}
~ {\cal G}\partial_{p_\mu} {\cal G}^{-1}
{\cal G}\partial_{p_\nu} {\cal G}^{-1} {\cal G}\partial_{p_\lambda}  {\cal
G}^{-1}~,
\label{TopInvariantMatrix}
\end{equation}
where as before the integral is about the Fermi point in the 4D
momentum-energy space. Let us consider the expression
\begin{equation} 
({\cal N},{\cal Y}) = {\bf tr} \left[{\cal N Y}\right]~,
\label{NY}
\end{equation}
where ${\cal Y}$ is the generator of the $U(1)_Y$ group, the hypercharge matrix. It is clear
that the Eq.(\ref{NY}) is robust to any perturbation  of the Green's function, which does not
violate the $U(1)_Y$ symmetry, since in this case the
hypercharge matrix ${\cal Y}$ commutes with the Green's function $ {\cal G}$. The same occurs
with any power of ${\cal Y}$, i.e. $({\cal N},{\cal Y}^n)$ is also invariant under symmetric
deformations. That is why  one can introduce the generating function for all the topological
invariants containing powers of the hypercharge
\begin{equation} 
 \left( e^{i\theta_Y {\cal Y} } ,{\cal N} \right) ={\bf tr}  \left[ e^{i\theta_Y {\cal Y} }
{\cal N} \right]~.
\label{GeneratingFunction}
\end{equation}
All the powers $({\cal N},{\cal Y}^n)$, which are topo0logical invariants, can
be obtained by differentiating of Eq.(\ref{GeneratingFunction}) over the group
parameter
$\theta_Y$. Since the above parameter-dependent invariant is robust to
interactions between the fermions, it can be calculated for the noninteracting
particles.  In the latter case the matrix ${\cal N}$ is diagonal with the
eigenvalues $C_a=+1$ and $C_a=-1$ for right and left fermions correspondingly. 
The trace of this matrix
${\cal N}$ over given irreducible fermionic representation of the gauge group
is (with minus sign) the symbol $N_{(y/2,\underline a,I_W)}$ introduced by
Froggatt and Nielsen in Ref.\cite{FroggattNielsen}.  In their notations
$y/2(=Y)$ , $\underline a$ , and $I_W$ denote hypercharge, colour
representation and the weak isospin correspondingly. 

For the Standard Model with hypercharges for 16 fermions given in Eq.(\ref{SU42}) one has
the generating function: 
\begin{equation} 
 \left( e^{i\theta_Y {\cal Y} } ,{\cal N} \right) =\sum_a C_a e^{i\theta_Y
Y_a}=2\left(\cos {\theta_Y\over 2} -1\right) \left( 3e^{i \theta_{Y}/6} +  
e^{-i
\theta_{Y}/2}\right) ~.
\label{GeneratingFunctionY}
\end{equation}
The factorized form of the generating function reflects the factorization in
Eq.(\ref{Terazawa}) and directly follows from this equation: The generating function for the
momentum space topological invariants for ``holons'' is $\left(  e^{i\theta_{BL} ({\cal
B-L})},{\cal N}_{\rm holon}
\right)=\left( 3e^{i \theta_{BL}/3} +   e^{-i
\theta_{BL}}\right)$, which must be multiplied by the ``spinon'' factor 
$2(\cos {\theta_R\over
2}- \cos {\theta_L\over 2})$. 

In addition to the hypercharge the weak charge is also conserved in the
Standard model above the electroweak transition. The generating function for the
topological invariants which contain the powers of  
both the hypercharge $Y$ and the weak charge $W_3$ also  has the factorized form:
\begin{equation} 
  \left( e^{i\theta_W {\cal W}_3^L } e^{i\theta_Y
{\cal Y}  },{\cal N}
\right)~  = 2\left(\cos {\theta_Y\over 2} -\cos {\theta_W\over 2}\right) \left( 3e^{i
\theta_{Y}/6} +   e^{-i
\theta_{Y}/2}\right) ~.     
\label{GeneratingFunctionWY}
\end{equation}
The generators of the
$SU(3)_C$ colour group, which is left-right symmetric, do not change the  form of the
generating function in  Eq.(\ref{GeneratingFunctionWY}).

The nonzero values of Eq.(\ref{GeneratingFunctionWY}) show that the Green's function
is singular at ${\bf p}=0$ and $p_0=0$, which means that some fermions must be
massless.

\subsubsection{Discrete symmetry and massless fermions}
\label{DiscreteSymmetryEffect}

Choosing the parameters $\theta_Y=0$ and $\theta_W=2\pi$ one obtains
the maximally possible value of the generating function:
\begin{equation} 
  \left( e^{2\pi i {\cal W}_3^L },{\cal N} \right)~  = 16 ~.     
\label{16}
\end{equation}
which means that all 16 fermions of one generation are massless above the
electroweak scale 200 GeV. This also shows that in many cases only the discrete
symmetry group, such as the $Z_2$ group $ e^{2\pi i {\cal W}_3^L}$, is enough
for the mass protection. 

In the planar state of $^3$He in Eq.(\ref{BogoliubovNambuHamPlanarPhase}) each
of the two Fermi points at ${\bf p}=\pm p_F\hat{\bf l}$ has zero topological
charge, $N_3=0$. Nevertheless the  gapless fermions in the planar state are
supported by the topological invariant $\left( P,{\cal N}\right)$ containing
the discrete $Z_2$ symmetry ($PP=1$) of the planar state vacuum. This symmetry
is the combination of discrete gauge transformation and spin rotation by $\pi$:
$e^{pi}C_{2}^z\Delta=\Delta$ (on discrete symmetries of superfluid phases of
$^3$He see Ref. \cite{VollhardtWolfle}). Applying to the Bogoliubov-Nambu
Hamiltonian (\ref{BogoliubovNambuHamPlanarPhase}), for which the generator of
the $U(1)$ gauge rotation is $\tau_3$, this symmetry operation has the form
$P=e^{\pi i\tau_3/2}e^{\pi i\sigma_z/2}=-\tau_3\sigma_z$: $P{\cal H}={\cal H}P$.
The nonzero topological invariants, which support the mass (gap) protection for
the Fermi points  at ${\bf p}=\pm p_F\hat{\bf l}$, are correspondingly
$\left(P,{\cal N}\right)=\pm 2$.
Thus at each Fermi point there are gapless fermions. They acquire the
relativistic energy spectrum in the low energy corner.  

In this relativistic limit the discrete symmetry $P$, which is responsible for
the mass protection, is equivalent to the $\gamma^5$-symmetry for Dirac
fermions. If the $\gamma^5$ symmetry is obeyed, i.e. it commutes with the Dirac
Hamiltonian, $\gamma^5{\cal H}\gamma^5={\cal H}$, then the Dirac fermion has
no mass. This is consistent with the nonzero value of the topological invariant:
it is easy to check that for the massless Dirac fermion one has
$\left(\gamma^5,{\cal N}\right)= 2$. This connection between  topology
and mass protection looks trivial in the relativistic case, where the absence
of mass due to $\gamma^5$ symmetry can be directly obtained from the Dirac
Hamiltonian. However the equations in terms of the topological charge, such as  
Eq.(\ref{16}), appears to be more general, since they remain valid even if
the Lorentz symmetry is violated at higher energy and the  Dirac
equation is not applicable any more. In the non-relativistic case even the
chirality is not a good quantum number at high energy (this in particular means
that transitions between the fermions with different chirality are possible at
high energy, see Sec. \ref{AndreevReflectionInterface} for an example). The
topological constraints, such as in Eq.(\ref{16}), protect nevertheless the
gapless fermionic spectrum in non-Lorentz-invariant Fermi systems.

\subsubsection{Nullification of topological invariants
below electroweak transition and massive fermions}

When the electroweak symmetry $U(1)_Y\times SU(2)_L$  is
violated to $U(1)_Q$, the only remaining charge -- the electric charge $Q=Y+W_3$ -- produces
zero value for the whole generating function according to Eq.(\ref{GeneratingFunctionWY}):
\begin{equation} 
 \left( e^{i\theta_Q {\cal Q} }, {\cal N} \right) =  \left(
e^{i\theta_Q {\cal Y} }  e^{i\theta_Q {\cal W}_3^L },{\cal N} \right) =
0~.
\label{GeneratingFunctionQ}
\end{equation}
The zero value of the topological invariants implies that even if the singularity in the
Green's function exists it can be washed out by interaction. Thus each elementary fermion in
our world must have a mass after such a symmetry breaking.

What is the reason for such a symmetry breaking pattern, and, in particular, for such
choice of electric charge $Q$? Why the nature had not chosen the more natural symmetry
breaking, such as 
$U(1)_Y\times SU(2)_L
\rightarrow U(1)_Y$, $U(1)_Y\times SU(2)_L
\rightarrow SU(2)_L$ or $U(1)_Y\times SU(2)_L
\rightarrow U(1)_Y\times U(1)_{W_3}$? The possible reason is provided by
Eq.(\ref{GeneratingFunctionWY}), according to which the nullification of all the
momentum-space topological invariants occurs only if the  symmetry breaking scheme
$U(1)_Y\times SU(2)_L
\rightarrow U(1)_Q$ takes place with the charge $Q=\pm Y\pm W_3$.
Only in such cases the topological mechanism for the mass protection
disappears. This can shed light on the origin of the electroweak transition. It is possible
that the elimination of the mass protection is the only goal of the transition.  This is
similar to the Peierls transition in condensed matter: the formation of mass (gap) is not the
consequence but the cause of the transition. It is energetically favourable to have
masses of  quasiparticles, since this leads to decrease of the energy of the fermionic
vacuum. Formation of the condensate of top quarks, which generates the
heavy mass of the top quark, could be a relevant scenario for that (see
review \cite{TaitPhD}).

In the $G(224)$ model the  electric charge $Q={1\over
2}(B-L) +W_{L3}+W_{R3}$  is left-right symmetric. That is why, if only the electric charge is
conserved in the final broken symmetry state, the only relevant topological
invariant $\left( e^{i\theta_Q {\cal Q} }, {\cal N} \right)$ is always
zero, there is no mass protection and the Weyl fermions must be paired into Dirac fermions.
This fact does not depend on the definition of the hypercharge, which appears at the
intermediate stage where the symmetry is
$G(213)$. It also does not depend much on the definition of the
electric charge $Q$ itself: the only condition for the nullification of the
topological invariant is the symmetry (or antisymmetry) of $Q$ with respect to
the parity transformation.

\subsubsection{Relation to Axial Anomaly}

The momentum-space topological invariants determine the
axial anomaly in fermionic systems. In particular the
charges related to the gauge fields cannot be created from vacuum, this requires the
nullification of some inavriants (see  Sec.~\ref{ChiralAnomaly}):
\begin{equation} 
 \left(   {\cal Y} , {\cal N} \right) = \left(   {\cal Y}^3, 
{\cal N} \right)  =   \left(   ({\cal
W}_3^L)^2 {\cal Y}, {\cal N}
\right) =     \left(   {\cal Y}^2 {\cal
W}_3^L, {\cal N}
\right) =...=0~.
\label{AnomalyCancellation}
\end{equation}
Nullification of all these invariants is provided by the general
Eq.(\ref{GeneratingFunctionWY}), though in this equation it is not assumed that the groups
$U(1)_Y$ and $SU(2)_L$ are local.

\section{Effective relativistic quantum field theory emerging in a system with Fermi point.}

The correspondence between field theory and $^3$He-A is
achieved by replacing the gauge field and the metric
by appropriate $^3$He-A observables \cite{Exotic}.  To establish the
correspondence, the free energy or Lagrangian should be expressed in a covariant and
gauge invariant form and should not contain any
material parameters, such as ``speed of light''.  Then it can be equally
applied to both systems, Standard Model and $^3$He-A. Note that the effective quantum field
theory, if it does not contain the high energy cut-off, should not contain the speed of
light $c$ explicitly: it is hidden in the metric tensor.

\subsection{Collective modes of fermionic vacuum -- electromagnetic and gravitational fields.}

Let us consider the collective modes in the system with Fermi points. The effective fields
acting on a given particle due to interactions with other moving particles cannot destroy the
Fermi point. That is why, under the inhomogeneous perturbation of the fermionic vacuum the
general form of Eqs.(\ref{GeneralPropagator}-\ref{GeneralEnergy}) is preserved. However the
perturbations lead to a local shift in the position of the Fermi point
$p_{\mu}^{(0)}$ in momentum space and to a local change of the vierbein $e^\mu_b $ (which in
particular includes slopes of the energy spectrum.  This means that the low-frequency
collective modes in such Fermi liquids are the propagating collective oscillations of the
positions of the Fermi point and of the slopes at the Fermi point (Fig.
\ref{CollectiveModesFPFig}). The former
is felt by the right- or the left-handed quasiparticles as the dynamical gauge
(electromagnetic) field, because the main effect of the electromagnetic field
$A_\mu=(A_0,{\bf A})$ is just the dynamical change in the position of zero in the energy
spectrum: in the simplest case
$(E-eA_0)^2=c^2 ({\bf p}-e{\bf A})^2$. 

The collective modes related to a local change of the vierbein $e^\mu_b$ correspond to the
dynamical gravitational field. The quasiparticles feel the inverse tensor
$g_{\mu\nu}$ as the metric of the effective space in which they move along the geodesic curves
\begin{equation}
ds^2=g_{\mu\nu}dx^\mu dx^\nu
\label{InverseMetric}
\end{equation}
Therefore, the collective modes related to the slopes
play the part of the gravity field. 

Thus near the Fermi point the quasiparticle is the chiral massless fermion moving in the
effective dynamical electromagnetic and gravitational fields.

\subsection{Physical laws in vicinity of Fermi point: Lorentz invariance, gauge invariance,
general covariance, conformal invariance.}

In the low energy corner the fermionic propagator in Eq.(\ref{GeneralPropagator}) is gauge
invariant and even obeys the general covariance near the Fermi point.  For example, the local
phase transformation of the wave function of the fermion, $\chi \rightarrow
\chi e^{ie\alpha({\bf r},t)}$  can be compensated by the shift of the
``electromagnetic'' field
$A_\mu \rightarrow A_\mu + \partial_\mu \alpha$. These attributes of the electromagnetic
($A_\mu$) and gravitational ($g^{\mu\nu}$) fields arise spontaneously  as the low-energy
phenomena. 

Now let us discuss the dynamics of the bosonic sector -- collective modes of $A_\mu$ and
$g^{\mu\nu}$. Since these are the effective fields their motion equations do not
necessarily obey gauge inariance and general covariance. However, in some special
cases such symmetries can arise in the low energy corner.  The particular model with the
massless chiral fermions has been considered by Chadha and Nielsen \cite{Chadha}, who found 
that the Lorentz invariance becomes an infrared fixed point of the renormalization group
equations. What are the general conditions for such symmetry of the bosonic fields in the
low energy corner?

The effective Lagrangian for the collective modes is obtained by integrating over the
vacuum fluctuations of the fermionic field.  This principle was used by Sakharov and
Zeldovich to obtain an effective gravity
\cite{Sakharov} and effective electrodynamics
\cite{Zeldovich}, both arising from fluctuations of the fermionic vacuum. If the main
contribution to the effective action comes from the vacuum fermions whose momenta
${\bf p}$ are concentrated near the Fermi point, {\it i.e.} where the fermionic
spectrum is linear and thus obeys the ``Lorentz invariance'' and gauge inariance of
Eq.(\ref{GeneralPropagator}), the result of the integration is necessarily invariant under
gauge transformation, $A_\mu \rightarrow A_\mu + \partial_\mu \alpha$, and has a covariant
form. The obtained effective Lagrangian then gives the Maxwell equations for
$A_\mu$ and the Einstein equations for $g_{\mu\nu}$, so that the propagating bosonic
collective modes do represent the gauge bosons and  gravitons. 

Thus two requirements must be fulfilled -- (i) the fermionic system has a Fermi point and (ii)
the main physics is concentrated near this Fermi point. In this case the system
acquires at low energy all the properties of the modern quantum field theory: chiral
fermions, quantum gauge fields, and gravity. All these ingredients are actually  
low-energy (infra-red) phenomena. 

In this extreme case when the vacuum fermions are dominatingly relativistic, the bosonic
fields acquire also  another symmetry obeyed by massless relativistic Weyl fermions,  the
conformal invariance -- the invariance under transformation $g_{\mu\nu}\rightarrow
\Omega^2({\bf r},t)g_{\mu\nu}$. The gravity with the conformly invariant effective action,
the so-called Weyl gravity, is still a viable rival to Einstein gravity  in modern cosmology
\cite{Mannheim,Edery}:  The Weyl gravity (i) can explain the galactic rotation curves without
dark matter; (ii) it reproduces the Schwarzschild solution at small distances; (iii) it can
solve the cosmological constant problem, since the cosmological constant is forbidden if the
conformal invariance is strongly obeyed; etc. (see \cite{Mannheim2}).

\subsection{Effective electrodynamics.}

\subsubsection{Effective action for ``electromagnetic'' field}

Let us consider what happens in a practical realization of systems with Fermi points in condensed
matter -- in $^3$He-A. From Eqs.(\ref{MetricAPhaseGeneral}) and (\ref{4Potential}) it follows
that the fields, which act on the ``relativistic'' quasiparticles as electromagnetic and
gravitational fields, have a nontrivial behavior. For example, the same  texture of the
${\hat{\bf l}}$-vector is felt by  quasiparticles as the effective magnetic field
${\bf B}=p_F
\vec\nabla\times{\hat{\bf l}}$ according to Eq.(\ref{4Potential})  and simultaneously it
enters the metric according to Eq.(\ref{MetricAPhaseGeneral}). Such field certainly cannot
be described by the Maxwell and Einstein
equations together.
Actually the gravitational and electromagnetic variables coincide in $^3$He-A only
when we consider the vacuum manifold: Outside of this manifold they split.  
$^3$He-A, as any other fermionic system with Fermi point, has enough number
of collective modes to provide the analogs for the independent gravitational and
electromagnetic fields. But some of these modes are massive in $^3$He-A. For
example the gravitational waves correspond to the modes, which are different from
the oscillations of the  ${\hat{\bf l}}$-vector. As distinct from the photons
(orbital waves -- propagating oscillations of the ${\hat{\bf l}}$-vector)  the
gravitons are massive (Sec.\ref{VacuumPressureSection}). 

All these troubles occur because in $^3$He-A the main contribution to the
effective action for the most of the bosonic fields come from the
integration over vacuum fermions at the ``Planck'' energy scale,   $E \sim
\Delta_0$. These fermions are far from the Fermi points and their spectrum is
nonlinear. That is why in general the effective action for the bosonic fields is
not symmetric.

\subsubsection{Running coupling constant: zero charge effect.}\label{RunningCouplingConstant}

There are, however, exclusions.  For example, the action
for the  ${\hat{\bf l}}$-field, which contains the term with the
logarithmically divergent factor $\ln (\Delta/\omega)$ (see ref. \cite{Exotic}
and Sec. \ref{Improved3He}). It comes from the zero charge effect, the
logarithmic screenig of the ``electric charge'' by the massless fermions, 
for whom the
${\hat{\bf l}}$-field acts as electromagnetic field. Due to its logarithmic divergence this
term is dominanting at low frequency $\omega$: the lower the frequency the larger is the
contribution of the vacuum fermions from the vicinity of the Fermi point and thus the more
symmetric is the Lagrangian for the ${\hat{\bf l}}$-field. 

This happens, for example, in the physically important case discussed in
Sec.\ref{MagnetogenesisChiralFermions}, where the  Lagrangian  for the 
${\hat{\bf l}}$-texture is completely equivalent  to the conventional Maxwell Lagrangian for
the (hyper-) magnetic and electric fields. In this particular case the equilibrium state is
characterized by the homogeneous  direction of the ${\hat{\bf l}}$ vector, which is fixed by
the counterflow: ${\hat{\bf l}}_0
\parallel  {\bf v}_{\rm n}-{\bf
v}_{\rm s}$. The effective electromagnetic field is simulated by the small deviations of the
${\hat{\bf l}}$ vector from its equilibrium direction, 
${\bf A}=p_F\delta{\hat{\bf l}}$.  Since ${\hat{\bf l}}$ is a unit vector, its variation
$\delta{\hat{\bf l}}\perp {\hat{\bf l}}_0$. This corresponds to the gauge choice
${\bf A}^3=0$, if $z$ axis is chosen along the background orientation, ${\hat{\bf
z}}={\hat{\bf l}}_0$. In the considered case only the dependence on $z$ and $t$ is relevant.
As a result in the low-energy limit the effective Lagrangian for the
$A_\mu$ becomes gauge invariant, so that in this regime the
$A_\mu$ field does obey the Maxwell equations coming from the Lagrangian:  
\begin{eqnarray}
L={{\sqrt{-g}}\over 4\gamma^2}
g^{\mu\nu}g^{\alpha\beta}F_{\mu\alpha}F_{\nu\beta}=\label{EMLagrangian}\\ 
{\rm ln}
\left ( {\Delta_0^2\over T^2 } \right ) {{p_F^2v_F}\over {24\pi^2\hbar}}~
\left((\partial_z \delta \hat {\bf l})^2 -{1\over v_F^2}\left(\partial_t \delta \hat {\bf
l} +({\bf v}_{\rm s}\cdot\nabla) \delta \hat {\bf
l}\right)^2\right)~~.
\label{EMLagrangianAPhase}
\end{eqnarray}
Here $g^{\mu\nu}$ is the effective metric in Eq.(\ref{MetricAPhaseGeneral}) with the
background direction ${\hat{\bf l}}_0$;  $\gamma^2$ is a running
coupling constant, which is logarithmically divergent because of vacuum
polarization:
\begin{equation} 
\gamma^{-2} ={1\over 12\pi^2} {\rm ln} \left ({\Delta_0^2\over T^2 } \right
) ~.
\label{RunningCoupling}
\end{equation}  
This is in a complete analogy with the logarithmic  divergence of the fine
structure constant $e^2/4\pi\hbar c$ in quantum electrodynamics, which is
provided by polarization of the fermionic vacuum with two species of Weyl
fermions (or  with one Dirac fermion if its mass $M$ is small compared to $T$,
otherwise
$T$ is substituted by $M$).  The gap amplitude
$\Delta_0$, constituting  the ultraviolet cut-off of the logarithmically
divergent coupling, plays the part of the Planck energy scale, while the
infrared cut-off is provided by temperature.  

To extend  the  
Eq.(\ref{RunningCoupling}) to the moving superfluid it must be written in
covariant form introducing the four-temperature and the cut-off four vector
$\Theta_\mu=\Delta_0u_\mu$  where the four-temperature $\beta^\nu$ and four-velocity
$u_\nu$ are determined in Sec. \ref{SectionEnergyMomentum}:
\begin{equation} 
\gamma^{-2} ={1\over 6\pi^2} {\rm ln} \left (\beta^\mu \Theta_\mu \right
) ~.
\label{RunningCoupling2}
\end{equation}
At $T=0$ the infrared cut-off is provided by the magnetic field itself:
\begin{equation} 
\gamma^{-2} ={1\over 12\pi^2} {\rm ln} \left ({(g^{\mu\nu}  \Theta_\mu
\Theta_\nu)^2\over g^{\mu\nu} g^{\alpha\beta} F_{\mu\alpha}F_{\nu\beta}}
\right ) ~.
\label{RunningCoupling3}
\end{equation}
 
 Note that
$\Delta_0$ has a parallel with the Planck energy in some other
situations, too (see Sec.\ref{MagnetogenesisChiralFermions}).
Another example is the analogue of the 
cosmological constant, which arises in the effective
gravity of $^3$He-A and has the value
$\Delta_0^4/6\pi^2$ (see Sec.\ref{VacuumPressureSection})  and also Ref.
\cite{Volovik1986gravity}). This parameter also determines the gravitational
constant
$G \sim
\Delta_0^{-2}$ (see Secs.\ref{ConicalSpaces}, 
\ref{MassHyperphoton} and also Sec.\ref{Improved3He}, where it was found that
$G^{-1}=(2/9\pi) \Delta_0^{2}$.  

\subsection{Effective $SU(N)$ gauge fields from degeneracy of Fermi point.}

In $^3$He-A the Fermi point (say, at the north pole ${\bf p}=+p_F\hat{\bf l}$)
is  doubly degenerate owing to the ordinary spin ${\bf\sigma}$ of the $^3$He atom
(Sec.\ref{SpinIsospin}). This means that in equilibtium the two zeroes, each with the
topological invariant $N_3=-1$, are at the same point in momentum space. Let us find out what
can be the consequences of such double degeneracy of the Fermi point. It is clear that the
collective motion of the vacuum can split the Fermi points: positions of the two  points can
oscillate separately since the total topological charge of the Fermi points $N_2=2$ is
conserved at such oscillations. Since the propagator describing the two fermions is the
$4\times 4$  matrix there can be the cross terms coupling the fermions. If we neglect the
degrees of freedom related to the vierbein then the collective variables of the system with
the doubly degenerate Fermi point enter the fermionic propagator as
\begin{equation}
{\cal G}^{-1}=\check\tau^b e^\mu_b (p_{\mu} - e_+A_{\mu} - e_+\sigma_{\alpha}W^{\alpha}_\mu) 
~.
\label{Nonabelian FieldPropagator}
\end{equation} 
The new effective field $W^{\alpha}_\mu$ acts on the chiral quasiparticles as
a ``weak'' $SU(2)$  gauge field. Thus in this effective field theory the
ordinary spin of the $^3$He atoms  plays the part of the weak isospin
\cite{VolovikVachaspati,Exotic}. 

The ``weak'' field $W^{\alpha}_\mu$ is also dynamical and in the leading
logarithmic order obeys the Maxwell (actually Yang-Mills) equations. It is
worthwhile to mention that the ``weak'' charge is also logarithmically
screened by the fermionic vacuum, that is why one has the zero charge effect
for the $SU(2)$ gauge field instead of the asymptotic freedom in the Standard
Model, where the antiscreening is produced by the bosonic degrees of freedom
of the vacuum \cite{Gribov}. 

Appearance of the local $SU(2)$ symmetry in the low energy physics of
$^3$He-A implies that the higher symmetry groups of our vacuum can, in
principle, arise as a consequence of the Fermi point  degeneracy.  For
example, in the Terazava decomposition of 16 fermions into 4 spinons and 4
holons (Sec.\ref{FermionsStandardModel}) the 4-fold degeneracy can produce
both the
$SU(4)_C$ and $SU(2)_L\times SU(2)_R$ gauge groups. In particle physics the
collective modes related to the shift of the 4-momentum are also discussed in
terms of the ``generalized covariant derivative''
\cite{Martin,Sogami}. In this theory the gauge fields, the Higgs fields, and
Yukawa interactions, all are realized as shifts of positions of the degenerate
Fermi point, with degeneracy  corresponding to different quarks and leptons.

In the Eq.(\ref{Nonabelian FieldPropagator}) we did not take into account
that dynamically the vierbein can also oscillate differently for each of the
two elementary Fermi points. As a result the number of the collective modes
could increase even more. This is an interesting problem which must be
investigated in detail. If the degenerate Fermi point mechanism has really
some connection to the dynamical origin of the non-Abelian gauge fields, we
must connect the degeneracy of the Fermi point (number of the fermionic
species) with the symmetry group of the gauge fields. Naive approach leads to
extremely high symmetry group. That is why there should be some factors which
can restrict the number of the gauge and other bosons. For example there can be
some special discrete symmetry between the fermions of the degenerate point,
which restricts the number of massless bosonic collective modes. 

In principle the discrete symmetry being combined with the momentum space
topology can play a decisive role in formation of the effective $SU(N)$ gauge
fields. In Sec. \ref{DiscreteSymmetryEffect} we considered the planar phase of
superfluid $^3$He where  momentum-space topology is nontrivial only due to the
discrete $Z_2$ symmetry. The resulting momentum-space invariant is responsible
for the degenerate Fermi points, which in turn give rise to chiral fermions and
to the effective $SU(2)$ gauge field in the low energy corner.  On the
importance and possible decisive role of the discrete symmetries in
relativistic quantum fields see  Refs.
\cite{HarariSeiberg,Adler2,Peccei}.

Another source of the reduction of the number of the effective field has been
found by Chadha and Nielsen
\cite{Chadha}. They considered the massless electrodynamics with different
metric (vierbein) for the left-handed and right-handed fermions. In this 
model the Lorentz invariance is violated. They found that the two metrics
converge to a single one as the energy is lowered. Thus in the low-energy
corner the Lorentz invariance becomes better and better, and at the same time
the number of independent bosonic modes decreases.

There is however an open question in the Chadha and Nielsen approach: If the
correct covariant terms in action are provided only by the logarithmic
selection, then the logarithm is too slow function to account for the high
accuracy with which symmetries are observed in nature \cite{Iliopoulus}. 
As the $^3$He-A analogy indicates, the noncovariant terms in effective
action appear due to integration over fermions far from the Fermi
point, where the ``Lorentz'' invariance is not obeyed. Thus to obtain the
$SU(N)$ gauge field (and Einstein gravity) with high precision the Lorentz
invariance in the large range of the transPlanckian region is needed. In this
sense the Lorentz invariance appeares to be more fundamental, since it
established the local gauge invariance and general covariance of the
effective theory.

\subsubsection{Mass of $W$-bosons, flat directions and supersymmetry.}
\label{supersymmetry}

In $^3$He-A the $SU(2)$ gauge field acquires mass due to the non-renormalizable terms, which
come from the ``Planckian'' physics. However, in the BCS theory the mass of the ``$W$-boson''
is exactly zero due to the hidden symmetry of the BCS action, and becomes nonzero only due
to the non-BCS corrections: $m_W \sim \Delta_0^2/v_F p_F$. It is interesting that in
the BCS theory applied to the $^3$He-A state the hidden symmetry is extended up to the $SU(4)$
group. The reason for such enhancement of symmetry is still unclear. Probably this can be
related with the flat directions in the Ginzburg-Landau potential for superfluid $^3$He-A 
obtained within the BCS scheme (some discussion of that can be found in \cite{VolovikKhazan}
and in Sec. 5.15 of  the book \cite{Exotic}) or with the supersymmetry in the BCS systems
discussed by Nambu \cite{Nambu}. In any case the natural appearance of the
groups $SU(2)$ and $SU(4)$ in condensed matter effective quantum field theory
reinforces the $G(224)$ group, discussed in Sec. \ref{FermionsStandardModel},
as the candidate for unification of electroweak and strong interactions.

\section{Chiral anomaly in condensed matter systems and Standard Model.}
\label{ChiralAnomaly}

Massless chiral fermions give rise to a number of anomalies in the
effective action. The advantage of $^3$He-A is that this system is
complete: not
only the ``relativistic'' infrared regime is known, but also the behavior
in the
ultraviolet ``nonrelativistic'' (or ``transplanckian'') range is calculable, at
least in principle, within the BCS scheme. Since there is no need for a cut-off, all subtle
issues of the anomaly can be resolved on physical grounds. The measured quantities
related
to the anomalies depend on the correct order of imposing limits, i.e. on what
parameters of the system tend to zero faster: temperature $T$;
external frequency $\omega$; inverse
quasiparticle lifetime due
to collisions with thermal fermions $1/\tau$; inverse volume; the distance
$\omega_0$ between the energy levels of fermions, etc.  All this is very
important
for the $T\rightarrow 0$ limit, where   $\tau$ is formally infinite. 
An example of
the crucial difference between the results obtained  using
different limiting procedures is
the so called  ``angular momentum paradox''  in
$^3$He-A, which is also related to the anomaly: The orbital
momentum of the fluid at $T=0$ differs by several orders of magnitude,
depending on
whether the limit is taken while keeping $\omega\tau\rightarrow 0$ or
$\omega\tau\rightarrow \infty$. The ``angular momentum paradox'' in
$^3$He-A has
possibly a common origin with the anomaly in the spin structure of hadrons
\cite{Troshin}. 

\subsection{Adler-Bell-Jackiw equation.}

\begin{figure}[t]
\centerline{\includegraphics[width=\linewidth]{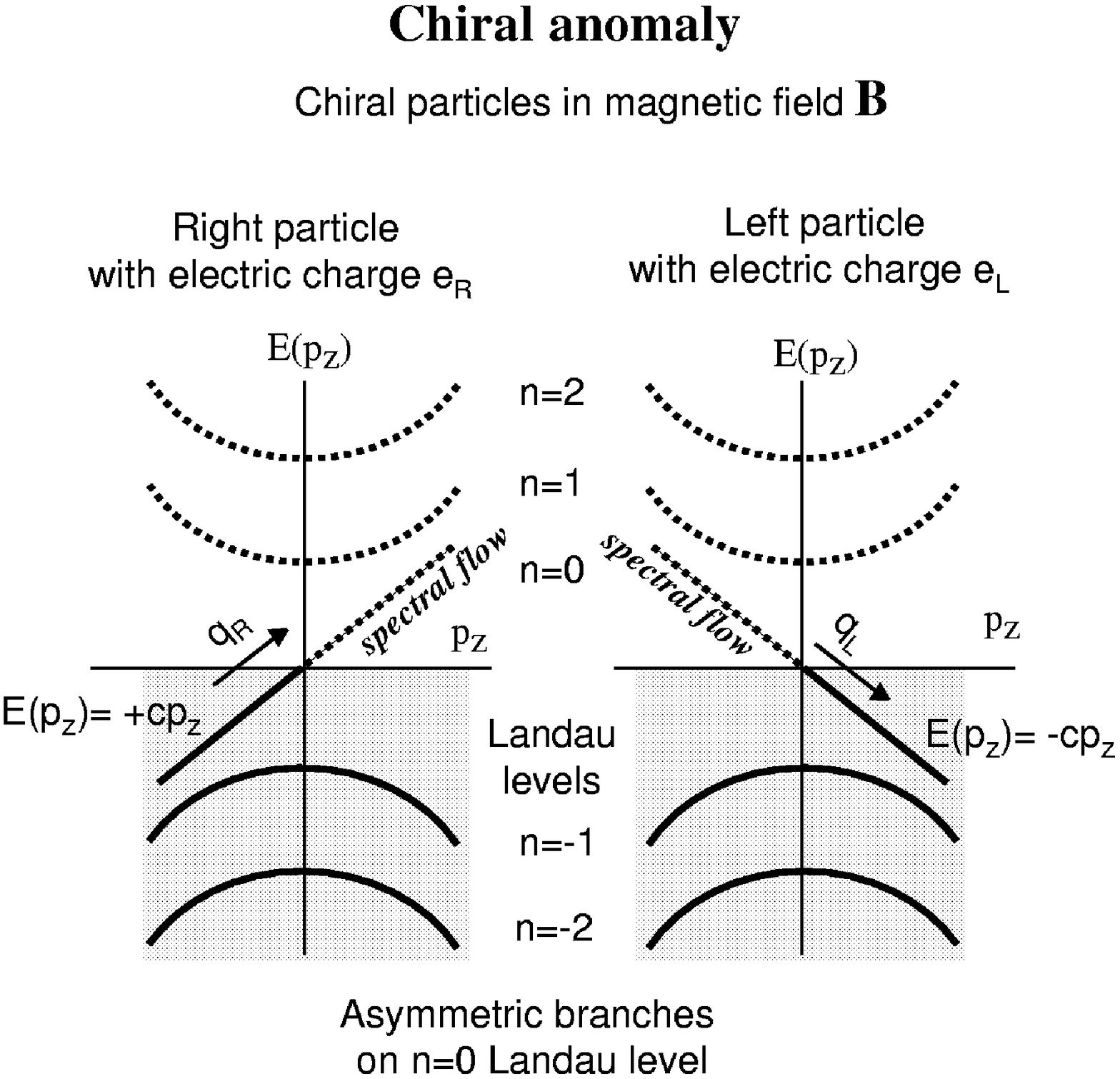}}
\medskip
\caption{Spectrum of massless right-handed and left-handed particles with
electric charges $e_R$ and $e_L$ correspondingly in a magnetic field
${\bf B}$ along $z$;
the thick lines show the occupied negative-energy states. Motion of the
particles in the plane perpendicular to ${\bf B}$ is
quantized into the Landau levels shown.  The free motion is thus effectively
reduced to one-dimensional motion along ${\bf B}$ with momentum $p_z$.
Because of the chirality of the particles the lowest ($n=0$) Landau
level, for which $E=cp_z$ if the particle is right-handed or $E=-cp_z$ if the
particle is left-handed, is asymmetric: it crosses zero only  in one
direction. If we now apply an electric field ${\bf E}$ along $z$, the spectral
flow of levels occurs: the right-handed particles are pushed together with
the energy levels from negative to positive energies according to the
equation of motion
$\dot p_z=e_RE$. The whole Dirac sea of the right-handed particles moves
up,  creating particles and the fermionic charge $q_R$ from the vacuum into
the positive energy continuum of matter.  The same electric field pushes the
Dirac sea of the left-handed particles down, annihilating the fermionic charge
$q_L$. There is a net production of the fermionic charge from the vacuum, if the
left-right symmetry is not exact, i.e. if the charges of left and right
particles are different.
 The rate of particle production is
proportional to  the density of states at the Landau level,
which is  $\propto e\vert {\bf B}\vert$, so that the rate of
production of fermionic charge $q$ from the vacuum is
$\dot q=(1/4\pi^2) (q_Re_R^2 -q_Le_L^2){\bf E} \cdot {\bf B}$.}
\label{ChiralAnomalyFig}
\end{figure}

The chiral anomaly is the phenomenon which allows the nucleation of the
fermionic charge from the vacuum \cite{Adler,BellJackiw}. Such nucleation
results from the spectral flow of the fermionic charge through the Fermi point
to high energy. Since the flux in the momentum space is conserved, it can be
equally calculated in the infrared or in the ultraviolet limits. In $^3$He-A
it is much easier to use the infrared regime, where the fermions obey all the
``relativistic'' symmetries.  As a result one obtains the same anomaly
equation, which has been derived by  Adler and by Bell and Jackiw for the
relativistic systems. The rate of production of some quasiparticle charge
$q$ from the vacuum in applied electric and magnetic fields is (see Fig.
\ref{ChiralAnomalyFig})
\begin{equation} 
\dot q=\partial_\mu J^\mu ={1\over {8\pi^2}} \sum_a C_a q_a e_a^2 F^{\mu\nu}F^{*}_{\mu\nu}~,
\label{ChargeParticlProduction}
\end{equation} 
Here $q_a$ is the charge carried by the $a$-th fermion which is nucleated together with the
fermion; $e_a$ is the charge of the $a$-th fermion with respect to the gauge
field $F^{\mu\nu}$; $C_a=\pm 1$ is the chirality of the fermion; and $F^{*}_{\mu\nu}$ is the
dual field strength. Note that the above equation is fully ``relativistic''. 

In a more general case when the chirality is not readily defined the above equation
can be presented in terms of the momentum-space topological invariant 
\begin{equation} 
\dot q=  {1\over {8\pi^2}}
\left( {\cal Q}{\cal E}^2, {\cal N} \right) F^{\mu\nu}F^{*}_{\mu\nu}~,
\label{ChargeParticlProductionGeneral}
\end{equation} 
where ${\cal Q}$ is the matrix of the charges $q_a$ and ${\cal E}$ is the matrix
of the ``electric'' charges $e_a$.

The Adler-Bell-Jackiw equation has been verified in  $^3$He-A
experiments (see Sec. \ref{VerificationAdlerBellJackiw}),  where the
``magnetic''  ${\bf B}=p_F\vec\nabla\times{\hat{\bf l}}$  and ``electric''
${\bf E}=p_F
\partial_t{\hat{\bf l}}$ fields have been simulated by the space and time dependent ${\hat{\bf
l}}$-texture. In particle physics the only evidence of axial anomaly is related to the decay of
the neutral pion $\pi^0\rightarrow 2\gamma$, although the anomalous nonconservation of the
baryonic charge has been used in different cosmological scenaria explaining an excess of
matter over antimatter in the Universe (see review \cite{Trodden}). 

\subsection{Anomalous nonconservation of baryonic charge.}
 
In the standard electroweak model there is an
additional accidental global symmetry $U(1)_B$  whose
classically conserved charge is the baryon number $B$.
Each of the quarks is assigned $B=1/3$  while the leptons (neutrino and
electron) have $B=0$. The baryonic number is not fundamental
quantity, since it is not conserved in unified theories, such as
$G(224)$ or $SO(10)$, where leptons and quarks are combined in the
same multiplet. At low energy the matrix elements for tramsformation
of quarks to leptons are extremely small and the baryomic charge can
be considered as a good quantum number with high precision. However, 
it can be produced due to the axial anomaly, in which it is generated from
the vacuum due to spectral flow. In the Standard Model there are two gauge
fields whose ``electric'' and ``magnetic'' fields become a source for
baryoproduction: The hypercharge field $U(1)_Y$ and the weak field $SU(2)_L$.
Let us first consider the effect of the hypercharge field. The production rate of baryonic
charge in the presence
of hyperelectric and hypermagnetic fields is 
\begin{equation}
{ N_F\over 4\pi^2} \left(  {\cal Y}^2  {\cal B},{\cal N} \right) ~{\bf B}_Y\cdot {\bf E}_{Y}
={ N_F\over 4\pi^2}(Y_{dR}^2 +Y_{uR}^2 - Y_{dL}^2 - Y_{uL}^2)~{\bf B}_Y\cdot {\bf E}_{Y},
\label{BarProdByHypercharge}
\end{equation} 
where $N_F$ is the number of families, $Y_{dR}$, $Y_{uR}$, $Y_{dL}$ and $Y_{uL}$ are
hypercharges of right and left $u$ and $d$ quarks. Since the hypercharges of
left and right fermions are different (see Eq.~(\ref{SU42})),  one obtains the
nonzero value of $ \left({\cal Y}^2  {\cal B},{\cal N} \right)=1/2$, and thus a
nonzero production of baryons by the hypercharge field
\begin{equation}
{N_F\over 8\pi^2}{\bf B}_Y\cdot {\bf E}_{Y}.
\label{BarProdByHypercharge2}
\end{equation} 
The weak  
field also contributes to the production of the baryonic charge:
\begin{eqnarray}
 { N_F\over 4\pi^2} \left(  {\cal W}_3^L {\cal W}_3^L{\cal B},{\cal N} \right) {\bf
B}^b_W\cdot {\bf E}_{bW}= - {N_F\over 8\pi^2}{\bf B}^b_W\cdot {\bf E}_{bW}.
\label{BarProdByHypercharge3}
\end{eqnarray}

Thus the total rate of baryon production in the Standard model takes
the form
\begin{eqnarray}
\dot B={ N_F\over 4\pi^2} \left[\left(  {\cal Y}^2  {\cal B},{\cal N} \right) ~{\bf
B}_Y\cdot {\bf E}_{Y}+\left(  ({\cal W}_3^L)^2 {\cal B},{\cal N} \right) {\bf B}^b_W\cdot {\bf
E}_{bW}\right]\nonumber \\
= {{N_F} \over {8 \pi^2}} \left ( {\bf B}_Y\cdot {\bf E}_{Y} -  {\bf B}^b_W\cdot {\bf
E}_{bW} 
   \right).
\label{TotalBaryoProduction}
\end{eqnarray}
The same equation describes the production of the leptonic charge $L$: one has $\dot L=\dot
B$ since $B-L$ is conserved due to anomaly cancellation.  This means that
production of one lepton is followed by production of three baryons.

The second term in Eq.(\ref{TotalBaryoProduction}), wich comes from
nonabelian
$SU(2)_L$ field, shows that the nucleation of baryons occurs when the topological charge
of the vacuum changes, say, by sphaleron or due to de-linking of linked loops of
the cosmic strings \cite{tvgf,jgtv,barriola}. The nontopological  term describes the
exchange of the baryonic (and leptonic) charge between the hypermagnetic field and the
fermionic degrees of freedom. 

It is important that the Eq.(\ref{TotalBaryoProduction}) is completely determined
by the invariants of the Fermi point and is valid even in the nonrelativistic systems.
That is why the same equation can be applied to $^3$He-A after being ajusted to the $^3$He-A
symmetry. 

\subsection{Analog of baryogenesis in $^3$He-A: Momentum exchange between superfluid vacuum and
quasiparticle matter.}\label{AnalogBaryogenesis}

In $^3$He-A the relevant fermionic charge, which is important for the
dynamics of superfluid liquid, is the linear momentum. The superfluid
background moving with velocty ${\bf v}_{\rm s}$ and the normal component
moving with velocity ${\bf v}_{\rm n}$ can exchange momentum. This exchange is
mediated by the texture of the $\hat{\bf l}$ field, which carries continuous
vorticity (see Eq.(\ref{Nermin-Ho})).  The momentum of the
flowing vacuum is transferred to the momentum carried by texture, and then
from texture to the system of quasiparticles. The force between the
superfluid and normal components arising due to this momentum exchange is
usually called the mutual friction, though the term friction is not very good
since some or essential part of this force is reversible and thus
nondissipative.  In superfluids and superconductors with curl-free superfluid
velocity ${\bf v}_{\rm s}$, the mutual friction is produced by the dynamics of
quantized vortices which serve as mediator. 

\begin{figure}[t]
\centerline{\includegraphics[width=\linewidth]{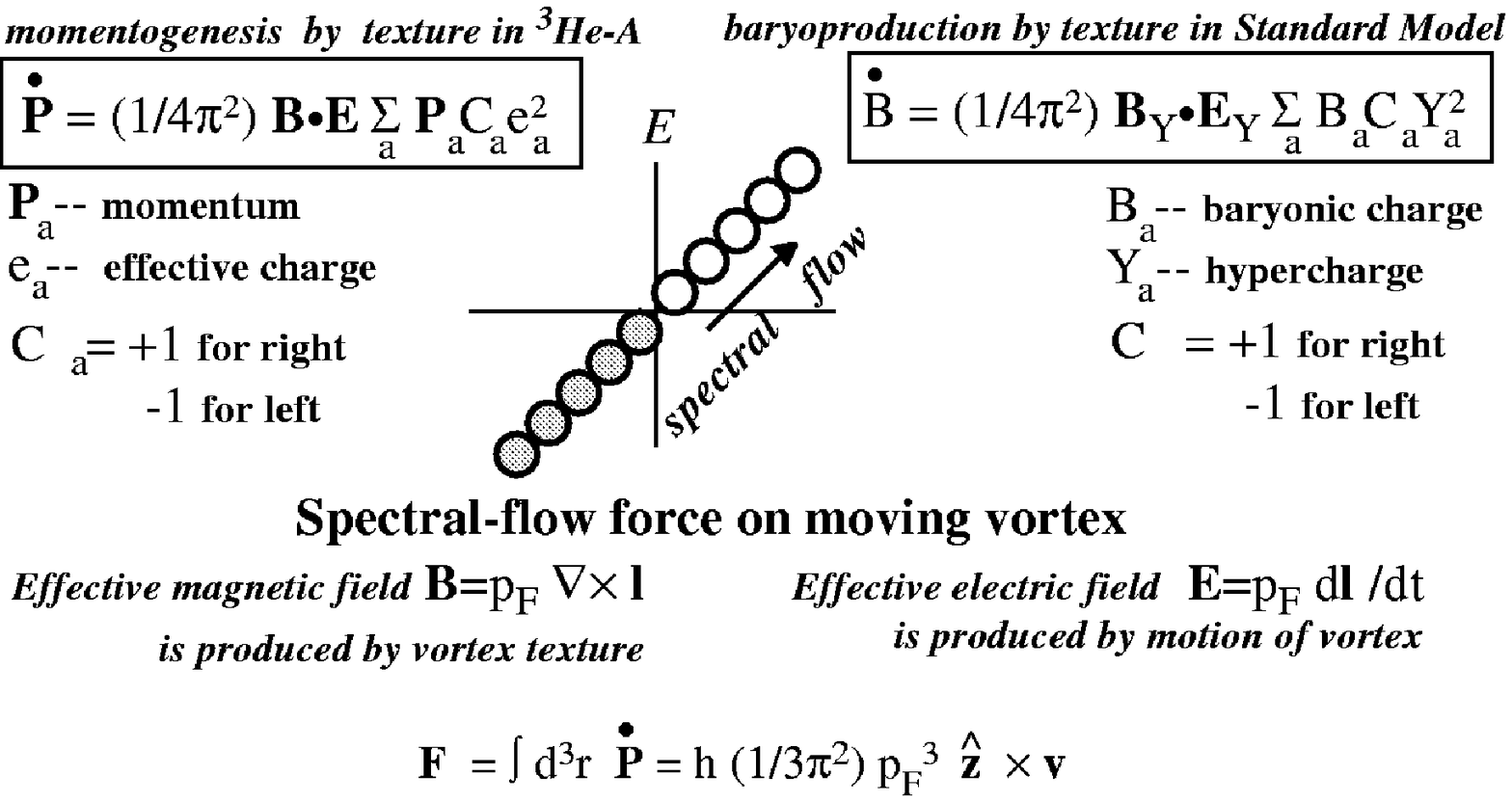}}
\medskip
\caption{Production of the fermionic charge in $^3$He-A (linear momentum) and
in Standard Model (baryonic number) are described by the same
Adler-Bell-Jackiw equation. Integration  of the anomalous momentum
production over the cross-section
of the moving continuous vortex gives the loss of linear momentum
and thus the additional force per unit length acting on the vortex  due to 
spectral flow.}
\label{momentogenesisFig}
\end{figure}

Here we are interested in the process of the momentum transfer from the
texture to quasiparticles. It can be described in terms of the chiral
anomaly, since as we know the $\hat{\bf l}$ texture plays the part of the
$U(1)$ effective gauge field acting on relativistic quasiparticles, and these
quasiparticles are chiral. Thus we have all the conditions to apply the
axial anomaly equation (\ref{ChargeParticlProduction}) to this process of
transformation of the fermionic charge carried by magnetic field (texture) to
the fermionic charge carried by chiral particles (normal component)
(Fig.\ref{momentogenesisFig}). When a chiral quasiparticle crosses zero energy
in its spectral flow it carries with it its linear momentum ${\cal{\bf P}}=\pm
p_F\hat{\bf l}$.  That is why this ${\cal{\bf P}}$ is the proper fermionic
charge $q$ which enters the Eq.(\ref{ChargeParticlProduction}),
 and the rate of
the momentum production from the texture is
\begin{equation}
{\dot{\bf P}}= { 1\over 4\pi^2} \left(  {\cal{\bf P}} {\cal E}^2,{\cal N} \right) 
{\bf B} \cdot {\bf E} = { 1\over 4\pi^2} 
{\bf B} \cdot {\bf E} \sum_a  {\bf P}_a C_a e_a^2 ~.
\label{MomentumProductionGeneral}
\end{equation} 
Here ${\bf B}=(p_F/\hbar)\nabla\times\hat{\bf l}$ and   ${\bf E}= (p_F/\hbar)
\partial_t
\hat{\bf l}$ are effective ``magnetic''  and ``electric'' fields;  ${\cal E}$
is the matrix of corresponding ``electric'' charges in
Eq.(\ref{RelatBogoliubovNambuEnergy}):
$e_a=-C_a$ (the ``electric'' charge is opposite to the chirality of the $^3$He-A
quasiparticle); and ${\bf P}_a=-C_a p_F\hat{\bf l}$ is the momentum
(fermionic charge) carried by the $a$-th fermionic quasiparticle. Using this
translation to the $^3$He-A language one obtains that the momentum production
from the texture per unit time per unit volume is
\begin{equation}
 {\dot{\bf P}}=- {p_F^3\over 2\pi^2 \hbar^2}\hat{\bf l}
\left(\partial_t\hat{\bf l}\cdot(\nabla\times \hat{\bf l})\right) 
    ~.
\label{MomentumProductionAPhase}
\end{equation} 

It is interesting to follow the hystory of this term in $^3$He-A. First the
nonconservation of the momentum of the superfluid vacuum at $T=0$ has been
found from the general consideration of the superfluid hydrodynamics of the
vacuum \cite{VolovikMineev1981}.  Later it was found that the quasiparticles
must be nucleated whose momentum production rate  is described by the same
Eq.(\ref{MomentumProductionAPhase}), but with the opposite sign
\cite{Combescot}. Thus the total momentum of the system has been proved to
conserve. In the same paper Ref.\cite{Combescot} it was first found that the
quasiparticle states in $^3$He-A in the presense of twisted texture of
$\hat{\bf l}$ has strong analogies with the eigenstates of a charged particle
in a magnetic field. Then it became clear \cite{Volovik1986} that the momentum
production is described by the same equation as the axial anomaly in
relativistic quantum field theory. Now we know why it happens: The spectral
flow from the texture to the ``matter'' occurs through the Fermi point and thus
it can be described by the physics in the vicinity of the Fermi point, where
the ``relativistic''  quantum field theory necessarily arises.

\subsection{Axial anomaly and force on  $^3$He-A vortices.}
\label{AxialAnomalyAndForce}

From the underlying microscopic theory we know that the total linear momentum
of the liquid is  conserved. The  Eq. (\ref{MomentumProductionAPhase}) thus
implies that in the presence of a time-dependent texture the momentum is
transferred from the  texture (the distorted superfluid vacuum or magnetic
field) to the heat bath of quasipartcles  (analogue of matter). The rate of
the momentum transfer gives an extra force acting on a moving $\hat{\bf
l}$-texture. This force influences the dynamics of the continuous textures,
which represents the vortex in $^3$He-A (analog of stringy texture in the
Standard Model \cite{AchucarroVachaspati}), and this force has been measured
in experiments on the rotating $^3$He-A \cite{BevanNature} (see Sec.
\ref{VerificationAdlerBellJackiw}).

\begin{figure}[t]
\centerline{\includegraphics[width=\linewidth]{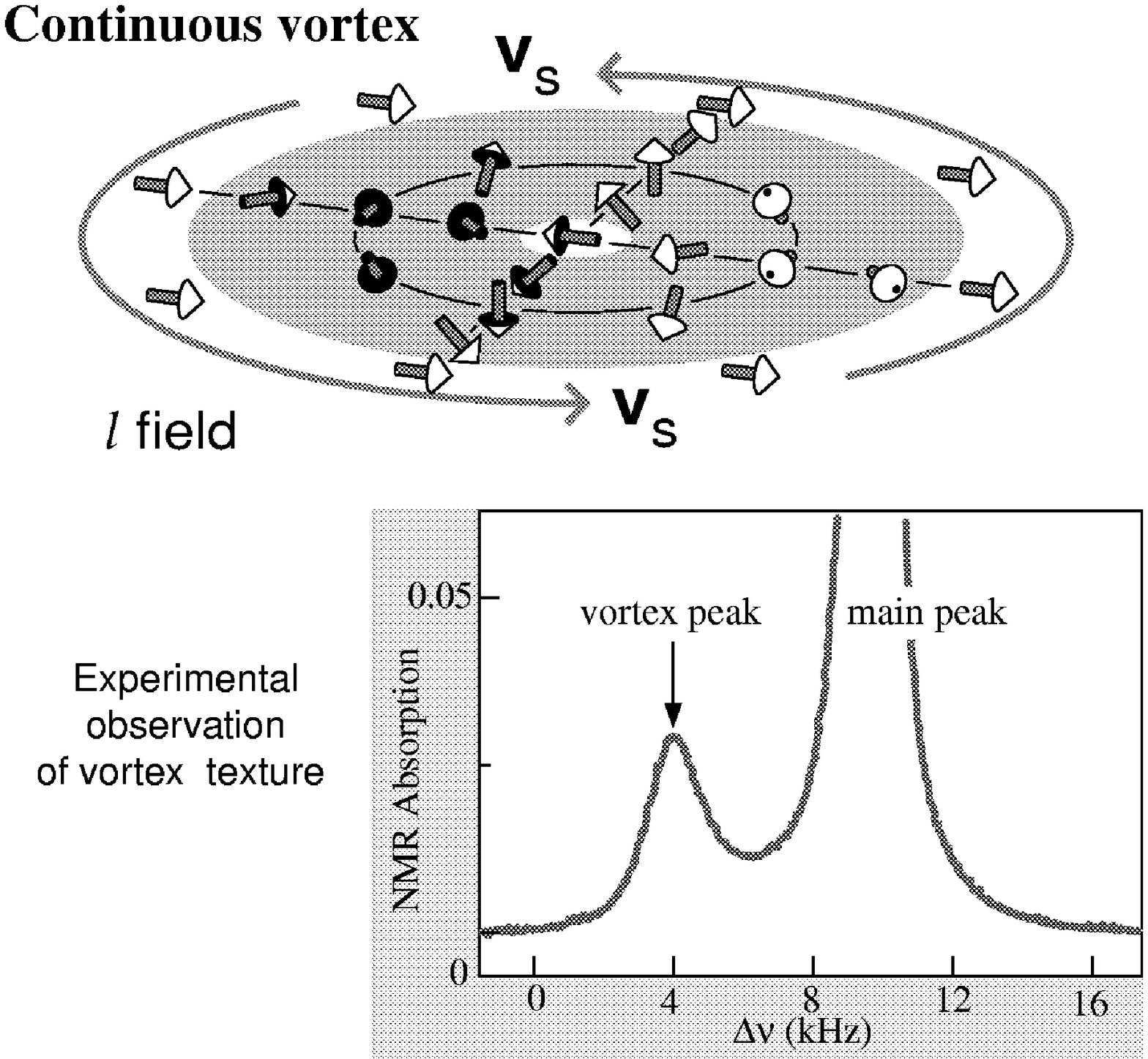}}
\medskip
\caption{{\it Top}: An $n_1=2$ continuous vortex in $^3$He-A. The arrows  
indicate the local direction of the order parameter vector ${\hat{\bf l}}$.
Under experimental conditions the direction of the ${\hat{\bf l}}$-vector
in the bulk liquid far from the soft core is kept in the plane perpendicular to
applied magnetic field far from the core. This does not change the topology of
the  Anderson-Toulouse-Chechetkin vortex: the  ${\hat{\bf l}}$-vector covers the
whole $4\pi$ sphere within the soft core. As a result there is 
$4\pi$ winding of the phase of the order parameter around the soft core,
which corresponds to $n_1=2$ quanta of anticlockwise circulation. {\it
Bottom}:  The NMR absorption in the characteristic
vortex satellite originates from the soft core where the ${\hat{\bf l}}$
orientation deviates from the homogeneous alignment in the bulk. Each soft core
contributes equally to the intensity of the satellite peak and gives a
practical tool for measuring the number of vortices.}
\label{ContinuousVortexFig}
\end{figure}

The continuous vortex texture, first discussed by Chechetkin \cite{Chechetkin} and
Anderson and Toulouse\cite{AT} (ATC vortex, Fig.
\ref{ContinuousVortexFig}), has in its simplest realization  the
following distribution of the ${\hat{\bf l}}$-field  ($\hat{\bf z}$,
${\hat{\bf \rho}}$ and ${\hat{\bf \phi}}$ are unit vectors of the cylindrical
coordinate system)
\begin{equation}
{\hat{\bf l}}(\rho,\phi)={\hat{\bf z}} \cos\eta(\rho) + {\hat{\bf
r}} \sin\eta(\rho)~,
\label{lTextureContVortex}
\end{equation}
where $\eta(\rho)$ changes from $\eta(0)=0$ to $\eta(\infty)=\pi$.  Such
${\hat{\bf l}}$-texture  forms the so called soft core of the vortex, since it
contains nonzero vorticity of superfluid velocity in Eq.(\ref{v_s}):
\begin{equation}
{\bf v}_{\rm s}(\rho,\phi)=  {\hbar\over 2 m_3 \rho}[1-\cos\eta(\rho)]{\hat {\bf
\phi}}~~,
\label{v_sContVortex}
\end{equation}
In comparison to a more familiar singular vortex, the continuous vortex has
a regular superfluid velocity field,  with no singularity on the vortex axis, but the
circulation of the superfluid velocity about the soft core is still
quantized:  
$\oint d{\bf x}\cdot {\bf v}_{\rm s}=\kappa$ with $\kappa=2\pi\hbar/m$, which
is twice the conventional circulation quantum number in the paircorrelated
system $\kappa_0=2\pi\hbar/2m$, where $2m$ is the mass of the Cooper pair.
Quantization of circulation in continuous vortex is related to the topology of
the ${\hat{\bf l}}$-field:  according to Mermin-Ho relation (\ref{Nermin-Ho})
the  ${\hat{\bf l}}$-vector covers the whole $4\pi$ sphere when the soft core
is swept. 

 The stationary vortex has nonzero effective ``magnetic'' field, ${\bf
B}=(p_F\hbar)  \nabla \times {\hat {\bf l}}$.  If the vortex moves with a
constant velocity ${\bf v}_{\rm L}$ with respect to the heat bath the moving
texture acquires the time dependence, ${\hat{\bf l}}({\bf r}-{\bf v}_{\rm
L}t)$, and this leads to the effective ``electric''  field 
\begin{equation}
 {\bf
E}={1\over \hbar}\partial_t {\bf A}=-{p_F\over \hbar}({\bf v}_{\rm L}\cdot
{\bf\nabla}){\hat {\bf l}}
\label{EfieldContVortex}
\end{equation}

The net production of the quasiparticle momenta by the spectral flow in the
moving vortex means: If the vortex moves with respect to the system of
quasiparticles (the normal component of liquid or matter, whose flow is
characterised by the normal velocity ${\bf v}_{\rm n}$),  that there is a force acting
between the normal component and the vortex. Integration of the anomalous
momentum transfer in Eq.(\ref{MomentumProductionAPhase}) over the cross-section of
the soft core of the moving ATC vortex gives the following force acting on the
vortex (per unit length) from the system of quasiparticles
\cite{Volovik1992}:
\begin{equation}
{\bf F}_{sf}=\int d^2 \rho{ p_F^3\over {2\pi^2\hbar^2}} \hat {\bf l} ~(
\partial_t \hat {\bf l} \cdot  ( \nabla\times \hat {\bf l} ))=
\int d^2 \rho{ p_F^3\over {2\pi^2\hbar^2}} \hat {\bf l} ~(
(({\bf v}_{\rm n}-{\bf v}_{\rm L})\cdot\nabla) \hat {\bf l} \cdot  (
\nabla\times
\hat {\bf l} ))=-2\pi \hbar
 C_0{\hat {\bf z}}  \times ({\bf v}_{\rm L}-{\bf v}_{\rm n}) ,
\label{SpFlowForce}
\end{equation}
where 
\begin{equation}
C_0= {p_F^3\over 3\pi^2\hbar^3}~.
\label{C0}
\end{equation}
Note that this spectral-flow force is transverse to the relative motion of the
vortex and thus is nondissipative (reversible). In this derivation it was
assumed that the quasiparticles and their momenta, created by the spectral
flow from the vacuum, are finally absorbed by the normal component. The
retardation in the process of absorption and also the viscosity of the normal
component lead to a dissipative (friction) force between the vortex and the
normal component:
${\bf F}_{fr}=-\gamma ({\bf v}_L-{\bf v}_{\rm n})$.  There is no momentum
exchange between the vortex and the normal component if they move with the
same velocity.  The result (\ref{SpFlowForce}) for  the spectral-flow force,
was also confirmed in a  microscopic theory \cite{Kopnin1993}.

The important property of the spectral-flow force (\ref{SpFlowForce}) is that
it does not depend on the details of the vortex structure: The result for
${\bf F}_{sf}$ is robust against any deformation of the ${\hat{\bf l}}$-texture
which does not change its asymptote, i.e. the topology of the vortex. In this
respect this force resembles another force, the force between the vortex
texture and the superfluid vacuum, which acts on the vortex moving with
respect to the superfluid  vacuum. This is the well-known Magnus force:
\begin{equation}
{\bf F}_{M}=  2\pi \hbar
 n {\hat {\bf z}}  \times ({\bf v}_L-{\bf v}_{\rm s}(\infty)) ~,
\label{MagnusForce}
\end{equation}
where again $n$ is the particle density -- the number density of $^3$He atoms;
${\bf v}_{\rm s}(\infty)$ is the uniform velocity of the superfluid vacuum far from
the vortex. 

Let us recall that the vortex texture (or quantized vortex in
$U(1)$ superfluids) serve as mediator (intermediate object) for the momentum
exchange between the superfluid vacuum and the fermionic heat bath of
quasiparticles. The momentum is transferred from the vacuum to texture (this
produces the Magnus force acting on the vortex texture) and then from the
texture to the ``matter'' (with minus sign the spectral-flow force in
Eq.(\ref{SpFlowForce}) acting on the vortex texture from the system of
quasiparticles). In this respect the texture (or vortex) corresponds to the
sphaleron or to the cosmic string in relativistic theories. If the other
processes are neglected then in the steady state these two forces acting on
the texture from vacuum and ``matter'' must compensate each other. From this
balance of the two forces one obtains that the vortex must move with the
constant velocity determined by the velocities
${\bf v}_{\rm s}$ and ${\bf v}_{\rm n}$ of vacuum and matter:    ${\bf
v}_L=(n{\bf v}_{\rm s}-C_0{\bf v}_{\rm n})/(n-C_0)$. However this is valid
only under special conditions. Firstly the dissipative friction must be taken
into account. It comes in particular from the retardation of the spectral
flow process. The retardation also modifies the nondissipative spectral flow
force.  Secondly the analogy with the gravity shows that there is one more
force of topological origin -- the so-called Iordanskii force in
Eq.(\ref{IordanskiiForce}). It comes from the gravitational analog of the
Aharonov-Bohm effect experienced by particles moving in the presence of the
spinning cosmic string (see Sec.\ref{IordanskiiForce}).

\subsection{Experimental verification of Adler-Bell-Jackiw equation in rotating $^3$He-A.}
\label{VerificationAdlerBellJackiw}

The spectral flow force acting on the vortex has been measured in experiments
on vortex dynamics in $^3$He-A \cite{BevanNature,BevanJLTP}. In such
experiments a uniform array of vortices is produced by rotating the whole
cryostat. In equilibrium the vortices and the normal component of the fluid
(heat bath of quasiparticles) rotate together with the cryostat.  An
electrostatically driven vibrating diaphragm  produces an oscillating
superflow, which via the Magnus force generates the vortex motion, while the
normal component remains clamped due to its high viscosity. This creates a
motion of  vortices with respect both to the heat bath (``matter'') and the
superfluid vacuum. The vortex velocity
${\bf v}_L$  is determined by the overall
balance of forces acting on the vortices.  This includes the spectral flow
force ${\bf F}_{sf}$ in Eq.(\ref{SpFlowForce}); the Magnus force ${\bf F}_{M}$
in Eq.(\ref{MagnusForce}); the friction force ${\bf F}_{fr}=-\gamma ({\bf
v}_L-{\bf v}_{\rm n})$;  and the Iordanskii force in
Eq.(\ref{IordanskiiForce}) coming from the gravitational analog of the
Aharonov-Bohm effect (see Sec.\ref{IordanskiiForce}). For the doubly quantized
vortex the Iordanskii force is
\begin{equation}
{\bf F}_{\rm Iordanskii}=  2\pi \hbar
 n_{\rm n} {\hat {\bf z}}  \times ({\bf v}_{\rm s}(\infty)-{\bf v}_{\rm n}) ~,
\label{IordanskiiForce1}
\end{equation}
where the momentum carried by quasiparticles is expressed in terms of the
normal component density: $\sum_{\bf p} {\bf p} f({\bf p})= mn_{\rm n}({\bf
v}_{\rm s}(\infty)-{\bf v}_{\rm n})$. Since for the steady state motion of
vortices the sum of all forces acting on the vortex must be zero, 
${\bf F}_{M}+{\bf F}_{sf}+{\bf F}_{\rm Iordanskii}+{\bf F}_{fr}=0$, one has
the following equation for ${\bf v}_L$:
\begin{equation}
\hat{\bf z}\times ({\bf v}_L-{\bf v}_{\rm s}(\infty))+
d_\perp \hat{\bf z}\times({\bf
v}_{\rm n}-{\bf v}_L)+d_\parallel({\bf v}_{\rm n}-{\bf v}_L) =0~~,
\label{ForceBalance}
\end{equation}
where 
\begin{equation}
d_\perp=1- {n-C_0\over n_{\rm s}(T)}~,
\label{d_perpAphase}
\end{equation}
$n_{\rm s}(T)=n-n_{\rm n}(T)$ is the density of the superfluid component; and
$d_\parallel=\gamma/2\pi n_{\rm s}$. 

\begin{figure}[t]
\centerline{\includegraphics[width=\linewidth]{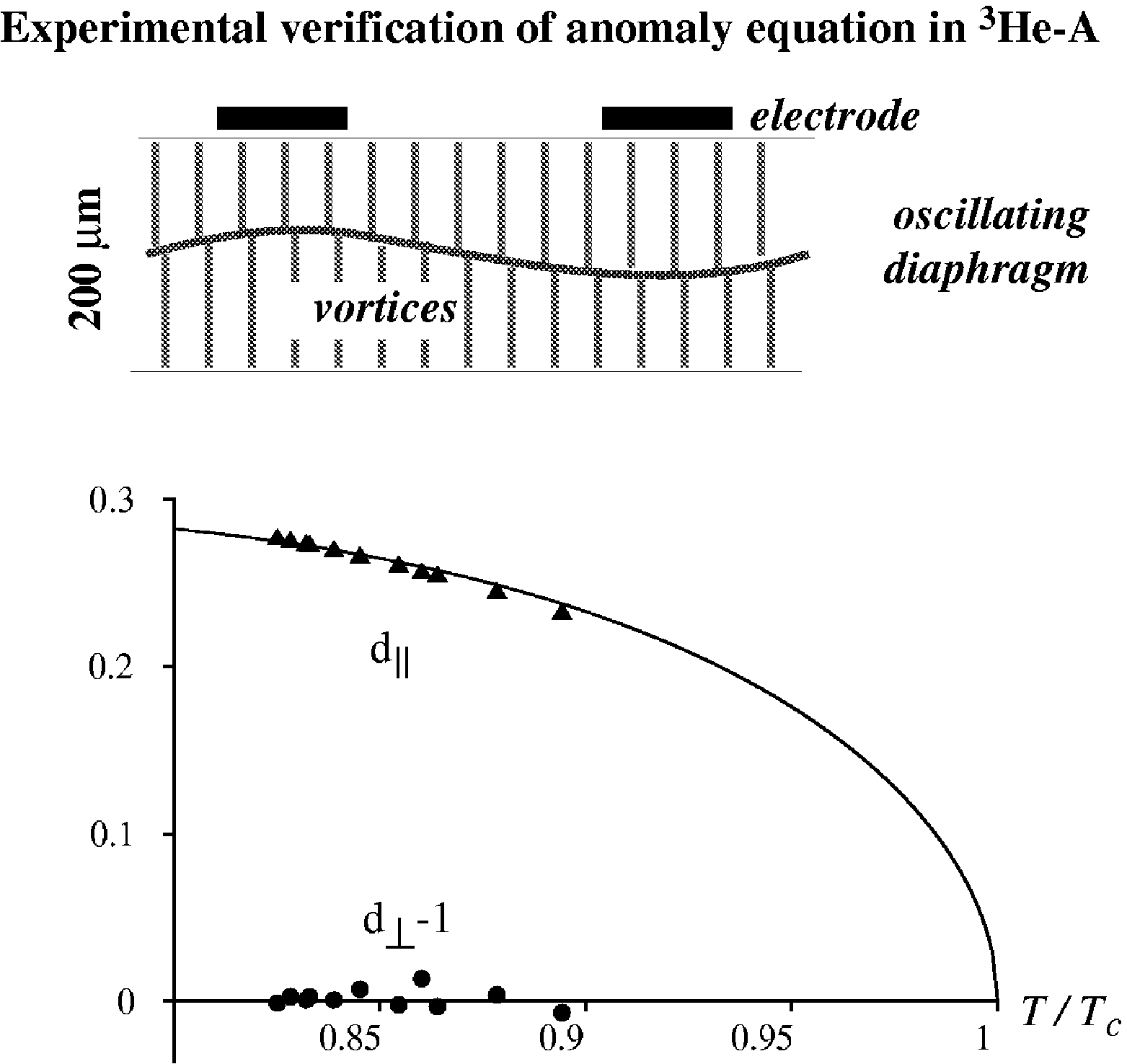}}
\medskip
\caption{A uniform array of vortices is produced
by rotating the whole cryostat, and oscillatory superflow perpendicular
to the rotation axis is produced by a vibrating diaphragm, while 
the normal fluid (thermal excitations) is clamped by viscosity, ${\bf v}_{\rm
n}=0$. The velocity
${\bf v}_L$ of the vortex array is determined by the overall
balance of forces acting on the vortices.  These vortices
produce additional dissipation proportional to $d_\parallel$ and
coupling between two orthogonal modes proportional to
$1-d_\perp$.}
\label{ExpmomentogenesisFig}
\end{figure}

Measurement of the damping of the diaphragm resonance and of the 
coupling between different eigenmodes of vibrations enables both dimensionless
parameters, $d_\perp$ and $d_\parallel$, to be deduced. The most important for us is the 
parameter $d_\perp$, which gives information on the spectral flow parameter $C_0$. The effect
of the chiral anomaly is crucial for $C_0$: If there is no anomaly then $C_0=0$ and
$d_\perp=n_{\rm n}(T)/n_{\rm s}(T)$; if the anomaly is fully realized the
parameter $C_0$ has its maximal value, $C_0=p_F^3/3\pi^2\hbar^3$, which
coincides with the particle density of liquid $^3$He in the normal state,
Eq.(\ref{LuttingerTheorem}). The difference  between the particle density of
liquid $^3$He in the normal state $C_0$ and the particle density of liquid
$^3$He in superfluid $^3$He-A state $n$ at the same chemical potential $\mu$
is determined by the tiny effect of superfluid correlations on the particle
density and is extremely small:
$n -C_0\sim n (\Delta_0/v_Fp_F)^2=n(c_\perp/c_\parallel)^2 \sim 10^{-6}
n$; in this case  one must have
$d_\perp\approx 1$ for all practical temperatures, even including the region
close to $T_c$, where the superfluid component $n_s(T) \sim n (1-T^2/T_c^2)$
is small. $^3$He-A experiments, made  in the whole temperature 
range where $^3$He-A is stable,  gave precisely
this value within experimental uncertainty,  
$|1-d_\perp|<0.005$\cite{BevanNature} (see Fig. \ref{ExpmomentogenesisFig}).  

This means that the anomaly is fully realized in the dynamics of the
$\hat{\bf l}$ texture and provides an experimental verification of the
Adler-Bell-Jackiw axial anomaly equation (\ref{ChargeParticlProduction}),
applied to $^3$He-A. This supports the idea that baryonic charge (as well as
leptonic charge)  can be generated by electroweak gauge fields through the
anomaly. 
 
In the same experiments with the $^3$He-B vortices the effect analogous to
the axial anomaly is temperature dependent and one has the crossover from 
$d_\perp\approx 1$ at high $T$ to $d_\perp=n_{\rm n}(T)/n_{\rm s}(T)$ at low
$T$ (see Sec. \ref{BPhaseExp} and Fig. \ref{CallanHarveyExpFig}). The reason
for that is that Eq.~(\ref{ChargeParticlProduction}) for axial anomaly and the
corresponding equation  (\ref{MomentumProductionAPhase}) for the momentum
production are valid only in the limit of continuous spectrum, i.e. when the
distance $\omega_0$ between the energy levels of fermions in the texture is
much smaller than the inverse quasiparticle lifetime: $\omega_0\tau \ll 1$. 
The spectral flow completely disappears in the opposite case $\omega_0\tau \gg
1$, because the spectrum becomes effectively discrete.  As a result, the
force acting on a vortex texture differs by several orders of magnitude for
the cases $\omega_0\tau \ll 1$ and $\omega_0\tau \gg 1$. The parameter
$\omega_0\tau$ is regulated by temperature.  In case of  $^3$He-A the vortices
are continuous and thus $\omega_0$ is extremely small and can be negelected.
This means that the spectral flow is maximally possible and thus the
Adler-Bell-Jackiw  anomaly equation is applicable there for all practical
temperatures and it was experimentally confirmed.

Note in conclusion of this Section, that the spectral flow in  
vortices realizes the momentum exchange between the 3+1 fermionic system
outside the vortex and the 1+1 fermions living in the vortex core. This
corresponds to the Callan-Harvey process of anomaly cancellation
\cite{CallanHarvey} between the systems of different dimension
\cite{CallanHarveyEffect,StoneSpectralFlow} (see also Sec.
\ref{SpectralFlowSingVort}).
 
\section{Macroscopic parity violating effects.}

\subsection{Helicity in parity violating systems.}

Parity violation, the asymmetry between left and right, is one of the
fundamental properties of the quantum vacuum. This effect is strong at high
energy of the order of electroweak scale, but is almost imperceptible in the
low-energy condensed matter physics. Since at this scale the left and right
particles are hybridized and only the left-right symmetric charges survive. For
example, Leggett's suggestion to observe the macroscopic effect of parity
violation using such macroscopically coherent atomic system as superfluid
$^3$He-B is very far from realization
\cite{Leggett,VollhardtWolfle}. On the other hand,   an analog of parity
violation exists in superfluid $^3$He-A alongside with the related phenomena,
such as chiral anomaly which we discussed in previous section and macroscopic chiral currents
(for a review see Refs. \cite{VollhardtWolfle,Exotic}). So, if we cannot investigate the
macroscopic parity violating effects directly we can simulate analogous 
physics in $^3$He-A.

Most of the macroscopic parity violating phenomena are related to helicity:
the energy of the system in which the parity is broken contains the helicity
term $\lambda{\bf A}\cdot ({\bf\nabla}\times {\bf A})$, where ${\bf A}$ is the
relevant collective vector field. To have such terms the parity $P$ must be
violated together with all the combinations containing other descrete
symmetries, such as $CP$, $PT$, $CPT$, $PU_2$ (where $U_2$ is the rotation by
$\pi$), etc. Such terms sometimes lead to the instability of the vacuum
towards the spacially inhomogeneous state, the helical instability. In nematic
liquid crystals, for example, the excess of the chiral molecules of one
preferred chirality leads to the helicity term, $\lambda\hat{\bf n}\cdot
({\bf\nabla}\times \hat{\bf n})$,  for the nematic vector (director) field
${\hat{\bf n}}$. This leads to formation of the cholesteric structure, the
helix. The same phenomenon occurs in superfluid
$^3$He-A, where at some conditions there is a helical instability of the
homogeneous counterflow ${\bf w}={\bf v}_{\rm n}-{\bf v}_{\rm s}$, which we
discuss in this Section. The interest to this instability arises because it
is the counterpart of the helical instability of an excess of the massless
right-handed electrons over the left-handed, which leads to
formation of the helical hypermagnetic field ${\bf B}_Y$
\cite{JoyceShaposhnikov,GiovanniniShaposhnikov}. The formed  
field ${\bf B}_Y$, after the electroweak transition occurs, is transformed to
electromagnetic magnetic field ${\bf B}(={\bf B}_Q)$. Thus the helical
instability can be the source of formation of primordial cosmological magnetic
fields (see also recent review paper on cosmic magnetic fields
\cite{CosmicMagneticReview} and references therein).

\begin{figure}[t]
\centerline{\includegraphics[width=\linewidth]{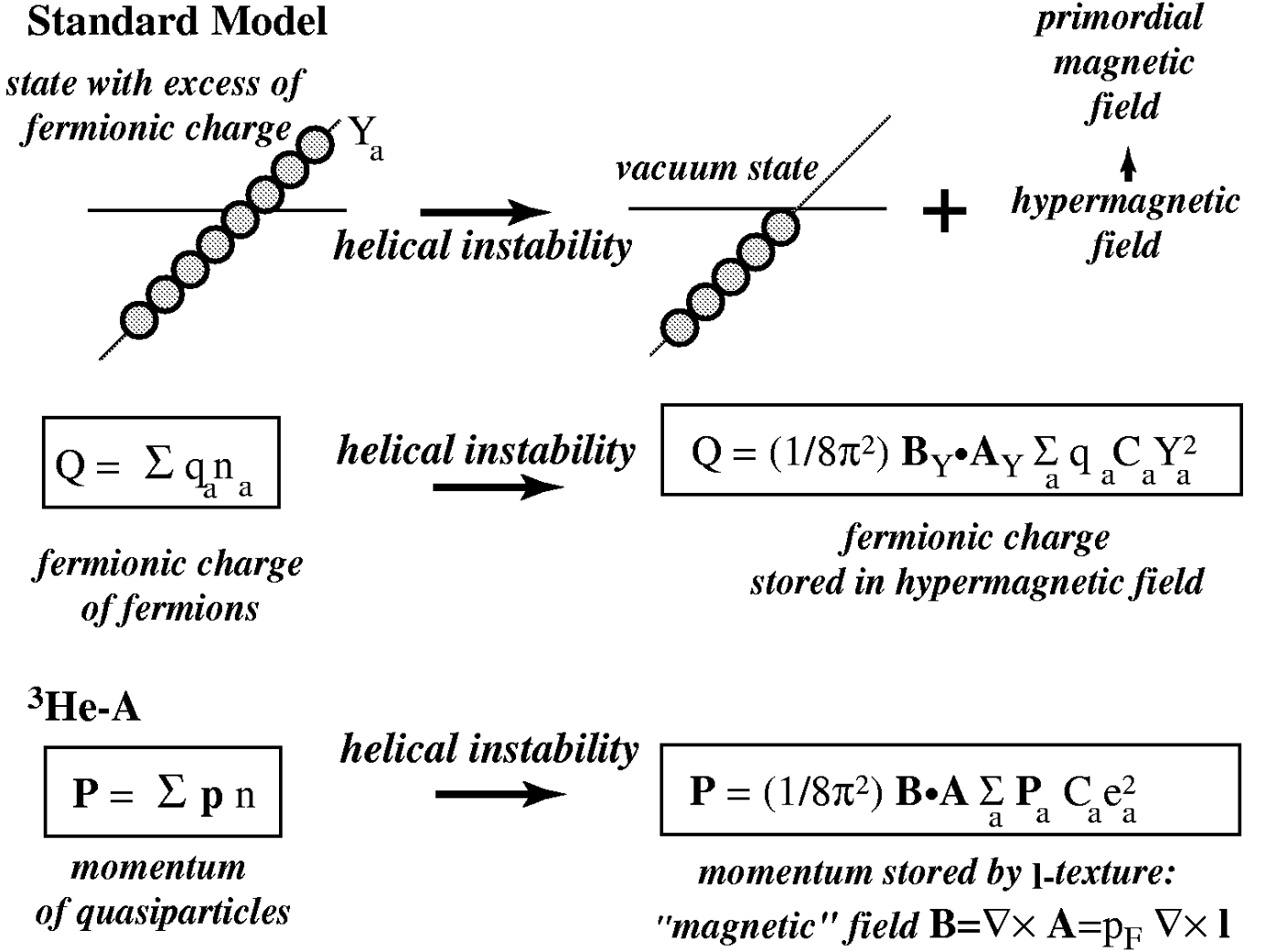}}
\medskip
\caption{The fermionic charge of right-handed minus that of
left-handed particles is conserved at the classical level but not if 
quantum properties of the physical vacuum are taken into account. This charge
can be transferred to the ``inhomogeneity'' of the vacuum via the axial
anomaly in the process of the helical instability.  The inhomogeneity which
absorbs the fermionic charge arises as a hypermagnetic field configuration in
the Standard Model and as the ${\hat{\bf l}}$-texture in $^3$He-A, which is
analogous to the magnetic field.}
\label{PrimordialMagnField1Fig}
\end{figure}

Here we show that the helical instability of the counterflow in $^3$He-A and
the helical instability of the system of the right electrons with nonzero
chemical potential are the same phenomena and thus are described by the
same effective action, which contains the Chern-Simons helical term
(Fig.{ref{PrimordialMagnField1Fig}).

\subsection{Chern-Simons energy term.}

\subsubsection{Chern-Simons term in Standard Model.}

Due to axial anomaly fermionic  charge, say, baryonic or leptonic, can be
transferred to  the ``inhomogeneity'' of the vacuum. This inhomogeneity, which
absorbs the fermionic charge, arises as a helix of magnetic field
configuration, say, hypermagnetic field.  According to axial anomaly
equation (\ref{ChargeParticlProductionGeneral}),  the fermionic charge $Q$
absorbed by the hypermagnetic field, is
\begin{equation}
Q \{{\bf A}_Y\}={1\over 2\pi^2} \left( {\cal Q}{\cal Y}^2, {\cal N}\right){\bf A}_Y\cdot
({\bf\nabla}\times {\bf A}_Y)~~.
\label{FermionicChargeGeneral}
\end{equation}
Let us recall that for the noninteracting fermions this equations reads
\begin{equation}
Q \{{\bf A}_Y\}={1\over 2\pi^2} {\bf A}_Y\cdot
({\bf\nabla}\times {\bf A}_Y) \sum_a C_a Q_a Y_a^2~~,
\label{FermionicCharge}
\end{equation}
where again $a$ marks the fermionic species; $C_a=\pm 1$ is the chirality of the fermion;  
$Y_a$ and $Q_a$ are correspondingly the hypercharge of the $a$-th fermion and its relevant
fermionic charge, whose absorption by the hyperfield we discuss. The fermionic charge
which we interested in is the fermionic number $Q_a=3B_a+L_a$.  

If the fermions are non interacting and their chemical potentials 
$\mu_a$ are nonzero, then the energy functional contains the term which describes the
conservation of
$3B_a+L_a$,
\begin{equation}
-\sum_a\mu_a n_a~~,~~n_a= n_a^{\rm fermion} + n_a  \{{\bf A}_Y\}     ~.
\label{ChemicalPotentialTerm}
\end{equation}
Here $n_a$ is the total charge density $3B_a+L_a$, which is the sum of the
charge density $n_a^{\rm fermion}$   stored by the system of fermionic
quasiparticles  and the  charge density $n_a  \{{\bf A}_Y\}$ stored by the
$U(1)_Y$ gauge field. The latter gives the
Chern-Simons energy of the hypercharge $U(1)_Y$ field in the presence of
nonzero $\mu_a$:  
\begin{equation}
F_{CS}\{{\bf A}_Y\}= -\sum_a\mu_a  n_a  \{{\bf
A}_Y\}= - {1\over 8\pi^2}  {\bf A}_Y\cdot ({\bf\nabla}\times {\bf A}_Y) \sum_a
C_a \mu_a Y_a^2 ~~.
\label{csenergy1}
\end{equation}

As an example let us consider the unification scale, where all the fermions have
the same chemical potential $\mu$, since they can transform to each other at
this scale. From the generating function for the Standard Model in
Eq.(\ref{GeneratingFunctionY}) one has
$\sum_a C_a Y_a^2=\left( {\cal Y}^2, {\cal N}\right)=2N_F$ and thus the Chern-Simons energy
of the hypercharge field at high energy becomes:
\begin{equation}
F_{CS}\{{\bf A}_Y\}=-{\mu N_F\over 4\pi^2}  {\bf A}_Y\cdot
({\bf\nabla}\times {\bf A}_Y)  ~~.
\label{ChernSimonsY}
\end{equation}

\subsubsection{Chern-Simons energy in $^3$He-A.}

Let us consider homogeneous counterflow ${\bf w}={\bf v}_{\rm n}-{\bf
v}_{\rm s}=w\hat{\bf z}$. As will be clear below in
Eq.(\ref{FermionKineticEnergy}), at nonzero $T$ the counterflow orients  the
${\hat{\bf l}}$ vector along the axis of the counterflow, so that the
equilibrium orientations of the ${\hat{\bf l}}$ field are  ${\hat{\bf
l}}_0=\pm {\hat{\bf z}}$.   Since ${\hat{\bf l}}$ is a unit vector,
its variation
$\delta{\hat{\bf l}}\perp {\hat{\bf l}}_0$. In the gauge field analogy, in which
${\bf A}=p_F\delta{\hat{\bf l}}$, this corresponds to the gauge choice
${\bf A}^3=0$. 

In the presence of counterflow the energy of quasiparticles, which enters the equilibrium
distribution function in Eq.(\ref{Equilibrium}), is Doppler shifted by an amount
$-{\bf p}\cdot{\bf w}$, which is $\approx C_a p_F({\hat{\bf l}}\cdot{\bf w})$ near the
two nodes. The distribution function of the low-energy quasiparticles, i.e. in the vicinity of
nodes, acquires the form
\begin{equation}
f_a({\bf p})=\left(\exp {  E({\bf p})- \mu_a\over
T} \pm 1\right)^{-1}~,
\label{EquilibriumChemPot}
\end{equation}
where the effective chemical potential produced by the counterflow:
\begin{equation}
\mu_a= - C_a p_F({\hat{\bf l}}\cdot{\bf w})~~.
\label{ChemicalPotentials}
\end{equation}
$\mu_a$ is the effective chemical potential for quasiparticles. This
 is distinct from the chemical potential $\mu$,
which is the true chemical potential of the original bare particles, atoms of
the underlying liquid, $^4$He atoms in Sec.\ref{LandauKhalatnikovSection} and
$^3$He atoms in Sec.\ref{UniversalityClassesOf}. The latter arises from
the microscopic physics as a result of the conservation of number of atoms. 
The effective chemical potential $\mu_a$ for quasiparticles appears only in the
low energy corner, i.e. in the vicinity of the $a$-th node, since in general
there is no conservation law for the quasiparticles.

The relevant fermionic charge of $^3$He-A, which is anomalously conserved and
which corresponds to the number of fermions, is as in
Eq.(\ref{MomentumProductionGeneral}) the  momentum of quasiparticles along
${\hat{\bf l}}$, i.e. ${\bf P}_a=-C_a p_F{\hat{\bf l}}$.  According to our
analogy the helicity of the effective gauge field ${\bf A}=p_F{\hat{\bf l}}$
must carry the following linear momentum
\begin{equation}
{\bf P} \{{\bf A}\}={1\over 8\pi^2} {\bf A}\cdot
({\bf\nabla}\times {\bf A}) \sum_a C_a {\bf P}_a  e_a^2=-{p_F^3\over 4\pi^2}
{\hat{\bf l}}\left(\delta {\hat{\bf l}}\cdot ({\bf\nabla}\times \delta
{\hat{\bf l}})\right)~~.
\label{FermionicChargeMomentum}
\end{equation}

The total linear momentum of quasiparticles stored both in the heat bath of
quasiparticles (``matter'') and in the texture (``hyperfield'') is thus
\begin{equation}
{\bf P}={\bf P}^{\rm fermion}+{\bf P}^{\rm texture}~= \sum_{\bf p}{\bf p}f({\bf
p}) +{\bf P} \{{\bf A}\}~~.
\label{TotalQuasiparticleMomentum}
\end{equation}
The kinetic energy of the liquid in the presence of the counterflow contains the
term which is equivalent to Eq.(\ref{ChemicalPotentialTerm})
\begin{equation}
-({\bf v}_{\rm n}-{\bf
v}_{\rm s}){\bf P}=-{\bf w} \sum_{\bf p}{\bf p}f({\bf p}) -{\bf w}{\bf P} \{{\bf
A}\}~~. 
\label{CounterflowEnergy}
\end{equation}
The second term in the rhs is precisely the analogue of the Chern-Simons energy
in Eq.(\ref{csenergy1}), which is now stored by the ${\hat{\bf l}}$ field in the
presence of the counterflow:
\begin{equation}
F_{CS}\{\delta {\hat{\bf l}}\}=  
-{1\over 8\pi^2}  {\bf A}\cdot ({\bf\nabla}\times {\bf A}) \sum_a C_a \mu_a
e_a^2 \equiv -{p_F^3\over 4\pi^2} ({\hat{\bf l}}\cdot{\bf w}) \left(\delta
{\hat{\bf l}}\cdot ({\bf\nabla}\times \delta {\hat{\bf l}})\right)~~.
\label{csenergy2}
\end{equation}

\subsubsection{Kinetic energy of counterflow in $^3$He-A and its analog for
chiral fermions.}

The first term in
the rhs of Eq.(\ref{CounterflowEnergy}) together with the quasiparticle energy
$\sum_{\bf p}E({\bf p})f({\bf p})$ gives:
\begin{equation}
  {7 \pi^2 \over 180}{m^*p_F\over \Delta_0^2}  T^4 - {1\over 2} m n_{{\rm
n}\parallel}({\bf w} \cdot {\hat {\bf l}})^2 ~.
\label{FermionKineticEnergy}
\end{equation}
While the first term is the thermal energy of the gapless fermions, the second
term in Eq.(\ref{FermionKineticEnergy}) is with the minus sign the kinetic
energy of the counterflow. Here $n_{{\rm n}\parallel}$ is the density of the
normal component for the flow along
${\hat {\bf l}}$ (see Eq.(\ref{EquilibriumCurrent}) in
Sec.\ref{NormalComponentSection}):
\begin{equation}
n_{{\rm n}ik}=n_{{\rm n}\parallel} \hat l_i \hat l_k + 
n_{{\rm n}\perp} (\delta_{ik} -\hat l_i \hat l_k)~~,~~  
n_{{\rm n}\parallel}
\approx { m^*\over 3m} p_F^3
{T^2\over \Delta_0^2} ~~.
\label{LowTnormalDensity}
\end{equation} 
It is the second term in Eq.(\ref{FermionKineticEnergy}) which provides the
preferred orientation of the ${\hat {\bf l}}$ field by the counterflow
velocity. As will be shown in Sec.\ref{MassHyperphoton}  below it is
responsible for the mass of the gauge field boson. 

The Eq.(\ref{FermionKineticEnergy}) can be written in the relativistic form
applicable both for $^3$He-A and the system of chiral fermions:
\begin{equation}
\Omega(T,\mu_a)= \Omega(T,0) + \Omega'(T,\mu_a)=  {7 \pi^2 \over 180}
\sqrt{-g}   T^4 \sum_a 1 - {\sqrt{-g}
\over 12} T^2 \sum_a \mu_a^2  ~,
\label{FermionEnergyRelativisticForm}
\end{equation}
where in $^3$He-A the determinant of the metric tensor $g_{\mu\nu}$  is
according to Eq.(\ref{MetricAPhaseGeneral}):
\begin{equation}
\sqrt{-g}={1\over c_\parallel
c_\perp^2}={m^*p_F\over \Delta_0^2}~~.
\label{DetG}
\end{equation}
In relativistic system the second term $\Omega'(T,\mu_a)$ in
Eq.(\ref{FermionEnergyRelativisticForm}) is the correction to the  
thermodynamic potential due to the chemical potential if $\mu_a^2\ll
T^2$.  In derivation of Eq.(\ref{FermionKineticEnergy}) we also used the
condition $\mu_a^2\ll T^2$, which corresponds to 
$p_Fw\ll T$.  Under this condition the thermal energy, which is   
$\propto \sqrt{-g}T^4$ in both systems, is dominating.  The minus sign in the
kinetic energy of quasiparticles in Eq.(\ref{FermionKineticEnergy}) occurs
since the counterflow velocity ${\bf w}$ is kept fixed. This corresponds to
the minus sign in the relativistic version,
Eq.(\ref{FermionEnergyRelativisticForm}), where the chemical potentials are
kept fixed.

\subsubsection{Mass of hyperphoton.}\label{MassHyperphoton}

Another term which is important for the consiferation of the helical
instability in $^3$He-A is that which gives the mass of  
``hyperphoton''.  This mass can be obtained
from Eq.(\ref{FermionKineticEnergy}) by expanding  the unit vector ${\hat{\bf
l}}$ up to the second order in deviations: ${\hat{\bf l}}= {\hat{\bf
l}}_0+\delta {\hat{\bf l}}-(1/2){\hat{\bf l}}_0(\delta {\hat{\bf l}})^2$.
Inserting this equation to Eq.(\ref{FermionKineticEnergy}) and neglecting the
terms which does not contain the $\delta{\hat {\bf l}}$ field one obtains the
term whose translation to the relativistic language gives the mass for the
$U(1)_Y$ gauge field:  
\begin{eqnarray} 
F_{\rm mass} ={1\over 4} m n_{{\rm n}\parallel}w^2 ( \delta{\hat {\bf l}}\cdot
\delta{\hat {\bf l}} )
\label{MassEnergy1}\\ \equiv {1\over
12}
\sqrt{-g} g^{ik} A_i A_k {T^2  \over
\Delta_0^2} \sum_a e_a^2\mu_a^2 ~~.
\label{MassEnergy2}
\end{eqnarray}
In $^3$He-A this mass of the  ``hyperphoton'' is physical and important for
the dynamics of the $\hat{\bf l}$-vector.  It is  the gap in the spectrum  of
orbital waves -- propagating oscillations of
$\delta \hat{\bf l}$. This mass appears due to the presence of the counterflow,
which orients $\hat {\bf l}$ and thus provides  the restoring force for 
oscillations of $\delta \hat{\bf l}$. 

In principle, the similar mass can exist for the real hyperphoton. If the Standard Model is
an effective theory, the local $U(1)_Y$ symmetry arises only in the low-energy
corner and thus is approximate. It can be violated (not spontaneously but
gradually) by the higher order terms, which contain the Planck cut-off. Let
us recall that the cut-off parameter $\Delta_0$ does play the part of the
Planck energy scale. The  Eq.~(\ref{MassEnergy2}) suggests that the mass of the
hyperphoton could arise if both the temperature  
$T$ and the chemical potential $\mu_a$ are finite. This mass disappears in the
limit of an infinite cut-off parameter or is negligibly small, if the cut-off is
of Planck scale $E_{\rm P}$.  The $^3$He-A thus provides an  illustration of
how  the renormalizable terms are  suppressed by small ratio of
the energy to the fundamental energy scale of the theory \cite{Weinberg} and
how the terms of order $(T/E_{\rm P})^2$ appear in the effective quantum
field theory \cite{Jegerlehner}.  

On the other hand, the mass term can be obtained in the effective theory
too, if one relates its to the gravity, which is also
determined by the Planck energy scale. The nonzero
chemical potential in Eq.(\ref{FermionEnergyRelativisticForm}) gives the nonzero trace of
energy momentum tensor for quasiparticles (matter) according to  
Eq.(\ref{QuasipStressTensorRel2}):
\begin{equation}
~T^\mu{}_\mu=2\Omega'(T,\mu_a)  ~.
\label{TraceStressTensor}
\end{equation}
Using the Einstein equation, $R=-8\pi G T^\mu{}_\mu$, the
Eq.(\ref{MassEnergy2}) can be transformed to
\begin{equation} 
F_{\rm mass} = {1 \over
16\pi G\Delta_0^2}
\sqrt{-g} g^{ik} A_i A_k R {\sum_a e_a^2\mu_a^2 \over \sum_a \mu_a^2 } ~~.
\label{MassEnergy3}
\end{equation}
The microscpic parameters $\Delta_0^2$ and $G\sim \Delta_0^{-2}$ cancel each other, so that
no cut-off parameter enters the mass term. This effective action is general covariant and
scale invariant but it violates the gauge symmetry. Applying this for the hypermagnetic
field  ${\bf A}_Y$ with charges $e_a=Y_a$, and identifying $G\Delta_0^2=GE_{\rm P}^2=1$
one obtains for the case of equal chemical potentials:
\begin{equation} 
F_{\rm mass} = {5 \over 24}{1\over
16\pi}
\sqrt{-g} g^{\mu\nu} A_{\mu Y}A_{\nu Y}  R   ~~.
\label{MassHyperphotonTerm}
\end{equation}

In principle the term $\sqrt{-g} T^4  g^{ik} A_i A_k /
E_{\rm P}^2  $ is also possible, which gives the hyperphoton mass of order $ T^2/E_{\rm
Planck}$.

\subsection{Helical instability and ``magnetogenesis'' by chiral
fermions.}\label{MagnetogenesisChiralFermions}

The Chern-Simons term in Eqs.~(\ref{csenergy1}) and (\ref{csenergy2}) is odd
under spatial parity tranformation and thus can have a negative sign for the
properly chosen field or texture.  Thus one can have an energy gain from the
transformation of the fermionic charge  to the $U(1)$ gauge field. This is the
essence of the Joyce-Shaposhnikov scenario for the generation of primordial
magnetic field
\cite{JoyceShaposhnikov,GiovanniniShaposhnikov}.  In
$^3$He-A language this process describes the collapse of the counterflow, where the relevant
fermionic charge is the momentum, towards the formation of ${\hat {\bf l}}$-texture.   Such
a collapse of quasiparticle momentum 
was recently observed in  the rotating cryostat of the Helsinki Low
Temperature Laboratory \cite{Experiment,LammiTalk} (Fig.
\ref{PrimordialMagnField2Fig}). 

\subsubsection{Helical instability condition.}

The instability can be found by investigation of the eigenvalues of 
the quadratic form describing the energy in terms of the
deviations from the homogeneous counterflow, which play the
part of the ``hypermagnetic'' field
${\bf A}=p_F\delta{\hat{\bf l}}$. The quadratic form contains Eqs. 
(\ref{EMLagrangianAPhase}), (\ref{csenergy2}), and  (\ref{MassEnergy1})
describing correspondingly the ``magnetic'' energy, the Chern-Simons energy, 
and the term giving the mass of the  ``hyperphoton''. In $^3$He-A notations
this is:
\begin{equation} 
{{12\pi^2}\over {p_F^2v_F}} F\{\delta {\hat{\bf l}}\}= 
(\partial_z \delta \hat {\bf l})^2 {\rm ln}
 {\Delta_0\over T }  -3 m^*w
\delta {\hat{\bf l}}\cdot (\hat{\bf z}\times \partial_z\delta {\hat{\bf l}}) 
 + 
\pi^2  ( m^*w)^2  
{T^2\over \Delta_0^2}  ( \delta{\hat {\bf l}} )^2
\label{EnergyHyperMagnField}
\end{equation} 
or after the rescaling of the coordinates $\tilde z=zm^*w/\hbar$
\begin{equation} 
{4F\{\delta {\hat{\bf l}}\}\over C_0 m^*w^2} = 
(\partial_{\tilde z} \delta \hat {\bf l})^2 {\rm ln}
 {\Delta_0\over T }  -3  
\delta {\hat{\bf l}}\cdot (\hat{\bf z}\times \partial_{\tilde z}\delta
{\hat{\bf l}}) 
 + 
\pi^2   
{T^2\over \Delta_0^2}  ( \delta{\hat {\bf l}} )^2
\label{EnergyHyperMagnField2}
\end{equation}

The quadratic form in Eq.(\ref{EnergyHyperMagnField2}) becomes negative and
thus the uniform counterflow becomes unstable towards the nucleation of the
$\hat {\bf l}$-texture if
\begin{equation}
 {T^2\over \Delta_0^2} \ln {\Delta_0\over T}< {9\over
4\pi^2  } ~~.
\label{StabilityCondition2}
\end{equation}
If this condition is fulfilled, the instability occurs for any value of the
counterflow.  

In relativistic theories, where the temperature is always smaller than the
Planck cut-off $\Delta_0$, the  condition corresponding to
Eq.(\ref{StabilityCondition2}) is always fulfilled. Thus the excess of the
fermionic charge is always unstable towards nucleation of the hypermagnetic
field, if the fermions are massless, i.e. above the electroweak transition. In
the scenario of the magnetogenesis developed by Joyce and Shaposhnikov
\cite{JoyceShaposhnikov,GiovanniniShaposhnikov}, this instability is
responsible for the genesis of the hypermagnetic field well above the
electroweak transition. The role of the subsequent electroweak transition is
to transform this hypermagnetic field  to the conventional 
(electromagnetic $U(1)_Q$) magnetic field due to the electroweak symmetry
breaking. 

\subsubsection{Experimental ``magnetogenesis'' in  rotating $^3$He-A.}

In $^3$He-A the helical instability is suppressed by another mass of the 
``hyperphoton'', which comes from the spin-orbit interaction $-g_D({\hat{\bf
l}}\cdot {\hat{\bf d}})^2$ between the orbital vector ${\hat{\bf l}}$
and the vector ${\hat{\bf d}}$, describing the spin part of the order
parameter. This gives an additional restoring force acting on ${\hat{\bf
l}}$, and thus the additional mass of the gauge field: 
$-g_D({\hat{\bf l}}\cdot {\hat{\bf d}})^2= -g_D({\hat{\bf l}}_0\cdot {\hat{\bf
d}})^2 +(1/2)g_D(\delta{\hat{\bf l}})^2$, which is independent on the
counterflow. As a result the helical instability occurs only if the counterflow 
(the corresponding chemical potential
$\mu_a$) exceeds the critical treshold, determined by the additional mass of
the ``hyperphoton'', which is of order $M_{\rm hp}\sim
10^{-3}\Delta_0$. This is observed  experimentally \cite{Experiment}  (see
Fig.\ref{PrimordialMagnField2Fig}). When the counterflow in the rotating
vessel exceeds this treshold, the intensive formation of the ${\hat{\bf l}}$
texture by helical instability is detected by NMR. This corresponds to the
formation of the hypermagnetic field according to our analogy. The only
difference from the Joyce-Shaposhnikov scenario is that the mass of the
``hyperphoton'' provides the treshold for the helical instability.

\begin{figure}[t]
\centerline{\includegraphics[width=\linewidth]{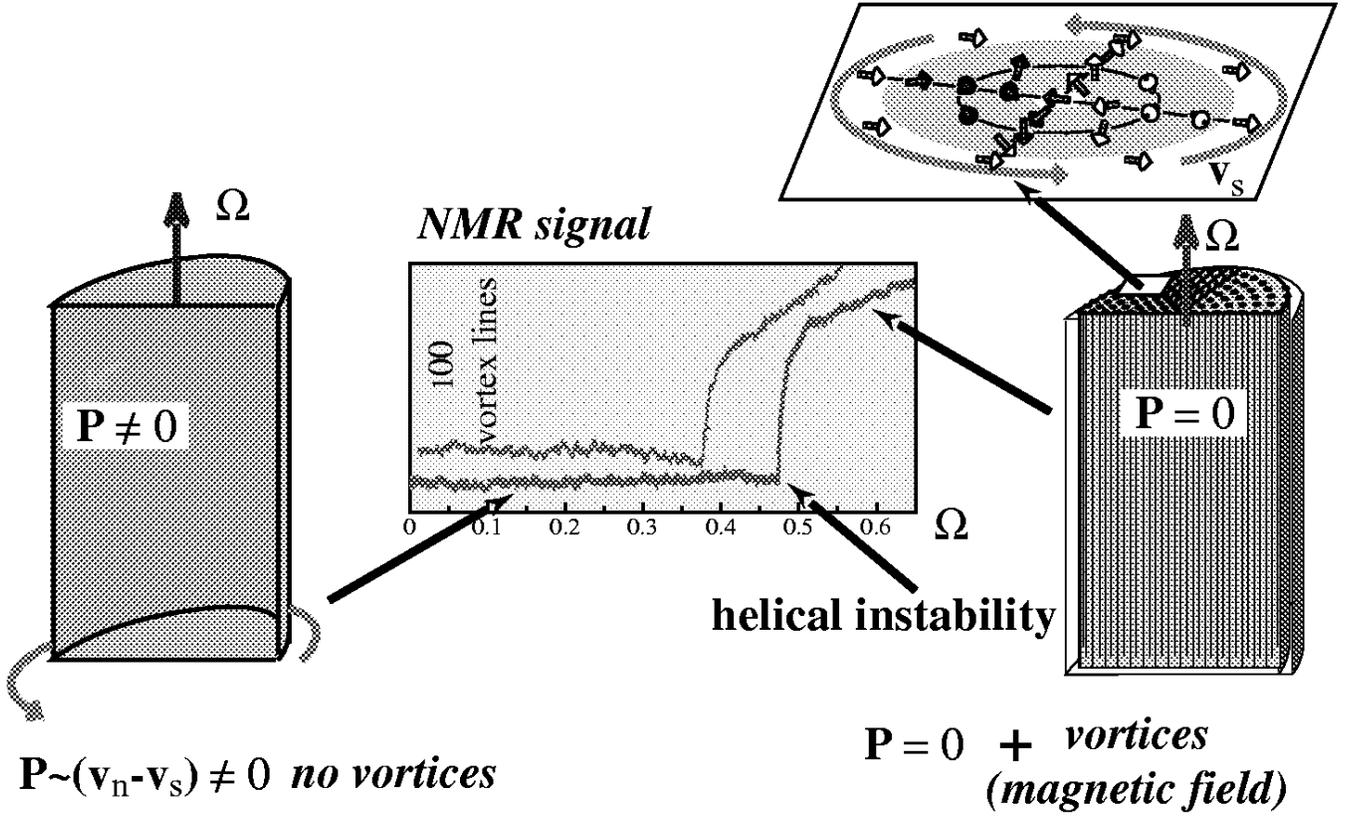}}
\medskip
\caption{{\it left}) The vortex-free state in the vessel rotating with
angular  velocity ${\bf \Omega}$ contains a counterflow, 
${\bf w}= {\bf v}_{\rm n}-{\bf v}_{\rm s} ={\bf \Omega}\times{\bf r}\ne 0$,  
since the average velocity of quasiparticles (the normal component)
is ${\bf v}_{\rm n}={\bf \Omega}\times {\bf r}$ and is not equal to 
the velocity of the superfluid vacuum, which is at rest, 
${\bf v}_{\rm s}=0$.  In the counterflow state, the quasiparticles have a 
net momentum $\propto n_{\rm n}{\bf w}$. In the presence of counterflow ${\bf
w}$, the energy of quasiparticles is Doppler shifted by the amount
${\bf p}\cdot{\bf w}\approx \pm p_F({\hat{\bf l}}_0\cdot{\bf w})$.
The counterflow therefore produces what would be an effective chemical
potential in particle physics. For right-handed particles, this is
$\mu_R= p_F({\hat{\bf l}}_0\cdot{\bf w})$ and for left-handed particles it
is $\mu_L=-\mu_R$. Thus rotation produces an excess of the
fermionic charge quasiparticle momentum, which is analogous to the excess of
the leptonic charge of chiral right-handed electrons if their  chemical
potential $\mu_R$ is nonzero. {\it middle}  When the excess of quasiparticle
momentum reaches the critical value, the helical instability occurs, which is
marked by abrupt jump of the intensity of the NMR satellite peak from zero,
which signals appearance of the 
${\hat{\bf l}}$-texture, playing the part of the magnetic field. {\it right}
The final result of the helical instability is a periodic array of continuous
$n_1=2$ vortices ({\it top}).  Formation of these vortices leads to
the effective solid body rotation of superfluid vacuum with 
$<{\bf v}_{\rm s}>\approx {\bf
\Omega}\times {\bf r}$. This essentially decreases the counterflow and thus the
fermionic charge. Thus a part of the fermionic charge is transformed into
``hypermagnetic'' field. }
\label{PrimordialMagnField2Fig}
\end{figure}

In principle, however the similar treshold can appear in the Standard Model if
there is a small ``non-renormalizable'' mass of the hyperphoton, $M_{\rm hp}$,
produced by the Planck-scale physics. In this case the decay of the fermionic
charge stops, and thus the excess of, say, the baryionic charge is not  washed
out any more, when the chemical potential of fermions becomes comparable with
$M_{\rm hp}$.  The observed baryon asymmetry would be achieved if the initial
mass of the hyperhoton at the electroweak temperature, $T\sim E_{\rm ew}$, is
$M_{\rm hp} \sim 10^{-9} E_{\rm ew}$. This is however too large compared with
the possible hyperphoton mass, discussed earlier, which is of order   
$ T^2/E_{\rm P}\sim E_{\rm ew}^2/E_{\rm P}$.

\subsection{Mixed axial-gravitational Chern-Simons term.}\label{MixedAxialGravitational}

\subsubsection{Parity violating current}

The chiral anomaly problem can be mapped to the angular momentum paradox in
$^3$He-A. To relate them let us consider the parity effects which occur for the rotating
chiral fermions. The macroscopic parity violating effects in a rotating system with chiral
fermions was first discussed by Vilenkin in Ref.
\cite{Vilenkin79}. The angular velocity of rotation ${\bf \Omega}$
defines the preferred direction of
polarization, and right-handed fermions move in the direction
of their spin.  As a result, such fermions develop a current
parallel to ${\bf \Omega}$.  Similarly, left-handed fermions
develop a current antiparallel to ${\bf \Omega}$.  The corresponding
current density was calculated in \cite{Vilenkin79}, assuming thermal
equilibrium at temperature $T$ and chemical potential of the fermions
$\mu$.  For right-handed fermions, it is given by
\begin{equation}
{\bf j}=\left({T^2\over{12}}+{\mu^2\over{4\pi^2}}\right){\bf\Omega}.
\label{PVC}
\end{equation}
The current ${\bf j}$ is a polar vector, while the angular velocity
${\bf\Omega}$ is an axial vector, and thus Eq.(\ref{PVC}) violates the
reflectional symmetry. 

If the current (\ref{PVC}) is coupled to a
gauge field $A^\nu$, the appropriate term in the Lagrangian density is
\begin{equation}
 L ={1\over c^2}{\bf\Omega}\cdot {\bf A}\left({T^2\over{12}}\sum_a
e_aC_a+{1\over{4\pi^2}}\sum_a e_aC_a\mu_a^2\right)  ~ ,
\label{calL}
\end{equation}
where $e_a$ are the corresponding couplings with the gauge field.
Since the rotation can be described in terms of metric, this represents the
mixed axial-gravitational Chern-Simons term in the effective action of
Standard Model
\cite{VolovikVilenkin}. After expressing it in the covariant form, this term
can be applied to
$^3$He-A too.

\subsubsection{Orbital angular momentum and free energy}

Let us consider a stationary liquid $^3$He-A in a vessel rotating with
angular velocity ${\bf\Omega}$ at a nonzero temperature.  We assume a
spatially homogeneous vector ${\hat{\bf l}}=\hat{\bf z}$ oriented along the
rotation axis. In $^3$He-A this can be achieved in the parallel-plane
geometry, while in the layered oxide
superconductor Sr$_2$RuO$_4$, which is believed to be a triplet superconductor
with a $^3$He-A-like order parameter,   the ${\hat{\bf
l}}$-vector is always fixed along the normal to the layers \cite{Rice,Ishida}.
The superfluid component (vacuum) is assumed to be at rest, ${\bf v}_{\rm
s}=0$, while the normal component  circulates in
the plane perpendicular to
$\hat{\bf l}$ performing the solid-body like rotation with the velocity ${\bf
v}_{\rm n}={\bf\Omega}
\times {\bf r}$. This state corresponds to the true thermodynamic equilibrium
in rotating vessel, though it is rather local than global 
minumum of the thermodynamic potential in rotating frame. The absolute
minimum would correspond to the system of quantized vortices in rotating
vessel, but since the energy barrier between the states with different
number of vortices is high, while the temperture is low, the probability
of the thermally activating transitions between the states (as well as
quantum tunneling) is less than
$\exp (- 10^{6})$. This is an advantage of superfluid $^3$He,
which allows us to support the states with given number of vortices,
including the vortex-free state \cite{SingleVortNucl}.

The value of the angular momentum of a rotating $^3$He-A has been a
subject of a long-standing controversy (for a review see
\cite{VollhardtWolfle,Exotic}).
Different methods for calculating the angular momentum give results
that differ by many orders of magnitude.  The result is also sensitive
to the boundary conditions, since the angular momentum in the liquid is not
necessarily the local quantity, and to whether the state is strictly
stationary or
has a small but finite frequency.  This is often referred to as the angular
momentum paradox.  The paradox is related to the axial anomaly induced by
chiral
quasiparticles and is now reasonably well understood. 

At $T=0$ the total angular momentum of the stationary liquid with
homogeneous ${\hat{\bf l}}=const$ is the same as obtained from the following
angular momentum density
\begin{equation}
 {\bf L} ={\hbar\over 2} ~\hat{\bf l}~ n ~.
 \label{MomentumDensityT=0}
\end{equation}
The physical meaning of the total angular momentum of
$^3$He-A is: each atom of the superfluid vacuum carries the angular
momentum $\hbar/2$ in the direction of $\hat{\bf l}$. This is in accordance
with the structure of the order parameter in Eqs. (\ref{PWave}) and
(\ref{APhase}) which state that the momentum dependence of the Cooper pair of
two $^3$He atoms is $\propto (p_x+ip_y)$, and this corresponds to the orbital
angular momentum $L_z=\hbar$ per Cooper pair. The
Eq.(\ref{MomentumDensityT=0}) is, however, valid only for the static angular
momentum. The dynamical angular momentum is much smaller:  ${\bf L}_{\rm dyn}
={\hbar\over 2} ~\hat{\bf l}~ (n-C_0)$ (let us recall that $(n-C_0)/n\sim
c_\perp^2/c_\parallel^2 \sim 10^{-6}$).  The presence of the anomaly parameter
$C_0$ in this almost complete cancellation of the dynamical angular momentum
reflects the same crucial role of the axial anomaly as in the ``baryogenesis''
by moving texture discussed in Sec. \ref{ChiralAnomaly}.

Now let us consider the nonzero temperature. According to Kita conjecture
\cite{Kita}, which was supported by his numerical calculations, the
extension of the total angular momentum of the stationary liquid with
${\hat{\bf l}}=const$ to $T\neq 0$ is 
\begin{equation}
 {\bf L}(T)={\hbar\over 2} ~\hat{\bf l}~ n_{{\rm s}\parallel}(T)~.
 \label{MomentumDensity}
\end{equation}
Here  $n_{{\rm s}\parallel}(T)$ is the temperature dependent density of the
superfluid component when it flows along   $\hat{\bf l}$; $n_{{\rm
s}\parallel}(T)=n-n_{{\rm n}\parallel}(T)$, where $n_{{\rm n}\parallel}(T)$ 
is given by  Eq.(\ref{LowTnormalDensity}). The Eq.(\ref{MomentumDensity})
suggests that $n_{{\rm s}\parallel}(T)/2$ is the effective number of the
``superfluid'' Cooper pairs which contribute to the angular momentum. 

The contribution of the angular momentum in Eq.(\ref{MomentumDensity}) to the
free energy density in $^3$He-A in the container  rotating with angular
velocity ${\bf\Omega}$ is
\begin{equation}
   \epsilon({\bf L})= -{\bf\Omega} \cdot {\bf L}(T)=-{\bf\Omega} \cdot {\bf L}(T=0)+
{\hbar\over 2} ({\bf\Omega}\cdot \hat{\bf l})  n_{{\rm n}\parallel}(T)~.
 \label{EnergyDensity}
\end{equation}
The first (zero-temperature) term on the right-hand side of
(\ref{EnergyDensity}) comes from the microscopic (high-energy) physics
and thus has no analogue in effective field theory, and we disregard it in
what follows. However the temperature correction, represented by the 2nd term, 
comes from the low-energy chiral quasiparticles, which comprise the normal
component in $^3$He-A, and thus is within the action of the effective field
theory. 

The numerical calculations in \cite{Kita} were made in the Fermi gas
approximation, i.e. under assumption that the Fermi liquid corrections are
absent and thus the Fermi-liquid mass $m^*=p_F/v_F$ is equal to the bare mass
$m$ of the $^3$He atom. In general case, when $m^*\neq m$, the second term in
Eq.(\ref{EnergyDensity}) at low $T$ must be modified
\begin{equation}
   \epsilon =   {\hbar\over 2} {m\over m^*}  n_{{\rm
n}\parallel}(T) ({\bf\Omega}\cdot \hat{\bf l})  ~.
 \label{EnergyDensityRelevant}
\end{equation} 
This modification follows from the comparison with the relativistic theory and
also from the fact that in the effective theory such microscopic parameter as
the mass $m$ of the $^3$He atoms never enters, since it is not contained in
the quasiparticle energy in Eq.(\ref{BogoliubovNambuEnergyAPhase})  (see
discussion in Sec. \ref{Improved3He}). As follows from
Eq.(\ref{LowTnormalDensity}) for the normal component density
$n_{{\rm n}\parallel}(T)$, the mass $m$ is cancelled in
Eq.(\ref{EnergyDensityRelevant}.

\subsubsection{Effective
Chern-Simons action for $\mu_a=0$ and  $T\neq 0$}

To translate Eq.(\ref{EnergyDensityRelevant}) to the language of
relativistic theories we use the dictionary in Eqs.(\ref{MetricAPhaseGeneral})
and (\ref{4Potential}). It is convinient to use the reference frame rotating
with the container, since in this frame all the fields including the effective
metric are stationary in equilibrium. In this rotating frame the velocity of
the normal component ${\bf v}_{\rm n}=0$, while the superflid velocity in this
frame is ${\bf v}_{\rm s} =-{\bf\Omega}
\times {\bf r}$. The mixed components of the metric tensor
$g^{0i}=({\bf\Omega} \times {\bf r})_i$, so that the angular
velocity is expressed through the effective gravimagnetic field
\cite{GravimagneticMonopole}:
\begin{equation}
{\bf B}_g={\bf\nabla}\times {\bf g}=2{{\bf\Omega}\over c_\perp^2}~,  ~{\bf
g}\equiv
g_{0i}= {v_{{\rm s}i}\over c_\perp^2}.
 \label{GravimagneticField}
\end{equation}
Here, we have made the following assumptions: (i) $\Omega  r <
c_\perp$ everywhere in the vessel, i.e. the counterflow velocity ${\bf
v}_{\rm n} -{\bf v}_{\rm s}$ is smaller than the pair-breaking critical
velocity $c_\perp=\Delta_0/p_F$ (the transverse ``speed of light''). This
means that there is no  region in the vessel where particles can have negative
energy (ergoregion). Effects caused by the ergoregion in rotating superfluids
\cite{CalogeracosVolovik} are discussed below in Sec.
\ref{VacuumUnderRotation.}. (ii) There are no vortices in the container. This
is typical for superfluid $^3$He, where the critical velocity for nucleation
of vortices is comparable to the pair-breaking velocity
$c_\perp=\Delta_0/p_F$ \cite{SingleVortNucl}. Even in the geometry when the
$\hat{\bf l}$-vector is not fixed, the observed critical velocity in $^3$He-A
was found to reach 0.5 rad/sec \cite{Experiment}. For the geometry with fixed
$\hat{\bf l}$, it should be comparable with the critical velocity in
$^3$He-B.   

Using Eq.(\ref{GravimagneticField}) one obtains that both the
Eq.(\ref{EnergyDensityRelevant}) in $^3$He-A and the first term in  
Eq.(\ref{calL}) for the Standard Model can be presented in a unified form:
\begin{eqnarray}
\nonumber L_{\rm Mixed~CS}(T, \mu_a=0)= {1 \over 24} \sum_a e_aC_a~ T^2{\bf
A}\cdot {\bf B}_g= \\={1 \over 24}
\sum_a e_aC_a~ ~T^2 e^{ijk} A_i
\nabla_j g_{k0}~.
 \label{ChernSimons1}
\end{eqnarray}
This equation (\ref{ChernSimons1}) does not contain
explicitly any material parameters of the system, such as the ``speeds of
light'' $c_\parallel$ and $c_\perp$ in $^3$He-A or the real speed of light $c$
in the Standard Model, and thus is equally applicable to any representative of
the Universality class of Fermi points.  

Eq.(\ref{ChernSimons1}) is not Lorentz invariant, but this is not
important here
because the existence of a heat bath
does violate the Lorentz invariance, since it provides a distinguished
reference frame. To restore the Lorentz invariance and also the general
covariance one must introduce the 4-velocity ($u^{\mu}$) and/or  4-temperature
($\beta^{\mu}$) of the heat bath fermions:
\begin{equation}
 L_{\rm Mixed~CS}(T, \mu_a=0)  =  {1 \over
24}
\sum_a e_aC_a~ ~(\beta_\mu \beta^\mu)^{-2} e^{\alpha\beta\mu\nu}\beta_\alpha
\beta^\beta A_\mu \partial_\nu g_{\gamma\beta}   ~.
 \label{ChernSimonsCovar}
\end{equation}
Note that this can be applied only to the fully equilibrium situations in
which $\beta_{\nu;\mu} +\beta_{\mu;\nu}=0$, otherwise the invariance
under the gauge transformation $A_\mu\rightarrow A_\mu +\partial_\mu\alpha$ is
violated.

\subsubsection{ Finite density of states and Chern-Simons term in the presence of
counterflow.}

To find the $^3$He-A counterpart of the second term in Eq.(\ref{calL}),  
let us consider the opposite case $T=0$ and $\mu_a\neq 0$.  According to
Eq.(\ref{ChemicalPotentials}) the counterpart of the chemical potentials
$\mu_a$ of relativistic chiral fermions is the superfluid-normal counterflow
in $^3$He-A.  The relevant counterflow, which does not violate the symmetry
and the local equilibrium condition, can be produced by superflow
along the axis of the rotating container.  Note that we approach the
$T\rightarrow 0$ limit in such a way that the rotating reference frame
is still active and determines the local equilibrium states. In the case of
rotating container this is always valid because of the interaction of the
liquid with the container walls. For the relativistic counterpart we
must assume that there is still a nonvanishing rotating thermal bath of
fermionic excitations. This corresponds to the case when the condition
$\omega\tau \ll 1$ remains valid, despite the divergence of
the collision time $\tau$. 

It follows from Eq.(\ref{ChemicalPotentials}) that at $T=0$ the
energy stored in the system of chiral fermions with the chemical potentials
$\mu_a$ and the energy of the counterflow along the $\hat{\bf l}$-vector are
described by the same thermodynamic potential: 
\begin{eqnarray}
\tilde\epsilon = \epsilon-\sum_a\mu_a n_a\equiv \sum_{\bf p} E({\bf p})f({\bf p}) -  {\bf
P}\cdot
  ({\bf v}_{\rm n}-{\bf v}_{\rm s}) =
\label{CounterflowEnergyEquiv1}
\\
 - { \sqrt{-g}\over 12\pi^2} \sum_a\mu_a^4 \equiv
-{m^*p_F^3\over 12 \pi^2 c_\perp^2} \left({\hat{\bf l}} \cdot({\bf v}_{\rm
s}-{\bf
v}_{\rm n})\right)^4  ~.
\label{CounterflowEnergyEquiv2}
\end{eqnarray}
The contribution in Eq.(\ref{CounterflowEnergyEquiv2})  comes from the
fermionic quasiparticles, which at $T\rightarrow 0$ occupy the negative
energy levels, i.e. the levels with $\tilde E<0$, where $\tilde E= E({\bf p}) + 
{\bf p}\cdot ({\bf v}_{\rm s}-{\bf v}_{\rm n})$.  In the relativistic
counterpart these are the energy states with $\tilde
E_a({\bf p})=cp-\mu_a <0$, i.e. the states inside the Fermi spheres 
$cp_{Fa}=\mu_a$. 

Variation of Eq.(\ref{CounterflowEnergyEquiv1}) with respect to 
${\bf v}_{\rm n}$ gives the mass current along the
${\hat{\bf l}}$-vector carried by these fermions:
\begin{equation}
mJ_{{\rm q}\parallel}\equiv {\bf P}\cdot {\hat{\bf l}} =  -{d  \tilde\epsilon \over
dv_{{\rm n}\parallel}}={m^*p_F^3\over 3 \pi^2 c_\perp^2} \left({\hat{\bf l}}
\cdot({\bf v}_{\rm s}-{\bf
v}_{\rm n})\right)^3~~.
\label{ZeroTCountercurrent}
\end{equation}
This shows that in the presence of a superflow with respect to the heat bath
the normal component density of superfluid $^3$He-A is nonzero even in the
limit $T\to 0$ 
\cite{Muzikar1983}:
\begin{equation}
 n_{{\rm n}\parallel}(T\to 0)={d  J_{{\rm q}\parallel}\over dv_{{\rm
n}\parallel}} =
  {m^*p_F^3\over 3\pi^2 mc_\perp^2}  ({\hat{\bf l}} \cdot({\bf v}_{\rm s}-{\bf
v}_{\rm n}))^2
  ~~.
\label{ZeroTnormalDensity}
\end{equation}
The nonzero density of the normal component at $T\rightarrow 0$ results from
the finite density of fermionic states  $N(\omega)=2\sum_a\sum_{\bf p}\delta
(\omega -\tilde E_a({\bf p}))$
 at $\omega=0$. This density of states has the same form for the system of
chiral relativistic fermions with nonzero chemical potential, where it is the
density of states on the Fermi surfaces, and for $^3$He-A, where also the
Fermi points are transformed to Fermi surfaces in the presence of
the counterflow:
\begin{equation}
N(0)=
{\sqrt{-g}\over \pi^2}\sum_a\mu_a^2\equiv  {p_Fm^*\over \pi^2 c_\perp^2}  
({\hat{\bf l}} \cdot({\bf v}_{\rm s}-{\bf v}_{\rm n}))^2
\label{DOSCounterflow}
\end{equation}

Since the counterflow leads to finite normal component density at
$T\rightarrow 0$ one can apply the Eq.(\ref{EnergyDensityRelevant}) for the
energy density describing the interaction of the orbital angular momentum with
rotation velocity. Then from Eq.(\ref{ZeroTnormalDensity}) and from the
$^3$He-A/relativistic-system dictionary it follows that the
Eq.(\ref{EnergyDensityRelevant}) at $T\rightarrow 0$ is nothing but the mixed
Chern-Simons term in the form
\begin{equation}
 L_{\rm Mixed~CS}(\mu_a,T=0)={1\over {8\pi^2}}\sum_a e_aC_a\mu_a^2 {\bf
A}\cdot {\bf B}_g~.
 \label{ChiralTerm2}
\end{equation}
This term is just the  second term in Eq.(\ref{calL}).

Thus the general form of the mixed Chern-Simons term, which is valid for
both systems of Fermi-point universality class and which includes both the
temperature $T$ and chemical potentials (in the Standard Model) or the 
counterflow velocity (in $^3$He-A) is
\begin{equation}
 L_{\rm Mixed~CS}(\mu_a,T)= {\bf
A}\cdot {\bf B}_g\left( {1\over {8\pi^2}}\sum_a e_aC_a\mu_a^2  
+ {T^2 \over 48}\sum_a e_aC_a~\right)~.
 \label{ChiralTermGeneral}
\end{equation}

\subsubsection{Unification of conventional and mixed CS terms.}

The form of Eq.(\ref{ChiralTermGeneral}) is similar to that of the induced
Chern-Simons term in Eq.(\ref{csenergy1}), which has been extensively
discussed both in the context of chiral fermions in relativistic theory
\cite{Vilenkin80,Redlich,JackiwKostelecky,Andrianov} and in $^3$He-A
\cite{LammiTalk}.  The main difference between (\ref{csenergy1}) and
(\ref{ChiralTermGeneral}) is that ${\bf B_g=\nabla\times g}$ is the
gravimagnetic field, rather than the magnetic field ${\bf B}$
associated with the potential ${\bf A}$.  Hence the name ``mixed
Chern-Simons term''.

Comparison of conventional and mixed  Chern-Simons  terms     
suggests that these two $CPT$-odd terms can be united if one uses the
Larmor theorem and introduces the combined fields:
\begin{equation}
{\bf A}_a=e_a {\bf A}+{1\over 2} \mu_a{\bf g}~,~{\bf
B}_a
=\nabla\times {\bf A}_a ~.
\label{MagneticGravimagnetic}
\end{equation}
Then the general form of the Chern-Simons $CPT$-odd term at $T=0$ is
\begin{equation}
{1\over 4\pi^2} \sum_a \mu_a C_a {\bf A}_a\cdot {\bf B}_a   ~.
\label{ChernSimonsGeneralGeneral}
\end{equation}

\subsubsection{Possible experiments in condensed matter.}

The parity-violating currents (\ref{PVC}) could be induced in
turbulent cosmic plasmas and could play a role in the origin of cosmic
magnetic fields \cite{Leahy}.  The corresponding $^3$He-A effects
are less dramatic but may in principle be observable.

Although the mixed Chern-Simons terms have the same form in
relativistic theories and in $^3$He-A, their physical manifestations
are not identical.
In the relativistic case, the electric current of chiral
fermions is obtained by
variation with respect to
$ {\bf A}$, while in $^3$He-A case the observable effects are obtained by
variation of the same term but with respect to $^3$He-A observables.
For example, the
expression for the current of $^3$He atoms is obtained by variation of
Eq.(\ref{ChiralTerm2}) over ${\bf
v}_{\rm n}$. This leads to
an extra particle current along the rotation axis, which is odd in
${\bf\Omega}$:
\begin{equation}
\Delta {\bf J}_{\rm q}({\bf\Omega})= {p_F^3\over \pi^2 }  {\hat{\bf
l}}~({\hat{\bf l}}\cdot({\bf v}_{\rm s}-{\bf v}_{\rm n})) ~ {{\hat{\bf l}
}\cdot
{\bf\Omega}\over m c_\perp^2}~.
 \label{Current}
\end{equation}
Eq.(\ref{Current}) shows that there is an ${\bf\Omega}$ odd contribution to
the normal component density at $T\to 0$ in $^3$He-A:
\begin{equation}
\Delta n_{{\rm n}\parallel}({\bf\Omega}) = {\Delta J_{\rm
q}({\bf\Omega})\over v_{{\rm n}\parallel}-v_{{\rm s}\parallel}}
=  {p_F^3\over \pi^2 }  {{\hat{\bf l}}\cdot
{\bf\Omega}\over  m c_\perp^2} ~.
\label{NormalDensityInRotation}
\end{equation}
The sensitivity of the normal component density to the direction of
rotation
is the counterpart of the parity violation effects in relativistic
theories with chiral fermions.
It should be noted though that, since ${\bf{\hat l}}$ is an axial
vector, the right-hand sides of (\ref{Current}) and
(\ref{NormalDensityInRotation}) transform, respectively, as a polar
vector and a scalar, and thus (of course) there is no real parity violation
in $^3$He-A. However, a nonzero expectation value of the axial vector of  the
orbital angular momentum ${\bf L} =(\hbar/2)n_{{\rm s}\parallel}(T){\bf {\hat
l}}$  does indicate a {\it spontaneously} broken reflectional symmetry, and an
 internal observer ``living'' in a $^3$He-A  background with a fixed
 ${\bf{\hat l}}$ would observe parity-violating effects.

The contribution (\ref{NormalDensityInRotation}) to
the normal component density can have arbitrary sign depending on
the sense of rotation with respect to ${\hat{\bf l}}$. This however does not
violate the general rule that the overall normal component density must be
positive:  The rotation dependent current $\Delta{\bf J}_{\rm
q}(\vec\Omega)$ was calculated as
a correction to the rotation independent current in
Eq.(\ref{ZeroTCountercurrent}). This means that we used the condition
$\hbar\Omega \ll m (v_{{\rm s}\parallel} - v_{{\rm n}\parallel})^2 \ll m
c_\perp^2$. Under
this condition the overall normal density, given by the sum of
(\ref{NormalDensityInRotation}) and
(\ref{ZeroTnormalDensity}), remains positive.

The ``parity'' effect in Eq.(\ref{NormalDensityInRotation})  is not very
small. The rotational contribution to the normal component density
normalized to the  density of the $^3$He atoms is $\Delta n_{{\rm
n}\parallel}/n =3\Omega/m c_\perp^2$ which is $ \sim 10^{-4}$ for
$\Omega \sim 3$ rad/s. This is within the
resolution of the vibrating wire detectors.

We finally mention a possible application of our results to the
 superconducting Sr$_2$RuO$_4$ \cite{Rice}.
An advantage of using superconductors is that the particle current
$\Delta{\bf J}_{\rm q}$ in Eq.(\ref{Current})  is accompanied by the
electric current $e\Delta{\bf J}_{\rm q}$, and can be measured directly.
An observation in Sr$_2$RuO$_4$ of the analogue of the parity violating
effect that we discussed here  (or of the other effects coming from the
induced Chern-Simons terms \cite{Goryo,Ivanov}), would be an unquestionable
evidence of the chirality of this superconductor.

 \section{Fermion zero modes and spectral flow in the vortex core.}
\label{FermionZeroModesOnVort}

As we discussed in Sec. \ref{ChiralAnomaly} the massless chiral fermions
influence the dynamics of the continuous vortex texture (stringy texture) due to
axial anomaly providing the flow of the momentum from texture to the heat bath
of fermions. Here we discuss the same phenomenon, which occurs in the core of
conventional singular vortices. Though the fermions can be massive in bulk
superfluid outside the core, they are gapless or have a tiny gap in the
vortex core (Fig. \ref{VortexVsStringFig}). These fermions living in the 
core of the vortex (or string) do actually the same job as chiral fermions in
the stringy texture. Though the spectral flow by fermion zero modes in the
core is not described by the Adler-Bell-Jackiw axial anomaly
equation (\ref{ChargeParticlProduction}), and is typically smaller than in
textures, in the regime of the maximal spectral flow one obtains the same
equations (\ref{SpFlowForce}-
\ref{C0}) for the generation of the momentum by moving vortex.

\begin{figure}[t]
\centerline{\includegraphics[width=\linewidth]{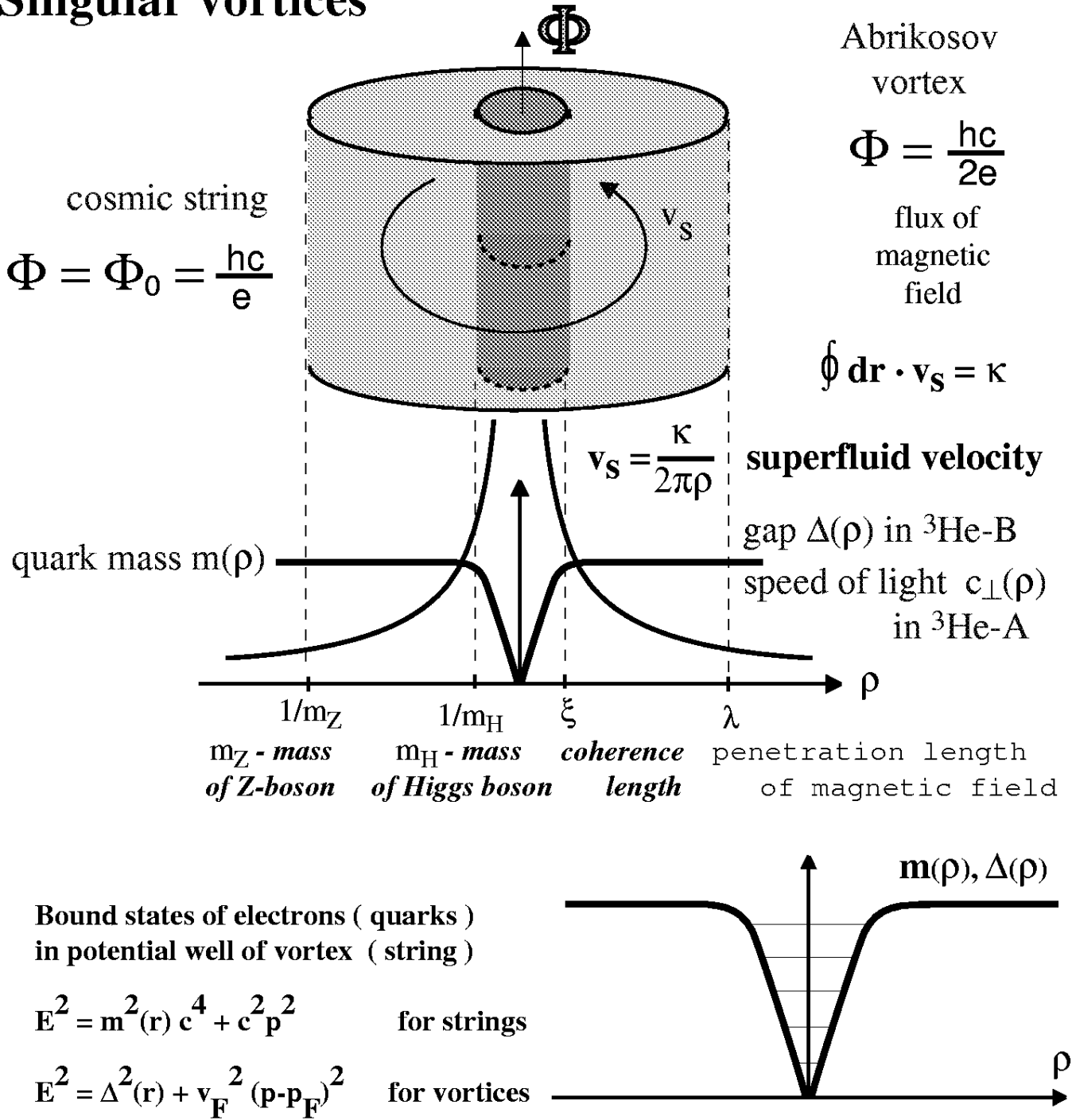}}
\medskip
\caption{Singular vortex and cosmic string. {\it top}: Abrikosov vortex in
superconductor is analog of the Nielsen-Olesen cosmic string. The role of the
penetration length $\lambda$ is played by the inverse mass of the $Z$-boson. If
$\lambda
\gg \xi$, the core size, within the region of dimension $\lambda$ the
Abrikosov vortex has the same structure as the vortex in neutral superfluids,
such as $^3$He-B, where the circulation of the superfluid velocity is
quantized. {\it bottom}: Masses of quarks and the gap of qusiparticles in
superconductors are suppressed in the vortex core. The core serves as potential
well for fermions which are bound in the vortex forming fermion zero modes.}
\label{VortexVsStringFig}
\end{figure}

The process of the momentum generation by vortex cores is similar to
that of generation of baryonic by the cores of cosmic strings
\cite{ewitten,tvgf,jgtv,barriola,gstv}. The axial anomaly is instrumental for
the   baryoproduction in the core of cosmic strings, but again the effect
cannot be described by the anomaly equation (\ref{ChargeParticlProduction}),
which was derived using the energy spectrum of the free massless fermions in
the presence of the homogeneous electric and magnetic fields. But in cosmic
strings these fields are no more homogeneous. Moreover the massless fermions
exist only in the vortex core as bound states in the 
potential well produced by the order parameter (Higgs) field.  Thus the
consideration of  baryoproduction by cosmic strings and momentogenesis by
singular vortices should be studied  using the spectrum of the massless (or
almost massless) bound states, the fermion zero modes.

\subsection{Fermion zero modes on vortices}

\subsubsection{Anomalous branch of chiral fermions}

The spectrum of the low-energy bound
states in the core of the axisymmetric vortex with winding number $n_1=\pm 1$
in the isotropic model of $s$-wave superconductor was obtained in microscopic
theory by Caroli, de
Gennes and Matricon
\cite{Caroli}:
\begin{equation}
E(L_z)=-n_1\omega_0 L_z~.
 \label{CaroliEq}
\end{equation}
This spectrum is two-fold degenerate due to spin degrees of freedom; $L_z$ is
the generalized angular momentum of the fermions in the core, which was found
to be half of odd integer quantum number (Fig.
\ref{FermionsOnVortexStringsFig}). The level spacing
$\omega_0$ is small compared to the energy gap of the quasiparticles outside
the core, $\omega_0\sim \Delta_0^2/E_F\ll\Delta_0$. This level spacing
$\omega_0$ is called the minigap, because $\omega_0/2$ is the minimal energy
of quasiparticle in the core. In the 3D systems the minigap depends on the
momentum $p_z$ along the vortex line. Here for simplicity we shall consider
the 2D case, when the vortex lines are reduced to point vortices.

\begin{figure}[t]
\centerline{\includegraphics[width=\linewidth]{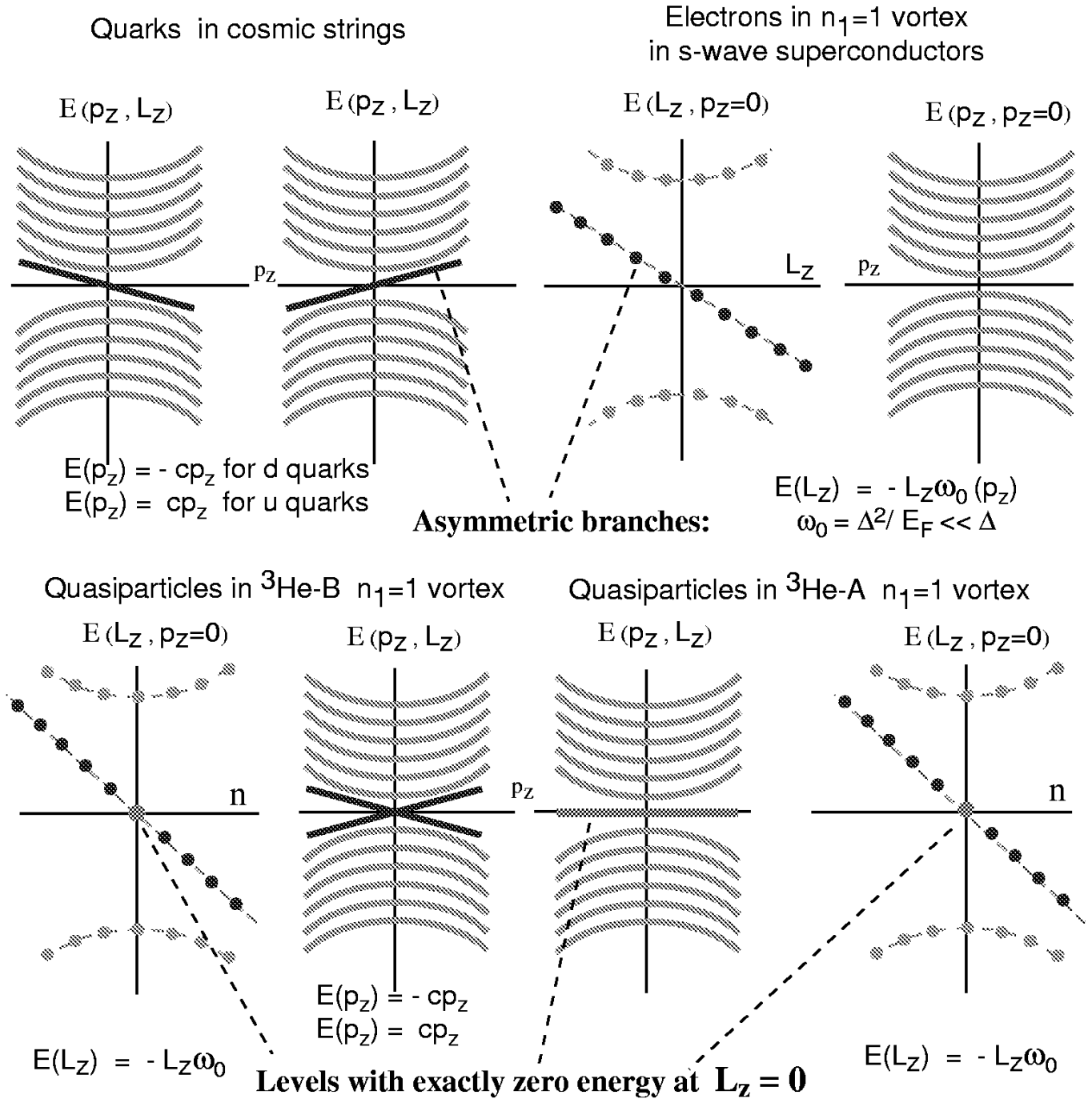}}
\medskip
\caption{Fermion zero modes on strings and on different condensed matter
vortices.}
\label{FermionsOnVortexStringsFig}
\end{figure}

Since the minigap is small, in many physical cases the discreteness of
$L_z$ can be neglected and in quasiclassical approximation one can consider
this quantum number as continuous.  Then from Eq.(\ref{CaroliEq}) it follows
that the spectrum as a function of continuous angular momentum
$L_z$ contains an anomalous branch, which crosses zero energy (Fig.
\ref{FermionsOnVortexStringsFig}). Thus in the quasiclasical approximation one
has  fermion zero modes on vortices. The fermions in this 1D ``Fermi liquid''
are chiral: the positive energy fermions have a definite sign of the angular
momentum $L_z$. In general case of arbitrary winding number
$n_1$, the number of fermion zero
modes, i.e. the number of branches crossing zero level as a function of $L_z$,
equals $-2n_1$ (see Ref.\cite{CallanHarveyEffect}). This represents an
analogue of the index theorem known for cosmic strings in the relativistic
quantum field theory (see Ref.\cite{GeneralizedIndexTheorem} and references
therein). The difference is that in strings the  spectrum of relativistic
fermions crosses zero energy as a function of $p_z$ (Fig.
\ref{FermionsOnVortexStringsFig}).  As a result the index
theorem discriminates between left-moving and right-moving
fermions, while in condensed matter vortices the index theorem  discriminates
between cw and ccw rotating fermions.

\subsubsection{Integer vs half-odd-integer angular momentum of fermion zero
modes}

The above properties of the fermions bound to the vortex core are
universal and do not depend  on the detailed structure of the vortex core. If
we now proceed to the non-$s$-wave superfluid or superconducting states, we
find that the situation does not change, with one exception: in some vortices
the quantum number $L_z$ is half of odd integer, in the others it is integer.
In the vortices of second type there is a true zero mode: at $L_z=0$ the
quasiparticle energy in Eq.(\ref{CaroliEq}) is exactly zero.  This true  
zero-energy bound state was first calculated in a microscopic theory
\cite{KopninSalomaa}  for the $n_1=\pm 1$ vortex in $^3$He-A. This difference
between two types of the fermionic spectrum becomes important at low
temperature $T<\omega_0$.

We consider here the representatives of these two types of vortices: the
traditional $n_1=\pm 1$ vortex in $s$-wave superconductor (Fig.
\ref{FermionsOnVortexStringsFig} {\it top right}) and the
simplest form of the $n_1=\pm 1$ vortex in $^3$He-A with $\hat{\bf l}$
directed along the vortex axis  (Fig.
\ref{FermionsOnVortexStringsFig} {\it bottom right}). Their order parameters
are
\begin{eqnarray}
\Psi({\bf r})= \Delta_0(\rho)e^{in_1\phi} ~,~~\oint d{\bf x}\cdot {\bf v}_{\rm
s}= n_1\pi\hbar/m~, 
\label{VorticesS}\\
A_{\mu
\,i}=\Delta_0(\rho)e^{in_1\phi}\hat z_\mu
\left(\hat x_{i} + i \hat y_{i}\right)~,~~\oint d{\bf x}\cdot {\bf v}_{\rm
s}= n_1\pi\hbar/m~,
 \label{VorticesP}
\end{eqnarray}
where $z$, $\rho$, $\phi$ are the coordinates of the cylindrical system
with the axis $z$ along the vortex line; and $\Delta_0(\rho)$ is the profile of
the order parameter amplitude in the vortex core with $\Delta_0(\rho=0)=0$.
The structure of the spectrum of the fermion zero modes does nor depend on
the profile of  $\Delta_0(\rho)$.

\subsubsection{Hamiltonian for fermions in the core}

After diagonalization over spin indices one finds that for each of two spin
components the Bogoliubov\,--\,Nambu Hamiltonian for
quasiparticles in the presence of the vortex has the form 
\begin{equation}
{\cal H}=\left(\matrix{M(p) &
\Delta_0(\rho)e^{in_1\phi} \left({p_x +ip_y\over p_F}\right)^{N_3}\cr
\Delta_0(\rho)e^{-in_1\phi} \left({p_x -ip_y\over
p_F}\right)^{N_3}&-M(p)\cr}\right)~.
 \label{MicroHamiltonian1}
\end{equation}
Here the integer index $N_3=0$ for the case of vortex in $s$-wave
superfluid in Eq.(\ref{VorticesS}) and $N_3=1$ for the $^3$He-A case (
Eq.(\ref{VorticesP})). For simplicity we assumed the 2D spatial dimension,
which is applicable for thin superfluid/superconducting films. In general  2D
system the integral index $N_3$ is the topological invariant in the momentum
space, Eq.(\ref{3DTopInvariant}),  which is responsible for the
Chern\,--\,Simons terms in the 2D superfluids/superconductors
\cite{VolovikYakovenko,VolovikEdgeStates,Senthill}. The two types of the
fermion zero modes, which we discuss, are determined by the parity
$W=(-1)^{n_1+N_3}$, which is constructed from the topological charges in real
and momentum spaces \cite{ZeroModesInChiral}.   

In what follows the fast radial motion of the fermions in the vortex core is
integrated out to obtain only the slow motion corresponding to the
low-energy fermion zero modes on anomalous branch. It is important that the
characteristic size $\xi$ of the vortex core is much larger than the wave
length $\lambda=2\pi/p_F$ of quasiparticle: $\xi p_F\sim E_F/\Delta_0 \sim
c_\parallel/c_\perp
\sim 10^3$. Thus for the radial motion we can use the quasiclassical
description in terms of trajectories, which are almost the straight lines
crossing the core. The description in terms of the trajectories are valid in
the quasiclassical region of energies between the two ``Planck'' scales,
$\omega_0\sim \Delta_0^2/v_Fp_F \ll E\ll \Delta_0$ discussed in Sec.
\ref{HierarchyEnergyScales}. 

\begin{figure}[t]
\centerline{\includegraphics[width=\linewidth]{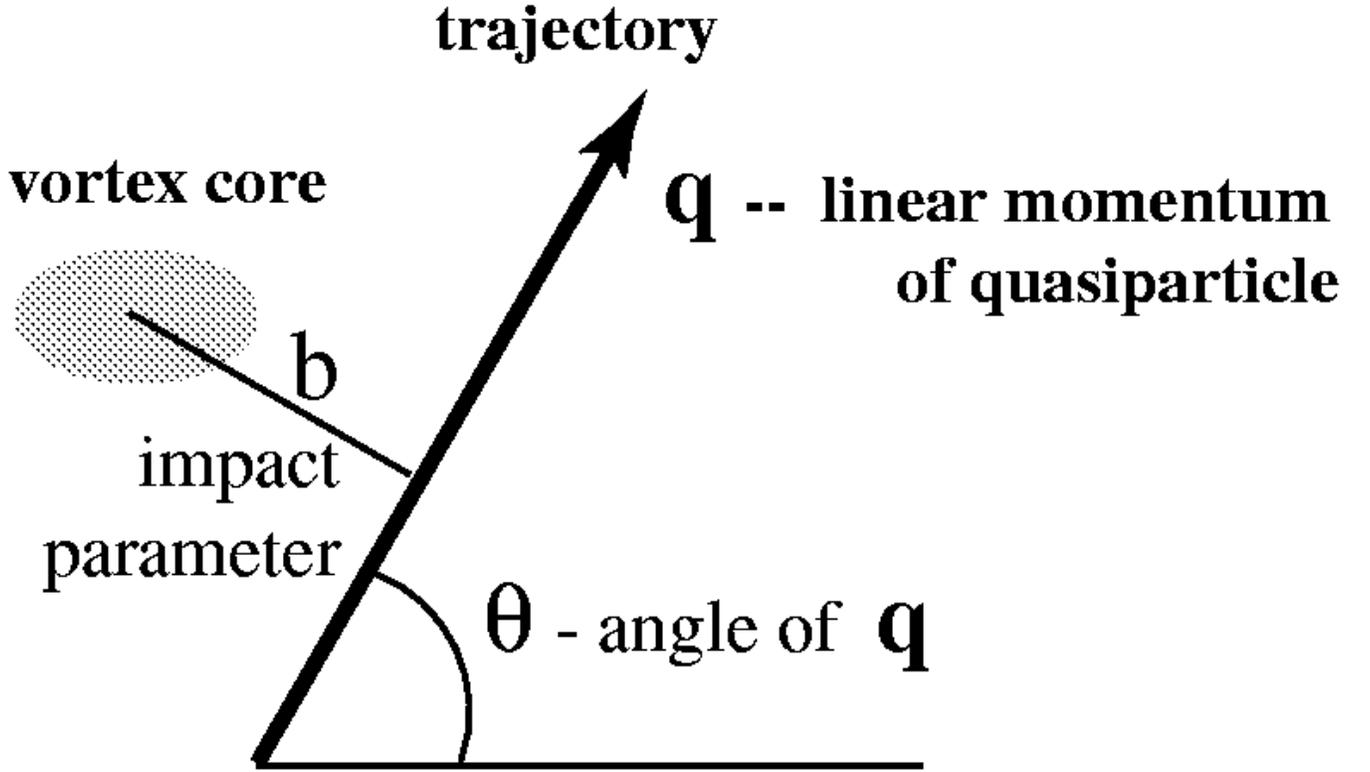}}
\medskip
\caption{In the quasiclassical approximation trajectories of quasiparticle are
straight lines.}
\label{TrajectoryFig}
\end{figure}

The low energy trajectories through the vortex
core are characterized by the direction
$\theta$  of the trajectory and the impact parameter $b$ (Fig.
\ref{TrajectoryFig}). The magnitude of the momentum of quasiparticle along
trajectory is close to
$p_F$, so that the momentum ${\bf p}$ is close to the value
\begin{equation}
{\bf q}=p_F(\hat {\bf x}\cos\theta + 
\hat {\bf
y}\sin\theta)~,
 \label{MomentumInPlane}
\end{equation}
and the velocity is close to ${\bf v}_F= {\bf q}/m^*$.
In terms of the angle $\theta$ the Eq.(\ref{MicroHamiltonian1}) can be
rewritten in the form
\begin{equation}
{\cal H}=\left(\matrix{M(p) &
\Delta_0(\rho)e^{i(n_1\phi + N_3\theta)}\cr \Delta_0(\rho)e^{-i(n_1\phi +
N_3\theta)}&-M(p)\cr}\right)~,
 \label{MicroHamiltonian2}
\end{equation}
which emphasizes the interplay between the  real-space and
momentum-space topologies.

Substituting $\chi\rightarrow e^{i{\bf
q}\cdot{\bf
r}}\chi$ and ${\bf p}\rightarrow {\bf q}-i{\bf \nabla}$, and
expanding in
small ${\bf \nabla}$, one obtains the quasiclassical Hamiltonian for
the
fixed trajectory $({\bf q},b)$:
\begin{equation}
{\cal H}_{{\bf q},b}=-i\check  \tau_3{\bf v}_F\cdot{\bf \nabla}  +
 \Delta_0(\rho)\left(\check  \tau_1     \cos(N_3\theta + n_1\phi)  -\check  \tau_2
\sin(N_3\theta  + n_1\phi)\right)~.
 \label{QuasiclassicalHamiltonian}
\end{equation}
Since the coordinates $\rho$ and $\phi$ are related by equation
$\rho\sin (\phi-\theta)=b$, the only argument is the coordinate 
along the trajectory
$s=\rho\cos (\phi-\theta)$ and thus the Hamiltonian in
Eq.(\ref{QuasiclassicalHamiltonian}) has the form
\begin{eqnarray}
 {\cal H}_{{\bf q},b} = -i{v}_F\check  \tau_3 \partial_s +
 \check  \tau_1  \Delta_0(\rho)   \cos \left(n_1 \tilde\phi
+(n_1+N_3)\theta\right)-\nonumber\\ -\check  \tau_2    \Delta_0(\rho)
  \sin \left(n_1 \tilde\phi +(n_1+N_3)\theta \right)  ~,
 \label{QuasiclassicalHamiltonian2}
\end{eqnarray}
where $\tilde\phi=\phi -\theta$ is expressed in terms of the coordinate $s$ as
\begin{equation}
\tilde\phi=\phi -\theta ~~,~~ \tan
\tilde\phi={b\over s} ~~.
 \label{tildephi}
\end{equation}

The dependence of the Hamiltonian on
the direction $\theta$ of the trajectory can be removed by the
following
transformation:
\begin{eqnarray}
\chi=
e^{i (n_1+N_3)\check  \tau_3 \theta/2}\tilde\chi,
\label{TransformationFunction}
\\
\tilde{\cal H}_{{\bf q},b}=e^{-i (n_1+N_3)\check  \tau_3 \theta/2}{\cal
H}e^{i (n_1+N_3)\check  \tau_3
\theta/2}=\nonumber\\
=
  -i{v}_F\check  \tau_3 \partial_s + \Delta_0(\sqrt{s^2+b^2})\left(\check 
\tau_1  \cos  n_1
\tilde\phi  -\check  \tau_2    \sin  n_1 \tilde\phi\right)~.
\label{QuasiclassicalHamiltonian3}
\end{eqnarray}

The Hamiltonian in Eq.(\ref{QuasiclassicalHamiltonian3}) does not depend on the
angle $\theta$ and on the topological charge $N_3$ and thus is the same for
$s$, $p$ and other pairing states. The dependence on $N_3$ enters only the
boundary condition for the wave function, which according to
Eq.(\ref{TransformationFunction}) is
\begin{equation}
\tilde\chi(\theta +2\pi)=(-1)^{n_1+N_3}\tilde\chi(\theta)
 \label{BoundaryCondition}
\end{equation}
With respect to this boundary condition, there are two classes of
systems: with  odd and even $n_1+N_3$. The
parity
$W=(-1)^{n_1+N_3}$  thus determines the spectrum of fermions in the vortex
core.

\subsubsection{Quasiclassical low-energy states on anomalous branch}

The state with the lowest energy corresponds to trajectories, which cross the
center of the vortex, i.e. with $b=0$. Along this trajectory one has
$
\sin
\tilde\phi=0$ and     $\cos   \tilde\phi = {\rm sign}~ s$. So that
the Eq.(\ref{QuasiclassicalHamiltonian3}) becomes
\begin{equation}
 \tilde{\cal H}_{{\bf q},b} = -i{v}_F\check  \tau_3 \partial_s +
 \check  \tau_1  \Delta_0(|s|)  {\rm sign}~ s ~.
 \label{SupersymmetriclHamiltonian}
\end{equation}
It is supersymmetric and thus contains the eigenstate with zero
energy. Let us write the corresponding eigen function  including all
the transformations:
\begin{eqnarray}
 \chi_{\theta, b=0}(s) =  e^{ip_F s} e^{i (n_1+N_3)\check  \tau_3
\theta/2} \left(\matrix{1 \cr -i\cr}\right)
\chi_0(s)~~,
\label{SupersymmetriclSolution1}
\\
\chi_0(s)=\exp{\left(-\int^s ds' {\rm sign}~
s'{\Delta_0(|s'|)\over v_F}
\right)}~.
 \label{SupersymmetriclSolution2}
\end{eqnarray}

Now we can consider the case of nonzero impact parameter.  When $b$ is small
the third term in Eq.(\ref{QuasiclassicalHamiltonian3}) can
be considered as perturbation and its average over the wave function in
Eq.(\ref{SupersymmetriclSolution1}) gives the energy levels in terms of
$b$ and thus in terms of the continuous angular momentum $L_z=p_Fb$:
\begin{equation}
E(L_z,\theta)=-n_1L_z \omega_0~,~\omega_0  = {\int_0^\infty d\rho
 {\Delta_0(\rho)\over p_F\rho}\exp{\left(-{2\over v_F}\int_0^\rho d\rho' 
\Delta_0(\rho')\right)}\over
\int_{0}^\infty d\rho \exp{\left(-{2\over v_F}\int_0^\rho d\rho' 
\Delta_0(\rho')\right)}}~.
 \label{QuasiclasicalEnergy}
\end{equation}
This is the anomalous branch of chiral fermions which crosses zero energy in
semiclassical approximation, when $L_z$ is continuous variable. For
nonaxisymmetric vortices the quasclassical energy depends also on $\theta$.
The minigap $\omega_0$ is of order $\omega_0\sim \Delta_0/(p_FR)$ where
$R$ is the radius of core. Typically $i$ is of order  coherence length
$\xi=v_F/\Delta_0$ and the minigap is  $\omega_0\sim \Delta_0^2/(p_Fv_F)\ll
\Delta_0$. In a large temperature region $\Delta_0^2/(p_Fv_F)<T<
\Delta_0$ these bound fermionc states can be considered as fermion zero modes.

\subsubsection{Quantum low-energy states and $W$-parity.}   

In exact quantum mechanical problem the generalized angular momentum $L_z$
has discrete eigen values. To find quantized energy levels we take into
account that the two remaining degrees of freedom, the angle $\theta$ and
the momentum $L_z$, are canonically conjugated
variables \cite{Stone,KopninVolovik}. That is why the next step is the
quantization of motion in the $\theta,L_z$ plane which can be obtained from the
quasiclassical energy in Eq.(\ref{QuasiclasicalEnergy}), if
$L_z$ is considered as an operator.    For the axisymmetric vortex, 
the Hamiltonian does not depend on $\theta$ 
\begin{equation}
H= in_1\omega_0\partial_\theta
 \label{HamiltonianTheta}
\end{equation}
and has the eigenfunctions $e^{-iE\theta/n_1\omega_0}$.
The boundary condition for these functions, the
Eq.(\ref{BoundaryCondition}), gives the quantized energy levels, which depend
on the $W$-parity:
\begin{eqnarray}
E(L_z)=-n_1L_z \omega_0~,
\label{SpectrumCore1}\\
L_z= n ~~,~~W=+1 ~~;
\label{SpectrumCore2}\\
L_z =  \left(n+{1\over 2}\right) ~~,~~W=-1~~,
\label{SpectrumCore3}
\end{eqnarray}
where $n$ is integer.

The phase $(n_1+N_3)\tau_3 \theta/2$ in
Eq.(\ref{TransformationFunction}) plays the part of Berry phase. It
shows
how the wave function of quasiparticle changes, when the trajectory is
adiabatically rotating by angle $\theta$. This Berry phase is instrumental for
the Bohr\,--\,Sommerfeld quantization of the energy levels in the vortex
core. It
chooses between the only two possible quantizations consistent with the
``CPT-symmetry'' of states in superconductors:   $E=n\omega_0$ and
$E=(n+1/2)\omega_0$. In both cases for each positive energy level  $E$ one can
find another level with the energy $-E$. That is why the above quantization
is applicable even to nonaxisymmetric vortices, though  the quantum
number $L_z$ is no more the angular momentum there.

Fig. \ref{FermionsOnVortexStringsFig}  shows the
quasiparticle spectrum in the core of vortex lines in 3D space. In the
top-right corner it is the spectrum in the core of vortices with $W=-1$. At the
bottom the spectrum of bound states in the core of vortices with
$W=1$ is shown. These are the real fermion zero
modes, since they cross zero energy. The most interesting situation occurs for
the $^3$He-A vortices  which contains a self conjugated
zero-energy level.  This $E=0$ level is
doubly degenerate due to spin.  Such exact zero mode is robust to any
deformations, which preserve the spin degeneracy.  This is the reason why the
branch with $L_z=0$ has zero energy for all momenta $p_z$   (right bottom
corner of Fig. \ref{FermionsOnVortexStringsFig})
as was first found in microscopic theory by Kopnin \cite{KopninSalomaa} (see
also \cite{ZeroModesInChiral}).

\subsubsection{Majorana fermion with $E=0$ on half-quantum vortex.}
\label{MajoranaFermion}

The most exotic situation will occur for the half-quantum
vortex \cite{VolMin}, which can exist in thin $^3$He-A films or in layered
chiral superconductors with the same order parameter.  The order parameter in
Eq.(\ref{APhase}) outside the core of $n_1=1/2$-vortex is
\begin{equation}
A_{\mu \,i}=\Delta_0\hat d_\mu \left(\hat e_{1i}  + i \hat
e_{2i}\right)= 
\Delta_0\left(\hat x_\mu\cos {\phi\over 2} +\hat y_\mu\sin {\phi\over 2}\right)\left(\hat x_{i}
+ i
\hat y_{i}\right)e^{i\phi/2}~.
\label{HalfQuantumVortex}
\end{equation}
The change of the sign of the vector $\hat{\bf d}$ when circumscribing around the
core is compensated by the change of the phase of the order parameter by
$\pi$ at which the sign of the exponent
$e^{i\phi/2}$ also changes. Thus the whole order parameter is smoothly
connected after circumnavigating. Because of the $\pi$ change of the phase
around the vortex, the circulation of the superfluid velocity around the vortex
$\oint d{\bf x}\cdot {\bf v}_{\rm s}=n_1\pi\hbar/m$ corresponds to $n_1=1/2$.
That is why the name half-quantum vortex. This vortex still had not been
observed in
$^3$He-A but its discussion has been extended to the flux quantization with
the half of conventional magnitude in superconductors
\cite{Geshkenbein}, which  finally led to the observation of the
fractional flux in high-temperature superconductors: the $n_1=1/2$ vortex
is attached to the tricrystal line, which is the junction of three grain
boundaries \cite{Kirtley1996}. 

For quasiparticles in the core of the  $n_1=1/2$ vortex, after spin
diagonalization one obtains the Hamiltonian in Eq.(\ref{MicroHamiltonian2}),
where  for one of the spin components one has  
$n_1=1$, while for the other component $n_1=0$. The $E=0$ level occurs only for
one spin component.  Since it is the only level with $E=0$ it cannot be moved
from the $E=0$ position by any perturbation: its shift is prohibited by the
``CPT''-symmetry. 

There are many interesting properties related to this $E=0$ level. Since the
$E=0$ level can be either filled or empty, there is a fractional  entropy 
$(1/2) ln~ 2$ per layer per vortex. The factor (1/2) appears because in
the pair correlated superfluids/superconductors one must take into account
that we artificially doubled the number of fermions introducing both the
particles and holes. On the $E=0$ level the particle excitation coincides with
its antiparticle (hole), i.e. the quasiparticle is a Majorana fermion
\cite{Read}. Majorana fermions at $E=0$  level lead to the non-Abelian
statistics of half-quantum vortices: the interchange of two point vortices
becomes identical operation (up to an overall phase) only on being repeated
four times \cite{Ivanov}.  This can be used for quantum computing
\cite{Kitaev}.

\subsubsection{Fermions on asymmetric vortices.}

Most of vortices in condensed matter are not axisymmetric. In superconductors
the rotational symmetry is violated by crystal lattice, while in superflud
$^3$He the axisymmetry is as a rule spontaneously broken in the core or
outside  the core (see e.g. \cite{NonaxisymmetricVortex}). The general form of
the  $a$-th branch of the low-energy spectrum of fermions, which cross zero in
quasiclassical approximation is (see \cite{KopninVolovik}) 
\begin{equation}
E_a(L_z,\theta)=\omega_a(\theta)
(L_z-L_{za}(\theta))
\label{SpectrumAsymmetricVortex}
\end{equation}
For given $\theta$ the spectrum crosses zero energy at some
$L_z=L_{za}(\theta)$.
The "CPT"-symmetry of the Bogoliubov-Nambu Hamiltonian requires that if
$E_a(L_zQ,\theta)$ is the energy of the bound state fermion, then
$-E_a(-L_z,\theta +\pi) $ also corresponds to the energy of quasiparticle in
the core.

Let us consider one pair of the conjugated branches related by the ``CPT''
symmetry. One can introduce the common gauge field $A_\theta(\theta,t)$  and  
the "electric" charge
$e=\pm 1$, so that $Q_\pm=eA_\theta$.
Then the quantum Hamiltonian becomes
\begin{equation}
{\cal H}= - {n_1\over 2} \left\{ ~\omega_0(\theta)~,~\sum_e \left(-i{\partial
\over
\partial\theta} - eA_\theta(\theta,t)\right)\right\}~~ ,
\label{HamiltonianAsymmetricVortex}
\end{equation}
where $\{~,~\}$ is anticommutator.
The Schr\"odinger equation for the fermions on the
$a$-th branch is
\begin{equation}
{i\over 2}(\partial_\theta \omega_0)\Psi(\theta)+
\omega_0(\theta)
\left(i\partial_\theta + eA(\theta)\right)  
\Psi(\theta)=n_1 E\Psi (\theta)
\label{SchrodEqAsymmetricVortex}
\end{equation}
 The normalized eigen functions are
\begin{equation}
\Psi(\theta)=\left<{1   \over
\omega_0(\theta)}\right>^{-1/2} ~ {1\over
\sqrt{2\pi \omega_0(\theta)}}\exp \left(i\int^\theta d\theta'\left( {n_1
E\over
\omega_0(\theta')} + eA(\theta')\right) \right) ~~.
\label{SolutionAsymmetricVortex}
\end{equation}
Here the angular brackets mean the averaging over the angle $\theta$.

For the self-conjugated branch  
according to the ``CPT''-theorem one has
$\int^{2\pi}_0   d\theta A(\theta)=0$, then using the boundary conditions in
Eq.(\ref{BoundaryCondition}) one obtains the equidistant energy levels:
\begin{equation}
E_n =\left(n +{1-W\over 4}\right) \left<{1   \over
\omega_0(\theta)}\right>^{-1}  ~~.
\label{QuantumSpectrumAsymmetricVortex}
\end{equation}
For axisymmetric vortices the integral index $n$ is determined by the
azimuthal  quantum number $L_z$. For the nonaxisymmetric vortex, $L_z$ is
not a good quantum number. Nevertheless the properties of the anomalous
branch of fermion modes on vortices are not disturbed by the nonaxisymmetric
perturbations of the vortex core structure, which lead only to
the renormalization of the minigap. These properties are dictated by real- and
momentum-space topology and are robust to perturbations.

\subsection{Spectral flow in  singular vortices: Callan-Harvey mechanism
of anomaly cancellation}\label{SpectralFlowSingVort}

Now let us consider again the  force, which arise when  the
vortex moves with respect to the heat bath. In Sec.
\ref{AxialAnomalyAndForce} we discussed this for the special case of the
continuous $^3$He-A vortex texture where the  macroscopic Adler-Bell-Jackiw 
anomaly equation could be used. Now we consider this effect using the
microscopic description of the spectral flow of fermion zero modes within the
vortex core. We show that the same force arises for any vortex in any
superfluid or superconductor under a special condition.

If the vortex moves with the
velocity
${\bf v}_L$ with respect to the heat bath, the coordinate ${\bf r}$ is
replaced by the
${\bf r}-{\bf v}_Lt$. The angular momentum $L_z$,  
which enters the quasiclassical energy of quasiparticle in
Eq.(\ref{QuasiclasicalEnergy}), shifts  with time, because the momentum
is ${\bf r} \times {\bf p}$. So the energy becomes
\begin{equation}
E(L_z,\theta)=-n_1\left(L_z- \hat{\bf z}\cdot({\bf v}_L\times {\bf
q})t\right)\omega_0~.
 \label{QuasiclasicalEnergyMovingVortex}
\end{equation}
Here $E_z=\hat{\bf z}\cdot({\bf v}_L\times {\bf
q})$ acts on fermions localized in the core in the same way that an
electric field $E_z$ acts on chiral fermions on an anomalous branch in magnetic
field or the  chiral fermion zero modes localized on a string in relativistic
quantum theory. The only difference is that under this ``electric'' field the 
spectral flow in the vortex occurs in the $L_z$ direction (${\dot L}_z=eE_z$)
rather than along
$p_z$ direction in strings where ${\dot p}_z=eE_z$. Since according to index
theorem, for each quantum number 
$L_z$ there are $-2n_1$ quasiparticle levels, the fermionic levels cross the
zero energy at the rate
\begin{equation}
{\dot n}= -2n_1{\dot L}_z=-2n_1 E_z({\bf q})=-2n_1\hat{\bf z}\cdot({\bf
v}_L\times {\bf q})~~.
 \label{LevelFlow}
\end{equation}

When the occupied level crosses zero, the quasiparticle on this level
transfers its fermionic charges from the vacuum (from the negative energy
states) along the anomalous branch into the heat bath (``matter''). For us the
important fermionic charge is linear momentum. The rate at which the momentum
${\bf q}$ is transferred from the vortex to the heat bath due to spectral flow 
\begin{eqnarray}
\partial_t{\bf P}={1\over 2} \int_0^{2\pi}{{d\theta}\over{2\pi}} ~{\bf q}\dot
n=
\label{MomentumFlow2Da}
\\
-n_1\int_0^{2\pi}{{d\theta}\over{2\pi}}~{\bf q}\left(\hat{\bf z}\cdot(({\bf
v}_L-{\bf v}_{\rm n})\times {\bf q})\right) =- n_1 {{p_F^2}\over
2}\hat{\bf z}\times({\bf
v}_L-{\bf v}_{\rm n}) ~~.
\label{MomentumFlow2Db}
\end{eqnarray}
The factor $1/2$ in Eq.(\ref{MomentumFlow2Da}) is to compensate the
double counting of particles and holes.

The extension to the 3+1 case is straightforward. 
The trajectory in the plane perpendicular to the vortex axis is
now characterized by the transverse momentum 
\begin{equation}
{\bf q}=\sqrt{p_F^2-p_z^2}~(\hat {\bf x}\cos\theta + 
\hat {\bf
y}\sin\theta)~,
 \label{MomentumInPlane2}
\end{equation}
so that the total momentum is on the Fermi surface: ${\bf q}^2+p_z^2=p_F^2$.
Due to that the momentum production is slightly modified:
\begin{eqnarray}
\partial_t{\bf P}={1\over 2} \int_0^{2\pi}{{d\theta}\over{2\pi}}
\int_{-p_F}^{p_F} {{dp_z}\over{2\pi}}~{\bf q}\dot n =
\label{MomentumFlow3Da}
\\
-n_1\int_{-p_F}^{p_F}
{{dp_z}\over{2\pi}}(p_F^2-p_z^2)\int_0^{2\pi}{{d\theta}\over{2\pi}}~\hat{\bf
q}\left(\hat{\bf z}\cdot(({\bf v}_L-{\bf v}_{\rm n})\times \hat{\bf q})\right)
=\nonumber
\\- n_1 {{p_F^3}\over 3\pi}\hat{\bf z}\times({\bf v}_L-{\bf v}_{\rm n})
~~.
\label{MomentumFlow3Db}
\end{eqnarray}
Thus the spectral flow force acting on a vortex from the system of
quasiparticles is
\begin{equation}
{\bf F}_{sf}=-\pi n_1\hbar
 C_0{\hat {\bf z}}  \times ({\bf v}_{\rm L}-{\bf v}_{\rm n}) .
\label{SpFlowForceSingular}
\end{equation}
This is in agreement with the result in Eq.(\ref{SpFlowForce}) obtained for
the $n_1=2$ continuous vortex using the Adler-Bell-Jackiw equation.
We recall that the parameter of the axial anomaly is $C_0=p_F^3/3\pi^2$ in 3D,
while in 2D case it is $C_0=p_F^2/2\pi$ in agreement with
Eq.(\ref{MomentumFlow2Db}).

In this derivation it was implied that all the quasiparticles, created from the
negative levels of the vacuum state, finally become part of the normal
component, i.e. there is a nearly reversible transfer of linear momentum from
fermions to the heat bath. This should be valid in the limit of large
scattering rate: $\omega_0\tau\ll 1$, where $\tau$ is the lifetime of the
fermion  on the $L_z$ level. This condition, which states that the
interlevel distance on the anomalous branch is small compared to the
life time of the level, is the crucial requirement for spectral flow to
exist. In the opposite limit $\omega_0\tau\gg 1$ the spectral flow is
suppressed and the corresponding spectral flow force is exponentially small
\cite{KopninVolovik} (see Sec. \ref{RestrictedSpectralFlow} below). This shows
the limitation for exploring the macroscopic Adler-Bell-Jackiw  anomaly
equation in the electroweak model and in $^3$He-A.

The process of transfer of linear momentum from the superfluid vacuum
to the  normal motion of fermions within the core is the realization of the
Callan-Harvey mechanism for anomaly cancellation \cite{CallanHarvey}.
In the case of the condensed matter vortices the anomalous nonconservation
of linear momentum in the 1+1 world of the vortex core
fermions and the anomalous nonconservation
of momentum in the 3+1 world outside the vortex
core  compensate each other. This is the same kind of  the Callan-Harvey
effect which has been discussed in Sec. \ref{AxialAnomalyAndForce} for the
motion of continuous textures in $^3$He-A. As distinct from  $^3$He-A, where
there are  gap nodes and chiral anomaly, the Callan-Harvey
effect for singular vortices occurs in any Fermi-superfluid: the anomalous
fermionic $L_z$ branch, which mediates the momentum exchange,  exists
in any topologically nontrivial singular vortex: the chirality of fermions
and the anomaly are produced by the notnrivial topology of the vortex. In the
limit $\omega_0\tau\ll 1$ this type of Callan-Harvey effect does  not depend on
the detailed structure of the vortex core and even on the  type of pairing, and
is determined solely by the vortex winding number $n_1$ and anomaly parameter
$C_0$. 

The first derivation of the spectral-flow force acting on a vortex was
made by Kopnin and Kravtsov in Ref. \cite{KopninKravtsov1976} who used the
fully microcopic BCS theory in Gor'kov formulation. It was developed further
by Kopnin and coauthors. That is why it is sometimes called the Kopnin force.

 \subsubsection{Restricted spectral flow in the
vortex core.}\label{RestrictedSpectralFlow} 

The semiclassical approach allows us to extend the derivation of the spectral
flow dynamics to the more complicated cases, when for example the core is not
symmetric or when the relation $\omega_0\tau\ll 1$ is not
fulfilled and the spectral flow is partially suppressed. Since the spectral
flow occurs through the zero energy, it is fully determined by the low-energy
spectrum. As in the cases of the universality classes in the 3+1 system, the
1+1 systems also have universality classes which determine the quasiparticle
spectrum in the low energy corner. In our case the generic spectrum of 1+1
fermions living in the vortex core is given by
Eq.(\ref{SpectrumAsymmetricVortex}). Note that the coordinate in the
``spatial'' dimension is the angle $\theta$, i.e. the effective space is
circumference $U(1)$.

Let us choose the frame of the moving vortex.
In this frame the energy of quasiparticle is well determined and the 
Hamiltonian for quasiparticles in the moving vortex is given by 
\begin{equation}
E_a(L_z,\theta)=\omega_a(\theta)
(L_z-L_{za}(\theta)) + ({\bf v}_{\rm s} - {\bf v}_L)\cdot{\bf q}~,
\label{SpectrumAsymmetricVortexFlow}
\end{equation}
where the last term comes from the Doppler shift.
We again consider the vortex in the 2D case where  ${\bf  q}=(p_F\cos\theta,
p_F\sin\theta)$. 
Kinetics of these low-energy quasiparticles is given by the 
Boltzmann equation for the distribution
function  $f_a(L_z,\theta)$ \cite{Stone}. Let us consider the simplest case
of axisymmetric vortex with one anomalous branch, then
$\omega_a(\theta)=-n_1\omega_0 L_z$, $L_{za}(\theta)=0$,  and the Boltzmann
equation becomes (in $\tau$ approximation):
\begin{equation}
\partial_t f-n_1\omega_0 \partial_\theta f -\partial_\theta (({\bf v}_{\rm s} -
{\bf v}_L)\cdot{\bf q}) 
~\partial_{L_z} f= -{f(L_z,\theta)-  f_{{\cal T}}(L_z,\theta)\over \tau} ~.
\label{BoltzmannEquation}
\end{equation}
The equilibrium  distribution $f_{{\cal T}}$ corresponds to the state, when the
vortex moves together with the heat bath, i.e.  when ${\bf v}_L = {\bf
v}_{\rm n}$:
\begin{equation}
f_{{\cal T}}(L_z,\theta)=\left(1 +\exp{-n_1\omega_0 L_z + ({\bf v}_{\rm s} -
{\bf v}_{\rm n})\cdot{\bf q} \over T}\right)^{-1}~.
\label{EquilibriumDistributionFunction}
\end{equation}
When ${\bf v}_L \neq {\bf
v}_{\rm n}$ the equilibrium is volated and the distribution function
evolves according to the Eq.(\ref{BoltzmannEquation}).

Introducing new variable 
$l=L_z - n_1({\bf v}_{\rm s} - {\bf v}_{\rm n})\cdot{\bf q}$ (we consider
here $n_1=\pm 1$) one obtains the equation for $f(l,\theta)$ which does not
contain
${\bf v}_{\rm s}$:
\begin{equation}
\partial_t f-n_1\omega_0 \partial_\theta f -\partial_\theta (({\bf v}_{\rm n} -
{\bf v}_L)\cdot{\bf q}) 
~\partial_{l} f= -{f(l,\theta)-  f_{{\cal T}}(l)\over \tau} ~.
\label{BoltzmannEquation2}
\end{equation}
Since we are interested in the momentum transfer only we can write 
equation for the net momentum of quasiparticles
\begin{equation}
{\bf P}=  {1\over 2}\int dl \int {d\theta\over 2\pi}  f(l,\theta) {\bf q}~,
\label{AverageMomentum}
\end{equation}
which is
\begin{equation}
\partial_t  {\bf P} -n_1\omega_0 \hat{\bf  z}\times  {\bf P}  +\pi C_0
\hat {\bf  z}\times ({\bf v}_{\rm n} - {\bf v}_L)
(f_{{\cal T}}(\Delta_0(T))-f_{{\cal T}}(-\Delta_0(T)))=  - { {\bf P}\over
\tau} ~.
\label{EvolutionAverageMomentum}
\end{equation}
Here we first consider only the bound
states below the gap $\Delta_0(T)$ and thus the integral  $\int dl \partial_l
n$ is limited by $\Delta_0(T)$. That is why the inegral gives
$f_{{\cal T}}(\Delta(T))-f_{{\cal T}}(-\Delta(T))=-\tanh (\Delta_0(T)/2T)$. 

In the steady state of ht evortex motion one has  $\partial_t {\bf
P}=0$ and the solution for the steady state momentum can be easily found
\cite{Stone}.  As a result  one  
obtains the following contribution to the spectral flow force due to  bound
states below $\Delta_0(T)$:
\begin{equation}
{\bf F}_{sf~bound}={ {\bf P}\over
\tau}=-{\pi  C_0\over 1+  \omega_0^2\tau^2}
\tanh {\Delta_0(T)\over 2T} [({\bf v}_L - {\bf v}_{\rm n})
\omega_0\tau +n_1\hat {\bf z}\times({\bf v}_L - {\bf v}_{\rm n})]  .
\label{FlowBoundStates}
\end{equation}
This equation contains both the nondissipative and friction forces. Now one
must add the contribution of unbound states above the gap $\Delta_0(T)$. The
spectral flow there is not suppressed,  since the distance between the levels
in the continuous spectrum is
$\omega_0=0$. This gives the spectral flow contribution from the thermal tail
of the continuous spectrum
\begin{equation}
{\bf F}_{sf~unbound}= -\pi n_1 C_0  \left(1- \tanh
{\Delta_0(T)\over 2T}\right)  \hat {\bf z}\times({\bf v}_L - {\bf v}_{\rm n}) 
~.
\label{FlowUnBoundStates}
\end{equation}

Finally the total  spectral-flow force is the sum of two contributions,  
Eqs.(\ref{FlowBoundStates}-\ref{FlowUnBoundStates}). The
nondissipative part of the force is
\begin{equation}
{\bf F}_{sf}=-\pi n_1 C_0\left[1 - {\omega_0^2\tau^2 \over 1+ 
\omega_0^2\tau^2}\tanh {\Delta_0(T)\over 2T} \right]\hat {\bf z}\times({\bf
v}_L - {\bf v}_{\rm n}) ~ ,
\label{FlowAllStates}
\end{equation}
while the contribution of the spectral flow to the friction force is
\begin{equation}
{\bf F}_{fr}= -\pi C_0  \tanh {\Delta_0(T)\over 2T}~{\omega_0\tau
\over 1+ 
\omega_0^2\tau^2}
 ({\bf v}_L - {\bf v}_n) ~.
\label{FlowFrictForce}
\end{equation}
 
 \subsubsection{Measurement of Callan-Harvey
effect in $^3$He-B}\label{BPhaseExp} 

The equations (\ref{FlowAllStates}) and (\ref{FlowFrictForce}) are
applied for the dynamics of singular vortices in
$^3$He-B, where the minigap $\omega_0$ is comparable with the inverse
quasiparticle lifetime and the parameter $\omega_0\tau$ is regulated by
temperature. Introducing again the Magnus and Iordanskii forces one obtains
the following dimensionless parameters $d_\perp$ and $d_\parallel$ in
Eq.(\ref{ForceBalance}) for the  balance of forces acting on a vortex:
\begin{equation}
d_\perp= {C_0\over n_{\rm s}}\left(1 -{n\over  n_{\rm s}} {\omega_0^2\tau^2\over
1+\omega_0^2\tau^2}  \tanh {\Delta_0(T)\over 2T}\right) - { n_{\rm n}\over
n_{\rm s}} ~,
\label{dPerp}
\end{equation}
\begin{equation}
d_\parallel= {C_0\over  n_{\rm s}} {\omega_0\tau\over
1+\omega_0^2\tau^2}  \tanh {\Delta_0(T)\over 2T} ~.
\label{dParallel}
\end{equation}

\begin{figure}[t]
\centerline{\includegraphics[width=\linewidth]{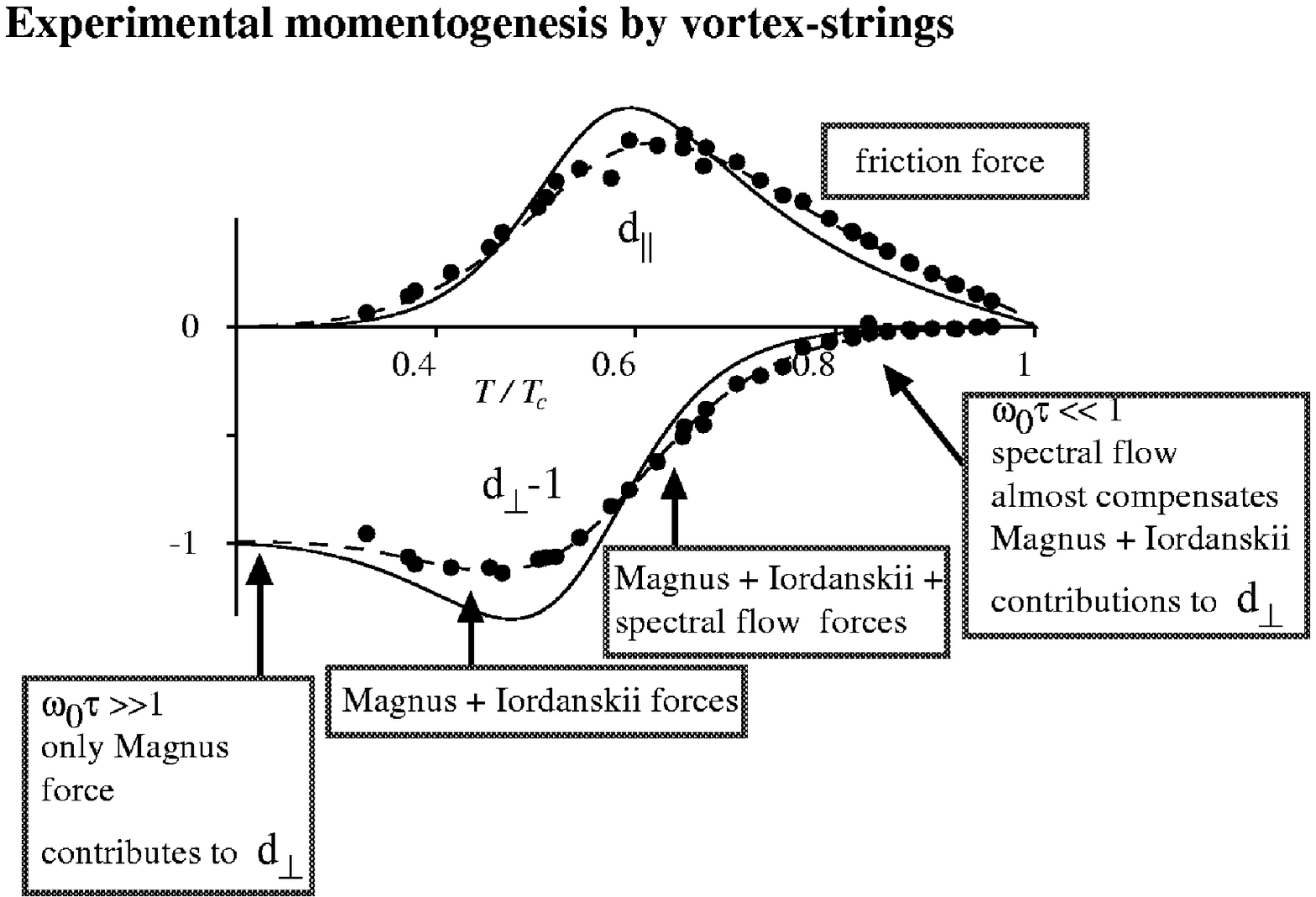}}
\medskip
\caption{Experimental verification of Callan-Harvey and gravitational
Aharonov-Bohm effects in $^3$He-B. Solid lines are Eqs.
(\protect\ref{dPerp}) and  (\protect\ref{dParallel}). The Callan-Harvey effect
is suppressed at low
$T$ but becomes maximal close to $T_c$, where the spectral flow transfers the
fermionic charge -- the momentum -- from the 1+1 fermionc in the core to the
3+1 bulk superfluid. The negative value of $d_\perp$ at intermediate $T$
demostrates the Iordanskii force, which comes from the analog of
the gravitational Aharonov-Bohm effect.}
\label{CallanHarveyExpFig}
\end{figure}

The  regime of the fully developed axial anomaly occurs when $\omega_0\tau
\ll 1$. This is realized close to $T_c$, since $\omega_0$ vanishes at $T_c$.
In this regime   $d_\perp =(C_0-n_{\rm n})/n_{\rm s}\approx (n-n_{\rm
n})/n_{\rm s}=1$. At lower $T$ both $\omega_0$ and
$\tau$ increase;  at $T\rightarrow 0$ one has an opposite regime,
$\omega_0\tau \gg 1$. The spectral flow becomes completely suppressed,
anomaly disappears and one obtains $d_\perp = -n_{\rm n})/n_{\rm s}
\rightarrow 0$. This negative contribution comes solely from the Iordanskii
force. Both extreme regimes and the crossover between them at
$\omega_0\tau \sim 1$ have been observed in experiments with $^3$He-B vortices
\cite{BevanNature,BevanJLTP,LammiTalk} (Fig. \ref{CallanHarveyExpFig}). The
friction force is maximal in the crosover region and disappears in the two
extreme regimes.  In addition the experimental observation of the
negative $d_\perp$ at low $T$ (Fig. \ref{CallanHarveyExpFig}) verifies the
existence of the Iordanskii force; thus the analog of the gravitational
Aharonov-Bohm effect (Sec. \ref{GravitationalAharonov-BohmEffect}) has been
measured in $^3$He-B.

\section{Interface between two different vacua and vacuum pressure in
superfluid $^3$He.}\label{Interface}

Here we proceed to the effects related to the nontrivial effective gravity 
of quantum vacuum in superfluid $^3$He-A. The main source of such metric are
the topological objects: (1) disgyration whose effective metric is analogous
to that of the cosmic string with the large mass; (ii) quantized vortices
which reproduce the effective metric of spinning cosmic string; (iii) solitons
and interfaces which simulate the degenerate metric and event horizon; etc.
Let us start with the interface between $^3$He-A and $^3$He-B, which appears
to be useful for the consideration of the vacuum energy of states with
different broken symmetry. We start with the interface between different
vacua (Fig. \ref{ABInterfaceFig}).

\begin{figure}[t]
\centerline{\includegraphics[width=\linewidth]{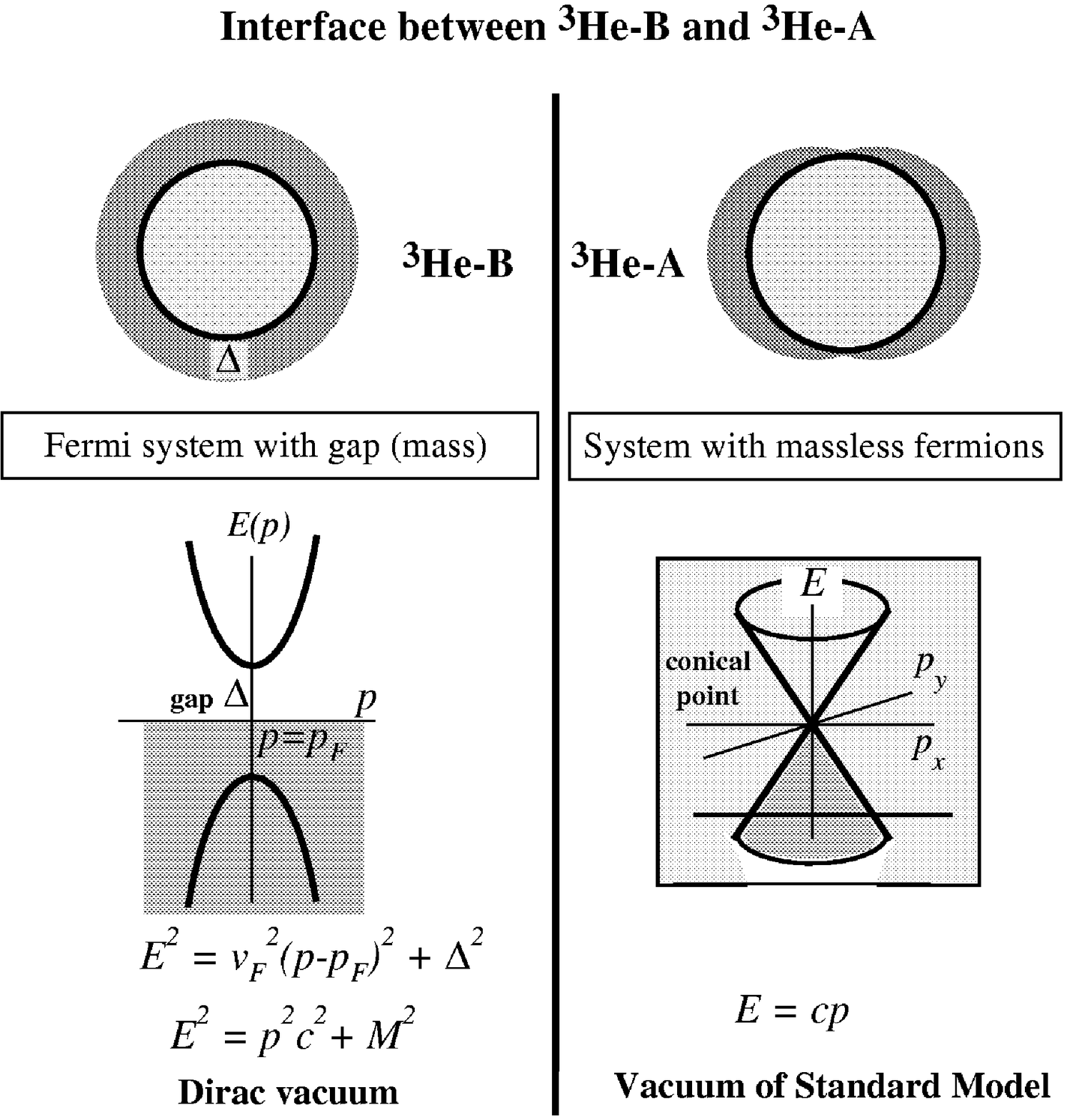}}
\medskip
\caption{Interface between two vacua of different universality classes.
Interface between $^3$He-A and $^3$He-B corresponds to the interface between
the vacuum of Standard Model and vacuum of Dirac fermions. 
Since the low-energy (quasi)-particles cannot penetrate from the right vacuum to
the left one, the interface is a perfect mirror. }
\label{ABInterfaceFig}
\end{figure}

\subsection{Interface between vacua of different universality classes and
Andreev reflection.}

\subsubsection{Fermions in two neighboring vacua.}

The vacua in $^3$He-A and $^3$He-B have different broken symmetries,
neither of them is the subgroup of the other, 
\cite{VollhardtWolfle} (Sec. \ref{BCS}). Thus the phase transition between
the two superfluids is of first order. The interface between them (the AB
interface) is stable and is stationary if the two phases have the same vacuum
energy. The difference between energies of $^3$He-A and $^3$He-B is regulated
by the magnetic field (see below) and temperature.  At $T\neq 0$ the
equilibrium and also the dynamics of the interface is determined by the
fermionic quasiparticles (Bogoliubov excitations). In the $^3$He-A vacuum the
fermions are chiral and massless, while in the $^3$He-B vacuum they are massive
according to Eq.(\ref{BogoliubovNambuEnergyBPhase}). At the temperature $T$
well below the temperature $T_c$ of the superfluid transition, when $T\ll
\Delta_0$, the thermal fermions are present only in the $^3$He-A. Close to the
gap nodes, i.e. at ${\bf p} \approx \pm p_F\hat {\bf l}$, the  energy
spectrum  of the  gapless $^3$He-A fermions  becomes ``relativistic''. Since
the $^3$He-B excitations are massive (gapped), the thermal excitations cannot
propagate from $^3$He-A through the AB interface (Fig.
\ref{ABInterfaceFig}). The scattering of the $^3$He-A fermions from the
interface, which in the BCS systems is known as Andreev reflection
\cite{Andreev}, is the dominating mechanism of the force experienced by the
moving AB interface. Due to the relativistic character of the $^3$He-A
fermions the dynamics of the interface becomes very similar to the motion of
the perfectly reflecting mirror in relativistic theories, which was heavily
discussed in the relation to the dynamic  Casimir effect (see eg
\cite{Neto,Law,Kardar}). So the investigation of the interface dynamics at
$T\ll T_c$ will give the possiblitity of the modelling of the  effects of
quantum vacuum.  On the other hand, using the relativistic invariance one can
easily calculate the forces acting on moving interface from the $^3$He-A heat
bath in the limit of low $T$ or from the $^3$He-A vacuum at $T=0$. This can be
done for any velocity of wall with respect to the superfluid vacuum and to the
heat bath. We discuss here the   velocities below the ``speed of light'' in
$^3$He-A. 

\subsubsection{Andreev reflection at the
interface.}\label{AndreevReflectionInterface}

The motion of the AB interface in the so called ballistic regime for the
quasiparticles has been considered in \cite{YipLeggett,KopninAB,LeggettYip} (see
also \cite{Palmeri}). In this regime the force on the  interface  comes from
the mirror reflection at the interface (Andreev reflection) of the
ballistically moving thermally distributed Fermi particles.
As in the case of moving vortex, three velocities are of importance in this
process: superfluid velocity of the vacuum ${\bf v}_{\rm s}$, normal
velocity of the heat bath (``matter'')
${\bf v}_{\rm n}$ and the velocity of the interface ${\bf v}_{\rm L}$. The
dissipation and thus the friction part of the force is absent when the wall is
stationary in the heat bath frame, i.e. when
${\bf v}_{\rm L}={\bf v}_{\rm n}$. 

Let us first recall the difference between the conventional and Andreev reflection in
superfluid $^3$He-A. For the low-energy quasiparticles their momentum is
concentrated in the vicinity of the Fermi momentum,
${\bf P}_a\approx -C_a p_F\hat{\bf l}$. In the conventional reflection   the
quasiparticle momentum is reversed, which in $^3$He-A means
that the quasiparticle acquires an opposite chirality $C_a$. The momentum
transfer in this process is $\Delta {\bf p}=2p_F\hat{\bf l}$, whose magnitude
$2p_F$ is well above the ``Planck'' momentum $m^*c_\perp$. The probability of
such process is exponentially small and becomes essential only if the
scattering center has an atomic size $a\sim 1/p_F$. Such atomic-size centers
are absent in the long wave limit of the effective theory. Even the AB
interface cannot produce such a process since the thickness $d$  of the
interface is about the coherence length
$d\sim \xi=\hbar v_F/\Delta_0= (c_\parallel/c_\perp)\hbar/p_F \sim
10^3~\hbar/p_F$, as a result the  usual scattering from the
AB interface is suppressed by huge factor $\exp(-p_Fd/\hbar)\sim
\exp(-c_\parallel/c_\perp)$. Thus the nonconservation of chirality of massless
chiral quasiparticle is possible, due to the ``transPlanckian'' physics,  but
it is exponentially suppressed. In relativistic theories such nonconservation
of chirality occurs in the lattice models, where the distance
between the Fermi points of opposite chiralities is of order of Planck
momentum \cite{NielsenNinomiya}. 

In Andreev reflection, instead of the momentum itself, the deviation of the
momentum from the Fermi point changes sign, ${\bf p}-{\bf P}_a\rightarrow
-({\bf p}-{\bf P}_a)$. The velocity of quasiparticle is reversed in this
process, but the momentum change can be arbitrary small, so that there is no
exponential suppression. In this process the chirality $C_a$ of quasiparticle
does not change. While in terms of the condensed matter observables, the
Andreev reflection is accompanied by the transformation of the particle to the
hole, it corresponds to the conventional reflection in relativistic theories.

We use the reference frame in which the interface is stationary and is
situated at the plane $x=0$. In this frame the order parameter (and thus the
metric) is time-independent and the energy of quasiparticles is well defined.
Let us first consider the simplified version when the quasiparticles at $x>0$,
where they are massless (Fig. \ref{ABInterfaceFig}), have an isotropic
``relativistic'' spectrum
$E=cp$.  The modification to the relevant $^3$He-A case will be obtained by
simple rescaling. The  spectrum of quasiparticles in the frame of the wall is
$\tilde E=E(p) + {\bf p}\cdot{\bf v}_{\rm s}$, where ${\bf v}_{\rm s}$ is the
stationary superfluid velocity in the wall frame. The equilibrium distribution
function of quasiparticles does not depend on the reference frame since it is
determined by the Galilean invariant counterflow velocity ${\bf w}={\bf
v}_{\rm n}-{\bf v}_{\rm s}$:  $f_{\cal T}({\bf p})=1/(1+e^{(E(p) -{\bf
p}\cdot{\bf w})/T})$ (compare with
Eq.(\ref{EquilibriumDistributionFunction}) for the case of the vortex).  

At $x<0$ fermions are massive. It is assumed that the temperature $T$ is much
less than the quasiparticle mass (or gap), so that there is practically the
vacuum state at $x<0$. The low-energy quasiparticles at $x>0$ cannot
penetrate through the wall; they are fully reflected and this provides
an analogue of perfect mirror. The reflection of thermal quasiparticles
leads to additional (``matter'') pressure on the wall from the right vacuum
and also, if the interface is moving with respect to the heat bath (${\bf
v}_L\neq {\bf v}_{\rm n}$), to the friction force experienced by the mirror.

The force acting on the wall from the massless quasiparticles living in the
half-space 
$x>0$ can be easily calculated in the ballistic regime, when the main
mechanism of the momentum transfer from the heat bath to the wall is the
scattering at the wall:
\begin{equation}
F_x=\sum_{\bf p} \Delta p_x v_{Gx} f_{\cal T}({\bf p})~~.
\label{AndreevForce}
\end{equation} 
Here ${\bf v}_G$ is the group velocity of incident  particles: 
\begin{equation}
v_{Gx}={d\tilde E \over dp_x}=c\cos\theta + v_{\rm s}~,
\end{equation}
$\Delta p_x$ is the momentum transfer after reflection
\begin{equation}
\Delta p_x=2p {\cos\theta+ v_{\rm s}/c\over  1 - 
v_{\rm s}^2/c^2}~~,
\label{AndreevMomentumTransfer}
\end{equation}  
and  $\theta$ is the angle between the  momentum ${\bf p}$ of incident
particle and the velocity of the wall
${\bf v}_{\rm L}$. Since the quasiparticle momentum $p$ is small the
momentum change $\Delta p_x$ is small compared to the cut-off
parameter $p_F$.  In pair-correlated condensed matter systems this correponds
to the Andreev reflection.

\subsection{Force acting on moving mirror from thermal relativistic
fermions.}

\subsubsection{Relativistic case.}

In the case of the thermal distribution of quasiparticles, it follows from
Eqs. (\ref{AndreevForce}) and (\ref{AndreevMomentumTransfer}) that the force
per unit area is ($c=1$):
\begin{equation}
{F_x(v_{\rm s},w)\over A}= - \hbar c {7 \pi^2 \over 60} {T^4\over (\hbar
c)^4}\alpha(v_{\rm s},w) ~~,
\label{AndreevForce1} 
\end{equation}
\begin{equation}
\alpha(v_{\rm s},w)= {1\over 1-v_{\rm s}^2}\int_{-1}^{-v_{\rm s}}  d\mu~ 
{(\mu +v_{\rm s})^2 \over (1-\mu w)^4}= {(1- v_{\rm s})^2\over 3(1+w)^3(1+ v_{\rm
s})(1+v_{\rm s}w)} ~~.
\label{AndreevForce2} 
\end{equation}
The force  disappears at $v_{\rm s}\rightarrow c$, because quasiparticles
cannot reach the interface moving from them with the ``speed of light''. The
force diverges at $v_{\rm s}\rightarrow -c$, when all quasiparticles become
trapped by the interface, so that the interface reminds the horizon of black
hole. 
 
In a global thermal equilibrium, which occurs when the normal component
is at
rest in the interface frame, i.e.  at $v_{\rm n}=0$ (see Sec.
\ref{GlobalThermodynamicEquilibrium}), the Eq.(\ref{AndreevForce1}) gives a
conventional pressure acting on the interface from the gas of (quasi)particles:
\begin{equation}
 {F_x(v_{\rm s},w=-v_{\rm s})\over A}=\Omega=\hbar c {7 \pi^2 \over 180} {T^4\over (\hbar
c)^4(1-v_{\rm s}^2)^2}=  {7 \pi^2 \over 180 \hbar^3} \sqrt{-g}  T_{\rm eff}^4    ~~
\label{AndreevForceEquilibrium} 
\end{equation} 
where again $T_{\rm eff}=T/\sqrt{-g_{00}}=T/\sqrt{1-v_{\rm s}^2}$,  see
Eq.(\ref{TolmanLaw}), and the real thermodynamic temperature $T$ again plays
the role of the Tolman's temperature in general relativity.

Now let us consider a small deviation from the equilibrium, $v_{\rm n}\neq 0$ but small.
Then the friction force appears, which is linear
in $v_{\rm n}$ in the interface frame and thus is proportional to the velocity of
the interface with respect to the normal component $v_L-v_{\rm n}$:  
\begin{equation}
{F_{\rm friction}\over A} =-\Gamma (v_L-v_{\rm n})~~,
~~\Gamma={3\over c}\Omega~~.
\label{FrictionForce} 
\end{equation}

\subsubsection{Force acting on moving AB interface.}

Now let us apply the obtained results to the $^3$He-A, which has an  anisotropic
velocity of light and also contains the vector potential ${\bf A}=p_F\hat {\bf
l}$.  Typically $\hat{\bf l}$ is parallel to the AB-interface, which
is dictated by boundary conditions. In such geometry the effective gauge
field is irrelevant: the constant vector potential ${\bf A}=p_F\hat {\bf
l}_0$ can be gauged away by shifting the momentum. The scalar potential
$A_0={\bf A}\cdot {\bf v}_{\rm s}$, which is obtained from the Doppler shift,
is zero since $\hat{\bf l}\perp {\bf v}_{\rm s}$ in the considered geometry.
In the same way the effective chemical potential $\mu_a=-C_ap_F(\hat{\bf
l}\cdot{\bf w})$ is also zero in this geometry. Thus if ${\bf p}$ is counted
from $e{\bf A}$ the situation becomes the same in relativistic case and one
can apply the equations (\ref{AndreevForceEquilibrium})
and (\ref{FrictionForce}). In the limit of small relative velocity $v_L-v_{\rm
n}$ one has
\begin{equation}
 {F_x\over A}=- {7 \pi^2 \over 180} \sqrt{-g}  T_{\rm eff}^4 \left(1+ 3{v_L-v_{\rm n}
\over c_\perp}\right)  ~~,~~T_{\rm eff}={T\over \sqrt{1-{{\bf v}_{\rm s}^2\over
c_\perp^2}}}~~.
\label{ForceOnAB} 
\end{equation} 
Here as before $\sqrt{-g}= 1/ c_\perp^2 c_\parallel$.

If one considers the steady state motion of the interface the conservation of
the particle current across the interface should be obeyed, i.e.  $n{\bf v}_{\rm s}
+\sum_{\bf p}({\bf p}/m)f_{\cal T}({\bf p})= const$. Since there are no
quasiparticles on the $^3$He-B side, the net quasiparticle momentum on the
$^3$He-A side must be also zero. This requires ${\bf w}=0$, i.e. ${\bf v}_{\rm
s}={\bf v}_{\rm n}$, and for this case one has
\begin{equation}
 {F_x\over A}=- {7 \pi^2 \over 180} \sqrt{-g}  T^4 {(1-v_{\rm s}/ c_\perp)^3\over
1-v_{\rm s}^2/ c_\perp^2}~.
\label{ForceOnAB2} 
\end{equation}

\subsection{Vacuum pressure and vacuum energy in $^3$He.}
\label{VacuumPressureSection}
 
\subsubsection{Interplay between vacuum pressure and pressure of matter.}

The pressure from the $^3$He-A side, which arises at finite $T$ can be compensated by the
difference in the vacuum pressure between $^3$He-A and $^3$He-B, so that the interface can be
stabilized. The energy difference between the two vacua is regulated by
external magnetic field $H$. The magnetic field distorts the $^3$He-B, so that
its energy depends on $H$ quadratically, $\tilde\epsilon_{\rm
vac~B}(H)=\tilde\epsilon_{\rm vac~B}(H=0)+
\beta H^2$, but it does not influence the $^3$He-A: with the same accuracy one has
$\tilde\epsilon_{\rm vac~A}(H)=\epsilon_{\rm vac~A}(H=0)$ \cite{VollhardtWolfle}. If $H_0$ is
the magnetic field at which the AB interface is in equilibrium at $T=0$, i.e.
$\tilde\epsilon_{\rm vac~B}(H_0)=\tilde\epsilon_{\rm vac~A}$, then at finite temperature the
magnetic field needed to equilibrate the interface is shifted:
$H^2(T)=H_0^2-{7 \pi^2 \over 180 \beta} \sqrt{-g}  T_{\rm eff}^4$. 

The interplay between the vacuum pressure and the pressure of the
``matter'' (quasiparticles)  is observed in experiments made in
Lancaster \cite{FisherPickett}. They measured the latent heat of transition 
-- the thermal energy released when the AB interface is pushed to the A-phase
side by decreasing magnetic field. While the total number of thermal
quasiparticles in $^3$He-A is conserved, their density increases with the
decrease of the volume of the $^3$He-A vacuum and the temperature rises. 

Similar situation could occur if both vacua have massless excitations, but
the ``speeds of light'' are different. In the acoustic case this can be the
interface between the Bose condensates with different density $n$ and speed of
sound $c$ on left and right sides of the interface. The simplest example is 
\begin{equation}
 g^{00}_L=-{1\over c_L^2} ~,~g^{00}_R=-{1\over c_R^2} ~,~  g^{ij}_L=g^{ij}_R=
\delta^{ij}~.
\label{MetricsAcrossinterface}
\end{equation}  
This effective space is flat everywhere except for the interface itself:
\begin{equation}
 R_{0101} =c\partial_x^2 c= {1\over 2}(c^2_R-c^2_L)\delta'(x) - (c_R-c_L)^2\delta^2(x)~.
\label{CurvatureAtinterface}
\end{equation} 
The situation similar to the AB interface corresponds to the case when $c_L\gg c_R$, so that
there are practically no quasiparticles on the left hand side of the interface, even if the
temperature is finite. The overpressure from the right hand side quasiparticles which arises
at the nonzero
$T$
\begin{equation}
 \Delta P=   {7 \pi^2 \over 180} \sqrt{-g} { T^4 \over g_{00}^2}~,
\label{OverPressure} 
\end{equation}
can be compensated by the difference in the vacuum pressures $P_{\rm vac~L}  - P_{\rm
vac~R}=\tilde\epsilon_{\rm vac~R}-\tilde\epsilon_{\rm vac~L}$, which is
determined by the ``transPlanckian physics'' (see Sec.\ref{VacuumEnergyAnd}).
In this way the order of magnitude coincidence between the vacuum energy (the
cosmological constant) and the energy of matter naturally arises.

\subsubsection{Vacuum energy in $^3$He.}
\label{VacuumEnergy3He}
 
Let us estimate the vacuum energy in $^3$He-A and $^3$He-B  using the BCS
theory. The  energy of
the superfluid, obtained from the diagonalization of the action in
Eqs.(\ref{BCSAction1}-\ref{BCSAction3}) by using the Bogoliubov
transformation, is
\begin{equation}
\tilde\epsilon_{\rm vac}= -{1\over 2}\sum_{{\bf p}\sigma} E ({\bf
p})~+~{1\over 2}\sum_{{\bf p}\sigma} M ({\bf p}) ~ +~{\Delta_0^2\over g}~.
\label{VacuumEnergyFermiSuper2}  
\end{equation}
Here 
\begin{eqnarray}
 E({\bf p})=  \sqrt{M^2({\bf p}) +  \Delta^2({\bf p})} ~,~M({\bf
p})={p^2\over 2m}-\mu\approx v_F(p-p_F)~,
\label{VacuumEnergyFermiSuper3} 
\\
\Delta^2({\bf p})=c_\perp^2({\bf p}\times \hat {\bf l})^2\approx
\Delta_0^2(\hat{\bf p}\times \hat {\bf l})^2 ~~{\rm in}~^3{\rm
He-A}~~,
\label{VacuumEnergyFermiSuper3A} 
\\
\Delta^2({\bf p})=c^2p^2\approx
\Delta_0^2~~{\rm in}~^3{\rm He-B}~~.
\label{VacuumEnergyFermiSuper3B} 
\end{eqnarray}
In Eqs.(\ref{VacuumEnergyFermiSuper3A}-\ref{VacuumEnergyFermiSuper3B}) we
took into account that the momentum $p$ is concentrated in the vicinity of
$p_F$. In Eq.(\ref{VacuumEnergyFermiSuper3}) we assumed for simplicity the
Fermi gas approximation for the underlying Fermi liquid; i.e. we
neglected the Fermi liquid corrections and put $m^*=m$ into the quasiparticle
energy spectrum
$M({\bf p})$ in the normal state of
$^3$He. As we shall see below the generalization to the real
liquid is straightforward. The last term in
Eq.(\ref{VacuumEnergyFermiSuper2}) is the Eq.(\ref{BCSAction3}) expressed in
terms of the amplitude of the order parameter $\Delta_0$; being written in this
form the coupling constant $g$ is different in $^3$He-A and $^3$He-B.

The first term in Eq.(\ref{VacuumEnergyFermiSuper2}) could be recognized as
the infinite energy of the ``Dirac vacuum'' of the occupied negative
energy states. The factor
$1/2$ in this term takes into account the  double counting of particles and
holes in the BCS theory. The energy of the ``Dirac vacuum''  catastrophically
diverges as
$-p^5/m$ for $p\gg p_F$, i.e at energies above the third ``Planck'' scale
$v_Fp_F$. This divergency is cancelled by the ``contr-term'' -- the second term
in Eq.(\ref{VacuumEnergyFermiSuper2}). Because of the cancellation the main
contribution to   Eq.(\ref{VacuumEnergyFermiSuper2}) becomes   of order of
third ``Planck'' energy per each particle:
$p_F^5/10m\pi^2 \sim n v_Fp_F$. This is the energy of the Fermi gas. For the
real liquid this would correspond to the energy of the liquid in the normal
state. Thus the further natural  regularization is achieved by consideration of
the difference between the energies of superfluid and normal liquids. This
effectively removes the contribution coming from  the region of energies above
the second  ``Planck'' energy scale $\Delta_0$.  Such regularization is
justified because we are interested in the energy difference between $^3$He-A
and $^3$He-B. At energies above
$\Delta_0$ there is no difference between these two phases, since they have the
same normal state. In the normal state the order parameter is zero and the
energy in the normal state is
\begin{equation}
\tilde\epsilon_{\rm norm}= {1\over 2}\sum_{{\bf p}\sigma}\left(M ({\bf p})- 
|M ({\bf p})| \right)=\sum_{{\bf p}\sigma} M ({\bf p})~\Theta
(-M ({\bf p}))
 ~.
\label{EnergyNormalState}  
\end{equation}
As a result one has 
for the energy difference:
\begin{eqnarray}
\nonumber\tilde\epsilon_{\rm vac}  -\tilde\epsilon_{\rm norm}= {1\over
2}\sum_{{\bf p}\sigma}\left(|M ({\bf p})|-E ({\bf p})\right)  
+ {\Delta_0^2\over g} \approx  -{1\over 4}\sum_{{\bf p}\sigma}{\Delta^2({\bf
p})\over E({\bf p})}   + {\Delta_0^2\over g}\\ \approx  -{p_F^2\over
4\pi^2v_F}\int_{p=p_F}{d\Omega\over 4\pi}\Delta^2({\bf p}) \ln
\left(\Lambda^2\over \Delta^2\right)   + {\Delta_0^2\over g}~.
\label{VacuumEnergyFermiSuper5a} 
\end{eqnarray} 
where the ultraviolet cut-off is $\Lambda=v_Fp_F$.
Now the remaining logarithmic divergency
can be removed, if the ground state is
considered. Minimizing the Eq.(\ref{VacuumEnergyFermiSuper5a})  
over the gap amplitude $\Delta_0$  by using equation  ${\rm min}(-x\ln
(1/x) + x/g)=-x_{\rm equilibrium}$,  one obtains  that in the ground state the 
logarithmic factor is cancelled. As a result the vacuum energy relevant for
the consideration of the relative energies of
$^3$He-A and
$^3$He-B is of order of the first ``Planck'' energy $\Delta_0^2/ v_Fp_F$
per particle (see also Eq.(3.52) in Ref.~\cite{VollhardtWolfle})  
\begin{equation}
\tilde\epsilon_{\rm vac}  -\tilde\epsilon_{\rm norm}
= -{p_F^2\over 4\pi^2v_F} \int_{p=p_F} {d\Omega\over 4\pi}\Delta^2({\bf
p})~â~\sim - n{\Delta_0^2\over v_Fp_F}~.
\label{VacuumEnergyFermiSuper5b}  
\end{equation}

The same equation (\ref{VacuumEnergyFermiSuper5b}) can be obtained using the
following Ansatz for the vacuum energy:
\begin{equation}
\tilde\epsilon_{\rm vac}= -{1\over
2}\sum_{{\bf p}\sigma}  E ({\bf p}) + \tilde\epsilon_{\rm norm}+ {1\over
2}\sum_{{\bf p}\sigma}\left( |M ({\bf p})| +{\Delta^2({\bf
p})\over 2E({\bf p})} \right)~.
\label{VacuumEnergyCounterterms} 
\end{equation}
Here the first term is the Dirac vacuum energy, while the other 3 terms
represent the ``counter-terms'' reflecting different stages of
``regularization''. Actually there was no real regularization, since
everything was obtained exactly within the microscopic BCS theory.
In principle, no cut-off parameters are needed for the BCS theory
which works at the ``transPlanckian'' scales too. All the integrals are
convergent and the ``cut-off'' parameters naturally arise within the theory:
they separate different regions of integration due to hierarchy of the
``Planck'' energy scales discussed in Sec. \ref{HierarchyEnergyScales}.    

\subsubsection{Cosmological term in $^3$He.}
\label{CosmologicalTerm3He}

Here we derived the Eq.(\ref{VacuumEnergyCounterterms}) 
assuming that  the normal state energy, the second term, is the energy of the
Fermi gas. However, the Eq.(\ref{VacuumEnergyCounterterms}) will hold for any
real liquid state too, if $\Delta_0\ll v_Fp_F$ and $\tilde\epsilon_{\rm norm}$
is understood the energy of the liquid in the normal state.  Let us recall that
the real liquid can be in equilibrium even without an external pressure. In a
full equilibrium at
$T=P=0$ one must have
$\tilde\epsilon_{\rm vac}=0$, as was discussed in Sec.\ref{VacuumEnergyAnd}.
On the other hand according to the same reasoning the normal state, which is
locally unstable equilibrium, must also have zero energy $\tilde\epsilon_{\rm
norm}=0$ at $T=0$, $P=0$. At first glance this is the paradox, since their
difference in Eq.(\ref{VacuumEnergyFermiSuper5a}) is certainly nonzero. But
there is no descrepancy since the superfluid and normal states at
$P=0$ and $T=0$ have different values of the chemical potential, $\mu_s$ and
$\mu_n$. Each of these values is ajusted to make the vacuum energy of the
corresponding state exactly zero. The difference between these chemical
potentials at $T=0$ and $P=0$ is $|\mu_n -\mu_s|\sim \Delta_0^2/v_Fp_F$. 

 In the $^3$He-A, where the angle average $\int{d\Omega\over
4\pi}\Delta^2({\bf p})=(2/3)\Delta_0^2$, the vacuum state energy in
Eq.(\ref{VacuumEnergyFermiSuper5a}) equals
\begin{equation}
\tilde\epsilon_{\rm vac}  -\tilde\epsilon_{\rm norm}=-{
\Delta_0^4\over 6
\pi^2}
\sqrt{-g}~~,~~\sqrt{-g}={1\over c_\perp^2c_\parallel}\equiv {p_F^2\over \Delta_0^2v_F}~.
\label{VacuumEnergyFermiSuper6}  
\end{equation}
This corresponds to the cosmological term with the Planck energy scale played by the parameter
$\Delta_0$. The same ``Planck'' energy determines
the ultraviolet cut-off of the logarithmically
divergent coupling in Eq.(\ref{RunningCoupling}). 

The cosmological term in Eq.(\ref{VacuumEnergyFermiSuper6}) cannot be obtained
using the effective field theory. But the order of magnitude of this vacuum
energy  can be estimated  by summation of the negative energies of the occupied
low-energy ``relativistic'' states of the Dirac vacuum in 
Eq.(\ref{SquareSpectrumAPhase}) using the cut-off at
$|E|=\Delta_0$:
\begin{equation}
\epsilon_{\rm vac~eff}=-\sum_{-\Delta_0<E<0} \sqrt{g^{ik}(p_i - eA_i)(p_k - eA_k)}=-{
\Delta_0^4\over 4 \pi^2}
\sqrt{-g}~.
\label{VacuumEnergyFermiSuper7}  
\end{equation}
Inspite of the close analogy between Eqs. 
(\ref{VacuumEnergyFermiSuper6}) and (\ref{VacuumEnergyFermiSuper6}) this 
vacuum energy is not ``gravitating'' since the total vacuum
energy in Eq.(\ref{VacuumEnergyCounterterms}) is zero for the real liquid. 

If the phase transition between $^3$He-A and $^3$He-B occurs at $T=0$ and
$P=0$ (this can be realized by applying the magnetic field) the energy density
$\tilde\epsilon$ of each vacuum remains zero: the ``vacuum is not
gravitating'' below and above transition. If $^3$He-A and
$^3$He-B coexist, the nonzero contribution to $\tilde\epsilon$ comes only from the
interface between them, i.e. the topological defects must ``gravitate''.   

Nonequilibrium deviations $\delta g_{\mu\nu}$ of the effective metric from their equilibrium
value $g^{(0)}_{\mu\nu}$ are also ``gravitating'', i.e. they enter the equation for the
collective modes related to the metric. Let us expand the  vacuum energy  in terms of
$\delta g_{\mu\nu}$. Since the term linear in $\delta g_{\mu\nu}$ is absent due to the vacuum
stability condition, only the quadratic terms are relevant. In $^3$He-A these are
\begin{eqnarray}
\tilde\epsilon(\delta g^{\mu\nu}) = { \Delta_0^4\over 24 \pi^2  }
\sqrt{-g^{(0)} }~ \left( {1\over 4}(\delta g^{11}g^{(0)}_{11} -\delta
g^{22}g^{(0)}_{22})^2  + (\delta
g^{12})^2 g^{(0)}_{11}g^{(0)}_{22}\right)
\label{GravMassTerm1} \\
 + { \Delta_0^4\over 48 \pi^2  }
\sqrt{-g^{(0)} } (\delta g^{11}g^{(0)}_{11} +\delta
g^{22}g^{(0)}_{22})^2
\label{GravMassTerm2}  
\end{eqnarray}
The Eq.(\ref{GravMassTerm1}), which gives rise to the mass of the conventional ``graviton'' --
perturbations with the momentum $L=2$, ie
$\delta g^{11} -\delta g^{22}$ and $\delta g^{12}$, describes the mass of the collective
clapping modes in $^3$He-A. The Eq.(\ref{GravMassTerm2}) is the mass term for the collective
pair-breaking mode, which corresponds to dilaton (see Sec. 5.17 in \cite{Exotic}). Both these
terms are determined by the ``transPlanckian'' physics and as a result the ``gravitons''
acquire the mass in $^3$He-A. It is interesting that in the BCS theory the
ratio of masses, $m_{\rm dilaton}/m_{\rm graviton} =\sqrt{2}$, is the same as
is given in the bi-metric theory of gravity in Ref.\cite{Logunov}.

\section{Vierbein defects.}\label{VierbeinDefects}

The field of the vierbein in general relativity can have topological defects, for example,
the point defects in 3+1 spacetime \cite{HansonRegge}. In $^3$He-A the relevant topological
objects in 3D space are line defects around which the dreibein rotates by $2\pi$
(Secs.~\ref{ConicalSpaces} and \ref{SpinningCosmicString}), and domain walls where one or
several vectors of the dreibein change sign across the wall. Let us start with the domain
walls.  At such wall in the 3D space (or at the 3D hypersurface in 3+1 space)  the vierbein is
degenerate, so that the determinant of the contravariant metric $g^{\mu\nu}$ becomes zero on
the surface.

\subsection{Vierbein domain wall.}\label{VierbeinDomainWall}

In gravity theory two types of the walls with the degenerate metric were considered: with
degenerate $g^{\mu\nu}$ and with degenerate $g_{\mu\nu}$\cite{Bengtsson}. The case of
degenerate  $g_{\mu\nu}$ was discussed in details in \cite{Bengtsson,BengtssonJacobson}.  
Both types of the walls could be generic. According to Horowitz
\cite{Horowitz}, for a dense set of coordinate transformations the generic
situation is the 3D  hypersurface where the covariant metric $g_{\mu\nu}$ has rank
3. The physical origin of the walls with the degenerate metric 
$g^{\mu\nu}$ in general relativity have been discussed by
Starobinsky \cite{Starobinsky}. They can arise after inflation, if the inflaton field has a
$Z_2$ degenerate vacuum. The domain wall separates the domains with  2 diferent
vacua of the inflaton field (in principle, the domains can have
different space-time topology, as is emphasized by Starobinsky \cite{Starobinsky}). The
metric $g^{\mu\nu}$ can everywhere satisfy the Einstein equations in vacuum, but at the
considered surfaces the metric $g^{\mu\nu}$ cannot be diagonalized as 
$g^{\mu\nu}={\rm diag}(1,-1,-1,-1)$. Instead, on such surface the metric is
diagonalized as $g^{\mu\nu}={\rm diag}(1,0,-1,-1)$ and thus cannot be inverted. 

Though the space-time can be flat everywhere, the coordinate transformation cannot remove
such a surface: it can only move the surface to infinity. Thus the system of such
vierbein domain walls divides the space-time into domains which cannot
communicate with each other. Each domain is flat and infinite as viewed by
 local ``inner'' observers who use the low energy ``relativistic'' quasiparticles
for communication. Such quasiparticles cannot cross the wall in the classical limit, so that
the observers living in $^3$He-A on different sides of the wall cannot communicate with each
other.  However, the ``Planck scale physics'' allows these worlds to communicate, since
quasiparticles with high enough energy are superluminal and thus can cross the wall.  This is
an example of the situation, when the effective spacetime, which is complete from the point of
view of the low energy observer, appears to be only a part of the more fundamental underlying
spacetime. 
 
In condensed matter the vierbein domain wall can be simulated by topological solitons in
superfluids and superconductors -- domain walls,  at which some of the three
``speeds of light'' crosses zero: in superfluid $^3$He-B
\cite{SalomaaVolovik,Volovik1990}; in chiral $p$-wave superconductors
\cite{MatsumotoSigrist,SigristAgterberg};   in $d$-wave superconductors
\cite{Volovik1997}; and in thin $^3$He-A films \cite{JacobsonVolovikThinFilm,VierbeinWalls}.
We consider the domain wall discussed in \cite{VierbeinWalls}, which simulates
the vierbein walls separating two flat space-time domains. When such vierbein
wall moves, it spilts into a black hole/white hole pair, which experiences the
quantum friction force due to Hawking radiation \cite{JacobsonVolovikThinFilm}.
We first discuss the stationary wall. Since the wall is topologically stable it
does not experience any dissipation.

\subsubsection{Vierbein wall in $^3$He-A film.}\label{VierbeinWallAPhase}
 
The simplest example of the vierbein walls we are interested in is provided by the
domain wall in superfluid $^3$He-A film which separates domains with opposite orientations of
the unit vector $\hat {\bf l}$ of the orbital momentum of Cooper pairs: $\hat {\bf l} =\pm 
\hat {\bf z}$.  Here the $\hat{\bf z}$ is along the normal to the film. 
The Bogoliubov-Nambu Hamiltonian for fermionic quasiparticles in
Eq.(\ref{BogoliubovNambuHamAPhase}) is
\begin{equation}
{\cal H}= \left({p_x^2 +p_y^2-  p_F^2\over 2m^*}\right)\check\tau^3 + {\bf e}_1\cdot {\bf
p}
\check\tau^1 +{\bf e}_2\cdot {\bf
p}
\check\tau^2
\label{HGeneral}
\end{equation}
Here as before $\check\tau^b$ are $2\times 2$ matrices for the Bogoliubov-Nambu spin, and we
neflected the conventional spin structure;
${\bf p}=
\hat{\bf x}p_x+ \hat{\bf y}p_y$ is the 2D momentum (for simplicity we
assume that the film is narrow so that the motion along the normal to the
film is quantized and only the motion along the film is free); the vectors ${\bf e}_1$ and
${\bf e}_2$  of the dreibein are given by Eq.(\ref{Dreibein}) in equilibrium $^3$He-A. But
within the domain wall the speeds of light can deviate from its equilibrium values
$c_\perp$. We assume the following order parameter texture in the wall:
\begin{equation}
{\bf e}_1(x)  =\hat{\bf x} c_x(x) ~,~{\bf e}_2 =\hat{\bf y} c_y(x)~.
\label{Texture}
\end{equation}
where the ``speed of light'' propagating along the axis $y$ is constant, while
the ``speed of light''  propagating along the axis
$x$ changes sign across the wall:
\begin{equation}
 c_y(x)=c_\perp~,~c_x(x)\equiv c(x)=   c_\perp\tanh {x\over d}  ~~,
\label{SpeedsLight}
\end{equation}
Across the wall the unit vector along the orbital momentum of Cooper pairs changes sign: 
\begin{equation}
\hat{\bf l}={{\bf e}_1\times{\bf e}_2\over 
|{\bf e}_1\times{\bf e}_2|}=\hat{\bf z}~{\rm sign}(x)~.
\label{lAcrossWall}
\end{equation}
At $x=0$ the dreibein is degenerate: the vector product ${\bf e}_1\times{\bf
e}_2=0$ and the third vector of the dreibein, the $\hat {\bf l}$ vector, is not
determined.
 
Since the momentum projection $p_y$ is the conserved quantity,  we come to
a pure 1+1 problem. Further we assume that  (i) $p_y= \pm p_F$; and (ii) the
parameters of the system are such that the thickness $d$ of the domain wall is
larger than the ``Planck'' legnth scale: $d
\gg \hbar/ m^*c_\perp$. This allows us to consider the ``relativistic'' range of
the momentum 
$\hbar/d \ll p_x \ll mc_\perp$, where the nonlinear correction $p_x^2$ can be
either neglected as compared to the relativistic term or considered in the
semiclassical approximation. Then rotating the Bogoliubov
spin and neglecting the noncommutativity of the $p_x^2$ term and $c(x)$ one has the
following Hamiltonian for the 1+1 particle:
\begin{eqnarray}
{\cal H}= M( {\cal P})\check\tau^3  +  {1\over 2} (c(x) {\cal P}  + {\cal P} 
c(x)) \check\tau^1~,\\    M^2( {\cal P}) ={{\cal P}^4 \over
4m^2}  + c_\perp^2p_y^2~.
\label{DiracCorrected}
\end{eqnarray}
where the momentum operator ${\cal P}_x
=-i\partial_x$ is introduced.
If the ${\cal P}^2$ term is completely neglected, one obtains the 1+1 Dirac
fermions
\begin{eqnarray}
{\cal H}=M\check\tau^3  +  {1\over 2} (c(x) {\cal P}  + {\cal P} 
c(x)) \check\tau^1~,\\   M^2 = M^2( {\cal P}=0)= c_\perp^2p_y^2~.
\label{Dirac}
\end{eqnarray}
The
classical spectrum of quasiparticles,
\begin{equation}
E^2 - c^2(x)p_x^2=M^2 ~,
\label{E}
\end{equation}
corresponds to the  contravariant metric
\begin{equation}
g^{00}=-1~,~g^{xx}= c^2(x)~~.
\label{metric}
\end{equation}
The line element of the effective space-time is
\begin{equation}
ds^2= -dt^2  + \bigl(c(x)\bigr)^{-2}\, dx^2  ~.
\label{LineElementWall}
\end{equation}

The metric element $g_{xx}$ is infinite at $x=0$, however the curvature is everywhere zero.
Thus the Eq.(\ref{LineElementWall}) represents a {\it flat} effective spacetime for any
function
$c(x)$. However, the singularity at $x=0$, where $g_{xx}=\infty$, cannot be removed
by the coordinate transformation. If at $x>0$ one introduces a
new coordinate
$\xi=\int dx/c(x)$, then the line element takes the standard  flat form 
\begin{equation}
ds^2=-dt^2 + d\xi^2  ~.
\label{LineElementFlatWall}
\end{equation}
However, the other domain -- the half-space with $x<0$ -- is completely removed by
such transformation.  The situation is thus the
same as discussed by Starobinsky for the domain wall in the inflaton field
\cite{Starobinsky}. 
 
These two flat spacetimes are disconnected in the relativisttic
approximation. However this approximation breaks down
near $x=0$, where the ``Planck energy physics'' becomes important and
nonlinearity in the energy spectrum appears in Eq.(\ref{DiracCorrected}):  The
two halves actually communicate due to the high-energy quasiparticles, which are
superluminal and thus can propagate through the wall. 

\subsubsection{Fermions across Vierbein Wall.}

In classical limit the low-energy relativistic quasiparticles do not
communicate across the vierbein wall, because the speed of light $c(x)$
vanishes at $x=0$. However the quantum mechanical connection can be
possible. 
There are two ways to treat the problem. In one approach one makes the
coordinate transformation first. Then in one of the domains, say, at $x>0$,
the line element is Eq.(\ref{LineElementFlatWall}), and one comes to the standard
solution for the Dirac particle propagating in flat space:
\begin{eqnarray}
\nonumber\chi(\xi)=
{A\over \sqrt{2}}\exp\left( i\xi\tilde E 
\right)\left(\matrix{Q
\cr
Q^{-1}  \cr}\right) +\\
{B\over \sqrt{2}}\exp\left(-
i\xi\tilde E \right)\left(\matrix{Q
\cr
-Q^{-1} \cr} \right) ~,
\label{DiracSolutionFlat}\\
\tilde E =\sqrt{E^2-M^2}~,~Q=\left({E+M\over E-M}\right)^{1/4}~.
\end{eqnarray}
Here $A$ and $B$ are arbitrary constants.
In this approach it makes no sense to discuss any connection to the other
domain, which simply does not exist in this representation.

In the second approach we do not make the coordinate transformation and work
with both domains. The wave function for the Hamiltonian Eq.(\ref{Dirac}) at
$x>0$ follows from the solution in Eq.(\ref{DiracSolutionFlat}) after restoring
the old coordinates:
\begin{eqnarray}
\nonumber\chi(x>0)=\\
\nonumber
{A\over \sqrt{2c(x)}}\exp\left( i\xi(x)\tilde E 
\right)\left(\matrix{Q 
\cr
Q^{-1}  \cr}\right) +\\
{B\over \sqrt{2c(x)}}\exp\left(-
i\xi(x)\tilde E \right)\left(\matrix{Q 
\cr
-Q^{-1}  \cr} \right),\\
\xi(x)=\int^x {dx\over
c(x)}
\label{DiracSolution+}
\end{eqnarray}
The similar solution exists at $x<0$.
We can now connect the solutions for the right and left half-spaces using (i) the
analytic cotinuation across the point $x=0$; and (ii) the conservation of the
quasiparticle current across the interface. The quasiparticle current e.g. at $x>0$
is
\begin{equation}
j =  c(x)  \chi^\dagger \check\tau^1 \chi=   |A|^2
-|B|^2  ~.
\label{QuasiparticleCurrent}
\end{equation}

The analytic cotinuation depends on the choice of the contour  around the $x=0$
in the complex $x$ plane. Thus starting from Eq.(\ref{DiracSolution+}) we
obtain two possible solutions at $x<0$. The first solution is obtained when the
point $x=0$ is shifted to the lower part of the complex plane:
\begin{eqnarray}
\nonumber\chi^{I}(x<0)=\\
\nonumber
{-iA e^{ -{\tilde E\over 2T_{\rm H}}}\over \sqrt{2|c(x)|}}\exp\left( i\xi(x)\tilde E 
\right)\left(\matrix{Q 
\cr
Q^{-1}  \cr}\right) +\\
{-iB e^{ { \tilde E\over 2T_{\rm H}}}\over \sqrt{2|c(x)[}}\exp\left(-
i\xi(x)\tilde E \right)\left(\matrix{Q 
\cr
-Q^{-1}  \cr} \right),
\label{DiracSolution-I}
\end{eqnarray}
where   $T_{\rm H}$ is
\begin{equation}
T_{\rm H}= {\hbar\over 2\pi}~ {dc\over dx} \bigg|_{x=0} ~.
\label{HawkingLikeT}
\end{equation}
The conservation of the quasiparticle current (\ref{QuasiparticleCurrent})
across the point $x=0$ gives the connection between parameters $A$ and
$B$:
\begin{equation}
|A|^2
-|B|^2 =|B|^2e^{ {\tilde E\over T_{\rm H}}} - |A|^2e^{ -{\tilde E\over T_{\rm H}}}~.
\label{CurrentConsI}
\end{equation}

The quantity $T_{\rm H}$ looks like the Hawking
radiation temperature determined at the singularity. As follows from
Ref.\cite{JacobsonVolovikThinFilm} it is the limit of the Hawking temperature
when the white hole and black hole horizons in the moving wall merge to form
the static vierbein wall (see Eq.(\ref{T(v)}) below). Note, that there is no
real radiation when the wall does not move.   The parameter $T_{\rm H}/\tilde E
\sim \partial \lambda /\partial x
$, where $\lambda= 2\pi/p_x=(2\pi /\tilde E) \partial c /\partial x$ is the de
Broglie wavelength of the quasiparticle. Thus the quasiclassical approximation holds
if $T_{\rm H}/\tilde E \ll 1$.

The second solution is obtained when the point $x=0$ is shifted to the upper
half-plane:
\begin{eqnarray}
\nonumber\chi^{II}(x<0)=\\
\nonumber
{iA e^{ {\tilde E\over 2T_H}}\over \sqrt{2|c(x)|}}\exp\left( i\xi(x)\tilde E 
\right)\left(\matrix{Q 
\cr
Q^{-1}  \cr}\right) +\\
{iB e^{ -{\tilde E\over 2T_H}}\over \sqrt{2|c(x)[}}\exp\left(-
i\xi(x)\tilde E \right)\left(\matrix{Q 
\cr
-Q^{-1}  \cr} \right),
\label{DiracSolution-II}
\end{eqnarray}
and the current conservation gives the following relation between parameters $A$
and
$B$:
\begin{equation}
|A|^2
-|B|^2 =|B|^2e^{ -{\tilde E\over T_H}} - |A|^2e^{ {\tilde E\over T_H}}~.
\label{CurrentConsII}
\end{equation}

Two solutions, the wave functions  $\chi^{I}$ and $\chi^{II}$, are connected by
the relation
\begin{equation}
\chi^{II}\propto\check\tau_3(\chi^{I})^* 
\label{parity}
\end{equation}
which follows from the symmetry
 of the Hamiltonian
\begin{equation}
H^*=\check\tau_3 H \check\tau_3 
\label{conjugation}
\end{equation}
The general solution is the linear combination of $\chi^{I}$ and $\chi^{II}$

Though on the classical level the two worlds on both
sides of the singularity are well separated, there is a quantum mechanical
interaction between the worlds across the vierbein wall.  The wave functions
across the wall are connected by the relation $\chi(-x)=\pm i\check\tau_3\chi^*(x)$
inspite of no possibility to communicate in the relativistic regime. 

\subsubsection{Communication across the wall via superluminal nonlinear dsipersion.}

In the above derivation we relied upon the analytic continuation and on the
conservation of the quasiparticle current across the wall. Let us justify this using
the nonlinear correction in Eq.(\ref{DiracCorrected}), which was neglected before.
We shall work in the quasiclassical approximation, which holds if $\tilde E\gg
T_H$. In a purely classical limit one has the dispersion
\begin{equation}
E^2= M^2 + c^2(x)p_x^2+{p_x^4\over 4(m^*)^2} ~,
\label{NonlineaClassical} 
\end{equation}
which determines two classical trajectories 
\begin{equation}
p_x(x)= \pm \sqrt{2m^*\left(\sqrt{\tilde E^2+(m^*)^2c^4(x)} -m^*c^2(x)\right)} ~.
\label{NonlineaClassicalTrajectories} 
\end{equation}
It is clear that there is no singularity at $x=0$, the two trajectories continuously
cross the domain wall in opposite directions, while the Bogoliubov spin continuously
changes its direction. Far from the wall these two trajectories give the two 
solutions,  $\chi^{I}$ and $\chi^{II}$, in the quasiclassical limit $\tilde E\gg
T_H$.  The function $\chi^{I}$ 
\begin{eqnarray}
\chi^{I}(x>0)={1\over \sqrt{2|c(x)|}}\exp\left( i\xi(x)\tilde E 
\right)\left(\matrix{Q 
\cr
Q^{-1}  \cr}\right),  \\
\chi^{I}(x<0)={-i\over \sqrt{2|c(x)|}}\exp\left( -i\xi(x)\tilde
E 
\right)\left(\matrix{Q 
\cr
-Q^{-1}  \cr}\right).
\label{QuasiclassicalSolutionI}
\end{eqnarray}  
describes the propagation of the quasiparticle from
the left to the right without reflection at the wall: in the quasiclassical
limit reflection is suppressed. The function $\chi^{II}$ desribes the propagation
in the opposite direction:
\begin{eqnarray}
\chi^{II}(x>0)={1\over \sqrt{2|c(x)|}}\exp\left( -i\xi(x)\tilde E 
\right)\left(\matrix{Q 
\cr
-Q^{-1}  \cr}\right),  \\
\chi^{II}(x<0)={i\over \sqrt{2|c(x)|}}\exp\left( i\xi(x)\tilde
E 
\right)\left(\matrix{Q 
\cr
Q^{-1}  \cr}\right),
\label{QuasiclassicalSolutionII}
\end{eqnarray}  
The quasiparticle
current far from the wall does obey the Eq.(\ref{QuasiparticleCurrent}) and is conserved across
the wall. This confirms the quantum mechanical connection between the spaces
obtained in previous section.

In the limit of small mass $M\rightarrow 0$, the particles become chiral with the
spin directed along or opposite to the momentum $p_x$. The spin structure of the
wave function in a semiclassical approximation is given by
\begin{equation}
\chi(x)= e^{i\check\tau_2{\alpha\over 2}}  \chi(+\infty) ~,~\tan \alpha= {p_x\over
2m^*c(x)}.
\label{SpinRotation} 
\end{equation}
Since $\alpha$ changes by $\pi$ across the wall, the spin of
the chiral quasiparticle rotates by $\pi$: the righthanded particle
tranforms to the lefthanded one when the wall is crossed. 

It appears that there is a quantum mechanic
coherence between the two flat worlds,  which do not interact classically across
the vierbein wall. The coherence is established by  nonlinear  correction 
to the spectrum of chiral particle: $E^2(p)=c^2p^2 + \gamma p^4$. In our
consideration the nonlinear dispersion parameter $\gamma$ was chosen, which 
allows the superluminal propagation across the wall at high momenta $p$. In
this case the result actually does not depend on the magnitude of
$\gamma$: in the relativistic low energy limit the amplitudes of the 
wave function on the left and right sides of the wall remain equal in
the quasiclassical approximation, though in the low
energy corner the communication across the wall is classically forbidden. 
Thus the only relevant input of
the ``Planck energy'' physics is the mere possibility of the superluminal
communication  between the worlds across the wall.   That is why the
coherence between  particles propagating in two classically disconnected
worlds can be obtained even in the relativistic domain, by using the
analytic continuation  and the conservation of the particle current across the
vierbein wall.

In principle, the two worlds across the vierbein wall can be smoothly connected by the 
path which does not cross the wall, say, in a M\"obius strip geometry
\cite{Silagadze}. In $^3$He-A the M\"obius geometry can be reproduced by the
half-quantum vortex (see  Sec.
\ref{MajoranaFermion} and review \cite{VolovikHalfQuantum}).  This vortex is an
analog of Alice string considered in particle physics by Schwarz
\cite{Schwarz}. A particle which moves around an Alice string continuously
flips its charge or parity or enters the ``shadow'' world.

\subsection{Conical spaces.}\label{ConicalSpaces}

\subsubsection{Antigravitating string.}\label{AntigravitatingString}

An example of the linear defects  in the vierbein field is the radial
disgyration in $^3$He-A (Fig.\ref{AntigravitatingStringFig}). Around this
defect one of the vectors of dreibein in Eq.(\ref{Dreibein}), say, ${\bf e}_1$
remains constant, ${\bf e}_2=c_\perp\hat{\bf z}$, while the other two are
rotating by $2\pi$ 
\begin{equation}
{\hat l}({\bf r})= {\hat\rho}~,~{\bf e}_1=c_\perp\hat{\bf \phi} ~~,
\label{RadialDisgiration}
\end{equation}
where as  before $z,\rho,\phi$ are the cylindrical coordinates with  the axis
${\hat {\bf z}}$ being along the defect line.
The interval corresponding to the metric in Eq.(\ref{MetricAPhase}) is
\begin{equation}
ds^2=dt^2 - {1\over  c_\perp^2} dz^2 -{1\over   c_\parallel^2}
\left(dr^2 +  { c_\parallel^2 \over  c_\perp^2  } r^2 d\phi^2\right)
~~,
\label{IntervalForRDisg}
\end{equation}

\begin{figure}[t]
\centerline{\includegraphics[width=\linewidth]{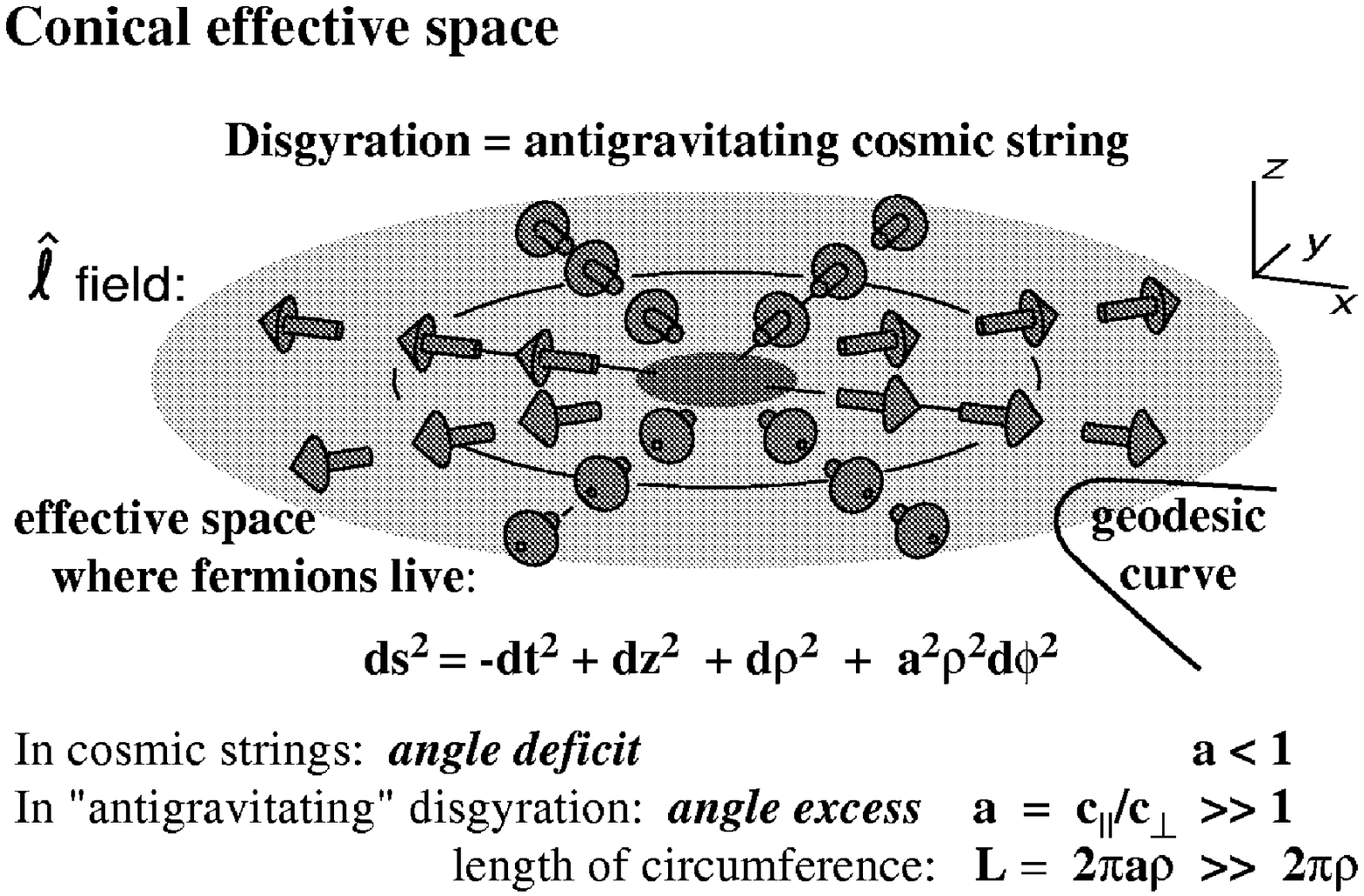}}
\medskip
\caption{The radial disgyration in $^3$He-A is equivalent to  
cosmic string with an excess angle. Since all the geodesic curves are
repelling from the string, the dysgyration serves as an example of the  
antigravitating string.}
\label{AntigravitatingStringFig}
\end{figure}

Rescaling the radial and axial coordinates
 $ \rho=c_\parallel R$ , $z= c_\perp Z $
one obtains
\begin{equation}
ds^2=dt^2 - dZ^2- dR^2 -a^2 R^2  d\phi^2~~,
~~a^2=c_\parallel^2/ c_\perp^2>1.
\label{RadialDysgMetric}
\end{equation}
In relativistic theories such conical metric, but with $a^2<1$, arises outside
the local strings.  The space outside the string core is
flat, but the proper length $2\pi R a$ of
the circumference of radius
$R$ around the axis is smaller than $2\pi R$, if $a<1$. This is so called
angle  deficit. In our case we have $a^2>1$, i.e. the  "negative angle 
deficit". The conical
singularity gives rise to the curvarture which is concentrated at the
axis of
disgyration ($R=0$)
\cite{SokolovStarobinsky,Banados}:
\begin{equation}
  {\cal R}^{R\phi}_ {R\phi}  =2\pi{a-1
\over  a }\delta_2({\bf R}) ~~,~ ~ \delta_2({\bf R})=\delta (X)\delta
(Y).
\label{CurvatureRDysg}
\end{equation}

Such metric can arise from the Einstein equations for the local
cosmic string with the singular energy density concentrated in the string core
\begin{equation}
 {\cal T}^{0}_ {0}  = {1-a
\over 4G a }\delta_2({\bf R})   ~~.
\label{NegativeEnergyDensity}
\end{equation}
where  $G$ is the gravitational constant. Since $a=c_\parallel /
c_\perp \gg
1$, this should be rather unusual cosmic string with a large negative
mass of
Planck scale, i.e. this string is antigravitating -- the trajectories of the
particles are repelled from the string (Fig.
\ref{AntigravitatingStringFig}).  

\subsubsection{Estimation of Newton constant.}\label{EstimationNewtonC
onstant}

If one finds such  singular  contribution to
the energy density of $^3$He-A in the presence of radial disgyration one can
extract the value of the effective gravitational constant in  $^3$He-A for this
particular case.

Let us consider the following Ansatz for the dreibein in the core of
the radial disgyration:
\begin{equation}
{\bf e}_2=c_\perp\hat{\bf z}~~,~~{\bf e}_1=f(\rho)c_\perp\hat{\bf \phi}
~~,~~f(\rho=0)=0~~,~~f(\rho=\infty)=1  ~~, 
\label{OrderParameter}
\end{equation}
which corresponds to the effective metric
\begin{equation}
ds^2=dt^2 - dZ^2- dR^2 -{a^2 \over f^{2}(Rc_\parallel)} R^2  d\phi^2~~ .
\label{GeneralRadialDysgMetric}
\end{equation}
The function $f(\rho)$ can be obtained from the Ginzburg-Landau free
energy
functional, Eq.(5.4) + Eq.(7.17) in \cite{VollhardtWolfle}, which for
the chosen
Anzats Eq.(\ref{OrderParameter}) has the form
\begin{eqnarray}
F=K{v_Fp_F^2\over 96\pi^2}  \int_0^{z_0} dz 
\int_{\rho<\rho_0} d^2\rho
\left[
\Lambda (1-f^2)^2 + {f^2\over
\rho^2} +
\left({df\over d\rho}\right)^2 \right]-
\label{DisgyrationFreeEnergy1}
\\
-  Kz_0{v_Fp_F^2\over 48\pi }  \int_0^{\rho_0} d\rho
{d(f^2)\over
d\rho}  ~~.
\label{DisgyrationFreeEnergy2}
\end{eqnarray}
Here $\rho_0$ and $z_0$ are the radius and the height of the
cylindrical
vessel with the disgyration on the axis;   $\Lambda \sim
\Delta_0^2(T)/v_F^2$; the overall dimensionless factor $K$ in the Ginzburg-
Landau
region close to the transition temperature $T_c$ is
\begin{equation}
 K(T) = 1-{T^2\over T_c^2} ~~,~~T\rightarrow T_c~.
\label{K(Tc)}
\end{equation}

The Eq.(\ref{DisgyrationFreeEnergy1}) is some kind of the
dilaton field.  The Eq.(\ref{DisgyrationFreeEnergy2}) is the pure divergence
and thus can be represented as the singular term, which does not depend on the
exact structure of the disgyration core, but nevertheless contributes the
core energy:
\begin{equation}
{\cal F}_{\rm div}= -2\pi K {v_Fk_F^2 \over 96 \pi^2}  \delta_2({\bf
\rho})
~~,~~F_{\rm div}=\int d^3x {\cal F}_{\rm div}= -  K{v_Fk_F^2 \over 48
\pi}
  ~~.
\label{SingularFreeEnergy}
\end{equation}
Now let us extract the ``Newton's constant'' $G$ for $^3$He-A by comparing this core energy
with the string mass
$M$  obtained by integration of ${\cal T}^0_0$:
\begin{equation}
 M= \int d^3X\sqrt{-g} {\cal T}^{0}_ {0}  =  {1-a
\over 4G} Z_0    ~~.
\label{NegativeMass1}
\end{equation}
Translating this to the $^3$He-A language, where the ``proper'' length
is
$Z_0=z_0
/c_\perp$,  and taking into account
that $a =c_\parallel/c_\perp \gg 1$ one has
\begin{equation}
 M=-   {c_\parallel
\over 4G c_\perp^2} z_0 ~~.
\label{NegativeMass2}
\end{equation}
Then from equation, $F_{\rm div}=M$, one obtains the ``gravitational
constant'' 
\begin{equation} G(T)=  {12 \pi \over  K(T)\Delta^2(T)}~.
\label{GravitationConstant}
\end{equation}
The same value is obtained from the energy-momentum
tensor for the analog of the graviton in $^3$He-A. $G(T)$ is inversely
proportional to the square of the ``Planck'' energy scale $\Delta_0(T)$ and depends on
$T$ increasing with $T$, which corresponds to the vacuum screening of
the
gravity. The temperature dependence of the gravitational
constant leads to its time dependence during the evolution of the
Universe. The latter has been heavily discussed starting with the Dirac
proposal (see Review\cite{Barrow}).

Though we cannot extrapolate the temperature dependence of $K(T)$ in
Eq.(\ref{K(Tc)}) to the low
$T$, we can expect that the overall temperature dependence of the Newton's
constant can be approximated
by
\begin{equation}
G(T)\sim G(T=0)  \left(1-{T^2\over T^2_c}\right)^{-2}~,~ G(T=0)\sim
{1\over
T_c^2} \sim {1\over
\Delta_0^2(0)}~.
\label{OverallTemperatureDependence}
\end{equation}
An exact value of $G$ in the low-$T$ regime will be discussed later in
Sec.\ref{Improved3He}. 

Note also that negative mass $M$ in Eq.(\ref{NegativeMass2}) does not mean that the vacuum
in $^3$He-A is unstable towards formation of the string: the  energy of the
radial disgyration is dominated by the positive energy term in
Eq.(\ref{DisgyrationFreeEnergy1}) which comes from the ``Planckian" physics:
\begin{equation}
 E_{\rm disg}(\rho_0)=K{v_Fk_F^2\over 48\pi } z_0\ln {\rho_0 \Delta_0
\over
c_\parallel}
  ~~.
\label{DisgyrationEnergy}
\end{equation}

\subsection{ Vortex vs spinning cosmic string.}\label{SpinningCosmicString}

Another example of the linear topological defects in the vierbein field is the quantized
vortex in $^3$He-A (Fig. \ref{SpinningStringFig}).  In its simplest
realization, which occurs, say, in
$^3$He-A films, where the $\hat{\bf l}$ is fixed by the boundary conditions,
the vortex structure is given by Eq.(\ref{VorticesP}). It can be laso written
in terms of the zweibein vectors
${\bf e}_1$ and
${\bf e}_2$, which are rotating by $2\pi$ around the origin
\begin{equation}
{\hat  {\bf l}}= {\hat  {\bf z}}~,~{\bf e}_1=f(\rho)c_\perp\hat{\bf \rho}~,~{\bf
e}_2=f(\rho)c_\perp\hat{\bf \phi} ~,~f(0)=0~,~f(\infty)=1.
\label{PlainVortex}
\end{equation}
From the definition of the superfluid velocity in Eq.(\ref{v_s}) one obtains
\begin{equation}
{\bf v}_{\rm s} ={\hbar\over 2 m\rho}\hat{\bf \phi}  ~,~\oint d{\bf x}
\cdot{\bf v}_{\rm s}\equiv \kappa= {\pi\hbar\over  m}.
\label{VortexField}
\end{equation}
For the general vortex in $^3$He-A in the parallel-plate geometry with fixed
$\hat{\bf l}$, the circulation $\kappa= n_1\pi\hbar/m$, where $n_1$ can be
integer or half-integer ($n_1=1/2$ for half-quantum vortex, Alice string  in
Sec. \ref{MajoranaFermion}\cite{VolovikHalfQuantum}). For
vortices in superfluid $^4$He one has $\kappa=n_1(2\pi\hbar /  m)$ with integer
$n_1$.

\begin{figure}[t]
\centerline{\includegraphics[width=\linewidth]{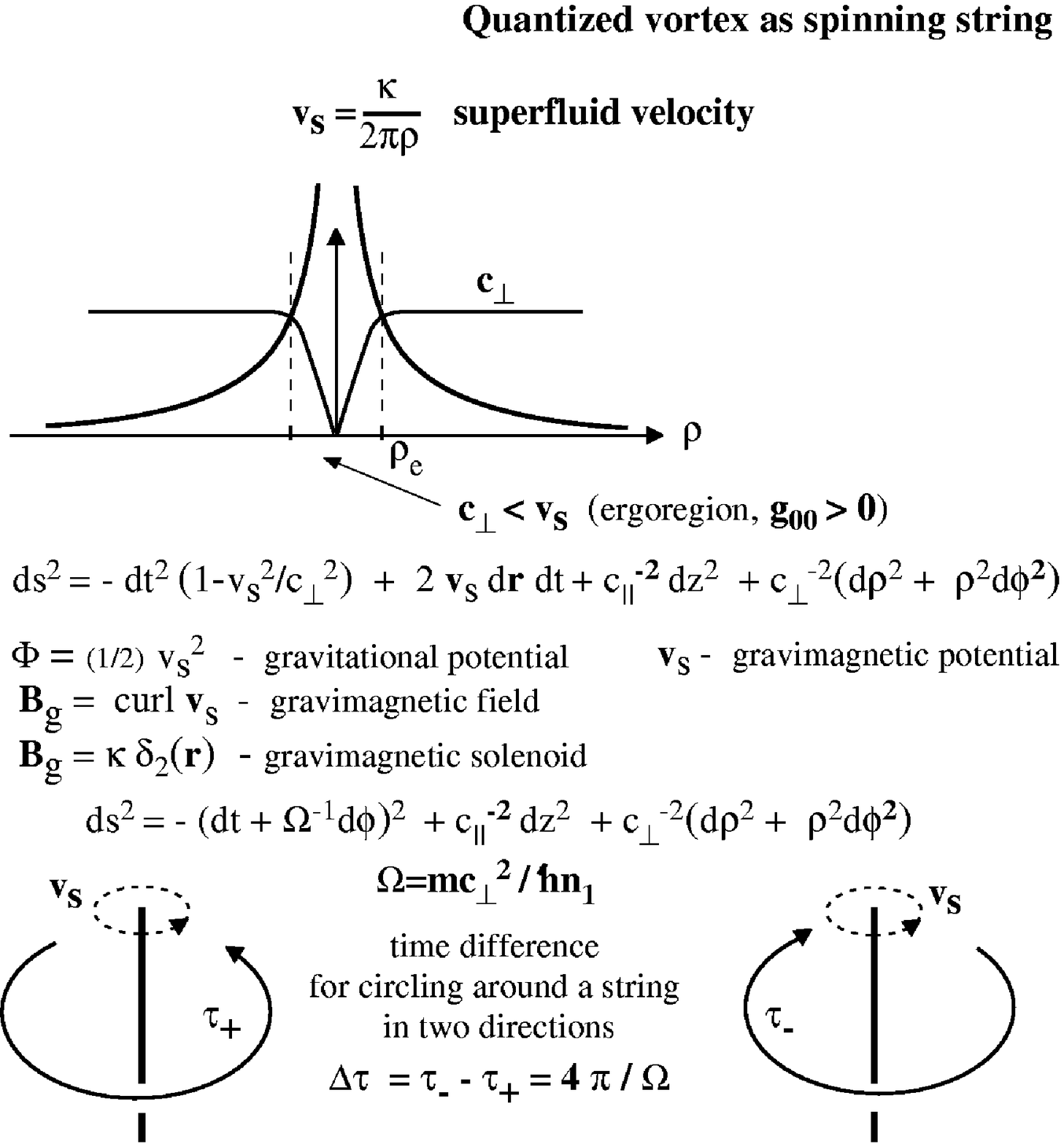}}
\medskip
\caption{The effective metric produced by superflow circulationg around
the vortex is similar to the metric of spinning string. {\it bottom}: As for
the spinning string, there is a constant time difference for the quasiparticle
circling with the ``speed of light'' around the vortex in clockwise and
anticlockwise directions. The string serves as a gravimagnetic solenoid and
quasiparticles experience the gravitational Aharonov-Bohm effects, which leads
to the additional Iordanskii force acting on a vortex. }
\label{SpinningStringFig}
\end{figure}

The superfluid velocity  field leads to Doppler shift of the quasiparticle energy which
in the low-energy limit modifies the effective metric according to
Eq.(\ref{MetricAPhaseGeneral}). The azimuthal flow around the core induces the effective
space, where quasiparticles  propagate along geodesic curves, with the interval 
\begin{equation}
  ds^2=-\left(1-{v_{\rm s}^2\over c_\perp^2}  \right)\left(dt +{\kappa
d\phi\over 2\pi( c_\perp^2-v_{\rm s}^2)} \right)^2 
+{d\rho^2\over c_\perp^2}
+{ \rho^2d\phi^2\over c_\perp^2-v_{\rm s}^2}+{dz^2\over c_\parallel^2}~.
\label{PhononInterval}
\end{equation}
The same metric is applicable for phonons propagating around the vortex in superfluid
$^4$He after the isotropic ``speed of light'' is introduced, $c_\perp=c_\parallel=c$.

Far from the vortex, where $v_{\rm s}^2/c_\perp^2$ is small and can be neglected, one has
\begin{equation}
ds^2=-\left( dt + {d\phi\over \omega}\right)^2 +{1\over c_\perp^2}(d\rho^2
+\rho^2d\phi^2)+{1\over c_\parallel^2} dz^2~~,~~\omega={2\pi  c_\perp^2 \over 
\kappa}
\label{IntervalVortexAsymp}
\end{equation}
The connection between the time and the azimuthal angle $\phi$  in
the interval suggests that there is a charactiristic angular
velocity $\omega$. For the phonons around the vortex in superfluid $^4$He 
$\omega= m c^2/n_1\hbar $, while for the $^3$He-A fermions $\omega= 2 m
c_\perp^2/n_1\hbar $.  The similar metric with rotation was obtained for the
so-called spinning cosmic string in $3+1$ space-time, which has the rotational
angular momentum 
$J$ concentrated in the string core, and for the spinning particle in the 2+1
gravity
\cite{CausalityViolation,MazurComment2,Staruszkievicz,Deser}:
\begin{equation}
ds^2=-\left( dt + {d\phi\over \omega}\right)^2 +{1\over c^2}(dz^2+ d\rho^2
+\rho^2d\phi^2)~~,~~\omega={1\over 4JG}
\label{IntervalSpinningStringAsymp}
\end{equation}
where $G$ is the gravitational constant. 
This gives the following correspondence between the circulation $\kappa$
around the vortex and the angular momentum $J$ of the spinning string 
\begin{equation}
\kappa = 8\pi JG~.
\label{JvsN}
\end{equation}
Thus vortices in superfluids simulate the spinning cosmic strings 
\cite{DavisShellard}. 

The   effect peculiar for the spinning string, which was modelled in
condensed matter, is  the gravitational AB topological effect
\cite{CausalityViolation}.  Outside the string the
metric, which enters the interval $ds$, is locally flat. But there is the time
difference for the particles propagating  around the spinning string  in
the opposite directions (Fig. \ref{SpinningStringFig}, bottom). Let us consider
the classical propagation of light along the circumference of radius $R$
assuming that there is confinement potential (mirrors). Then the trajectories
of phonons (null geodesics) at
$\rho=R$ and $z=0$ are described by the equation
$ds^2=0$, which at large distances from the core, $R\gg c/\omega$,  are the cw
and ccw  rotations with angular velocities 
\begin{equation}
\dot \phi_\pm={1\over \pm{R\over c} -{1\over \omega}}
\label{TimeDelay1}
\end{equation}
The difference in the periods $T_{\pm}=2\pi /\dot \phi_\pm$ for cw and ccw
motion of the phonon is
\cite{Harari}
\begin{equation}
2\tau={4\pi \over \omega}~~.
\label{Time Delay}
\end{equation}
The apparent ``speed of light'' measured by the internal observer is also
different for ``light'' propagating in opposite directions $c_\pm \approx
c(1 \pm c/\omega R)$.  

This asymmetry between the particles moving on different
sides of the vortex is the origin of the  Iordanskii force acting on the
vortex from the heat bath of quasiparticles (``matter''), which we
discuss in the next Section. On the quantum level, the connection between the
time variable
$t$ and the angle variable $\phi$ in the metric
Eq.(\ref{IntervalSpinningStringAsymp}) implies that the scattering cross
section of phonons (photons) on the vortex should be the periodic function of
the energy with the period equal to   $\hbar \omega$.  The asymmetric part of
this cross section gives rise to the  Iordanskii force.

It appears also, that even apart from the effective metric, the
condensed matter vortices and global spinning strings have the similar
properties. In particular, the spinning string generates the density of
the angular momentum in the vacuum outside the string
\cite{JensenKucera}. The angular momentum of the superfluid vacuum outside the vortex is also
nonzero and equals ${\bf L}=\int d^3x ~{\bf r}\times n{\bf v}_{\rm
s}$. For vortices in superfluid $^4$He this gives the density of the angular momentum
$n_1\hbar\hat{\bf z}$ per $^4$He atom.

\section{Gravitational Aharonov-Bohm effect and Iordanskii
force.}\label{GravitationalAharonov-BohmEffect}

\subsection{Gravitational Aharonov-Bohm effect.}

As we discussed in Secs.~\ref{AxialAnomalyAndForce} and \ref{VerificationAdlerBellJackiw}, in
superfluids, with their two-fluid hydrodynamics (for superfluid vacuum and quasiparticle,
which play the part of matter) there are 3 different topological contributions to the
force acting on the quantized vortex
\cite{3Forces}. The more familiar Magnus force arises when the vortex moves with respect to
the superfluid vacuum. For the relativistic cosmic string such force is absent since the
corresponding superfluid density of the quantum physical vacuum is zero. However the analog of
this force appears if the cosmic string  moves in the uniform background charge density
\cite{DavisShellard,Lee}. The other two forces of topological origin also have
analogs for the cosmic strings: one of them comes from the analog of the axial
anomaly in the core of electroweak string (see Sec.\ref{AnalogBaryogenesis}), and
another one -- the Iordanskii force \cite{Iordanskii} -- comes from the analog of the
gravitational Aharonov-Bohm effect \cite{VolovikGravAB,StoneIordanskii} experienced by the
spinning cosmic string discussed in the previous Section. The connection between the
Iordanskii force and conventional Aharonov-Bohm effect was developed in
\cite{Sonin1,SoninNew,Shelankov}. 

The Iordanskii force  arises when the vortex moves with respect to the
heat bath of excitations (``matter''). As we have seen from the
Eq.(\ref{IntervalSpinningStringAsymp}) there is a peculiar space-time
metric around the spinning string. This metric is locally flat, but its global properties
lead due to the time delay for any particle
orbiting around the string with the same speed, but in opposite directions, according to  
Eq.(\ref{Time Delay}). This gives rise to the quantum gravitational Aharonov-Bohm effect
\cite{CausalityViolation,MazurComment1,MazurComment2}.
We discuss here how the same effect leads to the asymmetry in the scattering of
particles on the spinning string and finally to the Iordanskii lifting force
acting on the vortex.  

In case of the superfluid $^4$He the equation for phonons with energy $E$
propagating in a curved space-time background created by the vortex follows
from the Lagrangian in Eq.(\ref{LagrangianSoundWaves} for scalar field. If one
neglects the change in the particle density around the vortex, one has
\begin{equation}
 {1\over c^2}\left(E  -  i{\bf v}_{\rm
s}\cdot\nabla\right)^2\alpha+\nabla^2\alpha=0~.
\label{EquationSoundWaves}
\end{equation} 
In the asymptotic region the quadratic terms ${\bf v}_{\rm s}^2/c^2$ can be neglected
and this equation can be rewritten as \cite{SoninNew}
\begin{equation}
E^2\alpha -c^2\left(-i\nabla + {E\over c}{\bf v}_{\rm s}({\bf r})\right)^2\alpha=0 ~~.
\label{ModifiedScalarField}
\end{equation}
This equation maps the problem under discussion to the Aharonov-Bohm (AB)
problem for the magnetic flux tube \cite{AB} with the vector potential ${\bf
A}={\bf v}_{\rm s}$, where the electric charge $e$ is substituted by the mass $E/c^2$
of the particle \cite{MazurComment1,JensenKucera,Galtsov}.

The symmetric part of the scattering cross section of quasiparticle with  energy
$E$ in the background of the vortex is \cite{SoninNew}:
\begin{equation}
{d \sigma_\parallel\over d\theta} = {\hbar c\over 2\pi E }\cot^2{\theta\over 2} ~  \sin^2  
{\pi E\over
\hbar \omega}~.
 \label{DiffCrossSectionVortex}
\end{equation}
This equation satsifies the periodicity of the cross section as a function of energy with the
period $\Delta E=\omega$ as is required by the spinning string metric in
Eq.(\ref{IntervalSpinningStringAsymp}). 
In superfluids the quasiparticle energy are typically small $E\ll \omega$. For small $E$ the
result in Eq.(\ref{DiffCrossSectionVortex}) was obtained by Fetter \cite{Fetter}. The
generalization of the Fetter result for the quasiparticles with  arbitrary spectrum
$E({\bf p})$ (rotons in $^4$He and the Bogoliubov-Nambu fermions in
superconductors) was recently suggested in Ref.\cite{Demircan}: In our notations
it is $d \sigma_\parallel/ d\theta=(\kappa^2 p/8\pi v_G^2) \cot^2(\theta/ 2)$, where
$v_G=dE/dp$ is the group velocity of quasiparticles. 

The equation for the particles scattered by the spinning string with zero mass gives
was suggested in Refs. \cite{MazurComment1,MazurComment2}:
\begin{equation}
{d \sigma_\parallel\over d\theta} = {\hbar c\over 2\pi E \sin^2(\theta/2)}~  \sin^2   {\pi
E\over
\hbar \omega}~,
 \label{DiffCrossSectionString}
\end{equation}
It preserves the most important properties of Eq.(\ref{DiffCrossSectionVortex}):
periodicity in $E$ and singularity at small scattering angle $\theta$.

This singularity at $\theta=0$ is the indication of the existence of the transverse
cross-section \cite{Shelankov}, which leads to the Lorentz force, which acts on the
magnetic flux tube in the presence of electric current carried by excitations, and to the
Iordanskii force, which acts on the vortex in the presence of the mass current carried by the
normal component of the liquid. The asymmetric part in the scattering of the
quasiparticles on the velocity field of the vortex has been calculated by Sonin
for phonons and rotons in $^4$He
\cite{SoninNew} and by Cleary \cite{Cleary} for the
Bogoliubov-Nambu quasiparticles in conventional superconductors. 
In the case of ``relativistic'' phonons the transverse cross section is periodic in the
phonon energy $E$ again with the period $\omega$
\cite{SoninNew,Shelankov}:
\begin{equation}
\sigma_\perp ={\hbar \over p} ~ \sin  {2\pi
 E\over \hbar \omega} 
\label{sigmaPerpVortex}
\end{equation}

At low $E\ll \omega$ this result was generalized for arbitrary excitations with the spectrum
$\tilde E({\bf p})=E({\bf p})+  {\bf p}\cdot{\bf v}_{\rm s}$ moving in the background of
the velocity field ${\bf v}_{\rm s}$ around the vortex, using a simple classical
theory of scattering
\cite{SoninNew}. Far from the vortex, where the circulating velocity is
small, the trajectory of the quasiparticle is almost the straight line
parallel, say, to the axis
$y$, with the distance from the vortex line being the impact
parameter
$x$. It moves along this line with the almost constant momentum
$p_y\approx p$ and  almost constant group velocity
$v_G=dE/dp$. The change in the transverse momentum during
this motion is determined by the Hamiltonian equation $dp_x/dt=-\partial
\tilde E/\partial x=-p_y \partial
v_{{\rm s}y}/\partial x$, or $dp_x/dy=-(p/v_G)\partial
v_{{\rm s}y}/\partial x$. The transverse cross section is obtained by
integration of $\Delta p_x/p$ over the impact parameter $x$:
\begin{equation}
\sigma_\perp = \int_{-\infty}^{+\infty}{dx\over
v_G}\int_{-\infty}^{+\infty}dy {\partial v_{{\rm s}y}\over \partial x} 
={
\kappa\over v_G}~.
\label{sigmaPerpLinear}
\end{equation}
For $g_G=c$ this is the result of Eq.(\ref{sigmaPerpVortex}) at $E\ll \omega$. Note that the 
result in Eq.(\ref{sigmaPerpLinear}) is pure classical: the Planck constant
$\hbar$ drops out.

\subsection{Iordanskii force on spinning string.}\label{IordanskiiForce}

 This
asymmetric part of scattering,   which describes the momentum transfer in
the transverse direction, after integration over the  distribution of
excitations  gives rise to the transverse force acting on the  vortex if
the vortex moves with respect to the normal component. This is  the
Iordanskii force:
\begin{eqnarray}
{\bf F}_{\rm Iordanskii}=\int {d^3p\over (2\pi)^3}\sigma_\perp(p) v_G f({\bf
p})
 {\bf p}\times {\hat{\bf z}} =\nonumber\\- \kappa  {\hat{\bf z}}\times  \int {d^3p\over
(2\pi)^3}  f({\bf p})
 {\bf p}
= \kappa  {\bf P}\times {\hat{\bf z}}
\label{IordanskiiForce2}
\end{eqnarray}
It depends only on the momentum density ${\bf
P}$ carried by excitations (matter) and on the circulation
$\kappa$ around the vortex. This confirms the topological origin of this
force. In the case of the equilibrium distribution of quasiparticles one has 
${\bf P}=mn_{\rm n}({\bf v}_n-{\bf v}_{\rm s})$.  
The same Iordanskii force must act on the spinning cosmic string, when it moves
with respect to the matter.

Iordanskii force has been experimentally identified in the rotating superfluid 
$^3$He-B (Fig. \ref{CallanHarveyExpFig}). According to the theory for the 
transport of vortices in 
$^3$He-B (Sec.\ref{VerificationAdlerBellJackiw}),  the Iordanskii force
completely determines the mutual friction parameter
$d_\perp \approx - n_{\rm n}/n_{\rm s}$ at low $T$ (see Sec.
\ref{BPhaseExp}), where the spectral flow is completely suppressed. This is in
accordance with the experimental data, which show that $d_\perp$ does approach
its negative asymptote at low $T$ \cite{Bevan}. At higher $T$ the spectral
flow becomes dominating which leads to the sign reversal of $d_\perp$.  The
observed negative sign of 
$d_\perp$  at low $T$   provides the experimental verification of
the analog of the gravitational Aharonov-Bohm effect on spinning cosmic string.

\section{Horizons, ergoregions,  degenerate metric, vacuum instability and all
that.}\label{HorizonsErgoregions}
 
\subsection{Event horizons in vierbein wall and Hawking
radiation.}\label{EventHorizonsVierbeinWwall} 

Let us consider the vierbein wall discussed in Sec. \ref{VierbeinDomainWall}, 
which is still stationary but there is a superflow across the wall with the
superfluid velocity
${\bf v}_{\rm s}=  v_{\rm s}\hat{\bf x}$ (Fig. \ref{HorizonVierbeinWallFig}).
The line element of the effective space-time in Eq.(\ref{LineElementWall})
becomes
\cite{JacobsonVolovikThinFilm}
\begin{eqnarray}
ds^2=- dt^2   + {1\over c^2(x)} (dx-v_{\rm s}dt)^2+{1\over c_\perp^2} dy^2+{1\over
c_\parallel^2} dz^2=
\label{LineElementMovingWall1}\\
 -\left(1-{v_{\rm s}^2\over c^2(x)}\right) \left(dt +{v_{\rm s}dx\over c^2(x)-v_{\rm
s}^2}\right)^2   + {dx^2\over c^2(x) -v_{\rm s}^2}  ~,~c(x)=   c_\perp\tanh {x\over d}  ~.
\label{LineElementMovingWall2}
\end{eqnarray}
In Eq.(\ref{LineElementMovingWall2}) we omitted the irrelevant metric of the space along the
wall. 

\begin{figure}[t]
\centerline{\includegraphics[width=\linewidth]{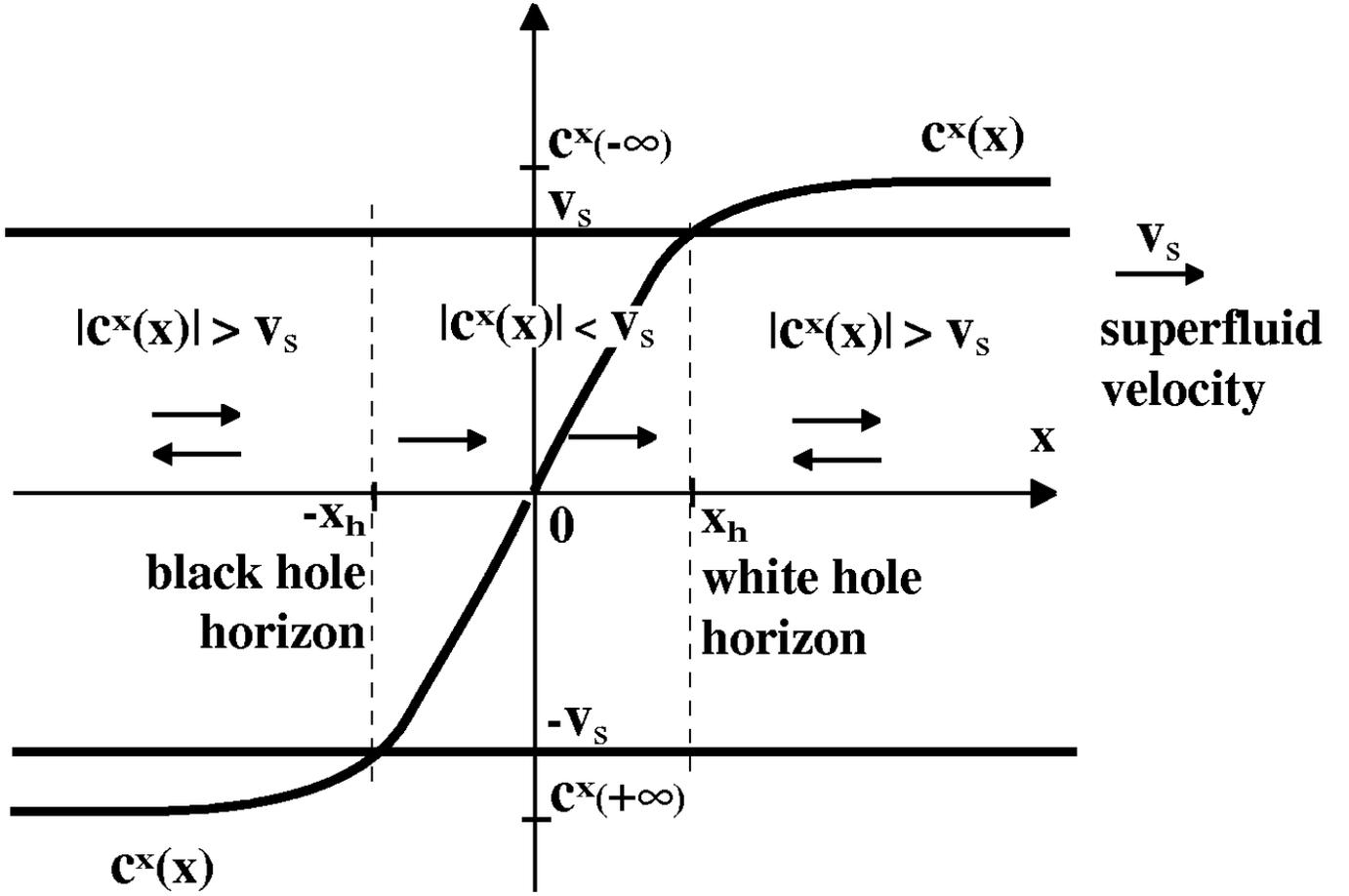}}
\medskip
\caption{When there is a superflow across the soliton, the vierbein wall
splits into a pair of horizons: black hole and white hole. Between the
horizons the superfluid velocity exceeds the ``speed of light'' and $g_{00}$
changes sign. Since their speed $c^x$ is smaller that the velocity of the
superfluid vacuum, the ``relativistic'' quasiparticles within the horizon can
move only along the streamlines. Horizon at $x=-x_h$ is the black-hole horizon,
since no information can be extracted from the region behind this horizon, if
the low-energy quasiparticles are used for communication.}
\label{HorizonVierbeinWallFig}
\end{figure}

The naive transformation 
\begin{equation}
\tilde t=t +\int^x {v_{\rm s}dx\over c^2(x)-v_{\rm s}^2}
\label{NaiveTrasnformation}
\end{equation}
in Eq.(\ref{LineElementMovingWall2}) gives 
\begin{equation}
ds^2= -\left(1-{v_{\rm s}^2\over c^2(x)}\right)  d\tilde t^2   + {dx^2\over c^2(x) -v_{\rm
s}^2}  ~.
\label{LineElementMovingWall3}
\end{equation}
This line element corresponds to the radial part of the Schwarzschild metric for the black
hole. The metric in Eq.(\ref{LineElementMovingWall3}) shows that there are two horizons in the
soliton: at planes $x_h$ where $c(x_h)=v_{\rm s}$ and at $x_-=-x_h$ where $c(-x_h)=-v_{\rm s}$.
This ``Schwarzschild'' metric has coordinate singularity at horizons and thus is not determined
globally, this is because the transformation in Eq.(\ref{NaiveTrasnformation}) is ill defined
at the horizons. The ``Schwarzschild'' metric can describe the effective space time
either outside the horizons or between them, at $-x_h<x<x_h$

The original metric in Eq.(\ref{LineElementMovingWall1}) is well determined
everywhere  except  for the physical singularity at $x=0$, where the
vierbein is degenerate). This is because in the effective gravity theory in
$^3$He-A (and also in superfluid $^4$He) the primary quantity is not the
metric, but the energy spectrum of quasiparticle, which in the low energy
corner becomes ``relativistic'' and acquires the Lorentzian form, thus giving
rise to the contravariant metric $g^{\mu\nu}$. Then from this contravariant
metric, if it is nondegenerate, the covariant metric $g_{\mu\nu}$ is obtained
which describes the effective space-time.  Thus only those space-times are
physical in these effective theories, which came from the physically
reasonable quasiparticle spectrum.  Since the spectrum of quasiparticles must
be determined everywhere, if the vacuum state is locally stable, the
contravariant effective metric $g^{\mu\nu}$ is well determined in the whole
underlying Galilean space time of the condensed matter. As a result the
covariant effective metric $g_{\mu\nu}$, which determines the line element of
the effective space-time, must be also determined everywhere except for the
places where the  the contravariant metric
$g^{\mu\nu}$ is degenerate.  The spectrum of the
$^3$He-A fermionic quasiparticles, which gives rise to the effective metric in
Eq.(\ref{LineElementMovingWall1}) is
\begin{equation}
\tilde E_a({\bf p})=  E_a({\bf p}) +p_x v_{\rm s}~,~E_a({\bf p})=\pm \sqrt{ c_\parallel^2 (p_z
-e_a p_F)^2 +c_\perp^2 p_y^2 + c^2(x) p_x^2}
\label{EDopplerShifted}
\end{equation}
which as we know can be written in the Lorentzian form $g^{\mu\nu}(p_\mu -
e_aA_\mu)(p_\nu - e_aA_\nu)=0$. 

Inspection of the energy spectrum of quasiparticles in Eq.(\ref{EDopplerShifted}) or of the
metric in Eq.(\ref{LineElementMovingWall1}) shows that the two horizons, at $x=x_h$ and at
$x=-x_h$, are essentially different. The horizon at $x=-x_h$ represents the black hole horizon,
while that at $x=x_h$ is the white hole horizon. 
This is because in the region between the horizons,  at $-x_h<x<x_h$, the group
velocity of the quasiparticle in the soliton frame $v_{Gx}=d\tilde E({\bf
p})/dp_x=d E({\bf p})/dp_x+ v_{\rm s}$ is positive for both directions of the
quasiparticle momentum $p_x$ (see Fig.
\ref{HorizonVierbeinWallFig}, where the superfluid velocity
$v_{\rm s}$ is chosen positive).  All the (low-energy)
quasiparticles inside the horizon will finally cross the plane $x=x_h$, which
means that this plane is the white hole horizon. But the quasiparticles cannot
cross the plane
$x=-x_h$ from inside, which indicates the black hole horizon: The ``inner'' observer living at
$x<-x_h$ cannot obtain the information from the region $x>-x_h$, if he uses 
the ``relativistic'' quasiparticles for communication.

Appearance of pairs of the white-hole/black-hole horizons is typical for the
condensed matter.

In the presence of horizons the notion of the vacuum state becomes subtle.
There are two important reference frames in which the vacuum state can be
defined: they give the same definition of the vacuum if there are no horizons,
but in the presence of the horizons the two vacua do not coincide. 

(i) The superfluid-comoving vacuum. This is
the vacuum as seen by the local ``inner'' observer, who is comoving with the
superfluid component, i.e. moving with the velocity ${\bf v}_{\rm s}$. This
vacuum is regulated by the energy $E({\bf p})$ in Eq.(\ref{EDopplerShifted}):
the states with the negative root in Eq.(\ref{EDopplerShifted}) are occupied.
This vacuum state can be also determined as the limit of the local
thermal state in Eq.(\ref{EffectiveT}) if $T\rightarrow 0$ at zero counterflow
velocity, $w=0$, or at any fixed ``subluminal'' counterflow velocity, $w<c$.
Such vacuum, however, cannot be defined globally, because in the reference
moving with the superfluid velocity, the metric is time dependent. 

(ii) The texture-comoving vacuum. This is the vacuum as determined in the frame
of the wall, where the metric does not depend on time (in general relativity
this means that the spacetime determined by such metric has a global timelike
Killing vector). Since there is no time dependence, the quasiparticle
energy is well defined globally  as $\tilde E({\bf p})$. The vacuum is
determined as the state in which the energy levels with
$\tilde E({\bf p})<0$ are occupied. 

In the presence of the horizons the superfluid-comoving vacuum and the
texture-comoving vacuum do not coincide in the region between the horizons. Some
energy states with the positive root in Eq.(\ref{EDopplerShifted}), which were
empty before, acquire in the texture-comoving frame the negative energy in the
presence of the horizons and must be occupied.  The process of the filling of
the negative energy levels is governed  by the spatial inhomogeneity, since for
the homogeneous system there is no preferred frame and the ``inner'' observer
does not know at $T=0$  whether the liquid is moving or not. The inhomogeneity is
provided by the spatial dependence of the metric within the soliton. Thus the
process of the dissipation of the superfluid-comoving vacuum is determined by
the derivatives of the texture, in a given case by the gradient of the ``speed
of light''
$c(x)$, or equivalently by the gradient of the metric, which plays the part of
the gravitational field.
 
An example of the dissipation, caused by the gradients of the metric, is the
Hawking radiation of quasiparticles \cite{hawkingnature}. This radiation is
characterized by the Hawking temperature, which depends on the gradient of the
``speed of light'' or, in terms of the gravity theory, by the ``surface
gravity'' $\kappa_S$ at the horizon:
\begin{equation}
T_{\rm H}={\hbar\over 2\pi} \kappa_S~,~\kappa_S= \left({dc\over dx}\right)_h
~~.
\label{HawkingT}
\end{equation}
In the case of Eq.(\ref{LineElementMovingWall2}) for the profile of the ``speed of light'' the
Hawking temperature depends on the velocity $ v_{\rm s}$:
\begin{equation}
T_{\rm H}(v_{\rm s})=T_{\rm H}(v_{\rm s}=0) \left( 1- { v_{\rm s}^2\over
c_\perp^2}\right)~~~,~~~T_{\rm H}(v_{\rm s}=0)={\hbar c_\perp
\over 2\pi  d}~.
\label{T(v)}
\end{equation}
Typically for this type of domain wall the Hawking temperature  $T_{\rm
H}(v_{\rm s}=0)$ is below  $1~\mu K$;  the Hawking flux of radiation could in
prinicple be detected by quasiparticle detectors.

The Hawking radiation leads to the energy dissipation and thus to the quantum
friction, which decreases the velocity $|v_{\rm s}|$ of the domain wall with respect to the
superfluid vacuum. Due to the deceleration of the wall motion, the Hawking temperature
increases with time. The distance between horizons,
$2x_h$, decreases until the complete stop of the domain wall when the two horizons
merge. The  Hawking temperature approaches its asymptotic value
$T_{\rm H}(v_{\rm s}=0)$ in Eq.(\ref{T(v)}); but when the horizons merge, the Hawking
radiation disappears: there is no more ergoregion (region with
negative energy states), so that the stationary domain wall is nondissipative,
as it should be.

Actually all this is valid when the quasiparticle spectrum near the
domain wall can be considered as continuous. The quasiparticle spectrum in
Eq.(\ref{EDopplerShifted}) takes place only if the horizons a far apart, i.e.
the distance between them is much larger than the superfluid coherence length,
$x_H\gg v_F/\Delta_0$. This is satisfied when the relative velocity between the
domain wall and the superfluid vacuum is close to $c_\perp$, so that $x_h\sim
(d/2) \ln [1/(1-v_{\rm s}^2/c_\perp^2)] \gg d$. Thus the dissipation stops when
the distance between the horizon becomes comparable to the ``Planck length"
$\xi=c_\parallel/\Delta_0$.  Also the Hawking radiation is only one of the
possible mechanisms of the dissipation: the real scenario of the relaxation of
the horizons depends on the details of the back reaction of the superfluid vacuum
to the filling of the negative energy states.

\subsection{Landau critical velocity and ergoregion} 

The horizons, discussed in the previous subsection, appeared as surfaces where the superfluid
velocity approached the ``speed of light'' $c_\perp$.  In general, i.e. for the
``nonrelativistic'' spectrum of quasiparticles, the quantum friction in
superfluids starts when the superfluid velocity exceeds the Landau critical velocity
\begin{equation} 
v_{\rm Landau}={\rm min}{E({\bf p})\over p}~.
\label{LandauVelocity1}
\end{equation}
Above this velocity the energy $\tilde E({\bf p})=E({\bf p}) +{\bf p}\cdot {\bf v}_{\rm s}$ of
some excitations, as  measured in the laboratory frame (or in the frame of the moving
soliton),  becomes negative. This allows for excitations to be nucleated from the vacuum. 
For a superfluid velocity field which is time-independent in the laboratory (or soliton)
frame, the region where 
$v_{\rm s}({\bf r})>v_{\rm Landau}$ and  quasiparticles can have
negative energy, is called the ergoregion. The surface $v_{\rm s}({\bf r})=v_{\rm Landau}$
which bounds the ergoregion, is called the ergosurface. 

In a given geometry discussed in previous subsection the ``speed of light'' coincides with the
Landau critical velocity, if only the ``relativistic'' quasiparticles are considered. 
Also the horizons coincide with the ergosurfaces, as it happens for the nonrotating black
hole. For the general superflow (or for the different type of the soliton
\cite{grishated})  horizon and ergosurface are separated from each other, as in the case of the
rotating black hole.  

In general, however, the behavior of  the system depends crucially on the
dispersion of the spectrum at higher energy. There are two possible cases. 

(1) The spectrum bends
upwards at high energy, i.e. $ E({\bf p})=cp +\gamma p^3$ with
$\gamma >0$.  Such dispersion can be realized for the fermionic 
quasiparticles in
$^3$He-A, for example from  Eq.(\ref{NonlineaClassical}) it follows that
$\gamma^{-1}=8(m^*)^2c_\perp$. Quasiparticles are ``relativistic'' in the low
energy corner but become ``superluminal'' at higher energy
\cite{JacobsonVolovikThinFilm,grishated}. In this case the Landau critical
velocity coincides with the ``speed of light'',
$v_{\rm Landau}=c$, so that the ergosurface is determined by
$v_{\rm s}({\bf r})=c$.  In the Lorentz invariant limit of the energy much
below the ``Planck'' scale, i.e. at $p^2
\ll \gamma/c$ (or $p\ll m^*c_\perp$) this corresponds  to the ergosurface at
$g_{00}({\bf r})=0$, which is just the definition of the ergosurface in
gravity.  In case of radial flow  of the  superfluid vacuum towards the origin
(Fig. \ref{UnruhSonicBlackHoleFig}), the ergosurface also represents the
horizon in the Lorentz invariant limit, and the region inside the horizon
simulates a black hole for low energy quasiparticles. Strictly speaking this is
not a true horizon for quasiparticles: Due to the nonlinear dispersion, their
group velocity $v_G=dE/d p =c+3\gamma p^2>c$, and thus the high energy
quasiparticles are allowed to leave the black hole region. It is, hence, a
horizon only for quasiparticles living exclusively in the very low energy
corner: they are not aware  of the possibility of ``superluminal'' motion.
Nevertheless, the mere possibility to exchange the information across the
horizon allows us to construct the thermal state on both sides of the horizon
and to investigate its thermodynamics, including the entropy related to the
horizon (see Sec.
\ref{ModTolmanLaw} and Ref. \cite{FischerVolovik}).

(2) In superfluid $^4$He the negative dispersion is realized, $\gamma<0$, with the group
velocity $v_G=dE/d p <c$ (if one neglects a small upturn of the spectrum at low
$p$). In such superfluids the ``relativistic'' ergosurface
$v_{\rm s}({\bf r})=c$ does not coincide with the true ergosurface, which is
determined by $v_{\rm s}({\bf r})=v_{\rm Landau}< c$.  In superfluid $^4$He, the
Landau velocity is related to the roton part of the spectrum, and is about four
times less than $c$. In case of radial flow inward, the ergosphere occurs at
$v_{\rm s}(r)=v_{\rm Landau}<c$, while the inner surface $v_{\rm s}(r)=c$ still
marks the horizon (Fig. \ref{UnruhSonicBlackHoleFig}).  This is in contrast to
relativistically invariant systems,  for which the ergosurface and the horizon
coincide for purely radial gravitational field of the chargless nonrotating
black hole.  The surface
$v_{\rm s}(r)=c$ stays a horizon even for excitations with very high momenta up
to some critical value, at which the group velocity of quasiparticle  again
approaches $c$.

\subsection{Painlev\'e-Gullstrand metric in effective gravity in
superfluids. Vacuum resistance to formation of horizon.}

Let us consider the spherically symmetric radial flow of the superfluid
vacuum, which is time-independent in the laboratory frame (Fig.
\ref{UnruhSonicBlackHoleFig}). For simplicity let us assume the isotropic
``speed of light'' as in superfluid
$^4$He. Then dynamics of the quasiparticles,
 propagating in this velocity field, is given by the line element provided by the effective
metric in Eq.(\ref{CovarianAcousticMetricReduced}):
\begin{equation} 
ds^2= -\left(1-{v_{\rm s}^2(r)\over c^2}\right)dt^2 + 2{v_{\rm s}(r)\over c^2}drdt
+ {1\over c^2} (dr^2 + r^2 d\Omega^2 )\,.
\label{PaileveInterval}
\end{equation}

\begin{figure}[t]
\centerline{\includegraphics[width=\linewidth]{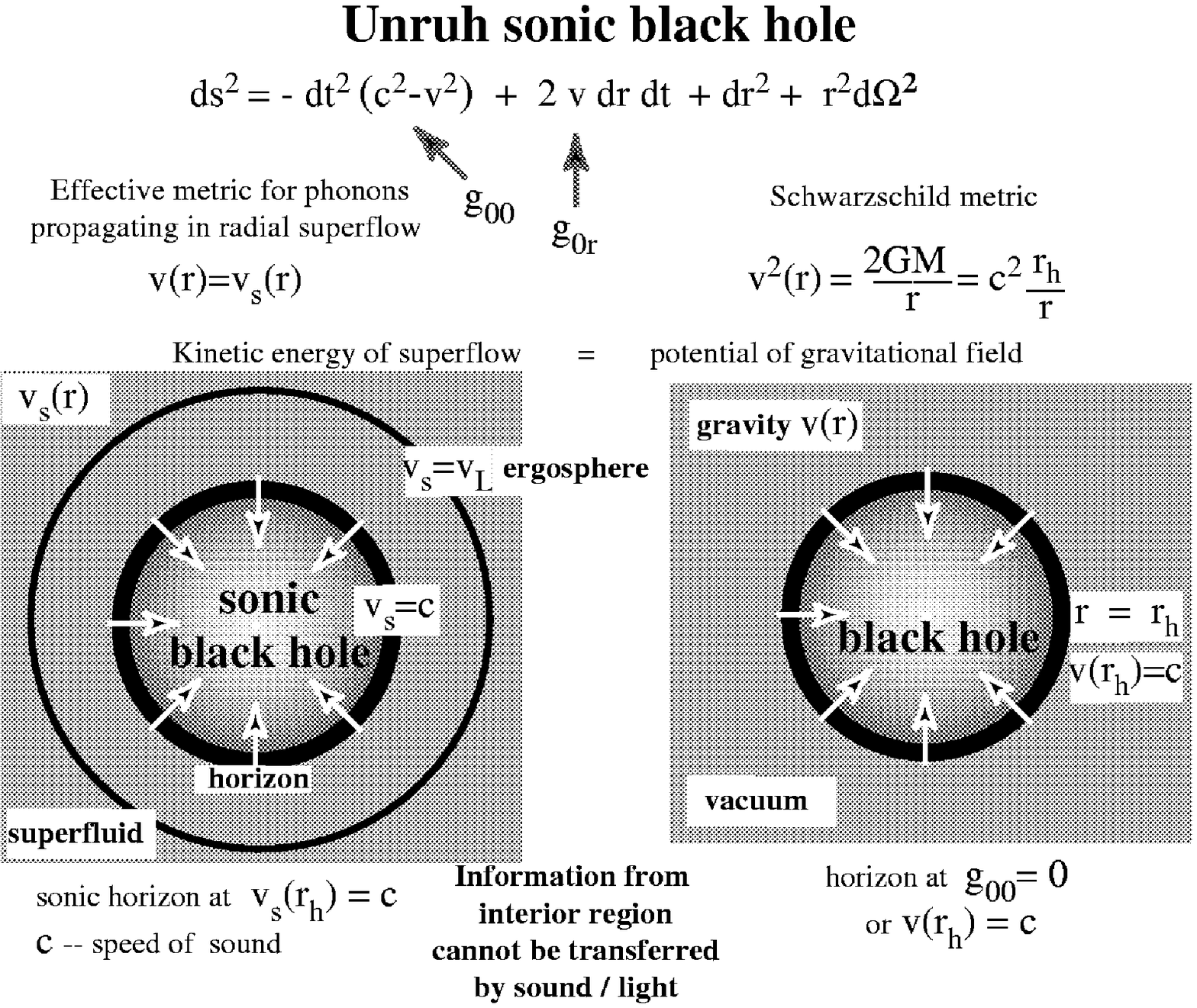}}
\medskip
\caption{Unruh sonic black hole in superfluid $^4$He. The horizon and
ergosphere coincide only for the ``relativistic'' phonons. For the real
quasiparticle spectrum the ergosphere occurs when the velocity of flow reaches
the Landau velocity for rotons.}
\label{UnruhSonicBlackHoleFig}
\end{figure}

This equation corresponds to the Painlev\'e-Gullstrand line elements. It
describes a black hole horizon if the superflow is inward
(see refs.\cite{unruh,vissersonic}; on the pedagogical review of
Panlev\'e-Gullstrand metric see \cite{Martel}; on the quantum vacuum effects
in gravity and in its condensed matter analogs see \cite{Liberati2}). If
$v_{\rm s}(r)=-c(r_{\rm h}/r)^{1/2}$ the flow simulates the  black hole in
general relativity. For the outward superflow with, say, 
$v_{\rm s}(r)=+c(r_{\rm h}/r)^{1/2}$ the  white hole is reproduced. For the general
radial dependence of the superfluid velocity, the Schwarzschild radius $r_{\rm h}$ is
determined as
$v_{\rm s}(r_{\rm h})=\pm c$; the  ``surface gravity''  at the Schwarzschild radius
is $\kappa_S  =(1/2c)dv^2_{\rm h}/dr|_{r_S}$;  and the Hawking temperature
$T_{\rm H}=\hbar\kappa_S/2\pi$.

It is not easy to create the flow in the Bose liquid, which exhbits the horizon for
phonons. This is  because of the  hydrodynamic instability which takes place behind the horizon
(see \cite{Liberati,Liberati2}).  From Eqs.(\ref{ContinuityEquation}) and
(\ref{LondonEquation}) of superfluid hydrodynamics at $T=0$ (which correspond to
conventional hydrodynamics of ideal curl-free liquid) it follows that for
stationary motion of the liquid one has the relation between $n$ and $v_{\rm s}$
along the streamline  \cite{FluidMechanics}:
\begin{equation} 
{\partial(nv_{\rm s})\over \partial v_{\rm s}}=n\left(1-{v_{\rm s}^2\over c^2}\right) \,.
\label{nvVSv}
\end{equation}
The current $J= nv_{\rm s}$ has a maximal value just at the horizon and
thus it must decrease behind the horizon, where $1-(v_{\rm s}^2/c^2)$ is negative. This is,
however, impossible in the radial flow since, according to the continuity equation
(\ref{ContinuityEquation}), one has $nv_{\rm s}=Const/r^2$ and thus the current must
monotonically increase across the horizon. This marks  the hydrodynamic instability behind
the horizon and shows that it is impossible to construct the time-independent flow with the
horizon without the fine-tuning of an external nonhydrodynamic
(``nongravitational'') force acting on the liquid
\cite{Liberati,Liberati2}.  Thus the liquid (vacuum) itself resists to the
formation of the horizon.  

Would the quantum vacuum always resist to formation of the horizon? Fortunately, not. In
the considered case of superfluid $^4$He, the same ``speed of light'' $c$, which  describes
the quasipartilces (acoustic waves -- phonons) and thus determines the value of the superfluid
velocity at horizon, also enters the hydrodynamic equations that establish the flow pattern of
the ``black hole''. For spin waves the ``speed of light'' can be less than the
hydrodynamic speed of sound, that is why the horizon can be reached before the
hydrodynamic instability.  In Fermi superfluid
$^3$He-A these two speeds can be well separated. The ``speed of light''
$c_\perp$ for quasiparticles, which determines the velocity of liquid flow at
the horizon, is  about $c_\perp
\sim 3$cm/sec, which is much less than the speed of sound
$s \sim 200$m/sec,  which determines the hydrodynamic instabilities of the
liquid. That is why there are no severe hydrodynamic constraints on the flow
pattern, the hydrodynamic instability is never reached and the surface gravity
at such horizons is always finite.

In  superfluids another instability can develop preventing the formation of
the horizon. Typically, the ``speed of light'' $c$ for
``relativistic'' quasiparticles coincides with the critical velocity,  at which
the superfluid state of the liquid becomes unstable towards  the normal state
of the liquid \cite{KopninVolovik1998}. In this case, when the superfluid
velocity with respect to the normal component or to the container walls exceeds
$c$, the slope
$\partial J/ \partial v_{\rm s}$ becomes negative and the superflow becomes
locally unstable.  The interaction between superfluid vacuum and
the walls leads to collapse of the superflow, so that the stationary superflow
with $w>c$ is impossible.   This instability, however, can be smoothened if the
container walls are properly isolated. In Ref.\cite{SimulationPainleve} it was
suggested to isolate the moving superfluid
$^3$He-A from the walls by the layer of superfluid $^4$He.  Then the
direct interaction and thus the momentum exchange between the flowing condensate
and the container walls is suppressed.  The momentum exchange occurs due to the gradients of the velocity field,
and can be made slow if the gradients are small, so that  the state with the
horizon can live long. Since the velocity gradient corresponds to the effective
gravitation field, the relaxation of the superflow in the presence of the
horizon at the initial stage of the flow instability becomes similar to the
Hawking process of the relaxation of the black hole. 

In principle, in Fermi superfluids, if the parameters of the system are
favourable, the ``speed of light'' can be made slightly less then the critical
velocity  at which the superfluid collapses. In this case the superflow can
remain stable in a supercritical regime, which means that the state with the
horizon can be stabilized: the fermionic quasiparticles formed in the Hawking
process finally occupy all the negative energy states and relaxation stops. 
Let us first consider this final state of stable horizon.  

\subsection{Stable event horizon and its momentum-space topology.}
\label{EventHorizonFermiSurface}

The horizon can be stabilized due to the nonlinear dispersion of the energy
spectrum. This can be illustrated on the simplest case of the motion through
the narrow place in the tube (Fig. \ref{HorizonsInFlowFig}). Let us consider
what happens with the system in the system, when we continuously increase the
velocity of the superflow from slightly below to slightly above the ``speed of
light''
$c_\perp$. Further we consider for simplicity the isotropic superfluid with
the nonlinear spectrum $E^2(p)=c^2p^2(1+p^2/p_{\rm P}^2)$. Here we assumed the
case of the superluminal dispersion, i.e. $\gamma >0$, and introduced the Planck
momentum, which enters the nonlinear dispersion, $p_{\rm P}
=\sqrt{c_\perp/2\gamma}$. In $^3$He-A it is of order
$ m^*c_\perp$.  We shall see that in some cases only this first nonlinear
correction does the whole job, so that the characteristic momenta are
concentrated well below the Planck scale,  $p\ll p_{\rm P}$, where the higher
order corrections are negligibly small.

\begin{figure}[t]
\centerline{\includegraphics[width=\linewidth]{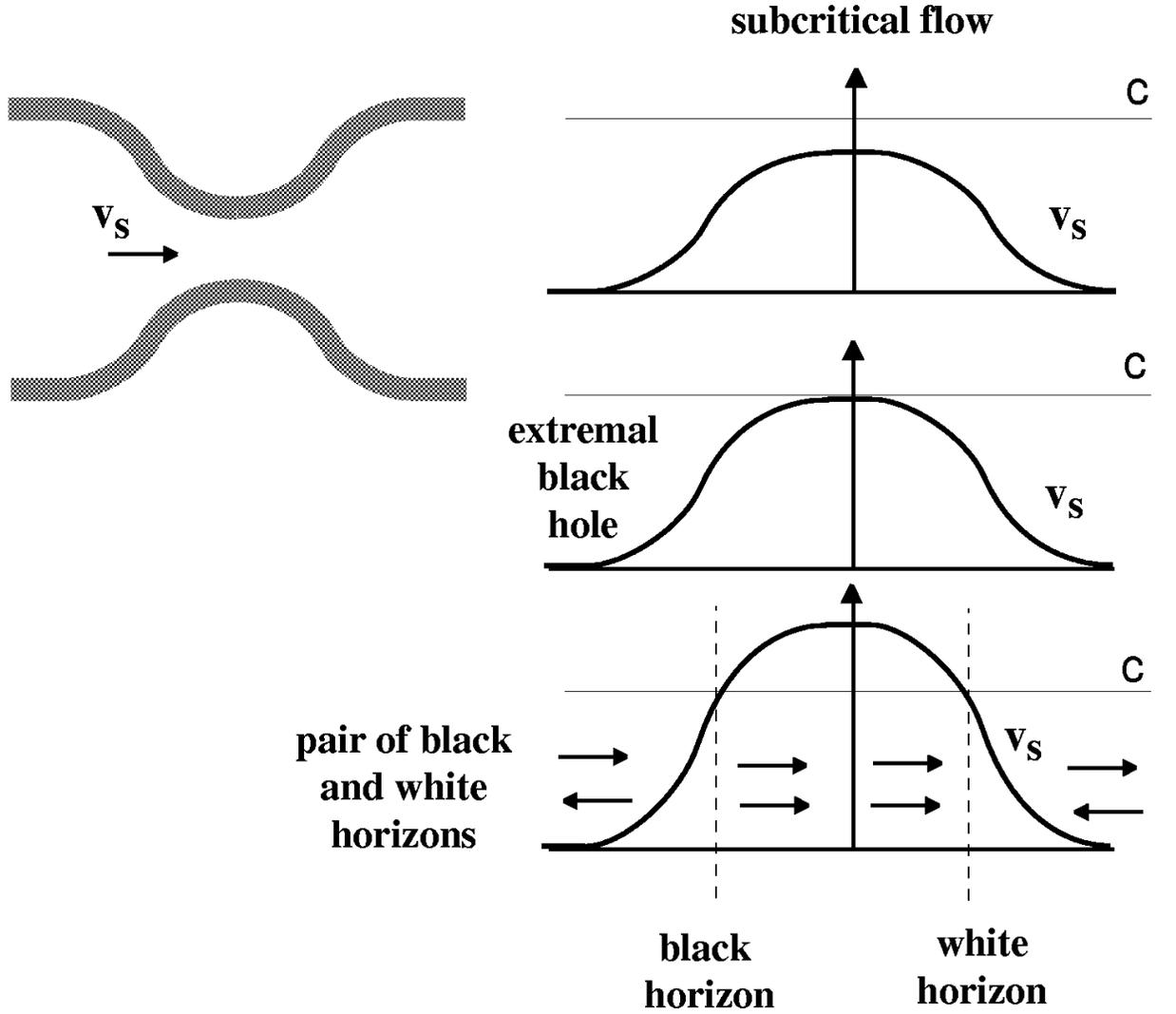}}
\medskip
\caption{The change of the effective space-time when the superfluid velocity
through the orifice continuously increases. When $v_{\rm s}$ exceeds the ``speed
of light'' the black-hole/white-hole pair appears. Arrows show possible
directions of the quasiparticles in the low-energy ``relativistic'' limit.
Between the horizons these quasiparticles can move only to the right. The
intermediate state when the velocity profile first touches the ``speed of
light'', has the effective metric which is equivalent to that in vicinity of the
horizon of extremal black hole.}
\label{HorizonsInFlowFig}
\end{figure}

When $v_{\rm s}$ exceeds the ``speed of light''
$c$ the pair of horizons, black and white, are formed similar to the case
discussed in Sec.
\ref{HorizonsErgoregions}. The only difference is that now the
superfluid velocity ${\bf v}_{\rm s}$ is space dependent, while the ``speed of
light'' $c$ is constant. The horizons can be considered as planes with
coordinate $x$ along the normal to the plane. The interval in the laboratory
reference frame, where the metric is time independent, has the form
\begin{equation} 
ds^2= -\left(1-{v_{\rm s}^2(x)\over c^2}\right)dt^2 + 2{v_{\rm s}(x)\over
c^2}dxdt + {1\over c^2} (dx^2 +dy^2+ dz^2)
\,.
\label{1DInterval}
\end{equation}
Since the superfluid velocity has a maximum at $x=0$, then in the region of
the crossover from subcritical flow to the horizons it can be approximated as 
\begin{equation} 
{v_{\rm s}(x)\over c} =1+ {\alpha\over 2} - {x^2\over
2x_0^2}~,~|\alpha|\ll 1~,
\label{VelocityTwoHor}
\end{equation}
where $x_0$ is of order of the dimension of the container.
The parameter $\alpha$ regulates the crossover: it changes from small negative
value when there are no horizons, to the small positive value, when the two
horizons appear. 

Within the relativistic limit the system with horizons, i.e. at $\alpha>0$,
does not possess the states of global thermodynamic equilibrium  (except
possibly at the Hawking temperature, which is determined by the quantum effect
and is very low). Since in such equilibrium the normal component velocity must
be zero in the laboratory frame, the counterflow velocity $w$ exceeds the
speed of light in the region between the horizons and thus the thermodynamic
potential in Eq.(\ref{EffectiveT}) is not determined. However, the
nonrelativistic corrections can restore the global thermodynamic equilibrium.
The laboratory frame energy spectrum of quasiparticles with the nonlinear
dispersion can be written for small $\alpha$ in the following form 
\begin{equation} 
\tilde E= p_x v_{\rm s}(x) + E(p)\approx {1\over 2}c |p_x|
\left({x^2\over x_0^2}-\alpha + {p_y^2+p_z^2\over
p_x^2}+{p_x^2\over p_{\rm P}^2}\right)      ~,~p_x<0~,~ p_y^2+p_z^2\ll
p_x^2~,~  |p_x|\ll p_{\rm P}~,
\label{SpectrumTwoHor}
\end{equation}
where we took into account that in the relevant low-energy region the
quasiparticle momentum is almost antiparallel to the superflow and thus   $|p_x|
\gg p_\perp$.

Inspection of the Eq.(\ref{SpectrumTwoHor}) shows the dramatic change in the
topology of the energy spectrum when $\alpha$ crosses zero. At $\alpha<0$,
where the superfluid velocity is everywhere subcritical, the energy of
quasiparticle is zero only 
at ${\bf p}=0$ (in real $¬3$He-A at
${\bf p}=\pm p_F\hat{\bf l}$). This is the topologically stable Fermi point
discussed in Secs. \ref{SystemsWithFermiPoints} and \ref{FermiPoints}. At
$\alpha>0$ the pair of horizons appears at $x=\pm x_h$, where
$x_h=x_0\sqrt{\alpha}\ll x_0$. As a result the topological invariant for the
Fermi point at
${\bf p}=0$ changes from nonzero value ($N_3=1$  or $N_3=-1$) to
$N_3=0$. Instead another topologically stable manifold of zeroes arises:
this is the Fermi surface, which appears in the region between the horizons,
i.e. at
$x^2<x_h^2$. The Fermi surface, the surface in momentum space where the
quasiparticle energy in Eq.(\ref{SpectrumTwoHor}) is zero, is determined by
equation
\begin{equation} 
 {p_y^2+p_z^2\over
p_x^2}+  {p_x^2\over p_{\rm P}^2} = {x_h^2-x^2\over
x_0^2}~~,~~ -x_h<x<x_h~.
\label{FermiSurfaceTwoHor}
\end{equation}
Thus the appearance of the horizons corresponds to the quantum phase
transition at which the topology of the quasiparticle spectrum
drastically changes. 

In the extended space, which includes the momentum ${\bf p}$ and the coordinate
$x$, the manifold of zeroes of the quasiparticle energy is the 3D Fermi
hypersurface in the 4D space $({\bf p},x)$.

At $\alpha>0$ the quasiparticles start, say by Hawking radiation, to fill the
negative energy states within the Fermi surface until all of them become
occupied. The state with the Fermi surface is thus the final
zero-temperature state of the Hawking radiation. The states of the global
thermodynamic equilibrium with nonzero $T$ are also determined by the Fermi
surface properties which, according to the conventional rules for the degenerate
Fermi system, are determined by the density of states at zero energy. If
$\alpha$ is small, the momenta  and energies of the occupied states are still
small compared to the Planck scale:
\begin{equation} 
 |p_x| \leq p_{\rm P}\alpha^{1/2} \ll p_{\rm P}~,~ |p_y| \leq
p_{\rm P}\alpha  \ll p_{\rm P}~,~E\sim c p_x\ll
E_{\rm P}~.
\label{EstimatonEnergyTwoHor}
\end{equation}
The density of states is thus $\sim p_xp_yp_z/E \sim (p_{\rm P}^2/c)\alpha^2$.

This is one of the examples when the effect of the Planck physics does not
require the knowledge of the details of transPlanckian physics, since only the
first nonlinear correction to the energy spectrum determines the phenomenon.

If $\alpha$ is small, the total number of the occupied states, $\sim A x_0
p_{\rm P}^3 \alpha^2$, where $A$ is the area of the horizon, is also small.
That is why one can expect that at $\alpha\ll 1$ the reconstruction of the
vacuum occurs without essential modification of the velocity profile, i.e.
without the considerable back reaction of the ``gravitational field''. 
However, this is not so simple. The detailed inspection of the backreaction of
the order parameter field to the filling of the negative energy levels shows
that the back reaction can be significant:  the supercritical
state with the Fermi surface is typically unstable towards the collapse of the
superfluidity \cite{KopninVolovik1998}. In such case,  there are no true
equilibrium states in the presence of the horizon. So that the final result
will be the merging and mutual annihilation of the two horizons. This occurs
either by the collapse of superfluidity, or, if the
collapse is smooothened, by thermal relaxations of the local equilibrium
states, which we discuss in Sec.\ref{ModTolmanLaw} and at very low temperature
by the Hawking radiation.

\subsection{Hawking radiation.}
\label{HawkingRradiationSection}

\begin{figure}[t]
\centerline{\includegraphics[width=\linewidth]{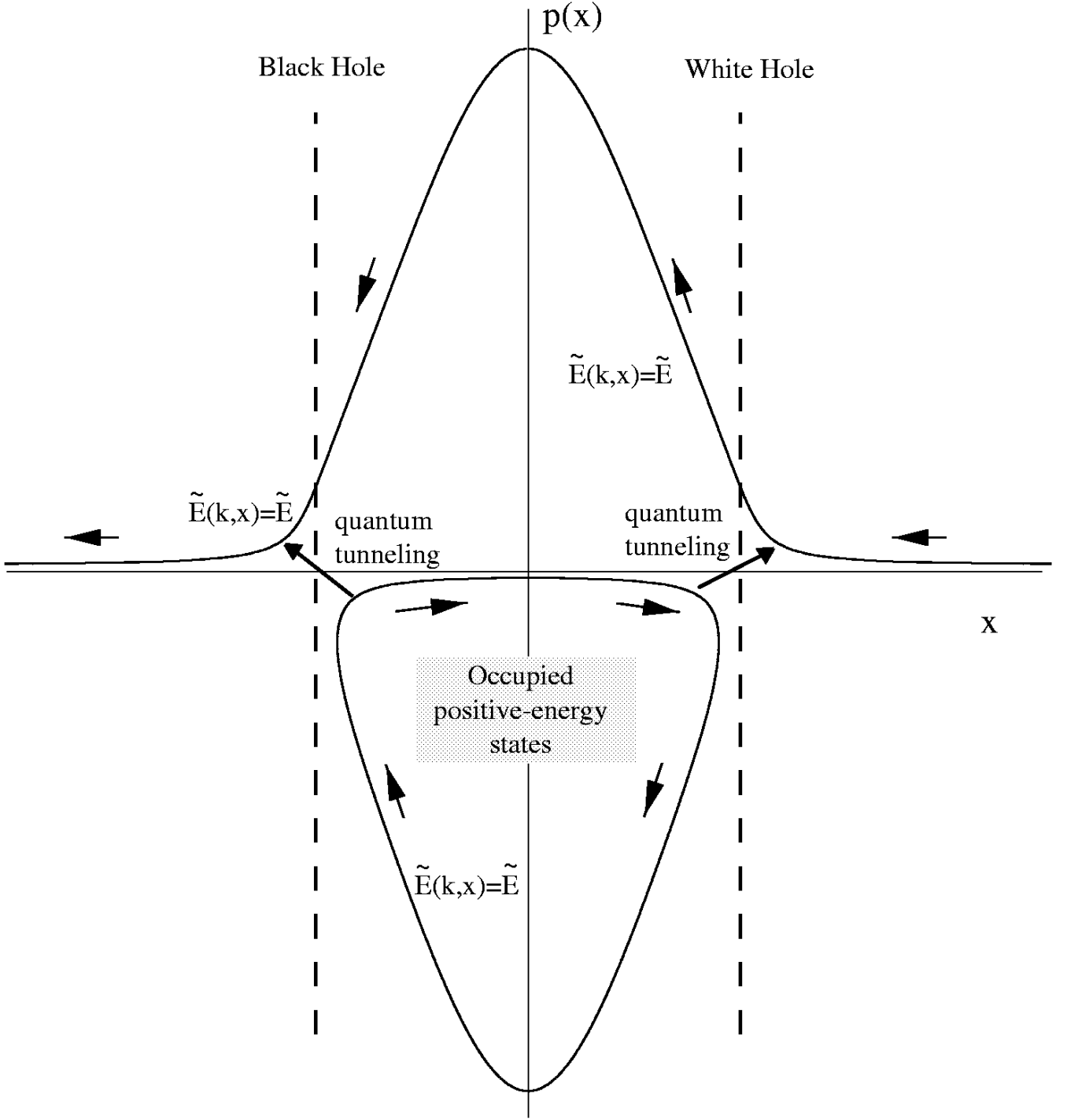}}
\medskip
\caption{Trajectories of quasiparticles in the presence of the
black-hole/white-hole pair in $1+1$ case. The nonlinear dispersion of the energy
spectrum at high momentum $p$ is taken into account. Only those positive-energy
trajectories are considered, which are responsible for the Hawking-like
radiation.  They are represented by curves of constant energy in the laboratory
(Killing) frame:
$\tilde E(p,x)=\tilde E >0$.  The trajectory between the horizons has negative
energy, $E<0$, in the frame comoving with the superfluid vacuum. That is why the
corresponding quantum states are initially occupied. But in the laboratory frame
these quasiparticles have positive energy, $\tilde E>0$, and thus can tunnel to
the mode with the same energy $\tilde E>0$, which is out-going from the black
hole.  The radiation from the black hole also occurs by tunneling to the mode
in-going to the white hole, which due to ``superluminal'' dispersion crosses both
horizons.}
\label{TrajectoriesHorizonsFig}
\end{figure}

Fig. \ref{TrajectoriesHorizonsFig} shows  the relevant classical trajectories of
quasiparticles in the presence of the horizons in simplest $1+1$ case. We
consider the positive energy states, as veiwed in the laboratory frame, where
the velocity field is time-independent. They are given by the equation
$\tilde E(p,x)=\tilde E >0$ with $\tilde E(p,x)$ given by
Eq.(\ref{SpectrumTwoHor}). Between the horizons the positive energy states in
the laboratory frame have the negative energy 
as viewed by the inner observer moving with the superfluid velocity ${\bf v}_{\rm
s}$: in the ``relativistic'' domain this energy is $E(p)=-cp<0$. The
corresponding quantum states belong to the quantum vacuum as viewed by
the comoving observer, that is why these states between the horizons are
initially occupied. This comoving vacuum is seen in the absolute
laboratory frame as highly excited state withe nonequilibrium distribution of
quasiparticles. The equilibrization occurs via tunneling of the quasiparticles
from the occupied excited states to the modes with the same positive energy
$\tilde E(p,x)=\tilde E >0$, which is out-going from the black-hole horizon. The
tunneling exponent is determined by the usual quasiclassical action
$2{\rm Im}\int p(x)dx$. At low energy, when the nonrelativistic corrections are
neglected, the momentum as a function of
$x$ on the classical trajectory has a pole at the horizon $p(x)\approx
\tilde E/ v'_{\rm s}(x-x_h)$. Here $v'_{\rm s}$ is the  derivative of the
superfluid velocity at the horizon, which is equivalent to the surface gravity.
Using the standard prescription for the treating of poles, one obtains for the
tunneling action
$2{\rm Im}\int p(x)dx=2\pi \tilde E/v'_{\rm s}$, which is similar to the thermal 
radiation with the Hawking temperature:
\begin{equation}
T_{\rm H}={\hbar\over 2\pi} \kappa_S~,~\kappa_S= \left({dv_{\rm s}\over
dx}\right)_h ~~.
\label{HawkingTvelocity}
\end{equation}
This can be compared with the Eq.(\ref{HawkingT}) for the Hawking temperature in
the case of moving soliton in terms of the
``surface gravity'' $\kappa_S$ at the horizon. The only difference is that in
the case of soliton the ``surface gravity'' is simulated by the gradient of the
``speed of light'' instead of the gradient of the superfluid velocity in
Eq.{\ref{HawkingTvelocity}).

There is another channel of dissipation. According to Fig.
\ref{TrajectoriesHorizonsFig} the radiation from the region between the horizons
also occurs by tunneling to the mode in-going to the white hole. This mode, due
to the ``superluminal'' dispersion, crosses both horizons and is transformed to
the mode out-going from the black-hole horizon. The tunneling exponent is given
by Eq.(\ref{HawkingTvelocity}) where it is now the velocity gradient at the
white-hole horizon, which determines the Hawking temperature of this tunneling.

\subsection{Extremal black hole.}\label{Bridge.}

Let us consider the transition point, $\alpha=0$. After rescaling
of the Eq.(\ref{1DInterval}) at $\alpha=0$ one obtains the following interval 
\begin{equation} 
ds^2= -x^2dt^2 + 2 dxdt +   dx^2 +dy^2+
dz^2=-x^2\left(dt-{dx\over x^2} \right)^2 +{dx^2\over x^2} + dy^2+dz^2
\,.
\label{BridgeInterval1}
\end{equation}
The plane $x=0$, where the black and white horizons merge, marks the bridge
between the two spaces. The metric in Eq.(\ref{BridgeInterval1}) coincides with
the metric in the vicinity of the horizon of an extremal black hole, whose mass
equal to the electric charge:
\begin{equation} 
ds^2= -\left(1-{r_h\over r}\right)^2dt^2  +\left(1-{r_h\over
r}\right)^{-2}dr^2 + r^2d\Omega
\,.
\label{ExtremalBlackHole}
\end{equation}
There is a strong divergency of the effective temperature at the horizon,
$T_{\rm eff}\propto T/x$, according to the Tolman's law in
Eq.(\ref{TolmanLaw}), where $T$ is the constant Tolman'temperature which is a
real temperature in superfluids. Because of that all the thermodynamic
quantities have singularity at the horizon where the Planck scale of energy
must intervene to cut-off the divergency. However, the first nonlinear
correction to the ``relativistic'' energy spectrum provides the necessary
cut-off already at low energy, which is much less than the Planck energy scale.

The energy spectrum in the vicinity of the horizon of our extremal
black hole is according to Eq.(\ref{SpectrumTwoHor})
\begin{equation} 
\tilde E= {1\over 2}c p_x
\left({x^2\over x_0^2}  + {p_y^2+p_z^2\over
p_x^2}+{p_x^2\over p_{\rm P}^2}\right)      ~,~p_x>0~,~ p_y^2+p_z^2\ll
p_x^2~.
\label{SpectrumExtremalHole}
\end{equation}
Thermodynamical quantities at the horizon, which follow from this spectrum can
be estimated using scaling relations. Since the characteristic energy $\tilde E$
is of order the temperature
$T$, the characteristic values of the coordinate $x$ and momentum ${\bf p}$ are
\begin{equation} 
 p_x \sim p_{\rm P}\left({T\over cp_{\rm P}}\right)^{1/3}~,~
|p_y|\sim |p_z| \sim  p_{\rm P}\left({T\over cp_{\rm P}}\right)^{2/3}
 ~,~ |x|\sim x_0\left({T\over cp_{\rm P}}\right)^{1/3}~.
\label{EstimatonEnergyExtremalHole}
\end{equation}
As a result the thermodynamic energy concentrated at the horizon is 
$E_{\rm hor}\propto AT|xp_xp_yp_z| \sim (1/c^2)Ap_{\rm P}x_0T^3$.
We can compare this with the thermodynamic energy in the bulk, which is of
order $E_{\rm bulk}\propto   (1/c^3)Ax_0T^4$, where 
$x_0$ is the dimension of the container (or the radius of the extremal black
hole). One has
$E_{\rm hor}/E_{\rm bulk}\propto   p_{\rm P}c/T\gg 1$. Thus the most of the
thermodynamic energy is concentrated in the vicinity of the horizon in the
region of thickness
$x_0\left(T/ cp_{\rm P}\right)^{1/3}$, which is very small, but still much
larger than the Planck legth scale.  All the characteristic energies and momenta
are also much less than the Planck scales. Thus the Planck physics provides only
the small nonlinear correction to the energy spectrum, and this is enough for
the complete determination of the thermodynamic properties of the extremal
horizon. The quantum gravity is not necessary to be introduced.

The most of the entropy of the system is also conncentrated at the horizon, 
$S_{\rm hor}(T)\sim (1/c^2)Ap_{\rm P}x_0T^2$. We considered the region of
temperatures $T_{\rm quantum}\ll T \ll cp_{\rm P}$, where $T_{\rm quantum}\sim
\hbar c/x_0$ is the temperature at which the quantization of the
quasiparticle levels becomes important and the scaling law for the entropy
becomes different. The entropy at this crossover temperature is $S_{\rm
hor}(T_{\rm quantum})\sim Ap_{\rm P}^2 (T_{\rm quantum}/cp_{\rm P})$. 
For the conventional black hole the quantum crossover temperature is of order
the Hawking temperature.

\subsection{Thermal states in the presence of horizons. Modified Tolman's
law.}\label{ModTolmanLaw}

In typical superfluid/superconducting systems with relativistic-like
quasiparticles the pair of black and white horizons does not allow the states
with global thermal equilibrium to exist because of the back reaction of the
superfluid vacuum.  However, if the walls of container are properly isolated
the system can evolve to the final state, when the horizons merge, through the
sequence of the local equilibrium states with inhomogeneous
$T$ and
${\bf v}_{\rm n}$. In a local equilibrium the
counterflow velocity $w$ in Eq.(\ref{EffectiveT} must be less than the ``speed
of light''. That is why in the presence of horizons, when the superfluid
velocity ${\bf v}_{\rm s}$ exceeds
$c$, the normal component velocity ${\bf v}_{\rm n}$ must be ajusted to produce
${\bf w}={\bf v}_{\rm n}-{\bf v}_{\rm s}$ below $c$ everywhere.
Thus the normal component velocity is necessarily inhomogeneous in the
presence of horizon and this gives rise to the dissipation due to viscosity of
the normal component.   Such local equilibrium
states were constructed in Ref.\cite{FischerVolovik} in the simplest case of
the 1+1 space-time with the flow profile in the right bottom part of Fig.
\ref{HorizonsInFlowFig}. The ``speed of light''
$c$ is kept constant, while the superfluid velocity $v_{\rm s}(x)$,  depends on
the coordinate $x$ and exceeds $c$ in the region between the horizons. Though
the superfluid velocity is ``superluminal'' between the horizon, $v_{\rm s}>c$,
the counterflow velocity $w$  appears to be everywhere ``subluminal'' reaching
the maximim value
$w=c$ at the horizon, with $w<c$ both outside and inside the horizons.  The
local equilibrium with the effective temperature $T_{\rm eff}$ in
Eq.(\ref{EffectiveT}) is thus determined on both sides of the horizon. 

These thermal states are obtained using the two-fluid hydrodynamics discussed
in Sec. \ref{LandauKhalatnikovSection}. We neglect quantum effects related to
gravity, including the  Hawking radiation process. This is permitted if all the
relevant energies are much higher than temperature of the crossover to the
quantum regime, which is of order of the Hawking
temperature, $T_{\rm H}=(\hbar/2\pi) \kappa_S$, where $\kappa_S=(\partial_x
v_{\rm s})_{\rm hor}\sim c/x_0$ is the ``surface gravity'' at the horizon.  
Thus we assume that 
$T\gg T_{\rm H}$, and
$\hbar/\tau \gg T_{\rm H}$, where $\tau$ is the relaxation time due to
collision of thermal quasiparticles. The latter relation also shows that the
mean free path $l=c\tau$ is small compared with the characteristic length,
within which the velocity (or the gravitational potential)
changes: $l (\partial v_{\rm s}/\partial x) \ll v_{\rm s}$. This is just the
condition for the applicability of the two-fluid hydrodynamic equations, where
the variables are the superfluid
velocity $v_{\rm s}$ which, when squared, plays the part of the gravitational
potential, as well as  temperature $T(x)$ and velocity $v_{\rm n}(x)$ of the
normal component, which characterize the local equilibrium states of
``matter''. The dissipative terms in the two-fluid equations can then be
neglected in zeroth order approximation, since they are small compared to the
reversible hydrodynamic terms by the above parameter $l (\partial v_{\rm
s}/\partial x) \ll v_{\rm s}$.

If the back reaction is neglected, and thus the superfluid velocity
(``gravity'') field is fixed, the other hydrodynamic variables,  temperature
$T(x)$ and velocity $v_{\rm n}(x)$ of ``matter'', are determined by the
conservation of energy and momentum. From Eq.(\ref{CovariantConservation2}) for the $\nu=0$
component, which corresponds to the energy conservation for the ``matter'', it follows that the
energy flux
$Q$ carried by the quasiparticles is constant (note that $c$ is constant here). In the
``relativistic'' approximation one then has ($c=1$)
\begin{equation}
Q=-\sqrt{-g} T^x{}_0
=  2\Omega\frac{v_{\rm n}(1 + wv_{\rm s})}{ { 1-w^2  }} ={\rm const} \,,
\qquad \Omega={(2s+1)\pi\over 12}T_{\rm eff}^2~.
\label{horizon1}
\end{equation}
From the same Eq.(\ref{CovariantConservation2}) but for $\nu=1$, which is the momentum
conservation equation,  there results the first order differential equation
 \begin{equation}
-\partial_x \left( 2\Omega~{v_{\rm n}w\over
1-w^2} +  \Omega\right )=    2\Omega {w\over
1-w^2}\partial_x v_{\rm s} ~ .
\label{horizon2}
\end{equation}

If the energy flux is zero, the Eq.(\ref{horizon1}) gives two
possible states. Eq.(\ref{horizon1}) is satisfied by the trivial solution
$v_{\rm n}=0$.  Then from Eq.(\ref{horizon2}) it follows that
$T={\rm const}$. This corresponds to a true equilibrium state, or global thermodynamic
equilibrium discussed in Sec.~\ref{GlobalThermodynamicEquilibrium}. Such equilibrium state,
however, can exist only in the absence of or outside the horizon. The effective temperature,
which satisfies the Tolman's law, $T_{\rm eff}(x)= T/\sqrt{-g_{00}(x)}\equiv T/\sqrt{1-v_{\rm
s}^2(x)}$, cannot be continued across the horizon:
The effective temperature $T_{\rm eff}$ diverges when the horizon
is approached and becomes imaginary inside the horizon, where  $|w|>1$.

There is, however, another solution of Eq.(\ref{horizon1}): $1+wv_{\rm s}=0$. Since
$w^2<1$, this solution can be valid only inside the horizon, where $v_{\rm
s}^2>1$. From Eq.(\ref{horizon2})  it follows that $\Omega(x)={\rm
const}/(v_{\rm s}^2(x)-1)$, and thus the temperature behaves as
$T^2(x) \propto \Omega(x) (1-w^2(x))={\rm const}/ v_{\rm s}^2(x)$,  or
$T(x)=T_{\rm hor~in}/|v_{\rm s}(x)|$, where $T_{\rm hor~in}$ is the temperature
at the horizon, when approached from inside.   Thus inside the horizon one has
 a quasiequilibrium state with inhomogeneous temperature. The effective
temperature behind the horizon follows a modified Tolman law:
\begin{equation}
T_{\rm eff}(x)={T_{\rm hor~in}\over \sqrt{v_{\rm s}^2(x)-1}}
\label{TolmannBehindHor}
\end{equation}

Since the superluminal dispersion provides an energy
exchange between the matter (quasiparticles) inside and outside the
horizon,  the temperature must be continuous across the
horizon. Thus
$T_{\rm hor~in}=T_{\rm hor~out}$, and assuming that far from the horizons the
superfluid velocity vanishes, one has  $ T(x>h_h)=T_\infty$,
$T(x<x_h)=T_\infty/|v_{\rm s}(x)|$. Thus one has the  modified form of
the Tolman's law, which is valid on both sides of the black hole horizon: 
\begin{equation}
T_{\rm eff}(x)={T_\infty\over \sqrt{ \big|1 - {v_{\rm s}^2(x)\over c^2}\big|}}={T_\infty\over
\sqrt{|g_{00}(x)|}}\,.
\label{ModifiedTolmanLaw}
\end{equation}
The effective temperature $T_{\rm eff}$, which determines the local
``relativistic''  thermodynamics,  becomes infinite at the horizon. The
cut-off is provided by the nonlinear dispersion of the quasiparticle spectrum
with $\gamma>0$. The real  temperature $T$ of the liquid is continuous across
the horizon: 
\begin{equation}
T= T_\infty~~~~{\rm at}~~v_{\rm s}^2(x)< c^2~~~~~,~~~~ T(x)= T_\infty  {c\over
|v_{\rm s}(x)|}~~~~{\rm at}~~v_{\rm s}^2(x)> c^2 \,.
\label{ModifiedTolmanLaw2}
\end{equation} 

In the presence of superluminal dispersion, all physical quantities
are continuous at the horizon. On the other hand,
in the limit of vanishing dispersion they
experience kinks. For example, the jump in the
derivative of the temperature is
\begin{equation}
\nabla T|_{x_h+0}-\nabla T|_{x_h-0}=2\pi T_\infty T_{\rm H}~.
\label{JumpT}
\end{equation} 
This jump does not depend on details of the high-energy dispersion. This means that
in the limit of a purely relativistic system, the presence of a nonzero temperature at
infinity implies a  singularity at the horizon. This  coordinate
singularity at the horizon cannot be removed, since in the presence of
``matter'' with nonzero temperature the system is not invariant under
coordinate transformations, and this produces a kink in temperature
at the horizon.

\subsubsection{Entropy related to horizon}

Let us consider the entropy of quasiequilibrium thermal state across the
horizon,
\begin{equation}
 {\cal S}=\int d^D r~  S\,,\qquad S={\partial \Omega\over \partial T}~.
\label{TotalEntropy}
\end{equation}
The ``relativistic'' entropy, which is measured by a local observer living
in the quasiparticle world is the effective entropy
\begin{equation}
S_{\rm
eff}={\partial\Omega\over \partial T_{\rm
eff}} = {\partial\Omega\over \partial T} {\partial T\over \partial T_{\rm
eff}} =S\sqrt{1-w^2}~.
\label{EffectiveEntropy}
\end{equation}
For our thermal state, in the presence of a horizon in 1+1 dimension, the
{\em total real entropy} can be divided into 3 contributions:
\begin{equation}
 {\cal S}={\cal S}_{\rm ext} + {\cal S}_{\rm int} +{\cal S}_{\rm
hor}~.
\label{TotalEntropyDivison}
\end{equation}
At $T_\infty \gg
T_{\rm H}$, the exterior entropy, which comes from the bulk liquid,
is proportional to the size $L_{\rm ext}$ of the external region:
${\cal S}_{\rm ext}
\propto T_\infty  L_{\rm ext}$. A similar estimate holds for the
entropy of the interior region, ${\cal S}_{\rm int}
\propto T_\infty  L_{\rm int}$. The entropy related to the
horizon is well separated from the bulk terms, since it contains the ``Planck''
energy cut-off due to the infinite red shift at the horizon.  In the 1+1 case
such entropy comes from the logarithmically divergent contribution at the
horizon:
\begin{equation}
{\cal S}_{\rm
hor} ={\pi \over 3} \int dx {T_\infty\over |1-v_{\rm s}^2|}~.
\label{HorizonEntropy}
\end{equation}
The infrared cut-off is determined by $x_0$, which characterizes the gradient of
the velocity field at horizon or the ``surface gravity'': $1-v_{\rm s}^2
\approx 2\kappa_S \Delta x \sim \Delta x/x_0$ where $\Delta x$ is the distance
from the horizon. 

The  ultraviolet cut-off is provided by the nonlinear dispersion of the
quasiparticle spectrum.   At the cut-off scale the nonlinear term becomes
comparable with the linear one. We use the same energy spectrum with
nonlinear dispersion as before, $\tilde E(p)=c|p|(1+p^2c^2/E_{\rm P}^2) +
pv_{\rm s}\approx
\kappa_S p\Delta x + c^3|p|^3/E_{\rm P}^2$, where the Planck energy scale is
$E_{\rm P}=cp_{\rm P} $. Then one has for the ultraviolet cut-off parameters the
equation $\kappa_S p x_c\sim c^3|p_c|^3/E_{\rm P}^2 \sim T_\infty$ and thus
the following estimation
\begin{equation}
 x_c=\kappa_S^{-1}\left({T_\infty\over E_{\rm P}}\right)^{2/3}~,~E_c\sim
cp_c\sim  T_\infty\left({E_{\rm P}\over T_\infty}\right)^{1/3}.
\label{ScalingCutOff}
\end{equation}
Again the cut-off energy $E_c$ appeared to be much smaller than the Planck
energy scale. This means that one does not need in the whole Planck scale
physics to discuss the horizon problem in the considered 1+1 space-time: only
the first nonlinear correction to the linear spectrum is important.
 
From Eq.(\ref{ScalingCutOff}) it follows that the entropy related to the horizon
is
\begin{equation}
{\cal S}_{\rm
hor} ={1  \over 9}   { T_\infty \over T_{\rm H}}
\ln \left({E_{\rm P}\over T_\infty}
\right)  ~ .
\label{HorizonEntropy2}
\end{equation}
This relation also means that the density of quasiparticle states diverges logarithmically
at the horizon 
\begin{equation}
N_{\rm hor}(E) =2\int {dxdp\over 2\pi} \delta(E-\tilde E(p,x))={1  \over
3\pi^2     T_{\rm H} }\ln
\left({E_{\rm P}\over E}\right)~,
\label{DOS}
\end{equation}
where $T_{\rm H}\ll E\ll E_{\rm P}$.

\subsection{Painlev\'e-Gullstrand vs Schwarzschild
metric in effective gravity. Incompleteness of space-time in effective gravity.}

As we have already discussed in Sec. \ref{EventHorizonsVierbeinWwall}, in the effective theory
of gravity, which occurs in condensed matter systems,   the   primary quantity is the
contravariant metric tensor 
$g^{\mu\nu}$ describing the energy spectrum. Due to this the 
two seemingly equivalent representations of the black hole metric,  
in terms of either the Painlev\'e-Gullstrand line element in Eq.(\ref{PaileveInterval}) and
the Schwarzschild line element
\begin{equation}
 ds^2=-\left(1- {v_{\rm s}^2\over c^2}\right)d\tilde t^2+{dr^2\over c^2-
v_{\rm s}^2} +{1\over c^2} r^2 d\Omega^2~.
\label{Schwarzschild}
\end{equation}
are not equivalent. 

The Eqs.(\ref{Schwarzschild}) and (\ref{PaileveInterval}) are related by the 
coordinate transformation. Let us for simplicity consider the abstract flow with the
velocity exactly simulating the Schwarzschild metric, i.e. $v^2_{\rm s}(r)=r_{\rm s}/r$ and we put
$c=1$. Then the coordinate transformation is
\begin{equation}
 \tilde t(r,t)=t +   \left({2\over v_{\rm s}(r)} + {\rm ln}~ {1- v_{\rm s}(r)\over
1+v_{\rm s}(r)}\right) ~,~d\tilde t=dt +{v_{\rm s}\over 1-v_{\rm s}^2}dr.
\label{Transformation}
\end{equation}
What is the difference between the Schwarzschild and Painlev\'e-Gullstrand space-times in
the effective gravity? The Painlev\'e-Gullstrand line elements directly follows from the
contravariant metric tensor $g^{\mu\nu}$ and thus is valid for the whole ``absolute'' 
Newton's space-time  $(t,{\bf r})$ of the laboratory frame, i.e. as is measured by the
external experimentalist, who lives in the real world of the laboratory and investigates the
dynamics of quasipartcles using the physical laws obeying the Galilean invariance of the
absolute space-time. 

The time $\tilde t$ in the Schwarzschild line element is the time as
measured by the ``inner''   observer at ``infinity''   (i.e. far from the black hole). The
``inner''  means that this observer `` lives''  in the superfluid background and uses
``relativistic''   massless quasiparticles (phonons in $^4$He or ``relativistic'' fermionic
quasiparticles  in $^3$He-A) as a light for communication and for sinchronization clocks. The
inner observer at some point $R\gg 1$ sends  quasiparticles pulse at the moment
$t_1$ which arrives at point $r$ at
$t=t_1+\int_{r}^Rdr/|v_-|$ of the absolute (laboratory) time, where  $v_+$
and
$v_-$ are absolute (laboratory) velocities of radially propagating quasiparticles,
moving outward and inward respectively
\begin{equation}
v_\pm ={dr\over dt}={dE\over dp_r}=\pm 1 +v_{\rm s}~.
\label{RadialVelocity}
\end{equation}
Since from the point of view of the inner observer the speed of light (i.e. the
speed of quasiparticles) is invariant quantity and   does not depend on direction
of propagation, for him  the moment of arrival of pulse to $r$ is not $t$
but  $\tilde t =(t_1+t_2)/2$, where
$t_2$ is the time when the pulse reflected from $r$ returns to observer at
$R$. Since $t_2-t_1=\int_{r}^Rdr/|v_-|+\int_{r}^Rdr/|v_+|$, one obtains
for the time measured by inner observer as
\begin{eqnarray}
 \tilde t(r,t)={t_1+t_2\over 2}=t +  {1\over 2} \left(\int^R_r{dr\over
v_+}+
\int^R_r{dr\over v_-}\right)=\nonumber\\
t +   \left({2\over v_{\rm s}(r)} + {\rm ln}~ {1-
v_{\rm s}(r)\over 1+v_{\rm s}(r)}\right)-   \left({2\over v_{\rm s}(R)} + {\rm ln}~ {1-
v_{\rm s}(R)\over 1+v_{\rm s}(R)}\right)  ~,
\label{InnerTime}
\end{eqnarray}
which is just the Eq.(\ref{Transformation}) up to a constant shift.

In the complete absolute physical space-time of the laboratory the external observer
can detect quasiparticles  radially propagating into (but not out of) the
black hole  or out of (but not into) the white hole. The energy spectrum
of the quasiparticles remains to be  well determined both outside and
inside the horizon.  Quasiparticles cross the black hole horizon with the
absolute velocity
$v_-= -1-v_{\rm s}=-2$ i.e.   with the double speed of light: $r(t) = 1  -
2(t-t_0)$. In case of a white hole horizon one has $r(t) = 1  +
2(t-t_0)$.  On the contrary, from the point of view of the inner observer
the  horizon cannot be reached and crossed: the horizon can be approached
only asymptptically for infinite time: $r(\tilde
t)=1+(r_0-1)\exp(-\tilde t)$.  Such incompetence of the local observer, who
"lives" in the curved world of superfluid vacuum,  happens because he is
limited in his observations by the ``speed of light'', so that the
coordinate frame he uses is seriously crippled in the presence of the
horizon and becomes incomplete.  

The Schwarzschild metric naturally arises for the inner observer, if the
Painlev\'e-Gullstrand metric is an effective metric for quasiparticles in
superfluids,  but not vice versa. The Schwarzschild metric
Eq.(\ref{Schwarzschild}) can in principle arise  as an effective metric in absolute
space-time; however, in the presense of a horizon such metric indicates an
instability of the underlying medium. To obtain   a line element 
of Schwarzschild metric as an effective metric for quasiparticles,  the
quasiparticle energy spectrum in the laboratory frame has to be
\begin{equation}
E^2=c^2\left(1-{r_{\rm s}\over r}\right)^2 p_r^2+c^2\left(1-{r_{\rm s}\over
r}\right) p_\perp^2~.
\label{SpectrumInSchwarzschild}
\end{equation}
 
In the presence of a horizon such spectrum has sections of the
transverse momentum $p_\perp$ with $E^2<0$. The imaginary
frequency of excitations  
signals the instability of the superfluid vacuum if
this vacuum  exhibits the Schwarzschild metric as an effective metric for
excitations:  Quasiparticle
perturbations may 
grow exponentially without bound in laboratory (Killing) time, as
$e^{t~{\rm Im} E}$, destroying the superfluid vacuum.  Nothing of this kind
happens in the case of the Painlev\'e-Gullstrand line element, 
for which the quasiparticle energy is real even behind the horizon.   
Thus the main difference between  Painlev\'e-Gullstrand and Schwarzschild
metrics as effective metrics is:  The first metric leads to the slow
process of the quasiparticle radiation from the vacuum at the horizon
(Hawking radiation), while the second one indicates a crucial instability
of the vacuum  behind the horizon. 

In general relativity it is assumed that the two metrics can be converted to each
other by the coordinate transformation in Eq.(\ref{Transformation}). 
In condensed matter the coordinate transformation
leading from one metric to another is not that innocent  
if an event horizon is
present.  The reason why the 
physical behaviour implied by the choice of metric  representation 
changes drastically is that the
transformation between the two line elements, $t\rightarrow t +\int^r
dr~v_{\rm s}/(c^2-v_{\rm s}^2)$, is singular on the horizon, and thus it can be applied
only  to a part of the absolute space-time.  In condensed matter, only such effective
metrics are physical which are determined everywhere in the real physical
space-time.  The two representations of the
``same'' metric cannot be strictly equivalent metrics, and we have
different classes of equivalence,  
which cannot be transformed to each other by everywhere regular coordinate
transformation.  Painlev\'e-Gullstrand metrics for black and white holes
are determined everywhere, but belong to two different classes.  The
transition between these two metrics occurs  via the singular
transformation 
$t\rightarrow t +2\int^r dr~v_{\rm s}/(c^2-v_{\rm s}^2)$ or via the Schwarzschild line
element, which is prohibited in condensed matter physics, as explained
above,    since it is pathological in the presence of a horizon: it is not
determined in the whole space-time and it is singular at horizon.

It is also important that in the effective theory there is no need for the
additional extension of space-time to make it geodesically complete. The effective space time
is always incomplete (open) in the presence of horizon, since it exists only in the low energy
``relativistic'' corner and quasiparticles escape this space-time to a nonrelativistic domain
when their energy increase beyond the relativistic linear approximation regime
\cite{grishated}. 

Another example of the incomplete space-time in effective gravity is provided by  vierbein
walls, or walls with the degenerate metric, discussed in
Sec.\ref{VierbeinDomainWall}. The ``inner'' observer, who lives in one of the domains
and measures the time and distances using the quasiparticles, his space-time is flat and
complete. But this is only half of the real (absolute)  space-time:  the other domains, which
do really exist in the absolute spacetime,  remain unknown to this ``inner''
observer. 

These examples show also the importance of the superluminal dispersion at high energy:
though many results do not depend on the details of this dispersion, merely the possibility
of the  information exchange by ``superluminal'' quasiparticles establishes the correct
continuation across the horizon or classically forbidden regions. The high energy dispersion
of the relativistic particles was exploited in recent works on black holes
\cite{BHdispers,Corley,BHlaser,origin,JacobsonTalk}.

\subsection{Vacuum under rotation.}\label{VacuumUnderRotation.} 

\subsubsection{Unruh and Zel'dovich-Starobinsky effects.}

An example of the quantum friction without the horizon is provided by a body rotating in
superfluid liquid at $T=0$ \cite{CalogeracosVolovik}. The effect is analogous to the
amplification of electromagnetic radiation and spontaneous emission by the body or black hole
rotating in quantum vacuum, first discussed by Zel'dovich and Starobinsky. The friction is
caused by the interaction of the part of the liquid, which is rigidly connected with the
rotating body and thus represents the comoving detector, with the ``Minkowski''  vacuum outside
the body. The emission process is the quantum tunneling of quasiparticles
from the detector to the ergoregion, where the energy of quasiparticles  is
negative in the rotating frame. The emission of quasiparticles, phonons and
rotons in superfluid $^4$He and Bogoliubov fermions in superfluid $^3$He, leads to the 
quantum rotational friction experienced by the body.

\begin{figure}[t]
\centerline{\includegraphics[width=\linewidth]{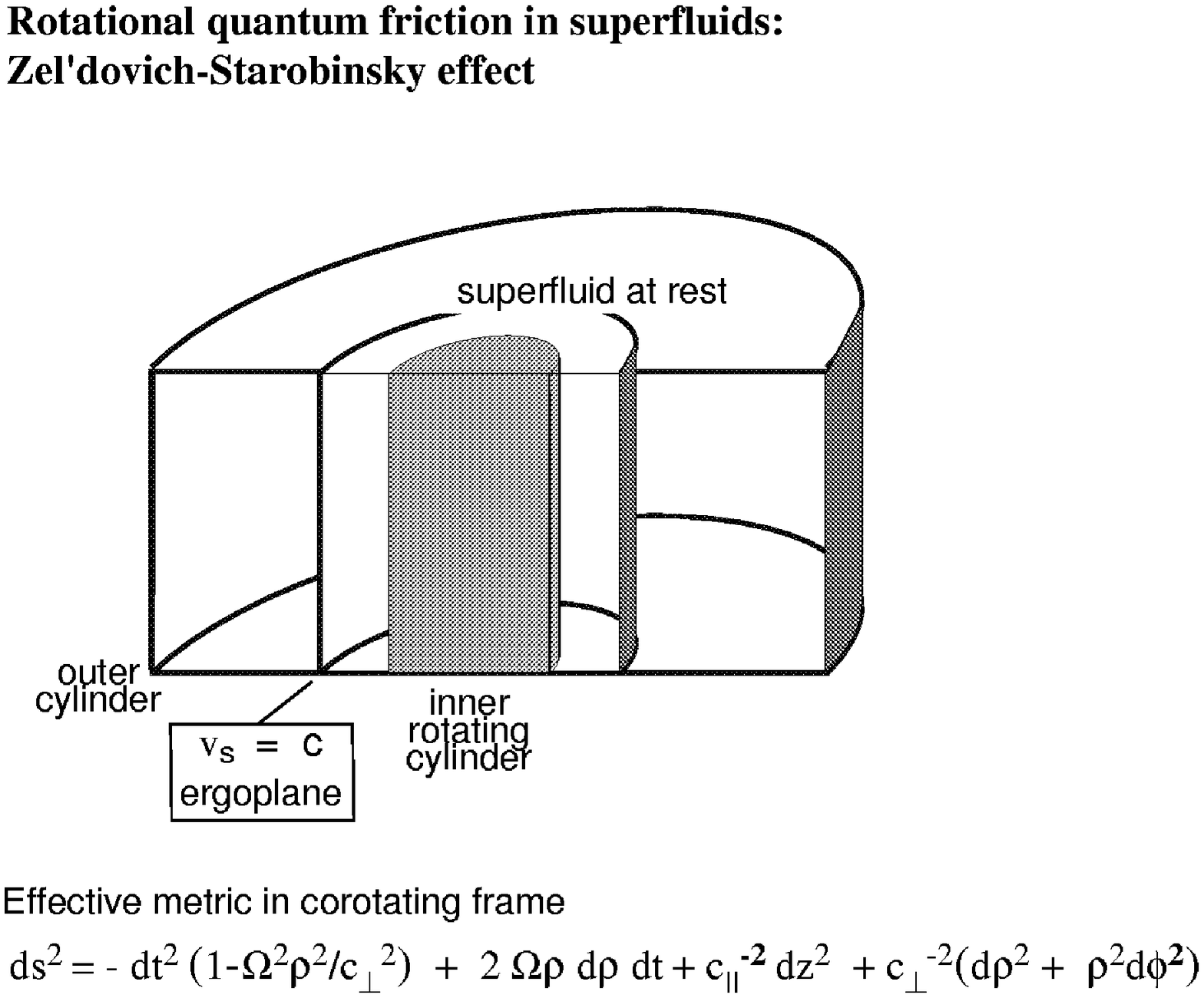}}
\medskip
\caption{Possible simulation of Zel'dovich-Starobinsky effect in superfluids.
The inner cylinder rotates forming the preferred rotating reference frame. In
this frame the effective metric has an ergoregion, where the negative
energy levels are empty. 
}
\label{RotVacuumFig}
\end{figure}

The motion with constant angular velocity is another realization of the Unruh
effect\cite{Unruh1}. In Uruh effect a  body moving in the vacuum with linear 
acceleration $a$ radiates the thermal spectrum of excitations with the Unruh temperature
$T_{\rm U}=\hbar a/2\pi c$. On the other hand the observer comoving with the body sees the
vacuum as a thermal bath with $T=T_{\rm U}$, so that the matter of the body gets heated to
$T_{\rm U}$ (see references in \cite{Audretsch}). It is difficult to simulate in condensed
matter the motion at constant proper acceleration (hyperbolic motion). On the other hand the
body rotating in superfluid vacuum simulates the uniform circular motion of the body with the
constant centripetal acceleration. Such motion in the
quantum vacuum was also heavily discussed in the literature (see the latest references in
\cite{RotatingQuantumVacuum,Leinaas,OrbitingUnruh}). The latter motion is stationary in the
rotating frame, which is thus a convenient  frame for study of the radiation and
thermalization effects for uniformly rotating body.

Zel'dovich \cite{Zeldovich1} was the first who predicted that the rotating
body (say, dielectric cylinder) amplifies those electromagnetic modes which
satisfy the condition 
\begin{equation}
 \omega - L \Omega<0 ~.
\label{ZeldovichCondition}
\end{equation}
Here $\omega$ is the frequency of the mode, $L$ is its azimuthal quantum
number, and $\Omega$ is the angular velocity of the rotating cylinder.
This amplification of the incoming radiation is referred to as 
superradiance \cite{BekensteinSchiffer}. The other aspect of this phenomenon is that due 
to quantum effects, the cylinder rotating in quantum vacuum
spontaneously emits the electromagnetic modes satisfying
Eq.(\ref{ZeldovichCondition}) 
\cite{Zeldovich1}. The same occurs for any rotating body, including the
rotating black hole \cite{Starobinskii}, if the above condition is satisfied.

Distinct from the linearly accelerated body, the radiation by a rotating
body does not look thermal. Also, the rotating observer does not see the
Minkowski vacuum as a thermal bath.  This means that the matter of the body,
though excited by interaction with the quantum fluctuations of the Minkowski
vacuum, does not necessarily acquire an intrinsic temperature  depending only on
the angular velocity of rotation. Moreover the vacuum of the rotating frame is not
well defined because of the ergoregion, which exists at the distance
$r_e=c/\Omega$ from the axis of rotation.

Let us consider a cylinder of radius $R$ rotating with angular
velocity $\Omega$ in the (infinite) superfluid liquid (Fig.
\ref{RotVacuumFig}). When the body rotates, the energy of quasiparticles is not
well determined in the laboratory frame due to the time dependence of the
potential, caused by the rotation of the body, whose surface is never perfect.
The quasiparticle energy is well defined in the rotating frame, where the
potential is stationary. Hence it is simpler to work in the rotating frame. If
the body is rotating surrounded by the stationary superfluid, i.e.
${\bf v}_{\rm s}=0$ in the laboratory frame, then in the rotating frame one has 
${\bf  v}_{\rm s}=-{\bf \Omega}\times {\bf r}$. Substituting this ${\bf
v}_{\rm s}=0$ in Eq.(\ref{CovarianAcousticMetricReduced}) one obtains that line
element, which determines the propagation of phonons in the frame of the body,
corresponds to the conventional metric of flat space in the rotating frame:
 \begin{equation}
ds^2=-(c^2-\Omega^2\rho^2)dt^2 - 2\Omega \rho^2d\phi dt +dz^2 +
\rho^2d\phi^2+d\rho^2 
 ~.
\label{Interval}
\end{equation}
The azimuthal motion of the quasiparticles in the rotating frame can be quantized in
terms of the angular momentum
$L$, while the radial motion can be treated in the quasiclassical approximation.
Then the energy spectrum of the phonons in the rotating frame is
\begin{equation}
\tilde E =E(p) + {\bf p}\cdot {\bf  v}_{\rm s}= c\sqrt{ {L^2\over \rho^2} + 
p_z^2 +  p_\rho^2}- \Omega L 
 ~.
\label{QuasiclassicalPhononSpectrum}
\end{equation}
For rotons and Bogoliubov fermions in $^3$He-B the 
energy spectrum in the rotating frame is  
\begin{eqnarray}
\tilde E(p)=\Delta + {(p-p_0)^2\over 2m_0} -\Omega L~,\\
\tilde E(p)=\sqrt{\Delta_B^2 +
v_F^2(p-p_F)^2} -\Omega L
 ~,
\label{RotonBogolonSpectrum}
\end{eqnarray}
where $p_0$ marks the roton minimum in superfluid $^4$He, while $\Delta$ is a roton gap.

\subsubsection{Ergoregion in superfluids.}

For the ``relativistic'' phonons the radius $\rho_e^{\rm rel}=c/\Omega$,  where
$g_{00}=0$, marks the position of the ergoplane. In the ergoregion, i.e. at 
$\rho>\rho_e^{\rm rel}=c/\Omega$, the energy of phonons  in
Eq.(\ref{QuasiclassicalPhononSpectrum}) can become negative for any rotation
velocity and $\Omega L> 0$ (Fig. \ref{RotatingVacuumFig}).  However, the real
ergoplane in superfluid $^4$He occurs at $\rho_e=v_{\rm Landau}/\Omega$, where
the Landau velocity in Eq.(\ref{LandauVelocity1}) is $v_{\rm Landau}\sim 
\Delta/p_0$. Let us assume that the angular velocity of rotation $\Omega$ is
small enough, so that the linear velocity on the surface of the cylinder  is
much less than the Landau critical velocity 
$\Omega R<v_{\rm Landau}$. Thus excitations can never be nucleated at the surface of cylinder.
However, in the ergoplane the velocity
$v_{\rm s}=\Omega \rho_e$ in the rotating frame reaches $v_{\rm Landau}$, so
that quasiparticle can be created in the ergoregion $\rho>\rho_e$. 

\begin{figure}[t]
\centerline{\includegraphics[width=\linewidth]{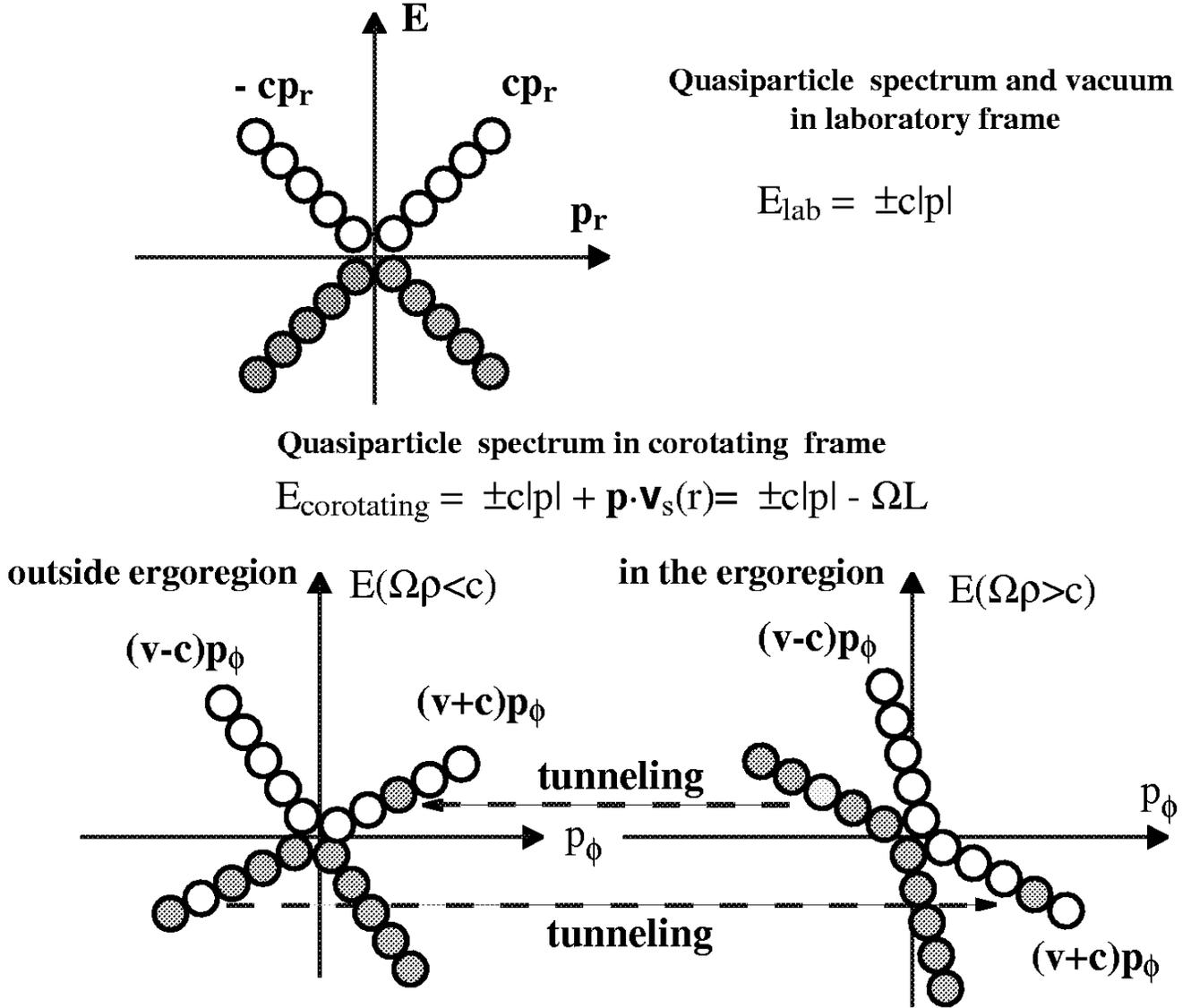}}
\medskip
\caption{Vacuum seen in the frame corotating with inner
cylinder is different from that viewed in the laboratory
frame.  The states which are occupied in the vacuum viewed in the laboratory
frame are shaded. If the ergoplane is close to the rotating inner
cylinder and is far from the outer cylinder, which is at rest in the laboratory
frame, the influence of the rotating cylinder on the quasiparticle
behavior is dominating. Thus the relevant frame for quasiparticles is the
rotating frame. In the ergoregion, some states, which are occupied in the
laboratory vacuum, have positive energy in the corotating frame. The 
quasiparticles occupying these levels must be radiated away. At $T=0$ the
radiation occurs via quantum tunneling from or to the region in the vicinity of
the surface of inner cylinder, where the interaction with the rotating cylinder
occurs.  The rate of tunneling reproduces the Zel'dovich-Starobinsky effect of
radiation from the rotating black hole.}
\label{RotatingVacuumFig}
\end{figure}

The process of creation of quasiparticles is determined by the
interaction with the rotating body; there is no radiation in the absence of the
body. If $\Omega R\ll v_{\rm Landau}=c$ one has $\rho_e\gg R$, i.e.   the
ergoregion is situated far from the cylinder; thus the interaction of the
phonons state in the ergoregion with the rotating body is small. This results
in a small emission rate 
 and thus in a small value of quantum friction, as will be discussed below. 

In superfluid $^4$He the Landau velocity for the emission of rotons is smaller than
that for the emission of phonons,
$v_{L}=c$. That is why the ergoplane for rotons, $\rho_e=v_{\rm
Landau}/\Omega$, is closer to the cylinder, than the plane $\rho_e^{\rm
rel}=c/\Omega$.  However, it appears (see below) that the emission of the
rotons is exponentially suppressed due to the big value of the allowed angular
momentum for emitted rotons: the Zel'dovich condition
Eq.(\ref{ZeldovichCondition}) for roton spectrum is satified only for
$L>\Delta/\Omega \gg 1$.

\subsubsection{Radiation to the ergoregion as a source of rotational quantum friction.}

For the Bose case  the radiation of  quasiparticles can be also considered 
as the process in which the particle in the normal Bose liquid in the surface
layer near surface of the rotating cylinder tunnels to the scattering state at
the ergoplane, where its energy is also
$\tilde E=0$ (Fig. \ref{RotatingVacuumFig}). In the quasiclassical
approximation the tunneling probability is
$e^{-2S}$, where
\begin{equation}
S={\rm {Im}} \int d\rho~p_\rho(\tilde E=0) ~.
\label{TunnelingGeneral}
\end{equation}
For the phonons with $p_z=0$ one has
\begin{equation}
S=L \int^{\rho_e}_{R}d\rho~\sqrt{{1\over \rho^2}-  {1\over
\rho_e^2}}\approx  L  \ln {\rho_e\over R}
 ~.
\label{TunnelingPhonons}
\end{equation}
Thus all the phonons with
$L>0$ are radiated, but the radiation probability decreases at higher $L$. 
If the linear velocity at the surface is much less than $c$, i.e. 
$\Omega R \ll c$, the probability of radiation of phonons with the energy
(frequency)
$\omega=\Omega L$ becomes
\begin{equation}
w \propto e^{-2S}=\left({R\over \rho_e}\right)^{2L}=\left({\Omega R\over
c}\right)^{2L} = \left({\omega R\over
cL}\right)^{2L}~,~\Omega R \ll c
 ~.
\label{PhononTunnelingProbability}
\end{equation}
If $c$ is substituted by the speed of light,  
Eq.(\ref{PhononTunnelingProbability}) is proportional to the superradiant
amplification of the electromagnetic  waves by rotating  dielectric
cylinder derived by Zel'dovich \cite{BekensteinSchiffer,Zeldovich2}.

The number of phonons  with the  frequency 
$\omega=\Omega L$  emitted per unit time can be estimated as 
$
\dot N= W e^{-2S} 
$,
where $W$ is the attempt frequency $\sim \hbar/ma^2$ multiplied by the number
of localized modes $\sim RZ/a^2$, where $Z$ is the height of the cylinder.
Since each phonon carries the angular momentum $L$, the cylinder  rotating in
superfluid vacuum (at $T=0$) is loosing its angular momentum, which means
the quantum rotational friction.

Let us consider the same process for the rotons.
The minimal $L$ value of the radiated quasiparticles, which have the gap
$\Delta$, is determined by this gap:
$L_{min}= \Delta/ \Omega p_0=v_{\rm Landau}/ \Omega$, where
$v_{\rm Landau}=\Delta/p_0$ is the Landau critical velocity for  emission of
rotons. Since the tunneling rate exponentially decreases with
$L$, only the lowest possible $L$ must be considered. In this case  the
tunneling trajectory with
$E=0$ is determined by the equation
$p=p_0$ both for rotons and Bogoliubov quaiparticles. For $p_z=0$ the classical
tunneling trajectory is thus given by
$p_\rho=i\sqrt{|p_0^2-L^2/\rho^2|}$. This gives for the tunneling exponent  
$e^{-2S}$ the equation
\begin{equation}
S={\rm {Im}} \int d\rho~p_\rho=L \int^{\rho_e}_{R}dr~\sqrt{{1\over \rho^2}- 
{1\over \rho_e^2}}\approx  L  \ln {\rho_e\over R}
 ~.
\label{RotonTunnelingAction}
\end{equation}
Here the position of the ergoplane is $\rho_e=L/p_0= v_{\rm Landau}/\Omega$. 
Since the rotation velocity $\Omega$ is always much smaller than the gap, 
$L$ is very big. That is why the radiation of rotons and Bogoliubov
quasiparticles with the gap is exponentially suppressed.

\section{How to improve helium-3.}\label{Improved3He}

\subsection{Gradient expansion}

Though $^3$He-A and Standard Model belong to the same universality class and
thus they have similar properties of the fermionic spectrum, the $^3$He-A cannot
serve as a good model for quantum vacuum. The effective action for bosonic
gauge and gravity fields obtained by integration over fermionic degrees of
freedom is contaminated by the terms which are absent in a fully
relativistic system. This is because the integration over fermions is not always
concentrated in the region where their spectrum is ``relativistic''. Thus the
question arises whether we can ``correct'' the $^3$He-A in such a way that
these uncomfortable terms are suppressed. 

If we neglect the spin degrees of freedom, the massless bosonic fields which
appear in $^3$He-A  due to the  breaking of symmetry $U(1)_N\times SO(3)_L$ 
is the triad field, which can be expressed in terms of the superfluid
velocity (torsion) and the $\hat{\bf l}$-field.  The quadratic energy in terms
of the gradients of the soft Goldstone variables (the London energy for
$^3$He-A)  in the reference frame where the heat bath is at rest, ${\bf v}_{\rm
n}=0$, can be written in the following general form \cite{VollhardtWolfle}:
\begin{eqnarray}
 F_{\rm London}=
{m\over 2}n_{{\rm s}\parallel}\left( \hat{\bf l}\cdot {\bf v}_{\rm s} \right)^2
+ {m\over 2}n_{{\rm s}\perp}\left( \hat{\bf l}\times {\bf v}_{\rm s} \right)^2
\label{London1} \\
+ {1\over 2}\left(C( \hat{\bf l}\times {\bf v}_{\rm s})\cdot(\hat{\bf
l}\times( \nabla\times\hat{\bf l})) -(C_0-C)( \hat{\bf l}\cdot {\bf
v}_{\rm s})(\hat{\bf l}\cdot( \nabla\times\hat{\bf l})) \right)
\label{London2}
\\
 +   (K_s ( \nabla\cdot\hat{\bf l})^2
+K_t(\hat{\bf l}\cdot( \nabla\times\hat{\bf l}))^2
+K_b(\hat{\bf l}\times( \nabla\times\hat{\bf l}))^2
\label{London3}
\end{eqnarray}
The temperatutre dependent coefficients can be obtained within the framework of
the BCS theory if one applies the gradient expansion \cite{Cross}. In its
simplest form  of BCS theory the gradient energy is determined by 4 parameters:

The `speeds of light'' $c_{\parallel}=v_F$ and $c_{\perp}$ characterize the
``relativistic'' physics of the low-energy corner. These are the parameters
of the quasiparticles in the ``relativistic'' low-energy corner below the
first ``Planck scale'',
$E\ll
\Delta_0^2/v_Fp_F$.  

The parameter $p_F$ is the property of the higher level in the hierarchy of
the energy scales -- the Fermi-liquid level. The related
quantity is the quasiparticle mass in the Fermi-liquid
theory, $m^*=p_F/c_{\parallel}$. This parameter also determined the
second ``Planck'' energy scale $\Delta_0=p_Fc_{\perp}$. The spectrum of
quasiparticles in $^3$He-A in the whole range $E\ll v_Fp_F$ is
determined by these 3 parameters, $c_{\parallel}$, $c_{\perp}$ and $p_F$.

Finally the bare mass $m$ of $^3$He is the parameter of the
underlying microscopic physics of interacting ``indivisible'' particles --
$^3$He atoms. This parameter does not enter the spectrum of quasiparticles in
$^3$He-A and thus the naive BCS theory. However the Galilean invariance of
the underlying system of $^3$He atoms requires that the kinetic energy of
superflow at $T=0$ must be $(1/2)mn{\bf v}_{\rm s}^2$, i.e. the bare mass
$m$ must be incorporated into the BCS scheme to maintain the Galilean
invariance. This is achieved in the Landau theory of Fermi liquid, where the
dressing occurs due to quasiparticles interaction. In the simplified approach
one can consider only that part of interaction which is responsible for the
renormalization of the mass and which restores the Galilean invariance of the
Fermi system.   This is the current-current interaction with the Landau
parameter $F_1=3(m^*/m-1)$, containg the bare mass $m$. This is the way how the
bare mass $m$ enters the Fermi-liquid and thus the BCS theory. 

In the real $^3$He liquid the ratio $m^*/m$ varies between about 3 and 6
depending on pressure. However, in the modified BCS theory this ratio can be
considered as a free parameter, which one can ajust to make the system more
close to the relativistic theories. As we discussed in Sec.
\ref{EinsteinActionAnd}, the superfluidity must be suppressed for the correct
Einstein term to prevail in the effective action for gravity. This must happen
if $m\rightarrow \infty$. In this limit the superfluid properties of the liquid
are really suppressed, since the superfluid velocity is inversly proportional to
$m$ according to Eq. (\ref{v_s}),  ${\bf v}_{\rm s}\propto 1/m$. In this limit
of heavy mass of atoms comprising the vacuum,  the vacuum becomes inert,  and
the kinetic energy of superflow $(1/2)mn{\bf v}_{\rm s}^2$, which is dominating
over the Einstein action in real $^3$He-A, vanishes as $1/m$. Thus we can
expect that in case of inert vacuum, the influence of the microscopic level on
the effective theory of gauge field and gravity is suppressed and the effective
action for the collective modes approaches the covariant and gauge invariant
limit of the Einstein-Maxwell action.   

For illustration that this really happens, let us consider how all these 4
parameters enter the gradient energy in Eq.(\ref{London1}-\ref{London3}) and
what happens when
$m\rightarrow
\infty$. According to Cross \cite{Cross} one has the following expression for
the coefficients in the gradient expansion (we are interested in a
low-temperature region $T\ll \Delta_0$). The normal (and thus superfluid)
component densities are given by Eq.(\ref{EquilibriumCurrent}), which for the
$^3$He-A quasiparticles are
\begin{eqnarray}
n_{{\rm n}\parallel}\approx {m^*\over
m}n^0_{{\rm n}\parallel}~,~n^0_{{\rm n}\parallel}=\pi^2 n{T^2\over
\Delta_0^2}~,~ n_{{\rm s}\parallel}=n-n_{{\rm n}\parallel}~,~n^0_{{\rm
s}\parallel}=n-n^0_{{\rm n}\parallel}~,
\label{ParallelNormal}\\
n_{{\rm n}\perp} = {m^*\over m}n^0_{{\rm n}\perp} ~,~n^0_{{\rm
n}\perp}={7\pi^4\over 15}n {T^4\over
\Delta_0^4} ~,~n_{{\rm s}\perp}=n-n_{{\rm n}\perp}~,~n^0_{{\rm
s}\perp}=n-n^0_{{\rm n}\perp}~. 
\label{PerpNormal}  
\end{eqnarray}
Here we introduced (with index $0$) the bare (nonrenormalized) values of
the normal and superfluid component densities, which correspond to the limit
of noninteracting Fermi-gas with $F_1=0$ and thus to
$m^*=m$.  In the derivation we considered
the low-temperature limit
$T\ll \Delta_0$; also the small terms of order of the anisotropy parameter
$c_{\perp}^2/c_{\parallel}^2\sim 10^{-6}$ have been neglected, so that the
particle density is the same in normal and superfluid states:
$n= {p_F^3\over 3\pi^2}$. The other parameters are accoring to
\cite{Cross}
\begin{eqnarray}
 C_0-C={1\over 2m}n_{{\rm s}\parallel}~,
~C={1\over 2m}n_{{\rm s}\perp}{n^0_{{\rm s}\parallel}\over n^0_{{\rm
s}\perp}}~,\\
K_s={1\over 32m^*}n^0_{{\rm s}\perp}~,\label{Splay} \\
K_t={1\over 96 m^*}\left(n^0_{{\rm s}\perp}+4n^0_{{\rm s}\parallel} +
3\left({m^*\over m}-1\right) {n_{{\rm s}\parallel} n^0_{{\rm s}\parallel}\over
n}\right)~â\label{Twist}  
\\
K_b={1\over 32 m^*}\left(2n^0_{{\rm s}\parallel} +
\left({m^*\over m}-1\right) {n_{{\rm s}\parallel} n^0_{{\rm s}\parallel}\over
n}\right)  +{\rm Log}~,\\
\label{Bend} 
~{\rm Log}={1\over 4 m^*}n
\int{d\Omega\over 4\pi}{(\hat{\bf l}\cdot \hat{\bf p})^4\over 
(\hat{\bf l}\times \hat{\bf p})^2} \left[1+2\int_{0}^\infty dM{\partial
f_{\cal T}\over
\partial E}\right]~.
\label{Log}  
\end{eqnarray}
Here ${\rm Log}$ is the term which contains $\ln (\Delta_0/T)$. The integral
in Eq.(\ref{Log}) is over the solid angle in momentum space; the energy spectrum
which enters the equilibrium quasiparticle distribution function $f_{\cal
T}=(1 +\exp{E/T})^{-1}$, with $E^2=M^2 +\Delta_0^2(\hat{\bf
p}\times\hat{\bf l})^2$ as is given by Eq.(\ref{BogoliubovNambuEnergyAPhase}).
This coefficient does not depend on the microscopic parameter $m$ and
represents the logarithmically diverging coupling constant in the Maxwell
effective action for the  magnetic field in curved space in
Eq.(\ref{EMLagrangian}).  

\subsection{Effective action in inert vacuum.}

In the limit $m\rightarrow\infty$ all the terms related to superfluidity
vanish since $v_s\propto 1/m$. Taking into account
that in this limit $n_{s\parallel}=n$ one obtains for the remaining ${\hat{\bf
l}}$-terms:
\begin{eqnarray}
 F_{\rm London}(m\rightarrow\infty)= 
{1\over 32m^*} n^0_{s\perp}  (\nabla\cdot\hat{\bf l})^2 +
 {1\over 96m^*}(n^0_{s\perp} + n^0_{s\parallel})  (\hat{\bf l}\cdot(
\nabla\times\hat{\bf l}))^2 
+  \left({1\over 32m^*}  n^0_{s\parallel} + {\rm Log}\right)(\hat{\bf
l}\times(
\nabla\times\hat{\bf l}))^2 ~.
\label{LondonInfty}
\end{eqnarray}
All three terms have the correspondence in QED and Einstein gravity. We have
already seen that the bend term, i.e. $(\hat{\bf
l}\times(
\nabla\times\hat{\bf l}))^2$, is exactly the energy of the the  magnetic field
in curved space with the logarithmically diverging coupling constant.
Let us now consider the twist term,  i.e. $(\hat{\bf
l}\cdot( \nabla\times\hat{\bf l}))^2$ and show that it corresponds to the
Einstein action.

\subsubsection{Einstein action in $^3$He-A.}\label{EinsteinActionAPhase}

The part of effective gravity, which is simulated by the superfluid velocity
field, vanishes in the limit of inert vacuum. The remaining part of
gravitational field is simulated by the inhomogeneity of the
$\hat {\bf l}$ field, which plays the part of the ``Kasner axis'' in the metric
\begin{eqnarray}
g^{ij}  =c_\parallel^2 \hat l^i\hat l^j +c_\perp^2(\delta^{ij}
-\hat l^i\hat l^j) ~,~g^{00}=-1,~g^{0i}=0,
~\sqrt{-g}={1\over c_\parallel
c_\perp^2} ,\label{MetricAPhaseGeneral2}
\\g_{ij}  ={1\over c_\parallel^2} \hat l^i\hat l^j +{1\over c_\perp^2}(\delta^{ij}
-\hat l^i\hat l^j),~g^{00}=-1~,~g^{0i}=0~.
\label{MetricAPhaseGeneralContra2}
\end{eqnarray}
The  curvature of the space with this metric is caused by spatial rotation of
the ``Kasner axis'' $\hat {\bf l}$. For the stationary metric, $\partial_t \hat
{\bf l}=0$, one obtains that in terms of the $\hat {\bf l}$-field the Einstein
action is 
\begin{equation}
-{1\over 16 \pi G}\int \sqrt{-g} R = {1\over 32 \pi G\Delta_0^2}
\left(1-{c_\perp^2\over c_\parallel^2}\right)^2{p_F^3\over m^*}\int   
((\hat{\bf l}\cdot(
\nabla\times\hat{\bf l}))^2 ~. 
\label{EinsteinActionAphase}
\end{equation}
It has the structure of the twist term in the gradient
energy (\ref{LondonInfty}) obtained in gradient expansion, which in the inert
vacuum limit is
\begin{equation}
 F_{\rm twist}={1\over 288}\left({2\over  \pi^2}-{T^2\over \Delta_0^2}
\right){p_F^3\over m^*}
\int   ((\hat{\bf
l}\cdot(
\nabla\times\hat{\bf l}))^2  
\end{equation}
We thus can identify the twist term with Einstein action. Neglecting
the small anisotropy factor ${c_\perp^2\over c_\parallel^2}$ one
obtains that the  Newton constant in the effective gravity of the
``improved'' $^3$He-A is:
\begin{equation}
 G^{-1} ={2\over 9\pi}\Delta_0^2- {\pi \over 9} T^2~.
\label{NewtonConstAphase}
\end{equation}
While the temperature independent part certainly depends on the details of
the transPlanckian physics, the temperature dependence of the Newton constant
pretends to be universal, since it does not depend on the parameters of the
system.  If one applies the regularization scheme provided by the transPlanckian
physics of $^3$He to the relativistic system one would suggest the
following temperature dependence of Newton constant in the vacuum with $N_F$
Weyl fermions ($N_F=2$ for
$^3$He-A): 
\begin{equation}
\delta [G^{-1}]=-   {\pi \over 18} N_F T^2~.
\label{NewtonConstTempCorrection} 
\end{equation}

\subsubsection{Violation of gauge invariance.}

Let us finally consider the splay term $(\nabla\cdot\hat{\bf l})^2$ in
Eq.(\ref{LondonInfty}). It has only the 4-th order temperature corrections,
$T^4/\Delta_0^4$.  This term has similar coefficient as the curvature term, but
it is not contained in Einstein action, since it cannot be written in
the covariant form. The structure of this term can be, however, obtained using
the gauge field presentation of the
$\hat{\bf l}$ vector, where ${\bf A}=p_F\hat{\bf l}$. It is known that similar
term can be obtained in the renormalization of QED in the leading order of
$1/N$ theory if the regularization is made by introducing the momentum cut-off:
\cite{Sonoda}:
\begin{equation}
 L'_{QED}= {1\over 96 \pi^2}(\partial_\mu A_\mu)^2~.
\label{divl1}  
\end{equation}
This term violates the
gauge invariance and does not appear in the dimensional
regularization scheme
\cite{WeinbergBook}, but it appears in the momentum cut-of procedure, which
violates the gauge invariance. Being written in covariant form the
Eq.(\ref{divl1})  can be applied to
$^3$He-A, where
${\bf A}=p_F\hat{\bf l}$ and $\sqrt{-g}=Const$:
\begin{equation}
 L'_{QED}= {1\over 96 \pi^2}\sqrt{-g}(\partial_\mu (g^{\mu\nu}A_\nu))^2=
~{1\over 96 \pi^2}{p_F^2c_\parallel^3\over c_\perp^2}(\nabla\cdot\hat{\bf
l})^2={c_\parallel^2\over c_\perp^2}{1\over 96
\pi^2}{p_F^3\over m^*}(\nabla\cdot\hat{\bf
l})^2.
\label{divl2}
\end{equation}
This term corresponds to  the splay term in
Eq.(\ref{LondonInfty}), but it contains an extra big factor of the vacuum
anisotropy $c_\parallel^2/ c_\perp^2$:
$F_{London~splay}= (c_\perp^2/ c_\parallel^2) L'_{QED}$. However, in the
isotropic case, where $c_\parallel = c_\perp$, they exactly coincide. This
suggests that the regularization provided by the ``transPlanckian physics'' of
$^3$He-A  represents the anisotropic version of the momentum cut-off
regularization of quantum electrodynamics.  

\section{Discussion}

Let us summarize the parallels between the quantum vacuum and superfluids, which were
touched upon in the review, and their possible influence on the quantum field theory.

Superfluid $^3$He-A and other possible representative of its
universality class provides an example of how the chirality, Weyl fermions,
gauge fields and gravity can emergently appear in the low emergy corner
together with the corresponding symmetries, which include the Lorentz symmetry
and local $SU(N)$ symmetry. This supports the ``anti-grand-unification'' 
idea that the quantum field theory, such as Standard Model or maybe GUT, is
an effective theory, which is applicable only in the infrared limit.  Most
of the symmetries of this effective theory are the attributes of the
theory: the symmetries gradually appear in the low-energy corner together
with the effective theory itself. 

The momentum space topology of the fermionic vacuum (Sec.
\ref{UniversalityClassesOf}) is instrumental in determination of the
universality class of the system. It provides the topological stability
of the low-energy properties of the systems of given class: the character of
the fermionc spectrum, collective modes and leading symmetries. The
universality class, which contains topologically stable Fermi points, is
common for superfluid $^3$He-A and Standard Model. This allowed us to provide
analogies between many phenomena in the two systems, which have the same
physics but in many cases are expressed in different languages and can be
visualized in terms of different observables. However, in the low-energy corner
they are described by the same equations if they are written in the covariant
and gauge invariant form. On this topological ground it appears that some of
the unification schemes of the strong and electroweak interactions is more
preferrable than the others: this is the $SU(4)_C\times SU(2)_L\times
SU(2)_R$ group (Sec. \ref{StandardModelAnd}).

The advantage of $^3$He-A is that this system is  complete being
described by the BCS model. This scheme incorporates not only the
``relativistic'' infrared regime, but also several successive scales of the
short-distance physics, which correspond  to different ultraviolet 
``transPlanckian'' ranges of high energy. Since in BCS scheme there is no
need for a cut-off imposed by hand, all subtle issues of the cut-off in
quantum field theory can be resolved on physical grounds. It appears,
however,  the $^3$He-A is not a perfect object for the realization of the
completely covariant effective gauge and gravity fields at low energy, because
the effect of the ``transPlanckian'' physics shows up even in the low-energy
corner. This in particular leads to many noncovariant terms
in the effective action, including the mass of the graviton.  On the other
hand, it is clear how to ``correct'' the $^3$He-A: the Lorentz invariance must
be extended far into the ``transPlanckian'' region to kill the noncovariant
non-renormalizable terms.  The killing is however never
complete, the ``nonrenormalizable'' terms are always there in effective
theory though they are small since contain the Planck energy cut-off
$E_{\rm P}$ in denominator. One example of such ``nonrenormalizable''
term in $^3$He-A, which is the remnant of the ``transPlanckian'' physics,
corresponds to the mass term for the $U(1)_Y$ gauge field of the hyperphoton
which violates the gauge invariance (Sec. \ref{MassHyperphoton}). 

Thus the condensed matter of the Fermi-point universality class shows one of
possibly many  routes from the low-energy ``relativistic'' to high energy
``transPlanckian'' physics.  Of course, one might expect the many routes to
high energy, since the systems of the same universality class become similar
only in the vicinity of the fixed point: they can diverge far from each other
at higher energies. Nevertheless, probably the first corrections could be
similar too.

Practically in all condensed matters of even different universality classes,
the effective action for some bosonic or even fermionic modes acquires
an effective Lorentzian metric. That is why the gravity is the field,  which
can be simulated most easily in condensed matter.  The gravity can be
simulated by flowing normal fluids, superfluids, and Bose-Einstein condensates;
by elastic strains, disclocations and disclinations in crystals, etc.  Though
the full dynamical realization of gravity takes place only in the
fermionic condensed matter with Fermi points, the acoustic type of gravity
are also useful for simulation of different phenomena related to marriage of
gravity and quantum theory. 

The analog of gravity in superfluids shows the possible way how to solve the
cosmological constant problem. The standard calculations of the energy density
of superfluid ground state in the framework of the effective
theory suggests that it is of the order of $E_{\rm P}^4$, if one
translates it into the language of the relativistic theories. It is of the same
magnitude as the field theoretical estimation for the vacuum energy and thus 
for the cosmological constant.  However, the stability analysis of the ground
state of the isolated superfluid liquid, which is certainly beyond the
effective theory, strongly forces the exact nullification of the appropriate
energy density of the liquid, which enters the Lagrangian at $T=0$ (Sec.
\ref{VacuumEnergyAnd}). In terms of the relativistic quantum field theory this
means that the equilibrium vacuum should not gravitate.  And this
conclusion cannot be obtained from the effective theory, while in the
underlying microscopic physics this result of the complete nullification
does not depend on the microscopic details. It is universal in terms of the
transPlanckian physics, but not in terms of the effective theory.

But what happens if the  phase transition occurs in which the symmetry of the
vacuum is broken, as is supposed to happen in early Universe when, say, the
electroweak symmetry was broken?  In the effectve theory, such transition must
be accompanied by the change of the vacuum energy, which means that the vacuum
has a huge energy either above or below the transition. However, in exact
microscopic theory of liquid the phase transition does not disturb the zero
value of the vacuum energy (Sec. \ref{VacuumPressureSection}), if the liquid
remains in equilibrium. The energy change is completely compensated by the
change of the chemical potential of the underlying atoms of the liquid which
comprise the vacuum state -- the quantity which is not known in effective
theory. Moreover, the analogy with the superfluids also shows that in
nonequilibrium case or at nonzero $T$
(Secs. \ref{GlobalThermodynamicEquilibrium} and \ref{VacuumPressureSection})
the effective vacuum energy must be of order of the energy density of matter.
This is in agreement with the mordern astrophysical observations.

Another object, where as in the problem of gravitating vacuum the marriage of
gravity and quantum theory is important, is the black hole.  Having
many objects for simulation of gravity,  we can expect in the nearest
future that the analogs of event horizon could be constructed in the
laboratory.  Most probably this will first happen in the laser trapped Bose
condensates \cite{Garay}. The condensed matter analogs of horizons may
exhibit Hawking radiation, but in addition the other, unexpected, effects
related to quantum vacuum could arise, such as instability
experienced by vacuum in the acoustic model of gravity.  Because the
short-distance physics is explicitly known in condensed matter this helps
clarify  the problem related to the vacuum in the presence of
the horizon, or in the other exotic effective metric, such as the degenerate
metric (Sec. \ref{VierbeinDomainWall}). 

At the moment only one of the exotic metrics has been experimentally
simulated. This is the metric induced by spinning cosmic string, which
produces the analog of the gravitational Aharonov-Bohm effect, experienced by
particles in the presence of such string. This type of Aharonov-Bohm effect has
been experimentally confirmed in superfluids by measurement of the
Iordanskii force acting on quantized vortices (Sec.
\ref{GravitationalAharonov-BohmEffect}).  

As for the other (nongravitational) analogies, the most interesting are
related to the interplay between the vacuum and the matter, and which can be
fully investigated in condensed matter, because of the absence of the cut-off
problem. These are the anomalies, which are at the origin of the exchange of
the fermionic charges between the vacuum and the matter. Such anomalies are
the attributes of the Fermi systems of universality class of Fermi points:
Standard Model and $^3$He-A. The spectral flow from the vacuum to the matter,
which carries the fermionic charge from the vacuum to matter, occurs just
through the Fermi point. Since in the vicinity of the Fermi point the equations
are the same for the two Fermi systems, the spectral flow in both systems is
described by the same  Adler, and Bell and Jackiw equation for axial anomaly
\cite{Adler,BellJackiw} (Sec. \ref{ChiralAnomaly}).  The
$^3$He-A provided the first experimental prove for the anomalous nucleation of
the fermionic charge from the vacuum. The Adler-Bell-Jackiw
equation was confirmed up to the numerical value of the factor $1/4\pi^2$ with
the precision of few percent. The anomalous nucleation of the baryonic or
leptonic charge is in the basis of the modern theories of the baryogenesis. 

The modified equation is obtained for the nucleation of the fermionic charge by
the moving string -- the quantized vortex. The transfer of the
fermionic charge from the ``vacuum'' to ``matter'' is mediated by the fermion
zero modes living on vortices. This condesed matter illustration of the 
cancellation of anomalies in 1+1 and 3+1 systems (the Callan-Harvey effect)
has been experimentally verified in $^3$He-B. This also means that the scenario
of the baryogenesis by cosmic strings has been experimentally probed. The other
effect related to axial anomaly --  the helical instability of the
superfluid/normal counterflow in $^3$He-A -- is also described by the same
physics and by the same equations as the formation of the (hyper) magnetic
field due to the helical instability experienced by the vacuum in the presence
of the  heat bath of the right-handed electrons  (Sec.
\ref{MagnetogenesisChiralFermions}). That is why its experimental observation
in $^3$He-A provided an experimental support for the Joyce-Shaposhnikov
scenario of the genesis of the primordial magnetic field.  In a future the
macroscopic parity violating effect suggested by Vilenkin
\cite{Vilenkin79} must be simulated in $^3$He-A (Sec.
\ref{MixedAxialGravitational}). In both systems it is desribed by the same
mixed axial-gravitational Chern-Simons action.

One may expect that the further theoretical and experimental exploration of
the vacuum / condensed matter analogies will clarify the properties of the
quantum vacuum.


\begin{thebibliography}{999}

\bibitem{FrogNielBook} C.D. Frogatt and  H.B. Nielsen, {\it Origin of
Symmetry}, World Scientific, Singapore - New Jersey - London - Hong Kong, 1991.

\bibitem{Chadha} S. Chadha,  and  H.B. Nielsen, Lorentz Invariance as a
Low-Energy Phenomenon, Nucl. Phys. {\bf B~217}, 125--144  (1983).

\bibitem{Weinberg} S. Weinberg,    What is quantum field theory, and what
did we think it is? hep-th/9702027

\bibitem{Jegerlehner} F. Jegerlehner, The ``ether-world'' and elementary
particles, hep-th/9803021.

\bibitem{parallel} G.E. Volovik, Field theory in superfluid $^3$He:  
What are the lessons for particle physics, gravity and high-temperature
superconductivity?,  Proc. Natl. Acad. Sci. USA {\bf 96},
6042 - 6047 (1999), cond-mat/9812381,;  G.E. Volovik, $^3$He and Universe
parallelism,  in ``Topological Defects and the Non-Equilibrium Dynamics of Symmetry
Breaking Phase Transitions'', Yu. M. Bunkov, H. Godfrin (Eds.), 
pp. 353-387 (Kluwer, 2000), cond-mat/9902171.

\bibitem{Rovelli} C. Rovelli, Notes for a brief history of quantum gravity,
gr-qc/0006061.

\bibitem{LaughlinPines} R.B. Laughlin and D. Pines, The Theory of Everything,
Proc. Natl. Acad. Sc. USA {\bf 97}, 28-31 (2000). 


\bibitem{Hu96}  B.L. Hu, Expanded version of an invited talk at 2nd International Sakharov
Conference on Physics, Moscow,  20 - 23 May 1996,  e-Print Archive: gr-qc/9607070.

\bibitem{Padmanabhan} T. Padmanabhan, Conceptual issues in combining general
relativity and quantum theory, hep-th/9812018.

\bibitem{LammiTalk} G.E. Volovik, Axial anomaly in $^3$He-A: Simulation of baryogenesis and
generation of primordial magnetic field in Manchester and Helsinki, Physica {\bf B~255}, 86
-- 107 (1998).

\bibitem{Khalatnikov} I.M. Khalatnikov: {\em An Introduction to the Theory of
Superfluidity}, (Benjamin, New York, 1965).

\bibitem{WessZumFerro} I.E. Dzyaloshinskii and G.E. Volovick, Poisson
brackets in condensed matter, Ann. Phys. {\bf 125}, 67 - 97 (1980); G.E.
Volovik,   Wess-Zumino action for the orbital dynamics of $^3He-A$," JETP Lett.
{\bf 44}, 185 - 189 (1986); Linear momentum in ferromagnets, J. Phys. C~ {\bf
20}, L83 - L87 (1987).

\bibitem{VollhardtWolfle} D. Vollhardt, and P. W\"olfle,  The
superfluid phases of helium 3,  Taylor and Francis, London - New York -
Philadelphia, 1990.

\bibitem{unruh} W. G. Unruh, Experimental black-hole
evaporation?, Phys. Rev. Lett. {\bf 46}, 1351-1354 (1981);   Sonic analogue of black holes and
the effects of high frequencies on black hole evaporation, Phys. Rev.
D {\bf 51}, 2827-2838 (1995).

\bibitem{vissersonic}
M. Visser, Acoustic black holes: horizons,
ergospheres, and Hawking radiation, Class. Quantum Grav. {\bf 15},
1767-1791 (1998)

\bibitem{StoneIordanskii} M. Stone,  Iordanskii force and the gravitational Aharonov-Bohm
effect for a moving vortex, Phys. Rev.    {\bf B~61}, 11780 -- 11786 (2000).

\bibitem{VisserBook} M. Visser, {\it Lorentzian Wormholes. From Einstein to
Hawking}, AIP Press, Woodbury, New York, 1995.

\bibitem{Woo} C.W. Woo, Microscopic calculations for condensed phases of helium, in: The
Physics of Liquid and Solid Helium, Part I, eds. K.H. Bennemann and J.B. Ketterson (John
Wiley \& Sons, New York, 1976).

\bibitem{Weinberg2} S. Weinberg,   Rev. Mod. Phys.  {\bf  61},
1  (1989).

\bibitem{Riess} A.G. Riess, A.V. Filippenko, M.C. Liu, P. Challis, A.
Clocchiatti, A. Diercks, P.M. Garnavich, C.J. Hogan, S. Jha, R.P.
Kirshner, B. Leibundgut, M. M. Phillips, D. Reiss, B.P. Schmidt, R.A. Schommer,
R.C. Smith, J. Spyromilio, C. Stubbs, N.B. Suntzeff, J. Tonry, P.
Woudt, R.J. Brunner, A. Dey, R. Gal, J. Graham, J. Larkin, S.C. Odewahn, and
B. Oppenheimer,
Tests of the accelerating universe with near-infrared
observations of a high-redshift type Ia supernova, astro-ph/0001384.

\bibitem{Sakharov} A. D. Sakharov:   Vacuum Quantum Fluctuations in Curved
Space and the Theory of Gravitation, Dokl. Akad. Nauk {\bf 177}, 70-71
(1967) [Sov. Phys. Dokl. {\bf 12}, 1040-41 (1968)]

\bibitem{FrolovFursaev} V. Frolov and D. Fursaev,
 Thermal fields, entropy, and black holes,  Class. Quant. Grav. {\bf 15},
2041-2074  (1998).

\bibitem{LandauLifshitz2} L.D. Landau and E.M. Lifshitz, {\em Classical Fields}, Pergamon
Press, Oxford, 1975.

\bibitem{Tolman} R.C. Tolman: {\it Relativity, Thermodynamics and Cosmology}
(Clarendon Press, Oxford, 1934).

\bibitem{VolovikMineev1982} G.E. Volovik and V.P. Mineev,  Current in
  superfluid Fermi liquids and the vortex core structure, Sov. Phys. JETP {\bf 56},
  579 - 586 (1982).

\bibitem{Grinevich1988} P.G. Grinevich and G.E.  Volovik, 
Topology of gap nodes in superfluid  $^3$He,   J. Low Temp. Phys.   {\bf 72}, 371-380 (1988).

\bibitem{LuttingerTheorem} J.M. Luttinger,    Phys. Rev. {\bf 119}, 1153
(1960). 

\bibitem{LuttingerTheoremTopology} M. Oshikawa, Topological Approach to Luttinger's Theorem
and the Fermi Surface of a Kondo Lattice,  Phys. Rev. Lett. {\bf 84}, 3370-3373 (2000). 

\bibitem{NewClass} G.E. Volovik,   A new class of normal Fermi
liquids,   JETP Lett. {\bf 53}, 222-225 (1991).

\bibitem{Blagoev} K.B. Blagoev and K.S. Bedell,   Luttinger theorem in
one dimensional metals,  Phys. Rev. Lett. {\bf 79},  1106-1109  (1997).

\bibitem{Wen} Wen, X.G.  (1990) Metallic non-Fermi-Liquid\index{sub}{Fermi
liquid} Fixed Point in Two and Higher Dimensions, {\it Phys. Rev.} {\bf
B~42},  6623-6630.

\bibitem{LuttingerLiquidReview} H.J. Schulz,  G. Cuniberti,  and P. Pieri,  
Fermi liquids and Luttinger liquids,  cond-mat/9807366.

\bibitem{Yakovenko} V.M. Yakovenko,   Metals in a high magnetic
field: a universality class  of marginal
Fermi liquid,  Phys. Rev. {\bf B~47},
8851-8857  (1993).

\bibitem{NambuJona-Lasinio} Y. Nambu  and G. Jona-Lasinio,    Dynamical model of
elementary particles based on an analogy with
superconductivity. I.,   Phys. Rev. {\bf
122}, 345-358 (1961); Dynamical model of elementary particles based on an
analogy with superconductivity\. II., Phys. Rev. {\bf 124},
246-254 (1961).

\bibitem{ColorSuperfluidity} M. Alford,  K.  Rajagopal and F. Wilczek,  
 QCD at Finite Baryon Density: Nucleon Droplets and Color
Superconductivity,  Phys. Lett. {\bf
B~422}, 247-256 (1998);F. Wilczek, F.  From Notes to Chords in QCD, Nucl.
Phys. {\bf A~642}, 1-13 (1998).

\bibitem{MiniBigBang} V.M.H. Ruutu, V.B. Eltsov, A.J. Gill,  T.W.B.  Kibble, M. Krusius,  
 Yu.G. Makhlin, B. Placais,  G.E.  Volovik and  Wen Xu, Vortex formation in
neutron-irradiated superfluid $^3$He as an analogue of cosmological defect
formation,   Nature {\bf  382},   334--336 (1996) . 

\bibitem{Kibble}  T.W.B.  Kibble, Topology of cosmic domains and strings,  J. Phys.  {\bf A~
9},   1387--1398 (1976) .

\bibitem{BevanNature}  T.D.C. Bevan, A.J. Manninen, J.B. Cook, J.R. Hook, H.E. Hall,
T. Vachaspati and G.E. Volovik, Momentogenesis by $^3$He vortices: an experimental 
analogue of primordial baryogenesis,  Nature,  {\bf 386}, 
689-692  (1997).

\bibitem{NielsenNinomiya} H.B. Nielsen and M. Ninomiya, Absence of
neutrinos on a lattice. I - Proof by homotopy theory, Nucl. Phys. {\bf
B~185}, 20 (1981), [Erratum - Nucl. Phys. {\bf B~195}, 541 (1982)];     Nucl.
Phys. {\bf B ~193} 173 (1981). 

\bibitem{Exotic} G. E. Volovik, {\it Exotic properties of superfluid
$^3$He}, World Scientific, Singapore - New Jersey - London - Hong Kong, 1992.

\bibitem{VolovikYakovenko} G.E. Volovik, and V.M. Yakovenko,  
 Fractional charge, spin and statistics of solitons in superfluid
$^3$He film, J. Phys.: Cond. Matter {\bf 1},  5263-5274  (1989). 

\bibitem{VolovikEdgeStates} G. E. Volovik, On edge states in superconductor
with time inversion symmetry breaking,  JETP Lett. {\bf 66 },  522-527 
(1997).

\bibitem{Ishikawa} K. Ishikawa, and T. Matsuyama, {\it Z. Phys. C} {\bf 33}, 41
(1986); {\it Nuclear Physics B}   {\bf 280}, 532 (1987).

\bibitem{Senthill} T. Senthil, J.B. Marston and M.P.A. Fisher, Phys.
Rev.   {\bf B~60}, 4245 (1999).

\bibitem{ReadGreen}  N. Read and D. Green,  Paired states of fermions in two
dimensions with breaking of parity and time-reversal symmetries, and the fractional quantum
Hall effect,Phys. Rev. {\bf B~61}, 10267-10297 (2000). 

\bibitem{Yakovenko2} V.M. Yakovenko,  Spin, statistics and charge
of solitons in (2+1)-dimensional theories,   Fizika (Zagreb) {\bf 21},
suppl. 3, 231 (1989) [cond-mat/9703195].

\bibitem{MomentumSpaceTopology} G.E. Volovik,  Momentum-space topology of
Standard Model, J. Low Temp. Phys.  {\bf 119}, 241 -- 247 (2000), hep-ph/9907456.

\bibitem{VolovikVachaspati} G.E. Volovik and T. Vachaspati, Aspects of $^3$He and the 
standard electroweak model, Int. Journ. Mod. Phys. B {\bf 10}, 471 - 521 (1996).

\bibitem{ColorSuperconductivity1}  S. Ying, The quantum aspects of relativistic
fermion systems with particle condensation, Annals Phys. {\bf 266},
295-350 (1998);  On the local finite density relativistic quantum field
theories,  hep-th/9802044.

\bibitem{Mermin-Ho}  N.D. Mermin and  T.L. Ho, Circulation and angular
  momentum in the A phase of superfluid $^3$He, Phys. Rev. Lett. {\bf 36}, 594-597 (1976).

\bibitem{Abrikosov}  A.A.  Abrikosov,    Phys. Rev. B {\bf  58},
2788  (1998).

\bibitem{PatiSalam} J.C. Pati and A. Salam, Is baryon number conserved?  Phys. Rev. Lett. 
{\bf 31}, 661-664 (1973);  Lepton number as the fourth color,
Phys. Rev.  {\bf D~10}, 275-289 (1974).

\bibitem{Foot} R. Foot, H. Lew, and R.R. Volkas,   Models of extended Pati-Salam
gauge symmetry, Phys. Rev.   {\bf D~ 44}, 859 (1991).

\bibitem{PatiNew} J.C. Pati, Discovery of Proton Decay: A must for theory, a challenge for
experiment, hep-ph/0005095.

\bibitem{Terazawa} H. Terazawa,  High Energy Physics in the 21-st Century, KEK
Preprint 99-46, July 1999, H.

\bibitem{Marchetti} P.A. Marchetti, Zhao-Bin Su, Lu Yu,  Dimensional reduction of
$U(1)\times SU(2)$ Chern-Simons bosonization: application to the $t-J$ model,
Nucl.Phys. {\bf B~482}, 731  (1996) and references therein.

\bibitem{FroggattNielsen}  C.D. Froggatt, and  H. B. Nielsen,   Why do
we have parity violation? Proceedings of the International Workshop on {\it What comes
beyond the Standard Model},  Bled, Slovenia, 29 June - 9 July 1998,   hep-ph/9906466.

\bibitem{TaitPhD} T.M.P. Tait, Signals for the electroweak symmetry breaking
associated with the top quark, hep-ph/9907462.

\bibitem{Zeldovich} Ya.B. Zel'dovich,  Interpretation of electrodynamics as a
consequence of quantum theory,  JETP  Lett. {\bf 6}, 345-347  (1967).

\bibitem{Mannheim} P.D.  Mannheim,   Implications of cosmic repulsion for
gravitational theory,     Phys. Rev. {\bf D~58}, 103511, pp. 1--12 (1998).

\bibitem{Edery} A. Edery and M.B. Paranjape,  Classical tests for Weyl gravity: Deflection
of light and time delay,     Phys. Rev.  {\bf D~58}, 024011, pp. 1--8 (1998).

\bibitem{Mannheim2} P.D.  Mannheim,   Cosmic acceleration and a natural solution to
the cosmological constant problem, e-Print Archive: gr-qc/9903005.

\bibitem{Volovik1986gravity} G.E. Volovik,  Analog of gravitation in superfluid
   $^3$He-A, JETP Lett. {\bf 44}, 498 - 501 (1986).

\bibitem{Gribov} I indebt to V.N. Gribov, who explained to me this point.

\bibitem{Martin}  C.P. Martin,  J.M.  Gracia-Bondia and  J.S.  Varilly,  The Standard
Model as a noncommutative geometry: The low-energy regime,   Phys. Rep. {\bf 294}, 
pp. 363--406 (1998).

\bibitem{Sogami} I.S. Sogami,  Generalized covariant derivative with gauge and
Higgs fields in the Standard Model,  Prog. Theor. Phys. {\bf 94},  117--123 (1995);
Minimal   $SU(5)$ Grand Unified theory based on generalized covariant
derivative with gauge and Higgs fields,  Prog. Theor. Phys.  {\bf 95},  637--655 (1996).


\bibitem{HarariSeiberg}  H. Harari and N. Seiberg,  Phys. Lett. {\bf
B~102}, 263 (1981).

\bibitem{Adler2}  S.L.  Adler,    fermion-sector frustrated $SU(4)$ as a
preonic precursors of the Standar Model,     Int. J. Mod. Phys. {\bf A~14},
1911-1934 (1999).

\bibitem{Peccei} R.D. Peccei,
Discrete and global symmetries in particle physics, hep-ph/9807516.

\bibitem{Iliopoulus} J. Iliopoulus, D.V. Nanopoulus and T.N. Tomaras,
Infrared stability or anti-grandunification, Phys. Lett.  {\bf B~55}, 141-144
(1980).

\bibitem{VolovikKhazan} G.E. Volovik and M.V. Khazan,  Dynamics of the A-phase of
$^3$He at low pressure, Sov. Phys. JETP {\bf 55}, 867 - 871 (1982).

\bibitem{Nambu} Y. Nambu, Fermion-boson relations in the BCS-type theories, Physica {\bf
D~15}, 147 (1985).

\bibitem{Troshin} S. M. Troshin, and N. E. Tyurin, Hyperon polarization in the constituent
quark model, Phys. Rev. {\bf D~55}, 1265 -- 1272 (1997).

\bibitem{Adler}  S. Adler,    Axial-vector vertex in spinor
electrodynamics,     Phys. Rev. {\bf 177}, 2426 - 2438 (1969).

\bibitem{BellJackiw}  J.S. Bell and R. Jackiw, A PCAC Puzzle: $\pi_0 \rightarrow
\gamma\gamma$ in the $\sigma$ Model,   Nuovo Cim. {\bf A~60},  47--61 (1969) .

\bibitem{Trodden} M. Trodden, Electroweak Baryogenesis,  Rev. Mod. Phys. 
{\bf 71},   1463-1500 (1999).

\bibitem{tvgf} T. Vachaspati and G.B. Field,  `Electroweak  
string configurations with baryon number, Phys. Rev. Lett. {\bf  73},
373--376 (1994);  {\bf 74}, 1258(E) (1995).

\bibitem{jgtv} J. Garriga and T. Vachaspati, Zero modes on  
linked strings', Nucl. Phys. {\bf B~438}, 161  (1995).

\bibitem{barriola}  M. Barriola,  Electroweak strings produce 
baryons, Phys. Rev. {\bf D~51}, 300 (1995).

\bibitem{VolovikMineev1981} G.E. Volovik and V.P. Mineev, $^3$He-A vs Bose
liquid:  Orbital angular momentum and orbital dynamics, Sov. Phys. JETP {\bf 54},
  524 - 530 (1981).

\bibitem{Combescot}  R. Combescot and T. Dombre, Twisting in superfluid $^3$He-A and
consequences for hydrodynamics at $T=0$, Phys. Rev.  {\bf B~33},  79-90 (1986).

\bibitem{Volovik1986} G.E. Volovik,  Chiral anomaly and the law of conservation
  of momentum in $^3$He-A, JETP Letters {\bf 43}, 551 - 554 (1986).

\bibitem{AchucarroVachaspati} A. Achucarro and T.   Vachaspati,  Semilocal and
electroweak strings,  Phys. Rep. {\bf 327}, 347-426 (2000).

\bibitem{Chechetkin} V.R. Chechetkin, Sov. Phys. JETP,  {\bf 44}, 706 (1976).

\bibitem{AT} P.W. Anderson and G. Toulouse,  Phase slippage without vortex cores: Vortex
textures in superfluid $^3$He, Phys. Rev. Lett. {\bf 38}, 508--511 (1977).

\bibitem{Volovik1992} G.E. Volovik,  Hydrodynamic action for   orbital and
superfluid dynamics of $^3$He-A at $T=0$',   JETP {\bf 75},  990-997 (1992).

\bibitem{Kopnin1993} N.B. Kopnin, Mutual friction in
superfluid  $^3$He. II. Continuous vortices in $^3$He-A at low temperatures, 
Phys. Rev.   {\bf B~47}, 14354 (1993)

\bibitem{BevanJLTP}  T.D.C. Bevan, A.J. Manninen, J.B. Cook, H. Alles, J.R.
Hook and H.E. Hall,  Vortex mutual friction in superfluid $^3$He vortices, J. Low Temp.
Phys. {\bf 109}, 423  (1997).

\bibitem{CallanHarvey} C.G. Callan, Jr. and J.A. Harvey,  Nucl. Phys.  {\bf
B~250}, 427  (1985).

\bibitem{CallanHarveyEffect} G.E. Volovik,   Vortex motion in fermi
superfluids and Callan-Harvey effect,  JETP Lett. {\bf 57}, 244 
(1993).

\bibitem{StoneSpectralFlow} M. Stone, Spectral flow, Magnus force and mutual
friction via the geometric optics limit of Andreev reflection, Phys. Rev.   
{\bf B~54}, 13222 (1996).

\bibitem{Leggett} A.J. Leggett,  Phys.
Rev. Lett. {\bf 39}, 587  (1977).

\bibitem{JoyceShaposhnikov}  M. Joyce, M. Shaposhnikov, Primordial
magnetic fields, right electrons, and the abelian anomaly, Phys.
Rev. Lett., {\bf 79}, 1193  (1997).

\bibitem{GiovanniniShaposhnikov} M. Giovannini and E.M.
Shaposhnikov, Primordial
hypermagnetic fields and triangle anomaly,  Phys.Rev. {\bf D~57} 2186 (1998).

\bibitem{CosmicMagneticReview}  O. Tornkvist, Cosmic Magnetic Fields from Particle Physics,
astro-ph/0004098.

\bibitem{Experiment}  V.M.H. Ruutu, J. Kopu, M. Krusius, U. Parts, B.
Placais, E.V. Thuneberg, and W. Xu , Critical velocity of 
vortex nucleation in rotating superfluid $^3$He-A, 
Phys. Rev. Lett.,{\bf 79}, 5058-5061 (1997).

\bibitem{Vilenkin79} A. Vilenkin, Phys. Rev.  {\bf D~ 20}, 1807  (1979); {\bf D~21}, 2260
(1980).

\bibitem{VolovikVilenkin} G.E. Volovik and A. Vilenkin, Macroscopic parity
violating effects and $^3$He-A,  Phys. Rev.  {\bf D~62}, 025014 (2000). 

\bibitem{Rice} M. Rice, Superconductivity: An analogue of superfluid 
$^3\!$He, Nature  {\bf 396}, 627-629 (1998);  

\bibitem{Ishida} K. Ishida, H. Mukuda, Y. Kitaoka 
{\it et al.}, Spin-triplet  superconductivity in Sr$_2$RuO$_4$ identified by $^{17}\!$O Knight
shift,  Nature  {\bf 396}, 658-660 (1998).

\bibitem{SingleVortNucl} \"U. Parts, V.M.H. Ruutu, J.H. Koivuniemi, Yu. N.
Bunkov, V.V. Dmitriev, M. Fogelstr\"om, M. Huenber, Y. Kondo, N.B. Kopnin,
J.S. Korhonen, M. Krusius, O.V. Lounasmaa, P.I. Soininen,   G.E. Volovik, 
Single-vortex nucleation in  rotating superfluid
$^3$He-B,  Europhys. Lett. {\bf  31},  449-454 (1995).

\bibitem{Kita} T. Kita, J. Phys. Soc. Jap. {\bf 67}, 216 (1998).

\bibitem{GravimagneticMonopole} D. Lynden-Bell and M. Nouri-Zonoz, Rev. Mod.
Phys.  {\bf 70}, 427 (1998).

\bibitem{CalogeracosVolovik} A. Calogeracos  and G.E. Volovik,  Rotational quantum friction
in superfluids: Radiation from object rotating in superfluid vacuum, JETP Lett. {\bf
69}, 281 -- 287 (1999).

\bibitem{Muzikar1983} P. Muzikar, and D. Rainer,    Phys.
Rev.  {\bf B~27},  4243 (1983); K. Nagai,    J. Low Temp. Phys.
{\bf 55},   233 (1984); G.E. Volovik, Symmetry in superfluid $^3$He, in: {\bf Helium Three}, 
eds. W.P. Halperin, L.P. Pitaevskii, Elsevier Science
   Publishers B.V., pp. 27 - 134 (1990).

\bibitem{Vilenkin80} A. Vilenkin,
Phys. Rev.  {\bf D~ 22}, 3080  (1980).

\bibitem{Redlich}  A.N. Redlich, and L.C.R.
Wijewardhana,  Phys. Rev. Lett. {\bf 54}, 970  (1985).

\bibitem{JackiwKostelecky}
R. Jackiw, and V. Alan Kosteleck$\rm{\acute y}$, Phys. Rev. Lett.
{\bf 82}, 3572 (1999).

\bibitem{Andrianov} A.A. Andrianov, R. Soldati, and L. Sorbo, Phys. Rev. D 59,
025002-1/13 (1999).

\bibitem{Leahy} A. Vilenkin, and D.A. Leahy, Ap. J. {\bf 254}, 77 (1982).

\bibitem{Goryo} J. Goryo and K. Ishikawa,  Phys. Lett {\bf A~260}, 294 (1999);
G.E. Volovik, Analog of quantum Hall effect in superfluid $^3$He film, Sov. Phys. JETP {\bf
67}, 1804 -- 1811 (1988).

\bibitem{Ivanov} D. A. Ivanov, Non-abelian statistics of half-quantum vortices in p-wave
superconductors, cond-mat/0005069.
 
\bibitem{ewitten} E. Witten, Nucl. Phys.  {\bf B249},
557 (1985).

\bibitem{tvgf} T. Vachaspati and G.B. Field,   Electroweak  
string configurations with baryon number, Phys. Rev. Lett. {\bf  73},
373--376 (1994);  {\bf 74}, 1258(E) (1995).

\bibitem{jgtv} J. Garriga and T. Vachaspati, Zero modes on  
linked strings', Nucl. Phys. {\bf B~438}, 161  (1995).


\bibitem{barriola}  M. Barriola,   Electroweak strings produce 
baryons', Phys. Rev. {\bf D~51}, 300 (1995).

\bibitem{gstv}  G.D. Starkman and T. Vachaspati, Galactic 
cosmic strings as sources of primary antiprotons, Phys. Rev. {\bf D~53},
6711  (1996). 

\bibitem{Caroli} C. Caroli, P.G. de Gennes, and J. Matricon, Phys.
Lett. {\bf 9}, 307  (1964).

\bibitem{GeneralizedIndexTheorem} S.C. Davis, A.C. Davis,
W.B. Perkins, Cosmic string zero modes and multiple phase transitions,
Phys. Lett. {\bf B~408}, 81-90 (1997).

\bibitem{KopninSalomaa} N.B.Kopnin and M.M.Salomaa,   Phys. Rev.
{\bf B44}, 9667 (1991).

\bibitem{ZeroModesInChiral} G.E. Volovik, Fermion zero modes on
vortices in  chiral superconductors,  JETP Lett. {\bf 70}, 609-614 (1999).

\bibitem{Stone} M. Stone, Phys. Rev. {\bf B} 54, 13222 (1996).

\bibitem{KopninVolovik}  N.B. Kopnin and G.E. Volovik, Flux-flow  in $d$-wave
superconductors: Low temperature universality and scaling,  Phys. Rev. Lett.
{\bf 79},  1377-1380  (1997); N.B. Kopnin and G.E. Volovik,
Rotating vortex core:  An instrument for detecting the core excitations, Phys.
Rev. {\bf B57}, 8526-8531 (1998).

\bibitem{VolMin} G.E. Volovik \& V.P. Mineev, 1976, JETP Lett. {\bf 24}, 561 -
563 (1976).

\bibitem{Geshkenbein} Geshkenbein, V., Larkin, A. \&   Barone, A. 
(1987) {\it Phys. Rev.} {\bf B ~36}, 235-238.

\bibitem{Kirtley1996} J.R. Kirtley,  C.C. Tsuei, M. Rupp,   et al.,
Phys. Rev. Lett., {\bf 76}, 1336 (1996).

\bibitem{Read}  N. Read \& D. Green,  Paired states of fermions in two
dimensions with breaking of parity and time-reversal symmetries,  and the
fractional quantum Hall effect,   Phys. Rev.  {\bf B~61}, 10267-10297
(2000). 

\bibitem{Kitaev}  S. Bravyi and A. Kitaev, Fermionic quantum
computation, quant-ph/0003137. 

\bibitem{NonaxisymmetricVortex} Y. Kondo,  J.S. Korhonen, M. Krusius, V.V.
Dmitriev, Yu. M. Mukharskiy, E.B. Sonin and G.E. Volovik,  Observation
of the nonaxisymmetric vortex in $^3$He-B, Phys. Rev. Lett.  {\bf 67}, 81 - 84
(1991).


\bibitem{KopninKravtsov1976}  N.B. Kopnin  and V.E. Kravtsov, 
  JETP Lett. {\bf 23}, 578 (1976).

\bibitem{Andreev} A. F. Andreev, The thermal conductivity of the
intermediate state in superconductors, Sov. Phys. JETP {\bf 19}, 1228 -- 1231 (1964).

\bibitem{Neto} P.A. Maia Neto and S. Reynaud,  Phys. Rev. {\bf A~47},
1639 (1993).

\bibitem{Law} C.K. Law,  Phys. Rev. Lett. {\bf 73}, 1931 (1994).

\bibitem{Kardar} M. Kardar and R. Golestanian, The ``friction'' of vacuum, and other
fluctuation-induced forces, Rev. Mod. Phys.  {\bf 71} 1233-1245 (1999). 

\bibitem{YipLeggett} S. Yip and A.J. Leggett,  Phys. Rev. Lett. , {\bf 57}, 345
(1986).

\bibitem{KopninAB} N.B. Kopnin,   Sov. Phys. JETP  {\bf 65}, 1187 (1987). 

\bibitem{LeggettYip}   A.J. Leggett and S. Yip, in: {\bf Helium Three}, eds.
W.P.Halperin, L.P.Pitaevskii, Elsevier Science Publishers B.V., p. 523 (1990).

\bibitem{Palmeri} J. Palmeri,  Phys. Rev. {\bf B~42}, 4010 (1990)

\bibitem{FisherPickett} M. Bartkowiak, S. W. J. Daley, S.N. Fisher, et al, Thermodynamics of
the A-B Phase Transition and the Geometry of the A-Phase Gap Nodes,  Phys. Rev. Lett. {\bf
83}, 3462-3465  (1999).

\bibitem{Logunov} A.A. Logunov, Teor. and Mat. Fizika {\bf 80}, 165 (1989).

\bibitem{HansonRegge} A.J. Hanson and T. Regge,  Torsion and
quantum gravity, in: Proceedings of the Integrative Conference on Group Theory
and Mathematical Physics, University of Texas at Austin, 1978;  R. d'Auria
and T. Regge, Nucl. Phys., {\bf B~195}, 308 (1982).

\bibitem{Bengtsson}
I. Bengtsson, Degenrate metrics and an empty black hole,
 Class. Quant. Grav. {\bf 8}, 1847-1858 (1991). 

\bibitem{BengtssonJacobson} I. Bengtsson, and T. Jacobson
Degenerate metric phase boundaries, Class. Quant. Grav. {\bf 14} (1997)
3109-3121; Erratum-ibid. {\bf 15}, (1998) 3941-3942.

\bibitem{Horowitz} G.T. Horowitz, Topology change in classical and quantum gravity,   Class.
Quant. Grav. {\bf 8}, 587-602 (1991). 

\bibitem{Starobinsky} A. Starobinsky, Plenary talk at Cosmion-99, Moscow, 17-24
October, 1999.

\bibitem{SalomaaVolovik} M.M. Salomaa,  G.E. Volovik,  Cosmiclike domain walls in
superfluid $^3$He-B: Instantons and diabolical points in (${\bf k}$,
$ {\bf r}\, $) space, Phys. Rev.  {\bf B~37}, 9298 - 9311 (1988);
Half-solitons in superfluid $^3$He-A: Novel $\pi /2$-quanta of phase slippage, 
J. Low Temp. Phys. {\bf 74}, 319 - 346 (1989).
 
\bibitem{Volovik1990}   G.E. Volovik, Superfluid $^3$He-B and gravity,
Physica  {\bf B~162}, 222-230 (1990). 

\bibitem{MatsumotoSigrist}  M. Matsumoto and M. Sigrist
Quasiparticle states near the surface and the domain wall in a $p_x\pm i
p_y$-wave superconductor, J. Phys. Soc. Jpn. {\bf  68}. 994 (1999);
cond-mat/9902265.

\bibitem{SigristAgterberg} M. Sigrist, D.F.
Agterberg,  The role of domain walls on the vortex creep dynamics in unconventional
superconducors, cond-mat/9910526.

\bibitem{Volovik1997} G. E. Volovik, On edge states in superconductor with time inversion
symmetry breaking,  JETP Lett. {\bf 66 },  522-527  (1997).

\bibitem{JacobsonVolovikThinFilm} T. A. Jacobson, G. E.  Volovik, Effective spacetime and
Hawking radiation from a moving domain wall in a thin film of $^3\!$He-A, JETP Lett. {\bf 68},
874-880 (1998).

\bibitem{VierbeinWalls} G.E. Volovik,  Vierbein walls in condensed matter, JETP Lett. {\bf
70}, 711-716 (1999).

\bibitem{Silagadze} Z.K. Silagadze, TEV scale gravity, mirror universe, and ...dinosaurs,
hep-ph/0002255.

\bibitem{VolovikHalfQuantum} G.E. Volovik,  Monopoles and fractional vortices in chiral
superconductors,  Proc. Natl. Acad. Sc. USA {\bf 97}, 2431-2436 (2000).

\bibitem{Schwarz}  A.S. Schwarz,      Nucl. Phys.   {\bf B~208}, 141-158  (1982).

\bibitem{SokolovStarobinsky} D.D. Sokolov, A.A. Starobinsky, Doklady
AN SSSR,
{\bf 234}, 1043 (1977) \lbrack Sov. Phys. - Doklady, {\bf 22}, 312
(1977)
\rbrack.

\bibitem{Banados} M. Banados, C. Teitelboim, J. Zanelli, Phys. Rev.
Lett., {\bf 69}, 1849 (1992).

\bibitem{Barrow} J.D. Barrow, Varying G and other constants, gr-
qc/9711084.

\bibitem{CausalityViolation} P.O. Mazur,  Phys. Rev. Lett.
{\bf 57}, 929 (1986).  

\bibitem{MazurComment2} P.O. Mazur, hep-th/9611206.

\bibitem{Staruszkievicz} A. Staruszkievicz, Acta Phys. Polon., {\bf 24}, 734
(1963).
 
\bibitem{Deser} S. Deser, R. Jackiw, and G. t'Hooft,  Ann. Phys. , 
{\bf 152}, 220 (1984).

\bibitem{DavisShellard} R.L. Davis and E.P.S Shellard,  Phys. Rev. Lett.
{\bf 63}, 2021 (1989).

\bibitem{Harari} D. Harari and A.P. Polychronakos,  Phys. Rev.
D, {\bf 38}, 3320 (1988).

\bibitem{JensenKucera} B. Jensen and J. Ku\v cera, J. Math. Phys.   {\bf
34}, 4975 (1993). 

\bibitem{3Forces} G.E. Volovik, Three nondissipative forces on a  moving vortex
line in superfluids and superconductors, JETP Lett. {\bf  62}, 65 - 71  (1995).

\bibitem{Lee} K.-M. Lee,  Phys. Rev. 
{\bf D~49}, 4265 (1994).

\bibitem{VolovikGravAB} G.E. Volovik, Vortex vs spinning string: Iordanskii force and
gravitational Aharonov-Bohm effect,    JETP Lett. {\bf 67},  881 - 887  (1998).

\bibitem{Iordanskii} S.V. Iordanskii,   Ann. Phys., {\bf 29}, 335
(1964);    ZhETF, {\bf 49}, 225 (1965);
\lbrack Sov. Phys. JETP, {\bf 22}, 160 (1966)\rbrack.

\bibitem{Sonin1} E.B. Sonin,    ZhETF, {\bf 69}, 921 (1975);
\lbrack  JETP, {\bf 42}, 469 (1976) \rbrack. 

\bibitem{SoninNew} E.B. Sonin, Phys. Rev. {\bf B~55}, 485 (1997).

\bibitem{Shelankov} A.L. Shelankov, Magnetic force exerted by the Aharonov-Bohm line, Europhys.
Lett., {\bf 43},   623-628 (1998).

\bibitem{AB} Y. Aharonov and D. Bohm, Phys. Rev. {\bf 115}, 485 (1959).

\bibitem{MazurComment1} P.O. Mazur,  Phys. Rev. Lett.
{\bf 59}, 2380 (1987). 

\bibitem{Galtsov} D.V. Gal'tsov and P.S. Letelier,   Phys. Rev. 
{\bf D~47}, 4273 (1993).

\bibitem{Fetter} A.L. Fetter, Phys. Rev. {\bf 136A}, 1488 (1964).

\bibitem{Demircan} E. Demircan, P. Ao and Q. Niu, Phys. Rev.  {\bf B~52}, 476
(1995).

\bibitem{Cleary} R.M. Cleary, Phys. Rev. {\bf 175}, 587 (1968).

\bibitem{Bevan} T.D.C. Bevan, A.J. Manninen, J.B. Cook, {\it et al}, Phys. Rev.
Lett.,  {\bf 74}, 750 (1995).

\bibitem{hawkingnature} S. W. Hawking, Black hole
explosions?, Nature {\bf 248}, 30-31 (1974).

\bibitem{grishated} T. A. Jacobson, G. E.  Volovik, Event horizons and ergoregions in
$^3\!$He, Phys. Rev. D {\bf 58}, 064021 (1998).

\bibitem{FischerVolovik} U.R. Fischer, and G.E. Volovik,  Thermal
quasi-equilibrium states across Landau horizons in the effective gravity of superfluids,
gr-qc/0003017.

\bibitem{Martel} K. Martel and E. Poisson,  Regular coordinate systems for
Schwarzschild and other spherical spacetmes, gr-qc/0001069.  

\bibitem{Liberati2} S. Liberati, Quantum vacuum effects in gravitational
fields: theory and detectability,  gr-qc/0009050.  

\bibitem{Liberati} S. Liberati, S. Sonego and M. Visser, Unexpectedly large surface gravities
for acoustic horizons? gr-qc/0003105  

\bibitem{FluidMechanics} L.D. Landau and E.M. Lifshitz,   Fluid Mechanics, p. 317, Pergamon
Press, 1989.

\bibitem{KopninVolovik1998} N.B. Kopnin and G.E. Volovik, Critical velocity and event
horizon in pair-correlated systems with ``relativistic'' fermionic quasiparticles,
JETP Lett. {\bf 67}, 140-145  (1998).

\bibitem{SimulationPainleve} G. E. Volovik, Simulation of
Painlev\'e-Gullstrand black hole in thin $^3$He-A film,
JETP Lett. {\bf 69}, 705-713 (1999).

\bibitem{BHdispers} S. Corley, T. Jacobson: {\em Hawking spectrum
and high frequency dispersion}, Phys. Rev. D {\bf 54}, 1568-1586
(1996)

\bibitem{Corley} S. Corley: {\em Computing the spectrum of black hole radiation
in the presence of high frequency dispersion: An analytical approach},
Phys. Rev. D {\bf 57}, 6280-6291  (1998)

\bibitem{BHlaser} S. Corley, T. Jacobson: {\em Black hole
lasers}, Phys. Rev. D {\bf 59}, 124011 (1999)

\bibitem{origin} T. Jacobson, On the origin of the outgoing
black hole modes, Phys. Rev. D {\bf 53}, 7082-7088 (1996)

\bibitem{JacobsonTalk}   T.A.  Jacobson, Trans-Planckian redshifts
and the substance of the space-time river,   hep-th/0001085.

\bibitem{Unruh1} W. G. Unruh,   Phys. Rev. {\bf D~14}, 870 (1976).

\bibitem{Audretsch} J. Audretsch and R. M\"uller,  Phys. Rev. {\bf A~50},
1755 (1994).

\bibitem{RotatingQuantumVacuum} P.C.W. Davies, T. Dray, C.A. Manogue,  
Phys. Rev. {\bf D~53}, 4382  (1996).

\bibitem{Leinaas} J.M. Leinaas, Accelerated electrons and the Unruh effect, 
Talk given at 15th Advanced ICFA Beam Dynamics Workshop on Quantum Aspects of Beam Physics,
Monterey, CA, 4-9 Jan 1998,
hep-th/9804179.

\bibitem{OrbitingUnruh}  W. G. Unruh, Acceleration radiation for orbiting
electrons, Phys. Rept. {\bf 307}, 163-171 (1998). 

\bibitem{Zeldovich1} Ya.B. Zel'dovich, Pis'ma ZhETF {\bf 14}, 270,
(1971) [JETP Lett. {\bf 14}, 180, (1971)].

\bibitem{BekensteinSchiffer} J.D. Bekenstein and M. Schiffer,  Phys.Rev.
{\bf D~58}, 064014  (1998).

\bibitem{Starobinskii} A.A. Starobinskii,   ZhETF, {\bf 64}, 48  
(1973) [JETP, {\bf 37}, 28   (1973)].

\bibitem{Zeldovich2} Ya.B. Zel'dovich,  ZhETF, {\bf 62}, 2076 
(1971) [JETP  {\bf 35}, 1085  (1971)]

\bibitem{Cross}   M.C. Cross, A generalized Ginzburg-Landau approach to the
superfluidity of $^3$He, J. Low Temp. Phys. {\bf 21}, 525-534 (1975).

\bibitem{Sonoda} H. Sonoda, Chiral QED out of matter,
hep-th/0005188; QED out of matter, hep-th/0002203.

\bibitem{WeinbergBook} S. Weinberg, The Quantum Theory of Fields,
Cambridge University Press (1995).


\bibitem{Garay}     L. J. Garay, J. R. Anglin, J.
I. Cirac, P. Zoller,     Black holes in Bose-Einstein
condensates, gr-qc/0002015.


\end{thebibliography}
\end{document}